\documentclass[11pt]{JHEP3}

\usepackage{amsmath,epsfig}
\usepackage{amssymb,amsfonts}
\usepackage{latexsym}
\usepackage[latin1]{inputenc}
\usepackage[T1]{fontenc}
 \usepackage{ae}
 \usepackage{aecompl}
\usepackage{subeqnarray}
\usepackage{xcolor}

\usepackage{graphicx}
\usepackage{longtable}
\usepackage[hang,footnotesize,bf]{caption}

\relax
\renewcommand{\theequation}{\arabic{section}.\arabic{equation}}
\def\be{\begin{equation}}
\def\ee{\end{equation}}
\def\bea{\begin{eqnarray}}
\def\eea{\end{eqnarray}}

\newcommand\fverb{\setbox\pippobox=\hbox\bgroup\verb}
\newcommand\fverbdo{\egroup\medskip\noindent%
                        \fbox{\unhbox\pippobox}\ }
\newcommand\fverbit{\egroup\item[\fbox{\unhbox\pippobox}]}

\newcommand{\bear}{\begin{eqnarray}}

\newcommand{\eear}{\end{eqnarray}}

\newcommand{\bsea}{\begin{subeqnarray}}
\newcommand{\esea}{\end{subeqnarray}}
\newbox\pippobox

\newcommand{\ga}{\gamma}

\newcommand{\da}{\delta}
\newcommand{\ka}{\kappa}
\newcommand{\ud}{\mathrm{d}}

\def\6{\partial}

\def\a{\alpha}
\def\g{\gamma}
\def\nn{\nonumber}

\def\half{\frac12}

\def\le{\left}
\def\ri{\right}

\def\pa{\partial}

\def\m{\mu}
\def\n{\nu}

\def\s{\sigma}

\def\sp{\;\;\;,\;\;\;}

\def\D{\Delta}
\def\sq
\def\a{\alpha}
\def\b{\beta}
\def\l{\lambda}

\def\d{\delta}

\def\ff{\phi}
\newcommand{\Figref}[1]{Fig.\ref{#1}}
\title{Effective Holographic Theories for low-temperature condensed matter systems}
\author{Christos Charmousis$^{a,b}$, Blaise Gout\'eraux$^a$, Bom Soo Kim$^{c,d}$, Elias Kiritsis$^d$\footnote{On leave of absence from Laboratoire APC,
Universit\'e Paris-Diderot Paris 7, CNRS UMR 7164, 10 rue Alice Domon et L\'eonie Duquet, 75205 Paris Cedex 13, France.}, Rene Meyer$^d$~\\
~\\
$^a$ \href{http://www.u-psud.fr}{Univ. Paris-Sud}, \href{http://www.th.u-psud.fr}{Laboratoire de Physique Th\'eorique}, CNRS UMR 8627, F-91405 Orsay, France;
~\\
$^b$ \href{http://www.phys.univ-tours.fr/}{LMPT}, Parc de Grandmont, Universit\'e Francois Rabelais, CNRS UMR 6083, Tours, France;
~\\
$^c$  \href{http://www.iesl.forth.gr}{IESL-FORTH}, 71110 Heraklion, Greece;
~\\
$^d$ \href{http://hep.physics.uoc.gr}{Crete Center for Theoretical Physics},
Department of Physics, University of Crete, 71003 Heraklion, Greece\\
}

\preprint{arXiv:1005.4690\\
CCTP-2010-5\\
LPT-10-34
}

\abstract{The IR dynamics of effective holographic theories capturing the interplay between charge density and the leading relevant scalar operator at strong coupling are analyzed. Such theories are parameterized by two real exponents $(\gamma,\delta)$ that control the IR dynamics.
By studying the thermodynamics, spectra and conductivities of several classes of charged dilatonic black hole solutions that include the charge density back reaction fully,  the landscape of such theories in view of condensed matter applications is characterized.
Several regions of the $(\gamma,\delta)$ plane can be excluded as the extremal solutions have unacceptable singularities.
The classical solutions have generically zero entropy at zero temperature, except when $\gamma=\delta$ where the entropy at extremality is finite.
The general scaling of DC resistivity with temperature at low temperature, and AC conductivity at low frequency and temperature across the whole $(\gamma,\delta)$ plane, is found. There is a codimension-one region where the DC resistivity is linear in the temperature.  For massive carriers, it is shown that when the scalar operator is not the dilaton, the DC resistivity  scales as the heat capacity (and entropy) for planar (3d) systems.
Regions are identified where the theory at finite density is a Mott-like insulator at T=0. We also find that at low enough temperatures the entropy due to the charge carriers is generically larger than at zero charge density.
}


\begin{document}

\section{Introduction and summary of  results}
\label{intro}

The AdS/CFT correspondence has provided a new look at the connection of gauge theories and string theories.
It also implements a new way to model strong-coupling physics using  string theory and its low energy approximations, namely gravitational theories.

Despite several attempts, the relevant (super)string theories, involving Ramond-Ramond (RR) backgrounds,
have so far defied solution, although important progress in this direction
has been achieved recently.
Most of this progress is related to the AdS/CFT correspondence and its avatars, and relies on effective low-energy string theories, emerging
when the strings can be approximated as point-like objects. In the most studied example, that of N=4 super Yang-Mills (sYM), this approximation becomes reliable when the 't Hooft coupling is large and the dual quantum field theory is strongly-coupled. The intuition developed in this limit suggests that the ``gap" between the gravitational masses  and the string excitation masses becomes large, making the gravity approximation of the string theory  reliable. This translates in the dual theory as the existence of a large gap in the spectrum of anomalous dimensions.

This new paradigm of strongly-coupled theories has been applied to understand the low-temperature dynamics of strongly-correlated electron systems (for recent progress, see \cite{s0,s1,s2,s3,s4}).
 The idea behind this approach is that of universality and non-trivial emergent behavior. We already know that the physics of condensed matter (CM)  systems is very varied, and involves many examples when emerging behavior (effective quasi-particles and associated interactions) may be described by very different-looking theoretical models. A class of such systems exhibits strongly-coupled degrees of freedom, in the border with magnetism, high-temperature ($T_c$) superconductors, heavy fermion systems, graphene and others, often with a behavior that is at odds with the traditional Fermi liquids.

The application of holographic ideas to such CM systems relies on the development and quantitative description of universality classes at low temperatures, which is often controlled by zero-temperature critical behavior (also known as quantum criticality, \cite{s3}). The selection criteria for the appropriate theories rest on the identification  and characterization of appropriate phases via their thermodynamic characteristics, their phase transitions and critical behavior, and most importantly,  their generalized transport coefficients which connect directly to what is measured by experiments. In particular,  in this paper we will search for  strange metal characteristics.

A basic and important ingredient of realistic CM systems is the presence of a finite density of charge carriers.
In holographic CFTs, this implies  the existence of a conserved global charge, and the duality between the associated conserved global current and a (massless) bulk gauge field. Finite density ground states therefore correspond to asymptotically AdS gravitational solutions  ``charged" under the appropriate bulk gauge field. This identification relies on an important assumption that real-life photons are weakly-coupled in strongly-correlated systems, a fact supported in most cases by data.

The initial study of a holographic finite density system involved charged Reissner-Nordstr\"om (RN) black holes of Einstein-Maxwell gravity with a cosmological constant. Such solutions proved to be the simplest laboratory for the study of phase transitions at finite density, indicating an interesting phase structure, \cite{RN1,RN2}.
Moreover, a study of the fermionic two-point function in this background suggested the presence of fermionic quasi-particles with non-Fermi liquid behavior, \cite{za,mc}, as well as an emerging scaling symmetry at zero temperature, \cite{mc2}, associated with the AdS$_2$ symmetry of the extremal RN black hole. The addition of a charged test scalar led to superfluid phase transitions, \cite{super1,super2,super3}, that upon weak gauging of the boundary U(1)  symmetry led to superconductivity.

The RN laboratory has however an important disadvantage from a CM point of view: it has a finite (and large) entropy at extremality. Such a fact is considered an anathema at zero temperature and seems to contradict the third law of thermodynamics.
In particular this is a signal of instability, since the Breitenlohner-Freedman bound, \cite{Breitenlohner:1982bm,Breitenlohner:1982jf}, for the IR AdS$_2$ is more stringent than that for the UV AdS$_4$. Furthermore, there are scalar operators that violate the bound and are expected to condense, destabilizing the RN saddle point, \cite{s2}.

The next step in modeling strongly-coupled holographic systems at finite charge density is to include the leading relevant (scalar) operator in the dynamics.
This is  generically uncharged, and it drives the renormalization group flow of the system from UV to IR. The corresponding holographic theory is an Einstein-Maxwell-Dilaton system with a scalar potential. In the case of zero charge density such theories have been analyzed in the recent past as they capture some essential features of pure YM theory in four dimensions, \cite{ihqcd1,ihqcd2,gubser2,gkmn1,gkmn2}. In the finite charge density case, solutions to such theories have been reconsidered in view of holographic applications, \cite{karch,gubser,kachru,kraus,gau,rey,hpst,cadoni,Edalati:2010hk,russo,chen,ulf,arean}.

The goal of the present paper is to describe a general framework for the discussion of the holographic dynamics of  Einstein-Maxwell-Dilaton systems with a scalar potential. This is a phenomenological approach based on the concept of Effective Holographic Theory (EHT), in analogy with Effective Field Theories low energy approximations to QFT.
This strategy has advantages and disadvantages. It allows a parametrization of large classes of IR dynamics and a survey of important observables. On the other hand, it is not obvious if concrete EHTs can be embedded in well-defined string theories, or if concrete classical solutions that are (mildly) singular are acceptable as saddle points of the dual CFTs, \cite{bad,ihqcd1,ihqcd2,gkmn2,pene}.

We will parametrize EHTs of the Einstein-Maxwell-Dilaton kind in terms of the IR asymptotics of the two scalar functions that enter in their two-derivative action: the scalar potential and the Maxwell (non-minimal) coupling. Firstly, we will investigate classical solutions \cite{char} that can be eventually completed into bona-fide asymptotically AdS solutions. This will be done in two different directions: by analyzing exact solutions in special cases and by studying the IR (extremal) asymptotics of the solutions, in order to classify the low-temperature asymptotics of the associated dual quantum field theories.

Secondly, using this holographic setup, we will compute the conductivity both at zero (DC) and
non-zero (AC) frequency, which along with the finite temperature entropy and specific heat are important experimental observables of finite charge density CM systems.
Indeed, some of the landmark experimental signals of strange metal behavior include resistivity linear in temperature and AC conductivity scaling as an inverse power of the frequency in some regimes.

The setup of the paper is as follows: For the rest of this section we will introduce and motivate the effective holographic theories we will study as well as some important definitions, giving at the end a consistent summary of the results and outlook of our work. In sections 2 and 3, we will study EHTs' dynamics at zero and non zero charge density. Conductivity will be treated in section 4 including the case of DBI dynamics. Then in sections 6-9 we will proceed to study in detail the equilibrium and non-equilibrium dynamics of EHT solutions.  Several complementary definitions and calculations can be found in the Appendix.

\subsection{Effective Holographic Theories}

The AdS/CFT correspondence is a correspondence between strongly-coupled adjoint theories and weakly-coupled string theories. It is expected to be valid more generally even when the dual field theory does not have an adjoint description. An important distinction is between ``color" (involving adjoint degrees of freedom that are gauged) with prototype examples SU(N) YM or N=4 sYM, and ``flavor" with prototype examples the Gross-Neveu model or flavor in QCD. In the first case all observables are singlets, and are described by closed strings in the dual theory. In the case of flavor, there are singlet and non-singlet observables. The singlet observables can be described by closed strings, the non-singlet ones by open strings.

The relevant string theories live in diverse dimensions, and have generically non-trivial RR backgrounds. They are currently beyond our computational control, although this may change in the future. In the best studied example, N=4 sYM,
the string theory at strong 't Hooft coupling $\l$ can be approximated by the effective supergravity theory as the curvature of the background is weak at strong coupling.
The signal in the dual QFT is that at large $\l$, there is a gap in the spectrum of anomalous dimensions. Indeed,  BPS operators have dimensions of ${\cal O}(1)$, while non-BPS operators (corresponding to string excitations) have dimensions of ${\cal O}(\l^{1\over 4})$.

Truncating a string theory to a finite spectrum of low-lying states is the central point of the EHT approximation. In a large-N CFT it can be a controllable approximation when the spectrum of dimensions has a gap that can be made parametrically large. This is analogous to the existence of mass gaps that make integrating out massive states a good approximation in usual QFTs.
In non-conformal cases the situation is more complicated.
At the intuitive level one can argue that a truncation of the string theory spectrum can be a reasonable approximation if none of the states that have been omitted can become relevant in the UV or the IR.

We can think about integrating-out massive string modes, in order to obtain an effective description of a small number of low energy modes. Whether this is a good approximation also depends on the particular classical solution that describes the dual theory ground state. If fluctuations have suppressed higher derivative interactions  then this has good chances of providing reliable information on the dual dynamics.

In a holographic string theory, string states that correspond to dual theory operators having non-zero couplings in the Lagrangian, will certainly have non-zero profiles in the string vacuum solution as their sources will be non-trivial. This may change however for string states that are non-sourced in the AdS boundary. This is decided by the dynamics of the theory, which determines their vevs, and therefore their non-trivial profiles in the string vacuum solution. Even if such vevs are non-trivial, neglecting them may not affect much the dynamics for other operators.

There may be cases where spectrum (and interaction) truncations give reliable results although there is no
 small parameter to justify their neglect. There are two complementary examples of such a state of affairs. One
 is the technique of level truncation in string field theory, used in order to calculate the effective tachyon potential
  in tachyon condensation studies, \cite{sen}. In these, the level truncation gives convergent results although it is not really understood why. A complementary example from QFT is YM. This is a theory with a single parameter, $\Lambda_{QCD}$ which is expected to provide a universal order of magnitude for vevs up to dimensionless numbers of order one. However, the phenomenological success of SVZ sum rules has provided unexpectedly small values for the relevant vevs, suggesting that (at least for some observables)
truncating the spectrum of string theory states may be a good approximation.

Our strategy will be to select a set of operators (or dual string fields) that are expected to dominate the dynamics, and parametrize their EHT in terms of a general two-derivative action. Higher-derivative corrections are expected if the interactions of other states cannot be neglected in the structure of the semi-classical vacuum solution. We will however assume that this is not the case for simplicity.

The next question  is to decide the set of string fields to keep in the discussion of the classical solution that should describe the ground state or the finite-temperature saddle points of the dual strongly-coupled field theory.
The minimal case includes only the metric. This is always important to consider as: it always encodes the energy distribution in the ground state; it is always sourced  by the boundary metric of the dual QFT.
In this minimal case, the two-derivative action is AdS gravity, with a single acceptable solution at zero temperature, namely AdS$_{p+1}$.

In the case of a system that contains a global U(1) symmetry (particle number conservation), a relevant operator that may be important is the associated conserved U(1) current $J_{\m}$ dual to a bulk U(1) massless gauge field\footnote{The gauge field may acquire a bulk mass if the global U(1) symmetry is spontaneously broken or anomalous, like $U(1)_A$ in QCD.} $A_{\m}$. At the two-derivative level, the action is that of Einstein-Maxwell with a cosmological constant. If the ground state of the theory contains a finite density of the global charge, then the gauge field must be non-trivial
and the relevant solution is the AdS-RN black hole.
The Einstein-Maxwell description is relevant for U(1) symmetries that are acting on adjoint fields, like the R-symmetries of N=4 sYM. They are closed string fields and they originate in the bulk.

There can be also U(1) symmetries that are generated by fundamental fields and are more like baryon number in large-N QCD. The origin of the associated U(1) gauge field is from flavor branes, and its effective action is given by the DBI action. The non-linearities of the DBI action implement interactions between charge carriers. Only in the dilute charge limit the action linearizes to a Maxwell action.
In this context, the relevant charged solution is the AdS-DBI-RN solutions first discussed in \cite{fer1,fer2}.

In many cases the important dynamics of strongly coupled theory is driven by a scalar leading-relevant operator. This is the case in YM, where the operator is $Tr[F^2]$ when the number of flavors $N_f\sim {\cal O}(1)$. On the contrary, in the Veneziano limit,  $N_f\sim {\cal O}(N_c)$ another scalar operator namely $\bar q q$ becomes important and it is the interplay of these two operators that determines the structure of the vacuum.

In such cases the minimal set of important bulk fields to consider for the vacuum solution are $g_{\m\n}$ controlling the energy, $A_{\mu}$ controlling the charge density and $\phi$ (dual to the relevant operator) controlling the coupling constant and its running in the IR.

For a CM system, $\phi$ maybe representing the strong interactions of the ion lattice, or or the effect of spins on the charge carriers. In a sense it could describe the $glue$ responsible for inducing interactions on the charge carriers. If the charge carriers are dilute, we expect that they do not backreact on the surrounding system  and  the dynamics can be treated in the probe approximation: solve first the equations for the Einstein-dilaton system, and in the background of this solution we can then solve the gauge field equations.

On the other hand, at sufficiently large doping, the back reaction cannot be neglected, and the full set of equations must be solved.

As mentioned earlier, we have also two options for the dynamics of the carriers. If they correspond to adjoints, then they arise from the closed string sector and a Maxwell action may be sufficient. If they are fundamentals then generically they originate on flavor branes and we should consider the DBI action.
In the latter case, at weak charge densities we expect also the charge dynamics to linearize and be expressed by the linear Maxwell action.

At the two-derivative level, the EHT in the Maxwell case is described in the Einstein frame by the action
\be
S=S_g+S_{Max}\,,\qquad S_g=M^{p-1}\int d^{p+1}x\sqrt{-g}\left[R-{1\over 2}(\partial\phi)^2+V(\phi)\right]
\label{ActionLiouville}\ee
$$
S_{Max}=-M^{p-1}\int d^{p+1}x\sqrt{-g}{Z(\phi)\over 4}F_{\m\n}F^{\m\n}
$$
Indeed, the most general two-derivative action containing a metric, a vector and a single scalar can be brought to this form via field redefinitions.

On the other hand, in the fundamental charge case the linear Maxwell action, $S_{Max}$ must be replaced by the DBI action as follows
\be
S_{DBI}= -M^{p-1}\int d^{p+1}x
e^{2k\phi} Z(\phi)\left(\sqrt{-\det(g+e^{-k\phi}F)}-\sqrt{-\det g}\right)
\label{2}\ee
Here $g = e^{-k \phi} g^\sigma$ is the Einstein frame metric, and $Z$ has been normalised such that for $F^2\ll 1$ the DBI action reduces to the Maxwell one. The parameter $k$ encodes the dilatonic nature of the scalar field $\phi$. If $\phi$ is the dilaton, $k={-\sqrt{2\over p-1}}$.

The dimension $p+1$ of the holographic bulk space-time has been left arbitrary here. Indeed, even for CM applications, $p$ can in principle take several values. For layered systems that are effectively $1+2$-dimensional, without any extra adjoint fields, we need $p=3$ to describe them. For real 3+1 dimensional systems without extra adjoint fields we would need p=4. However in the presence of extra adjoint fields more space-time dimensions are holographically generated and $p$ can take higher values (as in the case of N=4 sYM).

\subsection{The UV region}

It is standard in holographic theories to assume that there is a non-trivial (strongly coupled)  fixed point in the UV. This guarantees a globally controllable behavior, and an appropriate UV boundary where the theory is defined.

This assumption is however not necessary if our only interest is to understand the IR dynamics. There are two concurring reasons for this. The first is the Wilsonian definition of quantum field theory that requires a UV cutoff that can be anything, provided a full definition of dynamics is in effect at the cutoff energy. The second is that the non-trivial dynamical condition that defines
the ``vacuum" and effectively defines all expectation values, is imposed in the IR. Typically this is a regularity condition that that ties together the two linearly-independent solutions in the far IR. Even in mildly singular (and therefore acceptable) IR asymptotics such a regularity condition is in order.

To compute therefore correlators in the cutoff holographic theory, the only thing that remains is to evolve the solution containing the single integration constant from the IR to the cutoff using the equations of motion and then at the cutoff to distinguish the source from the vev contribution.
This is a priori non-trivial because although the source and the vev linearly-independent solutions start hierarchically different near the boundary, the eventually become similar and divergent in the IR. It is the linear combination that cancels the divergence and the proper and acceptable solution.

However, one can start from the IR with the two linearly independent solutions
and extend them toward the UV. If there is an intermediate scaling region then
 a cutoff there, can be used to separate the solution into a source and vev piece and eventually compute the correlator. Otherwise only the correlator at low frequencies and momenta can be computed.

 Note that the real parts of correlators typically suffer from scheme dependence because of renormalization choices whereas imaginary parts do not.

 In the rest of the paper we will pretend that there is an AdS completion to the geometries we are discussing. As argued above this is not necessary for computing IR data. We will assume though that the potential decreases toward the UV signaling a good IR approximation.

\subsection{The IR asymptotics}

Another important question concerns the form of the two scalar-dependent functions in the actions above, the scalar potential $V(\phi)$ and the gauge coupling constant function $Z(\phi)$.

The potential must be non-zero in all realistic examples as generic scalar operators have non-trivial n-point functions. Because of this the attractor mechanism that is playing an important role in spherically symmetric solutions in supergravity in the absence of the potential is not at work here\footnote{S. Kachru has suggested that even in this case one can place some constraints on the flows however.}.

What will be our main focus is the IR behavior of the functions, $V,Z$.
 Motivated by controlled examples
from string theory we will assume a pure exponential behavior at the IR,
\be
\lim_{\phi\to \infty} V(\phi)~~\sim~~  e^{-\delta\phi}\,.
\label{v}\ee
This is a generic behavior in the absence of IR fixed points with AdS$_{p+1}$ asymptotics.
It turns out that for generic values of the exponent $\d$, the asymptotic behavior in (\ref{v}) is enough to capture the important IR features of the theory, as shown in \cite{ihqcd2,gkmn2}. In the same references it was also shown that there are ``transition values" for $\delta$, that we will review in the next section, where important IR properties change. In the neighborhood of such transition values, a finer parametrization of the asymptotics is necessary in order to capture all the ramifications of the IR behavior.

In this work, we will take $V(\phi)=V_0 e^{-\delta\phi}$, remembering that this is an approximation valid in the IR. To enforce that our solutions and analysis is well grounded in the UV, we will demand that $e^{-\delta\phi}\to 0$ in the UV for our solutions. This will guarantee that simple modifications of the potential like $V\to {p(p-1)\over \ell^2}+V_0 e^{-\delta\phi}$ will lead to asymptotically AdS solutions without modification of the IR behavior. This will be the working assumption in this paper. As argued however above, our results are independent of the UV completion.,

We now focus on the gauge-coupling function $Z(\phi)$. Again experience from both gauge supergravity actions stemming from string theory and tachyon condensation indicates that we may take it to have exponential behavior in the IR:
\be
\lim_{\phi\to \infty} Z(\phi)~~\sim~~  e^{\gamma\phi}\,.
\label{z}\ee
In particular, in U(1)s arising in the adjoint (closed string) sector
$Z$ diverges in the IR, \cite{gubser}, indicating that the charge sector is driven to zero coupling in the IR.
On the other hand if the U(1) arises from the fundamental sector (open strings, $D-\bar D$ systems) then $\phi$ is the analogue of the tachyon and the general arguments of Sen, \cite{sen}, indicate that $Z\to 0$ in the IR and the gauge field is driven to strong coupling. This is triggered by tachyon condensation.
It has been suggested, \cite{tachyon,tachyon2}, that this is the mechanism for chiral symmetry breaking in holographic QCD.

Once we have the appropriate action as above we can address the question of the ground state solution of the system both at zero and finite temperature, by solving the appropriate equations of motion. Our main preoccupation in this paper is to unravel the dynamics in the cases where the presence of charge degrees of freedom backreact on the neutral sector, $g_{\m\n},\phi$.

From the solution one can study thermodynamics and in the presence of multiple candidates for the ground state of the system, the presence and nature of phase transitions. Moreover, the correlation function of currents can be computed at low energy and from this the AC and DC conductivities can be computed. They are important observables of the system, and can be used to characterize different models.

\subsection{On naked singularities}

As is well known by now, solutions to EHTs at zero temperature have  typically naked singularities. From the GR point of view naked singularities are anathema and are therefore  exorcized. In holography however it is known that not all naked singularities are unacceptable. Well-defined and controllable perturbations of
N=4 sYM lead to (mild) naked singularities in the bulk. Such singularities are generically resolved by lifting them to higher dimensions or eventually by the inclusion of the stringy states.

It was Gubser first who asked the question: when is a naked singularity in a solution of an EHT acceptable, \cite{bad}? One criterion that he introduced in scalar-tensor backgrounds can be phrased simply as follows:
\begin{itemize}
 \item{\em A naked singularity is unphysical if the scalar potential is {\em not bounded below} when evaluated in the solution}\footnote{Note that we are using the opposite sign for the potential compared to \cite{gubser2}.}.

However, as postulated in \cite{ihqcd2,gkmn2} there are further requirements that must be imposed in order to accept naked singularities in the context of an effective holographic setup.

 \item{\em A naked singularity describes holographic physics that is unreliable
 if the second-order equations describing the spectrum of small fluctuations around the solution are
  not well-defined Sturm-Liouville problems{\footnote{This has also been observed in tensorial
  perturbations of higher codimension braneworlds, \cite{Charmousis:2001hg}.}}}.
  The spectrum around a given vacuum solution is calculated by studying small fluctuations.
   Such fluctuations, once taken to be eigenstates of the $p$-dimensional Laplacian, satisfy a
   second-order equation in the holographic coordinate. Such an equation has two linearly
    independent solutions near the AdS boundary. Typically only one is normalizable.
    The same should also hold true in the IR. In the presence of naked singularities
    it may happen that both solutions are normalizable. Then an extra boundary condition
    is needed at the IR singularity in order to have a well-defined spectrum.
Such cases are deemed unreliable.

 \item{\em A naked singularity is unreliable if the minimal world-sheet of a string contributing
  to the calculation of Wilson loop expectation values descends arbitrarily close to the singularity.}
   Wilson loops expectation values at large-N are saturated by fundamental string world-sheets that start
   at the loop at the AdS boundary and end somewhere in the bulk. The ending point is determined dynamically
    by minimizing the Nambu-Goto action, which equivalently minimizes the world-sheet area given the boundary Wilson loop.
In common cases discussed in the literature for large Wilson loops, the endpoint of the string world-sheet is
 at a minimum of the string frame scale factor of the metric, \cite{kinar}.
 We demand that this remains a finite distance from the singularity.
\end{itemize}

Both the previous criteria, that we define to determine a {\em repulsive singularity}, resemble a state of affairs that has been studied in string theory
in a different context, namely the linear dilaton vacuum of string theory in two dimensions.
In that case, the effective string coupling constant, $e^{\Phi}$, diverges on one side of space.
 The theory however can be rendered well-defined and computable if a suitable potential (cosmological constant)
  is added which raises a barrier in front of the strong coupling singularity. Any finite energy process
  can approach only at a finite distance from the strong coupling singularity and physics is therefore well-defined.
This analogy is very close to what happens in Einstein-dilaton holographic models for YM, \cite{ihqcd1,ihqcd2,gkmn1,gkmn2}.

In EHTs the last criterion is ambiguous unless an embedding to string theory is known, or other information is available.
The reason is that the relevant metric scale factor is that of the string frame metric and this is related to the Einstein
metric via a dimension-dependent exponential for the dilaton, $\Phi$. Unless we know the dilaton dependence of our bulk
 scalar $\phi$ we cannot calculate the string frame metric from the Einstein metric.

There are two extreme situations:
\begin{itemize}
\item The scalar $\phi$ has nothing to do with the dilaton. In that case $g_{\m\n}^E=g_{\m\n}^{\s}$.
\item The scalar $\phi$ is the dilaton. In that case, taking account of the non-standard normalization of the kinetic term in (\ref{v}) we obtain
\be
g^{E}_{\m\n}=e^{-\sqrt{2\over p-1}~\phi}~g^{\s}_{\m\n}
\label{ein}\ee
\end{itemize}
In the general case when $\phi$ contains an admixture of the dilaton we have the relation
\be
g^{E}_{\m\n}=e^{-k~\phi}~g^{\s}_{\m\n}\,,\qquad 0\leq k\leq  \sqrt{2\over p-1}\,.
\label{ein1}\ee

\subsection{The conductivity}

An important observable in a theory with a conserved charge at finite density is the conductivity. It is defined as the response of the system to an external electric field, measured by the induced current proportional to electric field.
This is a linear response function and in general is a tensor. Since  the backgrounds that will be studied in this paper are homogeneous and isotropic, the conductivity $\s$ is parametrised by a scalar.

$\s$ can be calculated by linear response theory using a Kubo formula from the two-point function of the fluctuations of the spatial components of the gauge field in the appropriate background.
Here we will consider fluctuations with very long wavelengths compared to their frequency. This is a good approximation for optical CM experimental measurements of the conductivity. Therefore we will turn on a electric field with harmonic time dependence proportional to $e^{i\omega t}$.
We will define the frequency dependent (AC) conductivity in terms of the transverse retarded current-current correlator as is standard,
\be
\s(\omega)\equiv{G^R(\omega,\vec k=0)\over i\omega}
 \label{33}\ee
The DC conductivity on the other hand is the zero frequency limit of the AC conductivity. This limit may be ill-defined in several cases, \cite{rson}. A direct definition can be obtained by turning on a constant electric field, calculating the induced current and thus obtaining the ratio, in the linearized regime.
This was discussed in detail in \cite{karch} for probe solutions using the DBI action. For massive charge carriers, a string drag calculation is also possible as was pointed out in \cite{karch}.  We will come back to this in detail at sections 5 and 6.

\subsection{Setup and Results}

We consider a class of Effective Holographic Theories containing the metric, a U(1) gauge field, and a scalar (\ref{ActionLiouville}).
The action contains only second derivatives of the metric, and both the scalar potential and the gauge coupling function are parametrized by exponentials:

\be
Z(\phi)=e^{\ga\phi}\,,\qquad V(\phi)=-2\Lambda e^{-\da\phi}\,.
\label{00a}\ee
These parametrizations are string theory-motivated and describe the IR asymptotics of the dynamics, which depend on two real parameters $\gamma,\delta$.
Requiring that the IR potential $V(\phi)$ vanishes in the UV will allow to promote IR solutions to asymptotically AdS solutions.

We first study the bulk physics of uncharged solutions (see \cite{ihqcd2,gkmn2} for $p=4$) adding the dynamics of a test vector field. In particular, we provide power solutions for any $\d$, at zero and finite temperature.  We develop criteria for the generic naked singularities that appear in the extremal solutions based on previous works, \cite{bad,ihqcd2,gkmn2} and show that at zero temperature the following classification holds:
  \begin{itemize}

\item $0\leq |\delta|<\sqrt{2\over (p-1)}$. In this class of theories the extremal (T=0) solution has a good singularity and the spectra are continuous without a mass gap. At finite temperature there is a continuous phase transition at $T=0^{+}$, whose order is given by the parameter $\delta^2$. At any positive temperature the system prefers to be in the black hole phase. Moreover there is a single black hole solution at any temperature.

\item    $|\delta|=\sqrt{2\over (p-1)}$. This is a marginal case: the singularity is good and the spectrum is continuous with a mass gap. This is a case where the IR asymptotics of the potential can be further refined, \cite{ihqcd2}.  At finite temperature there is a minimum temperature $T_{\rm min}$ below which there is no black hole solution.
    At $T_{\rm min}$ the system undergoes a phase transition to the black hole phase. The order of the phase transition is continuous and depends on subleading terms in the scalar potential, \cite{gursoy1},\cite{gursoy2}.

\item     $\sqrt{2\over (p-1)}<\d\leq \sqrt{2(p+2)\over 3(p-1)}$. The IR singularity is stronger here but is still acceptable holographically.
     The spectrum is discrete with mass gap. The asymptotic behavior of the masses of the n-th excited state is $m_n\sim n$.
      At finite temperature there is a minimum temperature $T_{\rm min}$ below which there is no black hole solution.
      Above $T_{\rm min}$ there are generically two black hole solutions, a small and a large one. The small is unstable while the large one is thermodynamically stable. At a temperature $T_c>T_{\rm min}$ the system undergoes a first-order phase transition to the large black hole case. In special cases, there can be more than two black holes above $T_{\rm min}$ and more than one phase transition is possible.

\item $\d> \sqrt{2(p+2)\over 3(p-1)}$. Such cases are unreliable in the effective holographic description.

\item $\d> \sqrt{2p\over (p-1)}$. Such cases violate the Gubser criterion. They are not acceptable holographically.

    \end{itemize}

Always at zero charge density, we analyzed the spectra of probe vector fluctuations with the following conclusions:

  \begin{enumerate}

\item ${\gamma\over \d}+{(p-3)\over (p-1)\d^2}>{3\over 2}\,$. When $p>3$ or $p=3$ and and the UV dimension of the scalar $\D<1$
then the potential diverges both in the UV and the IR and the spectrum is discrete and gapped. This resembles to an insulator.

    On the other hand if    $p=3$ and and the UV dimension of the scalar $\D>1$ then the potential vanishes in the UV and the spectrum is continuous. This resembles a conductor.

\item $-{1\over 2}<{\gamma\over \d}+{(p-3)\over (p-1)\d^2}<{3\over 2}\,$. The spectral problem is unacceptable and therefore the holographic effective field theory unreliable.

\item  ${\gamma\over \d}+{(p-3)\over (p-1)\d^2}<-{1\over 2}\,$,   This is holographically acceptable and the same remarks apply here as in 1.

    \end{enumerate}

An important observable of such theories in view of applications to condensed matter physics is the conductivity, both AC and DC.
There are several techniques for calculating them depending on the theory.
At low charge density they can be computed in the probe approximation, where the charge density does not backreact on the system.
At finite charge densities which is the main preoccupation of this paper the backreaction cannot be neglected and therefore we must study the conductivity in the full setup. We will calculate AC conductivities at extremality (T=0). This is the regime  $\omega/T \gg 1$. The relevant background solution is the extremal solution that will provide the leading power behavior of the AC conductivity as a function of frequency.

The calculation of the DC conductivity is much trickier in the fully backreacted Maxwell case we study here. We will calculate it assuming massive carriers and we will perform a drag-force calculation as in \cite{karch}.
As we shall see the ``drag" DC conductivity calculation makes an interesting prediction in the case where the scalar operator is not the dilaton: the DC conductivity is tightly correlated to the entropy and therefore also to the heat capacity. In particular, in the planar case $p=3$, the DC resistivity $\rho$ verifies the very interesting relation
\be
\rho~~\sim~~ S~~\sim~~ C_V\,,
\label{a1}\ee
 that seems to be valid in the strange metal region.

At small charge densities we can use the probe approximation around the solutions mentioned above to study the conductivities:
\begin{itemize}
 \item  At the lowest densities the DC conductivity $\sigma=1/\rho$ behaves as
\be
\rho\sim T^{{2(p-1)\gamma\d+2(p-3)\over (p-1)\d^2-2}}\,.
\label{a2}\ee
In particular, in this dilute regime it is independent of the dilatonic nature of the scalar field, which can be parametrized by a  parameter $k$ in the passage between the Einstein and the string metric as
\be
g^{E}_{\m\n}=e^{-k~\phi}~g^{\s}_{\m\n}\,.
\label{a3}\ee

The conductivity  can be a linear function for
\be
\gamma=\gamma_{\rm linear}\equiv \frac\d2-\frac{(p-2)}{\d(p-1)}\,.
\label{a4}\ee
As the zero charge density system never exhibits an entropy linear in temperature, we conclude that when linear resistivity appears, it is due to the presence of charge carriers.

\item For higher charge density,
\be
\rho={1\over \sigma}\sim T^{2k(p-1)\d+4\over 2-(p-1)\d^2}\langle J^t\rangle\,.
\label{a5}\ee
In this  case, the resistivity can never be linear.

  \end{itemize}

We have then analyzed several classes of exact solutions with Maxwell finite (charge) density \cite{char}. For two families we studied the full class of solutions. The first is a two-parameter family with $\gamma\delta=1$ which contains two branches in the $\gamma,\delta$ plane. The second is a two-parameter family with $\gamma=\d$. In both cases the solutions have arbitrary boundary data (temperature, scalar charge  and charge density).
Finally we have also analyzed a family of scaling (near-extremal solutions)
for arbitrary $\gamma,\delta$ where the scalar charge is related to the charge density.

 We start with $\gamma\delta=1$ where the general solution is known \cite{char}. Black holes exist for $\delta^2<3$ and have zero entropy at extremality. We have analyzed the thermodynamics and phase structure of these solutions, which depend on the specific values of $\d$. We focus here on the canonical ensemble\footnote{The grand canonical ensemble is also analyzed in the text.}:

  \begin{itemize}

 \item $0\leq\delta^2\leq 1$. There is a single branch of black holes for any positive temperature. This is like the analogous zero charge case. The temperature rises without bound (except when $\delta^2=1$ where it has a maximal value) with the size of the horizon. At $T=0^+$ there is a continuous phase transition whose order depends on $\delta$ (either second or third order in this range). At any positive temperature the system thermodynamically prefers to be in the black hole phase.

\item $1<\delta^2 <1+{2\over \sqrt{3}}$.    In this case, apart from the extremal solution, at any finite temperature below a maximal temperature $T_{\rm max}$, there are two black hole solutions, a small and a large one. At any temperature $0<T<T_{\rm max}$, it is the small black hole that is thermodynamically dominant. It is also the stable one, while the large black hole is unstable. The transition at $T={0^+}$  between the extremal and the small black hole solution is continuous, with its order depending on $\delta$. In particular, all orders of critical behaviour can appear.
    Above $T_{\rm max}$ it is the extremal solution that dominates as it is the only one present. The transition is of zeroth-order  and suggests that in a UV-complete setup there will be also another black hole solution (of the RN type) that will dominate this region, changing the transition to first-order, and if appropriately tuned to higher-order. This is however not visible with our simple IR potentials.

\item $1+{2\over \sqrt{3}}<\delta^2 <3$. In this case  the extremal solution (T=0) is behaving differently. For $1+{2\over \sqrt{3}}<\delta^2<{5+\sqrt{33}\over 4}$, the spectrum of the charged carriers is gapped and discrete when the UV dimension if the scalar operator is $\D<1$ and continuous otherwise. These systems appear as {\em Mott insulators}. In the rest of the range ${5+\sqrt{33}\over 4}<\delta^2<3$ the spectrum is gapless.

    At $T>0$
     there is a single black hole solution at every non-zero temperature in the IR effective action. It is always unstable and its temperature diverges  at extremality, while it vanishes when the black hole becomes large. It is always thermodynamically subdominant to the extremal solution. This situation is similar to the chargeless case with  discrete spectrum. Indeed the black holes here correspond to the small black holes there. This correspondence suggests also the completion of this picture when one adds back the AdS asymptotics. In that case a large black hole branch will appear and will match to the small black hole case at some $T_{\rm min}$. Therefore for $T>T_{\rm min}$ there will be generically two black holes (plus the extremal solution). At some higher temperature a phase transition to the large black hole will take place. So the system will be a Mott-like insulator up to $T_c$ and then there will be a first order phase transition to a conducting phase.

  \end{itemize}
In the first two regimes  the  AC conductivity at extremality scales as
\be
	\sigma(\omega) \simeq \omega^n \,,\qquad n =  \frac{(3-\d^2)(5\d^2+1)}{|3\d^4-6\d^2-1|} - 1\,.
	\label{a6}
\ee
The exponent is always larger than 5/3 in the
 region,  $0 \leq \d^2 < 1+ \frac{2}{\sqrt{3}}$ and
 diverges at $\d^2 = 1+\frac{2}{\sqrt{3}}$ (see \Figref{gd1ACconductivity} for details). The system  behaves as a conductor.
 The situation in the case, $1+{2\over \sqrt{3}}<\delta^2 <3$, is different: for $\d^2<\frac{1}{4}(5+\sqrt{33})$, the Schr\"odinger potential diverges near the horizon at extremality, and hence the system is insulating, while it becomes conducting away from extremality (see the left plot in \Figref{Fig:gd1VSupper});
 for $\frac{1}{4}(5+\sqrt{33})<\d^2<3$, the system is conducting with a continuum of states at extremality.

The DC conductivity for massive charge carriers is depicted in \Figref{Fig:dg=1Resistivity}. In the range $0\leq\delta^2\leq 1$, the DC resistivity vanishes at zero temperature and then rises in the black hole phase. For a special value of $\d$ in this range the rise is linear at low temperatures. The validity of linear regime is determined by the IR dynamical scale, $\ell$.  In the intermediate range $1<\delta^2 <1+{2\over \sqrt{3}}$, the resistivity drops steeply with temperature in the unstable large black hole branch and rises slowly with temperature in the thermodynamically dominant small black hole branch. In this latter branch, the low-temperature resistivity is linear for a special value of $\delta$. Again  the validity of linear regime is determined by the IR dynamical scale. Finally, in the upper  branch $1+{2\over \sqrt{3}}<\delta^2 <3$, the resistivity is decreasing with the temperature.

In the special case $\delta^2=1$, the resistivity is finite at zero temperature, then increases, and finally diverges at a higher temperature signaling potentially a critical behavior.

The second class of solutions analyzed in full detail are the charged solutions of the theory with $\gamma=\delta$.
This class is special as it seems to be the only possible case on the $\gamma,\delta$ plane that have finite entropy at extremality (beyond AdS-RN black holes).
These are solutions with the full set of parameters (charge density and temperature). Unlike the previous case, they may not be the most general solutions. We have studied their thermodynamics and transport properties and the results are as follows.

 \begin{itemize}

 \item $0\leq\delta^2\leq 1$. There is a single branch of black holes for any positive temperature. This is like the analogous zero charge and $\gamma\delta=1$ case. The temperature rises without bound (except when $\delta^2=1$ where it has a maximal value) with the size of the horizon. There is no phase transition as $T\to0^+$. At any positive temperature the system thermodynamically prefers to be in the black hole phase. There is however a continuous transition from the charged black holes to the neutral ones at finite temperature, as the charge is sent to zero.

\item $1<\delta^2 <3$.   Apart from the extremal solution, at any finite temperature below a maximal temperature $T_{\rm max}$, there are two black hole solutions, a small and a large one. For all temperatures $0<T<T_{\rm max}$, the small black hole is thermodynamically dominant. It is  also the stable one, while the large black hole is unstable. Again, there is no phase transition as $T\to{0^+}$ is approached for finite charge, but there is a first order transition to the extremal solution as $Q\rightarrow 0^+$ for finite temperature. Above $T_{\rm max}$, the extremal solution dominates as it is the only one present.

\end{itemize}

The zero temperature AC conductivity for all such black holes behaves as $\omega^2$ independent of $\delta$.

The DC conductivity for massive charge carriers is depicted in \Figref{Fig:d=gResistivity}. In both ranges the resistivity is non-zero at zero temperature and has a regular low-temperature expansion in integer powers of the temperature starting with a linear term. Again the span of the linear regime is determined by the IR dynamical scale, $\ell$.

We have also investigated near extremal solutions for the whole $\gamma,\delta$ plane in the fully backreacted case. Such solutions agree for $\gamma\delta=1$ and $\gamma=\delta$ with the near extremal limits of the solutions mentioned above. They exist for a special value of the scalar charge and the charge density. They have a single parameter, the temperature.

Such solutions exist when
\be
2(p-1)+p\ga^2-2\ga\da-(p-2)\da^2>0 \sp \ga^2-\ga\da+2>0\sp 2-\da^2+\ga\da >0
  \ee
The condition above define the Gubser bound for the whole $(\gamma,\delta)$ plane, as shown in figure \ref{Fig41}.
The spin-2 IR spectra  of such solutions can be reliably calculated from the effective holographic action when
\be
2+\d(\d-\g) +{(\g-\d)^2\over 2(p-1)}>1
\ee
or
\be
2+\d(\d-\g) +{(\g-\d)^2\over 2(p-1)}<1 ~~~{\rm and}~~~
w_pu>4(p-1)
\ee
with $w_pu$ given in (\ref{wpu}).
The spin-1 spectra are always reliable except when
\be
2+\d(\d-\g) +{(\g-\d)^2\over 2(p-1)}<1 ~~~{\rm and}~~~
(\g-\d)^2>{2(p-1)\over p}
\ee
when the spectrum is reliable above a critical value of the charge density, (\ref{tan4}).

For $\delta=0, \gamma=-\sqrt{2(p-1)\over z-1}$, they provide Lifshitz solutions at extremality with Lifshitz exponent $z>1$. This suggests that generically Lifshitz solutions are generated by a scalar along a flat direction of the potential that drives the U(1) coupling constant to zero in the IR.

Analysing the phase structure, we find that, depending on $(\gamma,\delta)$, continuous phase transitions of all orders appear as zero temperature is approached. Furthermore, the near extremal solutions  have simple scaling thermodynamics functions. The entropy and heat capacity  in particular scale as
 \be
 S\sim C_Q\sim T^{(p-1)(\gamma-\d)^2\over 2(p-1)(1-\d(\d-\gamma))+(\d-\gamma)^2}
 \label{a7}\ee
near extremality. The entropy  is finite at extremality only if $\gamma=\delta$. When the exponent in (\ref{a7}) is negative this indicates an unstable black hole. In this region this black hole is never thermodynamically dominant.

When
\be\label{111}
\frac{(p-1)( \d^2 - \gamma^2 - 4) (8(1 -
    p) + (\d - \gamma) ( \gamma(1+p) + \d(5  p - 7)))}{4 (2(1 -
    p) + (\d - \gamma) (\gamma + \d (2 p - 3)))^2}>0
     \ee
     $$
     {(p-1)(\gamma-\d)^2\over 2(p-1)(1-\d(\d-\gamma))+(\d-\gamma)^2}<0
     $$
then the system has charged excitations with a mass gap and a discrete spectrum. In the special case of $p=3$ one more condition is needed for this: that the UV dimension of the scalar operator dual to $\phi$ is less than 1.
Such systems behave as Mott insulators.
In such cases there will a finite temperature phase transition from an insulating to a conducting phase with a continuous spectrum and no mass gap.

 The extremal AC conductivity scales as
\be
\sigma(\omega)\sim \omega^{\left|\frac{6(1-p)+(\d-\gamma)(p\gamma + (3p-4)\d)}{2(1 -
    p) + (\d - \gamma) (\gamma + (2 p - 3)\d )}\right| - 1}\,.
\label{a8}\ee
The exponent may become negative in a region of parameters, but in this same region the thermodynamics of such near extremal solutions is unstable.

The DC conductivity for massive charge carriers scales as
\be
\rho\sim T^{ {2k(p-1)
(\d-\gamma)+2(\d-\gamma)^2\over 2(p-1)(1-\d(\d-\gamma))+(\d-\gamma)^2}}\,.
\label{a9}\ee
As mentioned earlier, when the scalar is not the dilaton (k=0), we obtain $\rho\sim S^{{p-1\over 2}}$, and in $p=3$ in particular $\rho \sim S$.
These low-temperature scaling laws are expected to be valid throughout the $\gamma,\delta$ plane. When $p=3$, there are several cases that exhibit linear resistivity at low temperature. They have $\gamma$ given by $\gamma_{\pm}$
\be
\gamma_{\pm}=3\d+2k\pm 2\sqrt{1+(\d+k)^2}\,.
\label{a10}\ee
As the chemical potential scales in that case with a higher power of the temperature this behavior is valid in a regime where the chemical potential dominates the temperature  as in real strange metals.

The leading scaling behavior of thermodynamic functions and conductivities is expected to be universal and be valid for all solutions for this class of theories near extremality. Therefore, they provide a profile that might help select EHTs as descriptions of concrete CM systems.

We have also compared the entropy due to the charge carriers and compared it to the entropy in their absence. We found that generically the entropy at finite charge density dominates that in the absence of carriers as discussed in detail in the end of section \ref{secthermo}. Only in the gapped regions the both entropies are subleading and ${\cal O}(1)$ and therefore a comparison at this stage cannot be done.

Another interesting behavior concerns the zero charge density limit of the various solutions, in the context of the canonical ensemble.
This can be taken explicitly for the $\g=\d$ and $\g\d=1$ solutions and compared to the zero charge density solutions.

We find the following: In the $\g\d=1$ case  there seems to be no transition if we study the equilibrium thermodynamic functions.
In the $\g=\d$ case however there is a phase transition in the zero charge limit: it is  second-order if $\delta^2<1$ and  first-order if
$1<\delta^2<3$. The behavior of the conductivities in this zero charge limit is more interesting. The behavior of the AC conductivity at extremality  for the gapless charged cases is summarized in (\ref{a8}) and the exponent is independent of the charge density.
A calculation of the conductivity in the zero charge case, done in section \ref{sec:ACconductivity} gives instead for the conductivity
\be
\s\sim \omega^n\sp n={|\d^2-2\g\d-1|\over |1-\d^2|}-1
\label{a88}\ee
and this is generically different from (\ref{a8}).
This indicates that for all $\g,\d$, there is a transition as $Q\to 0$ that affects the scaling behavior of correlation functions.
This is also true in the case of the standard RN black holes.
Similar remarks apply to the DC conductivity at low temperature both at finite and zero charge density.

Finally we have found solutions to the bulk equations in the limit where there are strong self-interactions in the charge sector. This corresponds to the limit where the DBI action vanishes. In this limit the charge density has a limit which is independent of integration constants indicating that in flavor brane systems there is a saturation limit for the charge density, imposed by the DBI action. We have found two non-trivial solutions to the equations, one that is extremal, and another for $p=4$, that is near extremal but does not seem to fit the general pattern of Maxwell solutions described above.
This system is an insulator at zero temperature. Its DC resistivity for massive carriers behaves as $\rho\sim T^{1\pm {3\over 2}k}$ indicating linear resistivity when the scalar is not the dilaton ($k=0$).

\subsection{Outlook}

The set of holographic effective theories discussed here, parametrized by the exponents $\gamma,\delta$, is severely constrained by the nature of the IR singularity.  Imposing constraints so that the singularity is of the good kind removes part of the $\gamma,\delta$ plane from consideration.

The solutions fully analyzed, namely the uncharged solutions as well as $\gamma\delta=1$ and $\gamma=\delta$ cross sections, give a rather representative profile of the generic case.
The near extremal asymptotics are summarized by the scaling solutions that can be found for any value of $\gamma,\delta$.

Calculations of the transport coefficients, namely the conductivity, are important ingredients for the characterization of the associated physics.
We have given estimates for the scaling behavior of both the DC and AC conductivities. They suggest that there are EHTs that display  the hallmark behavior of strange metals, namely linear resistivity. Moreover, the DC resistivity is related in the planar case to the entropy, correlating two interesting aspects of strange metal behavior, linear heat capacity to linear resistivity.

There are however more features of the physics of the EHTs that need to be analyzed so that their suitability as theories of strange metal behavior can be assessed. An important general ingredient are the zero and low temperature spectra of fluctuations, that should be derived in order to completely characterize the low energy degrees of freedom, as well as the energy-energy and charge-charge correlators.

 In particular the interplay between  insulating versus conducting behavior must be further analyzed. It is a generic property of the holographic systems described here, in the planar case $p=3$, to favor conducting behavior. Indeed, with the exception of strongly relevant dynamics, the systems at small charge density seem to be conductors at any small but non-zero temperature.
 There is generically a continuous phase transition to the extremal solution which is insulating.

The phase structure also seems to be interestingly varying with the IR ``strength" of the scalar operator captured by the exponent $\delta$.
There are indications from our analysis that the appearance of discrete spectra at low temperature is ``delayed" by the finite charge density.
At zero charge density, such spectra appear when $\d^2>1$ while at finite density in the $\gamma=\delta$ case they never appear, and for $\gamma\delta=1$ their appearance is seemingly delayed until $\delta^2=1+{2\over \sqrt{3}}$. This structure needs verification from a more detailed analysis of low energy spectra.

The underlying fermionic nature of the holographic system is also a subject that needs clarification. Fermi surfaces and related observables are part of the phenomenology of strange metals and their cross-over behavior to Fermi liquids. A study of fermionic response functions along the lines of \cite{za,mc,mc2} is necessary in order to characterize the fermi surfaces and the presence of quasi-particles.

Last but nor least, the superconducting instability must be analyzed. For this the EHT used here needs to be extended by the inclusion of an extra charged scalar operator, $X$. At the quadratic level, and in the most general case, this introduces two extra functions of the neutral scalar into the model.

\section{The zero charge density dynamics\label{zero}}

This is a warm up case that has been studied extensively lately in \cite{ihqcd1,ihqcd2,gkmn1,gkmn2} in view of mapping the landscape of Einstein-dilaton gravity with a potential, with the intention to use such a model as a phenomenological model for large-N YM. Our aim here is to obtain constraints on $\gamma$ and $\delta$ in order to map the allowed region of parameters for the non-zero charge density exact solutions that will follow in the later sections.
The action is given by (\ref{ActionLiouville})
where we consider zero charge density and to start with a Liouville potential,
$$
V(\phi)=V_0 e^{-\delta\phi}, \qquad Z(\phi)=0\,.
$$
As argued earlier we will assume that the potential will be  modified eventually  in the UV so as to lead to asymptotically AdS solutions.
In the \emph{domain-wall} frame ansatz for the metric
\be
ds^2=e^{2A(r)}\left[-f(r)\ud t^2+\ud x^i\ud x_i\right]+{\ud r^2\over f(r)}\,,
\label{ToroidalMetrics2}
\ee
the equations are
\be
(p-1)A^{\prime\prime}+\half\phi^{\prime 2}=0\,,\qquad
g'(pA^{\prime}+g^{\prime})+{g^{\prime\prime}} =0\,,\qquad f=e^g\,,
\label{3}\ee

\be
(p-1)A'(g'+pA')-\half\phi^{\prime 2}-e^{-g}V(\phi)=0\,,
\label{4}\ee
where primes stand for derivatives with respect to $r$.

Since $A'(r)$ is monotonically decreasing asymptoting near the UV, so is $e^{A}$. There are two exclusive possibilities: either $e^{A}$ asymptotes to an AdS scale factor or it vanishes somewhere, and this is a naked singularity, \cite{ihqcd2}.
This property will persist also at finite density. The general solution to this system has been found in \cite{Charmousis:2001nq}.

For our purposes here it suffices to study zero temperature solutions and therefore we set $g=0$ in (\ref{3},\ref{4}). The background solution is
\be
e^A=r^{\frac{2}{(p-1)\d^2}}\,,\qquad e^{\d \phi}= \frac{V_0\d^4}{2\left(\frac{2p}{p-1}-\d^2\right)}~r^{2}\,,\qquad f=1\,.
\label{9a}
\ee
Defining \emph{general} coordinates,
\be
 	\ud s^2 = -D(r)\ud t^2 + B(r) \ud r^2 +C(r)\le(\ud x_i\ud x^i\ri),
	\label{ToroidalMetrics}
\ee
the background solution \eqref{9a} becomes
\be
D(r)=C(r)=r^{\frac{4}{(p-1)\d^2}}\,,\qquad B(r)=1\,,\qquad e^{\d \phi}=\frac{V_0\d^4}{2\left(\frac{2p}{p-1}-\d^2\right)}~r^{2}\,.
\label{9aGeneral}
\ee
This can also be written in ``AdS-like''-coordinates,
\bsea
	\ud s^2& = & r^2(-\ud t^2 + \ud x^2 + \ud y^2)+ r^{(p-1)\da^2-2}\ud r^2\,,\label{90}\\
	e^{\delta\phi} &=& \frac{V_0\d^4}{2\left(\frac{2p}{p-1}-\d^2\right)}r^{(p-1) \da^2} \,.
	\label{LiouvilleBackground}
\esea
This is not $AdS_{p+1}$ and explicitly breaks its $SO(p,2)$ symmetry group. This breaking corresponds to a non-zero value for $\da$, and goes together with a non-trivial scalar field profile. On the other hand, setting $\da=0$ restores the $SO(p,2)$ invariance and yields a constant scalar field. The potential $V(\phi)$ then is simply a constant, as expected. However, none of the analytic solutions presented below have both a non-trivial dilaton and $AdS_{p+1}$ asymptotics in the case of a pure cosmological constant.

By going through a conformal transformation of the metric $g_{\mu\nu}\to e^{-\frac\phi\da}g_{\mu\nu}$ in appropriate coordinates, one can show that this space-time is conformally flat, recovering Poincr\'e invariance. In the string case, $\gamma=\delta=\sqrt{2\over d-2}$, the coordinate transformation induces a logarithmic branch and then the background in the string frame is simply Minkowski space-time.

To carry out a general analysis, in the domain wall frame (\ref{ToroidalMetrics2}), it is convenient to introduce a ``superpotential" $W(\phi)$ as
\be
\phi'\equiv \frac{\ud\phi}{\ud r}=\frac{\ud W}{\ud\phi}\,,\qquad A'\equiv \frac{\ud A}{\ud r}=-\frac{W}{2(p-1)}\,.
\label{5}
\ee
Then the system of equations (\ref{3},\ref{4}) is equivalent to (\ref{5}) and
\be
2V=\frac{p ~W^2}{2(p-1)}-\left(\frac{\ud W}{\ud\phi}\right)^2\,.
\label{6}
\ee
Moreover, (\ref{3},\ref{4})
and (\ref{5},\ref{6}) have the same number of initial conditions, namely three.
Note also that each solution $W$ of (\ref{6}) provides two solutions to the equation of motion, one satisfying (\ref{5}) and another where the sign of both equations in (\ref{5}) is reversed.

Here we analyze the general solution of the zero-temperature superpotential equation,
eq. (\ref{6}), (below we set  $\l=e^\phi$).
\be\label{spot1}
-\l^2 (W'(\l))^2 + \frac{p}{2(p-1)}W^2(\l) = 2V(\l)\,,
\ee
To exclude the well known case of IR fixed points, that have been discussed extensively before, we assume that the potential is monotonic. It must be positive near the UV fixed point, and we will assume that it is positive,  $V(\l)>0$, everywhere.

First we observe some general properties:
\begin{enumerate}
\item The  solution can only exist as long as $|W(\l)|> \sqrt{{4(p-1)\over p} V(\l)}$.
\item The equation has a symmetry $W \to -W$, so we can
limit the analysis to $W>0$.
\item For any  value $\l_0\neq 0$ for the scalar, there are two solutions of
(\ref{spot1}), $W_+(\l),W_-(\l) $ passing through the point $\l_0$,
such that $W_+(\l_0) =  W_-(\l_0)$, and $W_+'(\l_0) = - W_-'(\l_0)$.
In other
words there are two branches of solutions: one where   $W$ and $W'$
have the same sign (i.e. $W_+'(\l)>0$) , another where they have opposite
sign ($W'_+(\l)<0$).

\item At any $\l_*\neq 0$ where  $|W(\l_*)|= \sqrt{{4(p-1)\over p} V(\l_*)}$, the derivative vanishes, $W'(\l_*)=0$.
\item A solution can go past such a point  $\l_*$ {\em only} if $V'(\l_*)=0$.
Indeed, suppose that $V'(\l_*)> 0$.
  If the solution exists for $\l<\l_*$, at the point $\l_*$ we have:
\be
W(\l_*) = \sqrt{{4(p-1)\over p} V(\l_*)}\,,\qquad W'(\l_*)=0\,,\qquad V'(\l_*) >0
\ee
$$ \Rightarrow
W(\l_* + \epsilon ) < \sqrt{{4(p-1)\over p}V(\l_*+\epsilon)}\,,	
$$
therefore the solution does not exists  for $\l>\l_*$.
\item By the same
argument, if $V'(\l_*)<0$, the solution does not exist for $\l<\l_*$.
\item It follows from points 3,4 and 5 that,  if $V(\l)$ is positive and monotonic, the two branches $W_+(\l)$ and $W_-(\l)$
(see point 4)  are completely
disconnected, since neither $W'$ nor $W$ can change sign. However two solutions belonging to different branches
can be glued together at a point $W'=0$.
\item All solutions that reach $\l=0$ have either $W(0) = \sqrt{{4(p-1)\over p}V(0)}$,
or $W'(0)=\infty$.
\end{enumerate}

In what follows we assume $V(\l) >0$ and without loss of generality we take $W(\l)>0$.

\begin{figure}[!ht]
\begin{center}
\begin{tabular}{cc}
	 \includegraphics[width=0.45\textwidth]{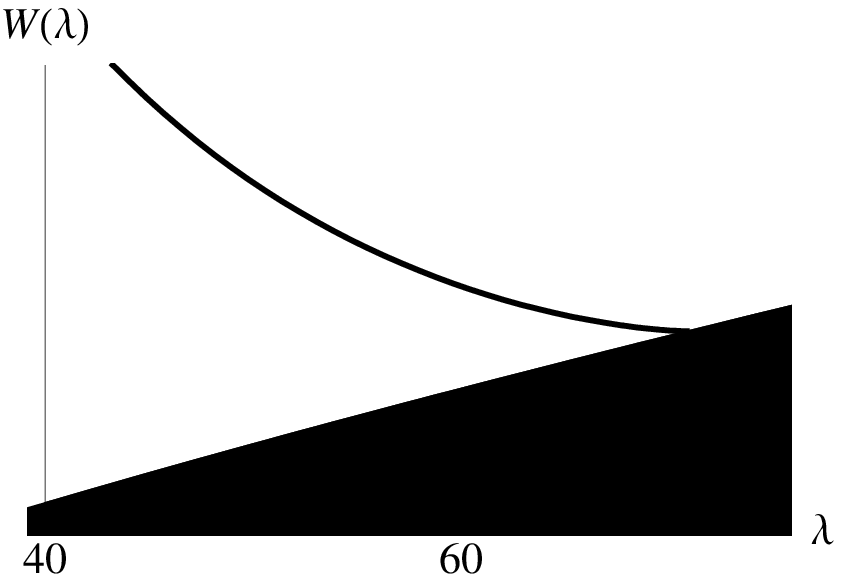}&	
	 \includegraphics[width=0.45\textwidth]{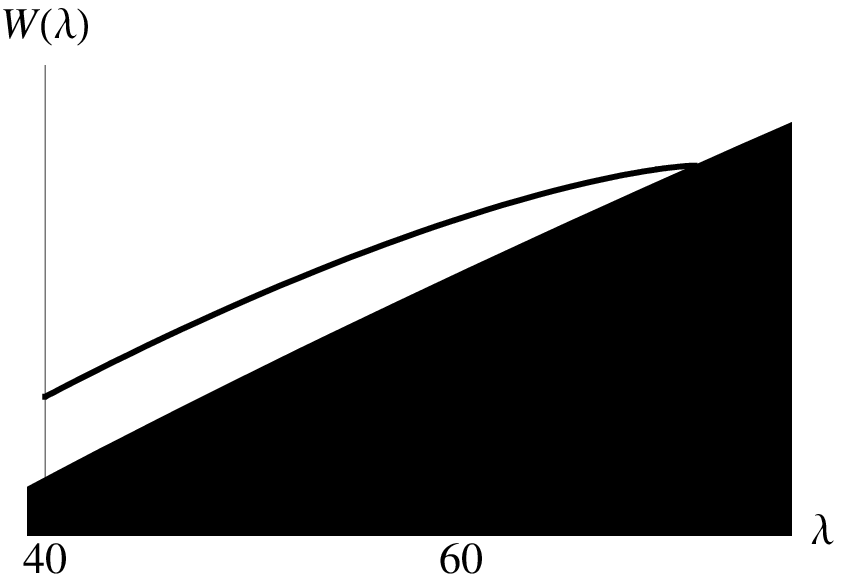}
 \end{tabular}
\caption{{Superpotential on (a) the $W_-$ branch  and (b) the
$W_+$ branch,  close to
a critical point. The black area is the ``forbidden'' region below
the critical curve $\sqrt{{4(p-1)\over p} V}$, where $W'$ would become imaginary. The solution stops where
it meets the critical curve.}}
\label{super5}
\end{center}
\end{figure}

An analysis on how solutions approach critical points for $p=4$ is described in detail in the appendix of \cite{gkmn2} and can be generalized to arbitrary $p>2$.
This behavior is summarized in \Figref{super5} for the $W_-$ branch.

\subsection{Solutions close to $\l\sim \infty$}

 We can analyze the solution of (\ref{spot1}) in the asymptotic region of large $\l$. We assume
for the potential a power-law behavior
\be\label{spot14}
V(\l) \underset{\l \to +\infty}{\sim} V_{0} \l^{|\delta|}\,,
\ee
where we took $\d$ to be negative so that the potential is increasing.\footnote{The potential has to increase at the AdS boundary. If it decreases in some intermediate range there there is an IR fixed point. We are not interested in this case here.}

There are two kinds of asymptotic solutions:
\begin{enumerate}
\item a continuous one-parameter family of  the form:
\be  \label{spot15}
W_C(\l) = C \l^{\pm k}  + {2V_0\over C\left({p\over p-1}\mp k(|\d|\mp k)\right)}\l^{|\d| \mp k } + \ldots\,,\qquad  k=\sqrt{p\over 2(p-1)}\,,
\ee
where $C$ is an arbitrary constant of integration;
\item a {\em single} solution that asymptotes as
\be\label{spot16}
W_s(\l)  = {W}_\infty \l^{|\d|\over 2}+\cdots, \qquad {W}_\infty = \sqrt{8 V_0 \over {2p\over p-1}-\d^2}\,.
\ee
\end{enumerate}
Notice that if $V_0>0$,   both types of solutions exist only if $|\d|<{2p\over p-1}$: for $|\d|>{2p\over p-1}$ the l.h.s. of the
differential equation is asymptotically negative. In this case there is no solution that reaches arbitrarily
large values of $\l$, but rather all solutions to (\ref{spot1}) are of the bouncing type: they reach a maximum value $\l_*$ where a $W_+$ and a $W_-$ solutions join.

\subsection{General Classification of the solutions} \label{general}

The general classification of solutions for the superpotential is as follows:
 for any positive and  monotonic potential $V(\l)$ with  AdS asymptotics in the UV and
\be
V(\l)\underset{\l \to +\infty}{\sim} V_\infty  \l^{|\d|}\,,\qquad  V_0 >0\,,\nn
\label{91}\ee
the zero-temperature superpotential equation has three types of solutions, that we name the {\em Generic},
the {\em Special}, and the {\em Bouncing} types:
:
\begin{enumerate}

\item  A continuous one-parameter family  that has a regular expansion near the boundary and
 reaches the asymptotic large-$\l$ region where it  grows as
\be
W \simeq C_b \l^{\sqrt{p\over 2(p-1)}}+\cdots  \qquad \l \to \infty\,,
\label{92}\ee
where $C_b$ is an arbitrary  positive real number.
These solutions lead to backgrounds with ``bad'' (i.e. non-screened) singularities at finite values of the radial coordinate $r_0$,
where $e^{A} \to 0$ and $\l \to \infty$ as
\be
e^A(r) \sim |r-r_0|^{1\over p}, \qquad \l(r)^{\sqrt{p\over 2(p-1)} } \sim {2(p-1)\over pC_b~(r_0-r)}\,.
\label{93}\ee
We call  this  solution {\em generic}. An example for $p=4$ is shown in the left part of  \Figref{superbad}
\begin{figure}[!ht]
\begin{center}
\begin{tabular}{ccc}
 \includegraphics[width=0.3\textwidth]{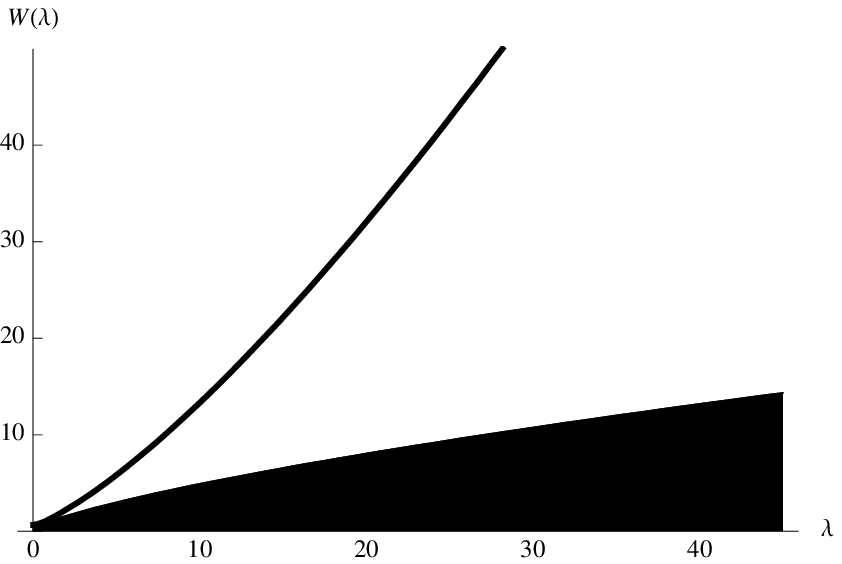}
\includegraphics[width=0.3\textwidth]{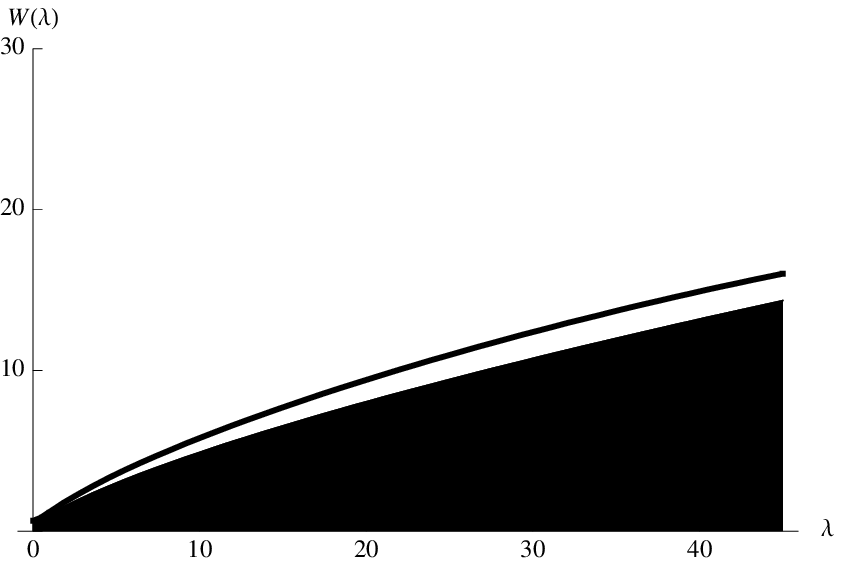}
\includegraphics[width=0.3\textwidth]{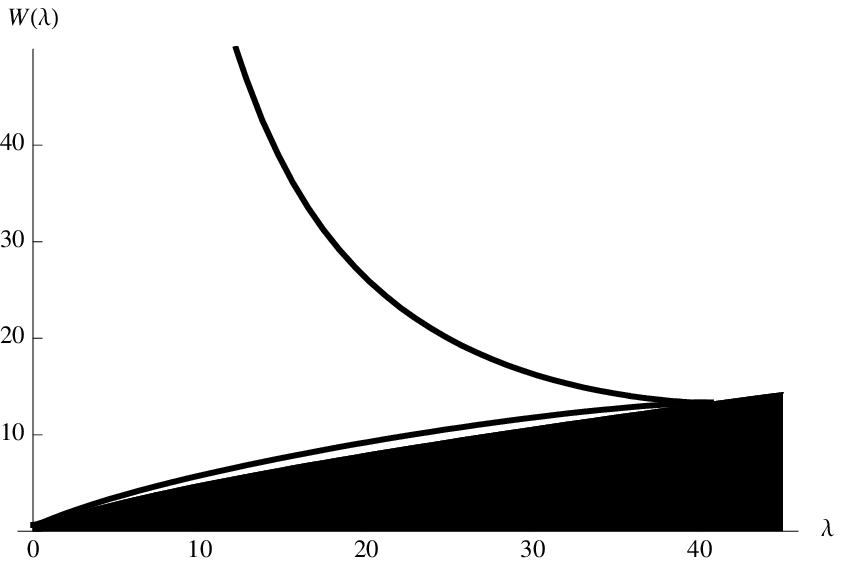}
\end{tabular}
\caption{From left to right, Superpotentials of the ``generic'', ``special'' and  ``bouncing'' type. The black area is the forbidden region below the curve $\sqrt{{4(p-1)\over p}V}$. The AdS boundary is at $\l=0$. } \label{superbad}
\end{center}
\end{figure}

\item A  unique   solution,    which also reaches the large-$\l$ region, but slower:
\be\label{singular}
W_s(\l)  = {W}_\infty \l^{|\d|\over 2}+\cdots, \qquad {W}_\infty = \sqrt{8 V_0 \over {2p\over p-1}-\d^2}\,.
\ee
We call  this  the {\em special} solution. An example is shown in the center part of  \Figref{superbad}.
We can subsequently find the dilaton and scale factor to be
\be
e^{|\d|\phi}={2\left({2p\over p-1}-\d^2\right)\over V_0\d^4}{1\over (r-r_0)^2}
\,,\qquad e^A=|r-r_0|^{2\over (p-1)\d^2}\,.
\label{95}\ee

\item A second continuous one-parameter family where $W(\l)$  does not reach the asymptotic large-$\l$ region.
 These solutions have two branches that both reach $\l=0$ (one in the UV, the
other in the IR)  and merge at a point $\l_*$
where $W(\l_*) = \sqrt{{4(p-1)\over p}V(\l_*)}$. The IR branch is again a  ``bad'' singularity
at a finite value $r_0$, where $W\sim \l^{-\sqrt{p\over 2(p-1)}}$,  and
\be
e^A(r) \sim |r-r_0|^{1\over p}, \qquad \l(r)^{-\sqrt{p\over 2(p-1)} } \sim {2(p-1)\over pC_b~(r_0-r)}\,.
\label{94}\ee
We call  this  solution {\em bouncing}. An example is shown at the right of \Figref{superbad}

\end{enumerate}

Notice that,  as two solutions with positive derivative cannot cross, the
special  solution (\Figref{superbad}) marks the
boundary between the generic (bad) solutions, that reach the asymptotic large-$\l$   region
as $\l^{\sqrt{p\over 2(p-1)}}$ (\Figref{superbad}) and the bouncing  ones,  that don't reach it,
\Figref{superbad}.
Notice that, if $|\d|>{2p\over p-1}$, only  bouncing  solutions exist.

Finally, all solutions end up in a naked singularity in the IR.
This can be seen as follows. We can evaluate the scalar curvature from the equations as
\be
R=-{p+1\over p-1}V+{\phi'^2\over 2}={p\over p-1}\left[W'^2-{p+1\over 4(p-1)}W^2\right].
\label{96}\ee
For $W\sim C\l^a$ in the IR
\be
R\simeq {p\over p-1}\left[a^2-{p+1\over 4(p-1)}\right]C^2\l^{2a}+\cdots
\ee

When $a<0$ such solutions violate Gubser's criterion. For positive $a>0$ there is an IR singularity.

\subsection{The spectra of spin-2 fluctuations\label{spectra}}

At zero temperature there are three types of perturbations, transverse-traceless fluctuations of the metric giving spin two excitations, a scalar excitation that corresponds to a gauge invariant combination of the trace of the metric and the scalar \cite{kn} and a vector fluctuation originating in the gauge field. The equations they satisfy were analyzed in detail for $p=4$ in \cite{kn,ihqcd2}. We will only consider the graviton fluctuation $h_{\m\n}$ here as the second order operator is the minimal Laplacian. By factorizing $h_{\m\n}(r,x^i)=h(r)\tilde h_{\m\n}(x^i)$ with $\square_4 \tilde h_{\m\n} =m^2 \tilde h_{\m\n}$ we obtain for the radial wave-function
\be
h''+p~A'h'+m^2e^{-2A}h=0\,.
\label{97}\ee
Changing variables and coordinates to the radial coordinates, we obtain a Schr\"odinger problem
\be
-\ddot\psi+V_s\psi=m^2\psi\,,\qquad {dz\over dr}=e^{-A}\,,\qquad h=e^{-{(p-1)A\over 2}}\psi
\,,\qquad V_s={p-1\over 2}\ddot A+{(p-1)^2\over 4}\dot A^2\,,
\label{8}\ee
where dots stand for derivatives with respect to $z$.
The Schr\"odinger potential always diverges $V_s\to+\infty$ near the AdS boundary.

We parameterize again the  superpotential as $W=C~\l^a$ in the IR (large values of $\l=e^{\phi}$). As negative values of $a$ violate Gubser's criterion  we restrict our discussion to $a>0$. We obtain
\be
A={1\over 2(p-1)a^2}\log|r-r_0|\,,\qquad e^{a\phi}={1\over Ca^2(r_0-r)}\,.
\label{98}\ee
The Schr\"odinger coordinate defined in (\ref{8}) is
\be
z=z_0+{(r-r_0)^{1-{1\over 2(p-1)a^2}}\over 1-{1\over 2(p-1)a^2}}
\,,\qquad A={1\over 2(p-1)a^2-1}\log|z-z_0|+\cdots
\label{99}\ee

\begin{itemize}

 \item $2(p-1)a^2<1\,$.
 In this case the IR limit $r\to r_0$ corresponds to $z\to \infty$.
    The Schr\"odinger potential from (\ref{8}) is in this case
    \be
 V_s\simeq {\zeta^2-\zeta\over z^2}\,,\qquad \zeta={1\over 4\left(a^2-{1\over 2(p-1)}\right)}<0\,,
 \ee
and vanishes near the IR singularity. The spectrum is therefore gapless and continuous. This is an acceptable system.

  \item $2(p-1)a^2=1\,$. In this case $z-z_0=\log|r-r_0|$ and $A={z-z_0}$
  The Schr\"odinger potential from (\ref{8}) asymptotes to a constant
    \be
 V_s\simeq {(p-1)^2\over 4}\,,
 \ee
 and the system has a gap and continuous spectrum above the gap. The spectral problem is well-defined and the system passes the  Gubser criterion.

\item $2(p-1)a^2>1\,$. In this case the IR limit $r\to r_0$ corresponds to $z\to z_0$.
    The Schr\"odinger potential from (\ref{8}) is
    \be
 V_s={\zeta^2-\zeta\over (z-z_0)^2}\,,\qquad \zeta={1\over 4\left(a^2-{1\over 2(p-1)}\right)}>0\,,
\label{100} \ee
 and rises steeply in the IR. This suggests  a discrete spectrum.
Near $z=z_0$ the two linearly independent solutions behave as
\be
\psi_1\sim (z-z_0)^{\zeta}\,,\qquad \psi_2\sim (z-z_0)^{1-\zeta}\,.
\label{101}\ee
$\psi_1$ is always normalizable near $z=z_0$ as $\int^{z_0}\ud z~ \psi_1^2<\infty$.

When $\zeta\geq {3\over 2}$, implying
\be
\sqrt{1\over 2(p-1)}<a\leq \sqrt{(p+2)\over 6(p-1)}\,,
\label{102}\ee
  the second solution $\psi_2$ is non-normalizable. This is an acceptable state
  of affairs as only one of the two solutions should be acceptable in the IR. In the opposite case,
  we must impose an extra boundary condition in the IR and this is against the dictums of the holographic correspondence.

For the remaining values of $a$, namely
\be
a> \sqrt{(p+2)\over 6(p-1)}\,,
\label{103}\ee
both solutions are square integrable in the IR and such superpotentials are therefore unacceptable.

  \end{itemize}

With the above results in stock, we now return to the solutions for the IR superpotential for a given potential described in \ref{general}.

  \begin{itemize}

  \item For any $\delta$, the bouncing solutions have $a<0$ and violate therefore Gubser's bound.

 \item When a generic solution exists ($|\d|<{2p\over (p-1)}\,$), $a=\sqrt{p\over 2(p-1)}$ and
 for $p>1$, this is always bigger than  $\sqrt{(p+2)\over 6(p-1)}\,$. Therefore, such solutions are always unreliable
  since they have a bad Sturm-Liouville problem for spin-2 fluctuations.

\item Therefore only the special solutions (that exist for $|\d|<\sqrt{2p\over p-1}\,$) with $a={|\delta|\over 2}$ have  a chance of being physical and reliable.
 They are acceptable for $\d\leq \sqrt{2(p+2)\over 3(p-1)}\,$.

    \end{itemize}

In the acceptable range the spectrum of spin-2 fluctuations is as follows:

  \begin{itemize}

\item $0\leq |\delta|<\sqrt{2\over (p-1)}\,$. In this class the spectra are continuous without a mass gap.

\item    $|\delta|=\sqrt{2\over (p-1)}\,$. This is a marginal case: the spectrum is continuous with a mass gap. The asymptotics in this case can be further refined \cite{ihqcd2} but we will not pursue this further here.

\item     $\sqrt{2\over (p-1)}<\d\leq \sqrt{2(p+2)\over 3(p-1)}\,$. Here the spectrum is discrete with mass gap. The asymptotic behavior of the masses of the n-th excited state is $m_n\sim n$.

\item $\d> \sqrt{2(p+2)\over 3(p-1)}\,$. Such cases are unreliable from a holographic point of view, as justified earlier.

\item  $\d> \sqrt{2p\over (p-1)}\,$. Such cases violate the Gubser criterion.

    \end{itemize}

\subsection{The spectrum of current excitations\label{specc}}

\label{current}

We now consider a small gauge field perturbation of the form $A_{\mu}(r,\omega,\vec k)e^{i\omega t+i\vec k\cdot \vec x}$, $\mu=0,1,2,...,p-1$,
 in the gauge $A_r=0$
 around the neutral background \eqref{9aGeneral}.
The effective action for the gauge field reads
\be
S_{F^2}=\int dr\int{d\omega d^{p-1}k\over (2\pi)^{p}}{Z\sqrt{DBC^{p-1}}\over 2}\left[{|A_0'|^2\over DB}-{|A_{i}'|^2\over BC}
-{|k_iA_j-k_jA_i|^2\over C^2}+{|k_iA_0-\omega A_{i}|^2\over DC}
\right],
\label{s2a}\ee
where we have used a general diagonal metric of the form \eqref{ToroidalMetrics}.
The perturbation equations are
\bea
A_{i}''+\partial_r\left[\log{Z\sqrt{{D\over B}C^{p-3}}}\right]A_{i}'+{B\omega\over D}(\omega A_i-k_iA_0)  -{B \over C}(\vec k^2 A_i-k_i(\vec k\cdot \vec A))&=&0\,,
\label{s3a}\\
A_{0}''+\partial_r\log{ZC^{p-1\over 2}\over \sqrt{D B}}A_{0}'-{B\vec k^2\over C}A_{0}+{B\omega \over C}\vec k\cdot \vec A&=&0\,,
\label{s5a}
\eea
along with the Gauss law
\be
\omega A_0'={D\over C}\vec k\cdot \vec A'\,.
\label{s6a}\ee

From   (\ref{s6a}) we may solve for the longitudinal component of $\vec A$ in terms of $A_0$.
the two independent degrees of freedom are therefore $A_{\perp}$ and $A_0$ and satisfy
\bea
A_{\perp}''+\partial_r\left[\log{Z\sqrt{{D\over B}C^{p-3}}}\right]A_{\perp}'+\left({B\omega^2\over D} -{B \over C}\vec k^2\right) A_{\perp}&=&0
\label{s7a}\,,\\
A_{0}''+\partial_r\log{ZC^{p-1\over 2}\over \sqrt{D B}}A_{0}'+\left({B\omega^2\over D} -{B \over C}\vec k^2\right) A_{0}&=&0
\label{s8a}\,.\eea
We first consider the IR asymptotics using  the extremal chargeless metrics (\ref{9a}),
\be
D=C=r^{4\over (p-1)\d^2}\,, \quad B=1 \,,
\ee
and
\be
 e^{\d \phi}={V_0\d^4\over 2\left({2p\over p-1}-\d^2\right)}~r^{2}\,,\qquad Z=e^{\gamma \phi}=Z_0r^{2\gamma\over \d}\,,\qquad
 Z_0^{\d \over \g}={V_0\d^4\over 2\left({2p\over p-1}-\d^2\right)},
\label{s9a}\ee
with the boundary at $r=\infty$ and the IR at $r\to 0$.
In the IR, we obtain
\be
A_{\perp}''+\left({2\gamma\over \delta}+{2(p-3)\over (p-1)\d^2}\right){1\over r}A'_{\perp}=-m^2 r^{-\frac{4}{(p-1)\d^2}} A_{\perp}\,,\qquad m^2=\omega^2-\vec k^2\,.
\label{s10a}\ee
We now define $\frac{dz}{dr} = r^{\frac{-2}{(p-1)\d^2}}$ and $A_{\perp}=\psi ~z^{-\beta},$ with $\beta=\left( {\gamma\over \delta}+{(p-3)\over (p-1)\d^2}\right) \left( \frac{(p-1)\d^2}{(p-1)\d^2 -2} \right)$ to obtain a Schr\"odinger problem
\be
-\psi''+\frac{\beta (\beta -1)}{z^2}\psi=m^2\psi\,,
\label{s11a}\ee
with the potential diverging to $+\infty$ in the far IR because $z \to 0$ as $r\to 0$ when $(p-1)\d^2 >2$ or with the
potential vanishing to $0$ in the far IR because $z \to \infty$ as $r\to 0$ when $(p-1)\d^2 <2$.

When $(p-1)\d^2 >2$, the two independent solutions are behaving near $z=0$ as
\be
\psi_{+}\sim z^{\beta}\,, \qquad
\psi_{-}\sim z^{-\beta+1}.
\label{s12a}\ee
If $\beta>{3\over 2}\,$, one of the solutions, $\psi_+$, is normalizable
in the IR and the other is not, while for $\beta<-{1\over 2}\,$,
$\psi_-$ is normalizable and $\psi_+$ is non-normalizable in the IR.
For $-{1 \over 2}<\beta<{3\over 2}\,$, both solutions are normalizable and this is holographically unacceptable.\footnote{The relevant quantum mechanical problem in treated by Landau and Lifshitz, and a regularization is presented that chooses one of the two solutions when the regulator is removed. However the end result seems to depend on the way to regularize, and is equivalent to choosing a boundary condition at the singularity.}
We deduce therefore that
\be
\left( {\gamma\over \delta}+{(p-3)\over (p-1)\d^2}\right) \left( \frac{(p-1)\d^2}{(p-1)\d^2 -2} \right)>{3\over 2}~~~~{\rm or}~~~~ \left( {\gamma\over \delta}+{(p-3)\over (p-1)\d^2}\right) \left( \frac{(p-1)\d^2}{(p-1)\d^2 -2} \right)<-{1\over 2}\,.
\label{s13a}\ee

In the UV, we assume the solution to become asymptotically  AdS with $D=B=C={\ell^2\over r^2}+\cdots$, $r\to 0$ and $Z\simeq 1+\cdots$.
The subleading terms marked with dots in the previous relations are associated to the scalar perturbation of UV scaling dimension $\D$ and are analyzed carefully for $p=3$ (where they are important) in appendix \ref{ap-uv}.

We can calculate therefore the UV asymptotics  of the Schr\"odinger potential first by considering the leading contributions due to the UV fixed point
\be
V_{\perp}={(p-1)(p-3)\over 4r^2}+\cdots
\ee
where the coordinate $r$ vanishes at the UV boundary.
We observe that the potential diverges $V_{\perp}\to\infty$ at the boundary
when $p>3$.
For $p=3$ we must calculate the subleading contributions as
\be
V_{\perp,p=3}=-{k\over 2}{\D(2\D-1)}r^{2\D-2}+\cdots
\label{s12}\ee
where the parameter $k$ is defined from the UV asymptotics of the gauge function $Z$ as described in (\ref{e1}).

Unitarity implies that $2\D-1\geq 0$.
When $\D<1$, the potential diverges in the UV.
 Regularity then implies that $k<0$, and the potential asymptotes to $V_{\perp,p=3}\to +\infty$.
If the perturbation is softer $\D\geq 1$ the potential vanishes in the UV.

From the above we conclude that,
\begin{itemize}

\item Condition (\ref{s13a}) must be satisfied in order for the spectrum to be holographically acceptable. In this range,  $\beta \left(\beta -1\right)>0$ always and the the effective potential diverges to $+\infty$ in the IR.

\item
For $p>3$
the potential is diverging to $+\infty$ on both the UV and the IR. The spectrum is therefore discrete and gapped. This describes an insulator
(although not a conventional one, as there is no charge density).

\item When p=3, and the UV dimension of the scalar $\D>1$ then the potential vanishes in the UV and the spectrum is continuous. This resembles a conductor.

\item When p=3, and the UV dimension of the scalar $\D<1$ then the potential diverges both in the UV and the IR and the spectrum is discrete and gapped. This resembles to an insulator.

\end{itemize}

When $(p-1)\d^2 <2$, the two independent solutions are behaving near $z=\infty$ as
\be
\psi_{+}\sim z^{\beta}\,, \qquad
\psi_{-}\sim z^{-\beta+1}.
\label{s12aa}\ee
If $\beta>{3\over 2}\,$, one of the solutions, $\psi_-$, is normalizable
in the IR and the other is not, while for $\beta<-{1\over 2}\,$,
$\psi_+$ is normalizable and $\psi_+$ is non-normalizable in the IR.
For $-{1 \over 2}<\beta<{3\over 2}\,$, both solutions are non-normalizable and this is holographically unacceptable.
Thus the same condition (\ref{s13a}) still holds for $(p-1)\d^2 <2$. In this range,
$\beta \left(\beta -1\right)>0$ always and the the effective potential vanishes to $0$ in the IR.
Thus the spectrum is continuous for $p \geq 3$. This resembles a conductor.

\begin{figure}[!ht]
\begin{center}
\begin{tabular}{cc}
	 \includegraphics[width=0.45\textwidth]{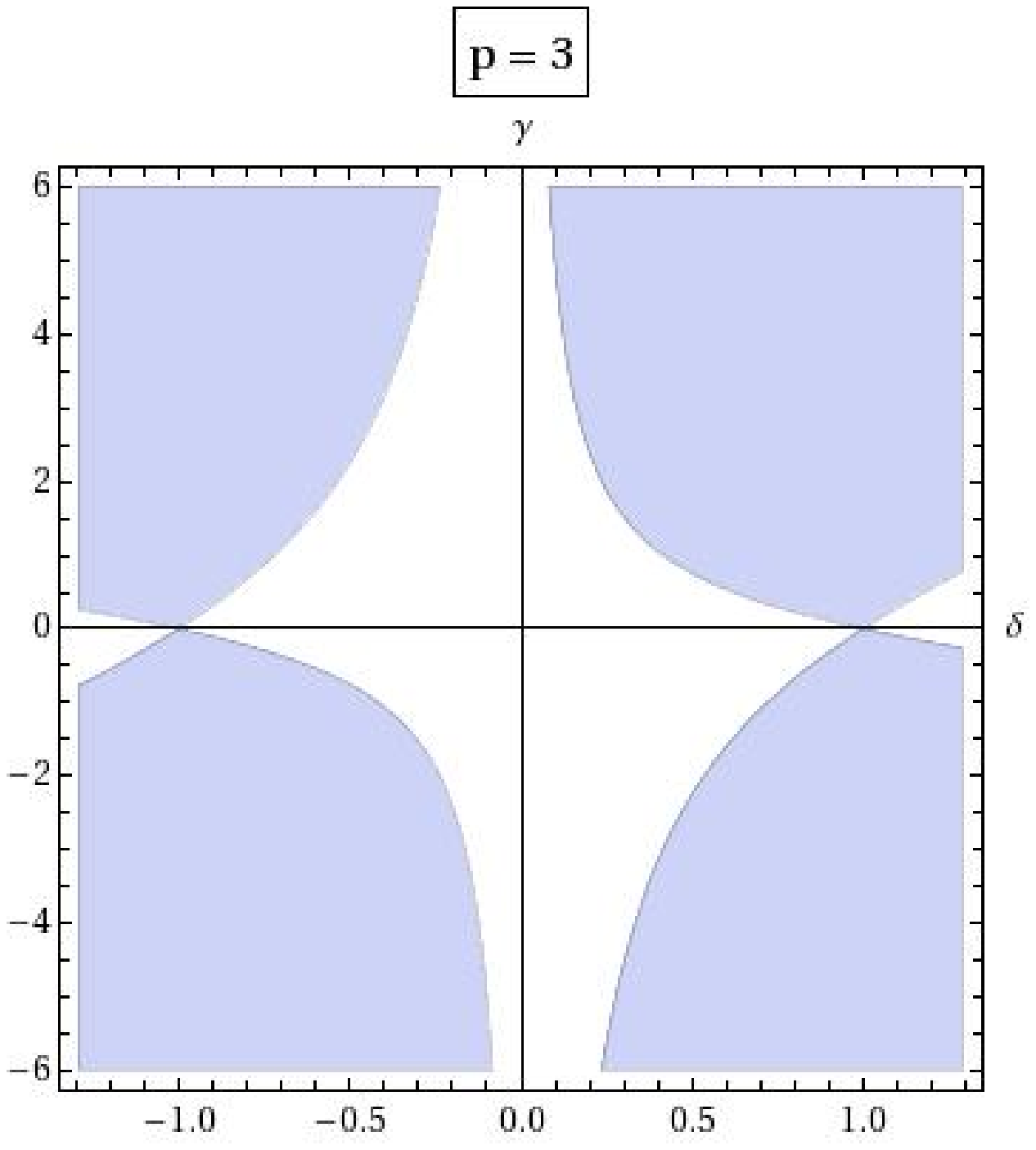}&	
	 \includegraphics[width=0.45\textwidth]{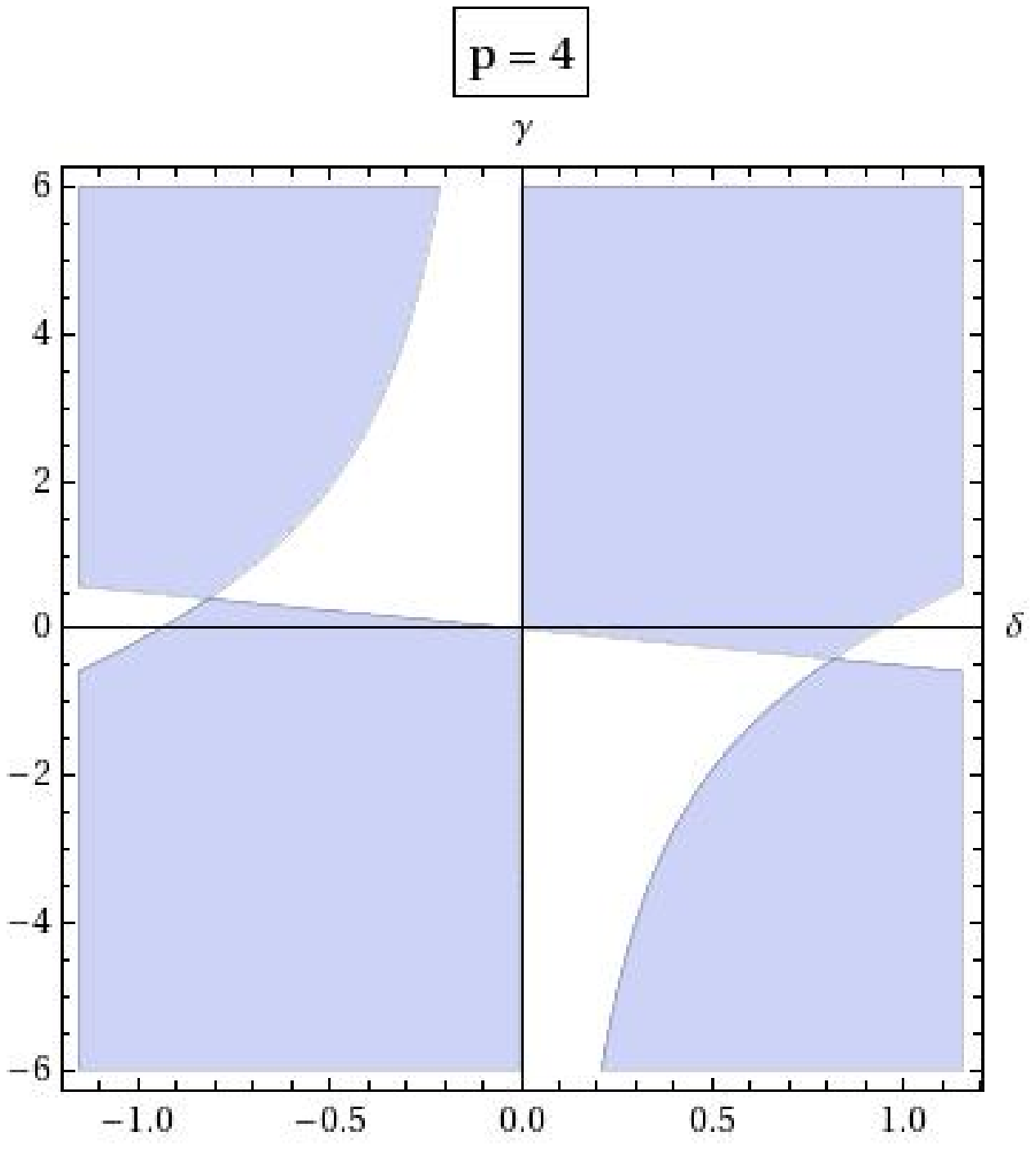}
 \end{tabular}
\caption{The region allowed on the $\gamma-\delta$ plane from the spin-two and spin-one spectra is displayed in blue for p=3 and p=4 dimensions.}
\label{gd}
\end{center}
\end{figure}

The spectrum of $A_0$ is similar.
The constraints above combined with the constraint on $\d$ derived in the previous section are displayed in \Figref{gd}.

\subsection{The phase diagram at finite temperature}

At finite temperature, the boundary conditions at infinity require (Euclidean) time to be a circle  with temperature $\beta$.
There are two solutions  to (\ref{3}), (\ref{4}) that are relevant. The first is the background solution with time compactified \eqref{9a}, for which the temperature is arbitrary.
This is the thermal gas solution. The other is the (small) black hole solution in domain-wall coordinates as in (\ref{ToroidalMetrics2})
 \be
e^A=r^{2\over (p-1)\d^2}\,,\qquad e^{\delta\phi}={V_0\d^4\over 2\left({2p\over p-1}-\d^2\right)}~r^{2}\,,\qquad f=1-\left({r_0\over r}\right)^{{2 p\over (p-1)\d^2}-1}\,,
\label{9}\ee
with temperature
\be
4\pi T=\left({2p\over (p-1)\d^2}-1\right)r_0^{{2\over (p-1)\d^2}-1}\,,
\label{10}\ee
For $|\d|<\sqrt{2\over p-1}\,$ the temperature vanishes at extremality $r_0\to 0$.
In the intermediate case, $|\d|=\sqrt{2\over p-1}\,$, the temperature is constant and independent of $r_0$. Finally in the case  $|\d|>\sqrt{2\over p-1}\,$ the black hole has infinite temperature at extremality.

Such solutions provide the IR asymptotics of solutions that are asymptotically AdS in the UV. For $p=4$ the complete phase diagram was drawn in \cite{gkmn2} for all $\d$. The situation for other dimensions is qualitatively similar and we describe it below.

\begin{itemize}

\item In the first case, $|\d|<\sqrt{2\over p-1}\,$, there is a single branch of asymptotically-AdS black holes qualitatively similar to (\ref{9}).
    Once $T>0$ they have lower free energy compared to the thermal gas solution. {Their free energy scales with temperature as
    $$W \sim T^{\frac{6 - (p-1) \delta^2}{2-(p-1)\delta^2}}\,,$$
    and hence the system generically shows continuous phase transitions to the dilatonic background at zero temperature. For the case $p=3$, which is interesting from the condensed matter point of view, there is a $n^\textrm{th}$-order phase transition for the range
    $$\frac{n-4}{n-2} <\delta ^2 < \frac{n-3}{n-1}\,,\quad n=4,5,6,\dots$$
    In particular, the transitions are fourth-order or higher.
    }

\item In the marginal case $|\d|=\sqrt{2\over p-1}\,$, there is again a single branch of asymptotically-AdS black holes that are qualitatively similar to (\ref{9}). Now however they have a minimum temperature
    given by (\ref{10}): $T>T_{\rm min}={p\over 4\pi}$.
    For $T<T_{\rm min}$ only the thermal gas solution exists. Once $T>T_{\rm min}$ the black hole solution has lower free energy compared to the thermal gas solution and therefore dominates. The transition at $T=T_{\rm min}$ is continuous. Its order depends crucially on subleading terms in the potential, \cite{gursoy1}, \cite{gursoy2}.

\item In the case $|\d|>\sqrt{2\over p-1}\,$, there are at least  two branches of   asymptotically-AdS black holes. In the generic case there are
exactly two branches. One branch corresponds to ``small" black holes  that are qualitatively similar to (\ref{9}).
 They are the only ones that appear as solutions to the IR potential.
 The large black hole branch appears because of the AdS asymptotics and is not visible when the IR potential is used. The large black-hole branch  merges with the small black hole branch at some critical value of the horizon.  There is also a  minimum temperature $T_{\rm min}$. At $T=T_{\rm min}$ there is a single black hole solution. At $T>T_{\rm min}$ there are two black hole solutions. The value of $T_{min}$ cannot be estimated from IR quantities alone but contains information from the full potential.

    For $T<T_{\rm min}$ only the thermal gas solution exists and it dominates the canonical ensemble. Even above $T_{\rm min}$ the thermal gas solution is dominant up to a temperature $T_c>T_{\rm min}$. At $T_c$ there is a first order phase transition, to the large black hole phase that dominates for all higher temperatures.

\end{itemize}

\begin{figure}
 \begin{center}
 \leavevmode \epsfxsize=12cm \epsffile{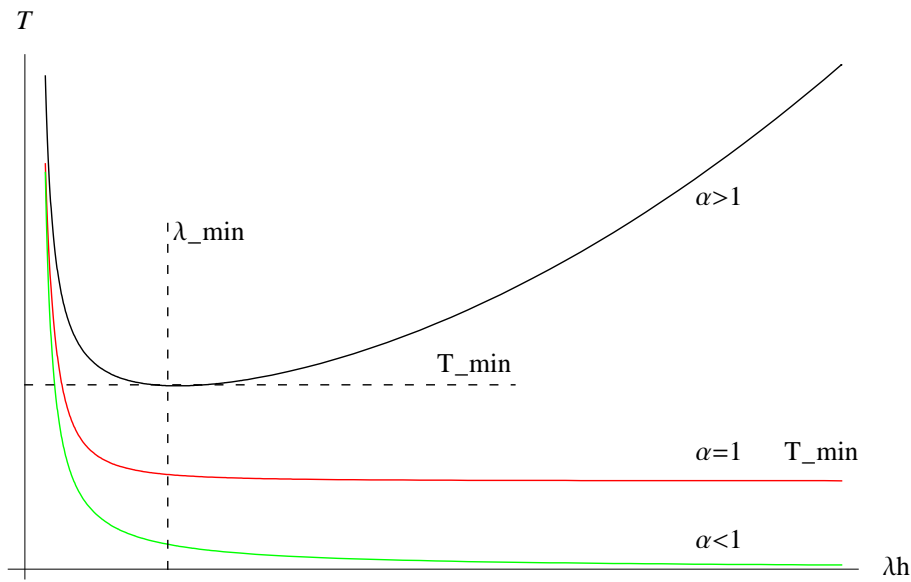}
 \end{center}
 \caption[]{Temperature as a function of the value of $\l$ at the horizon, $\l_h$ for
 the three different cases discussed in the text. Black holes exist only above $T_{min}$
 whose precise value depend on the particular zero-$T$ geometry.}
 \label{f1}\end{figure}

In \Figref{f1} the temperature of black holes as a function of the value of the scalar at the horizon is plotted. The three different behaviors described above are evident in the plot.

\begin{figure}
 \begin{center}
 \leavevmode \epsfxsize=12cm \epsffile{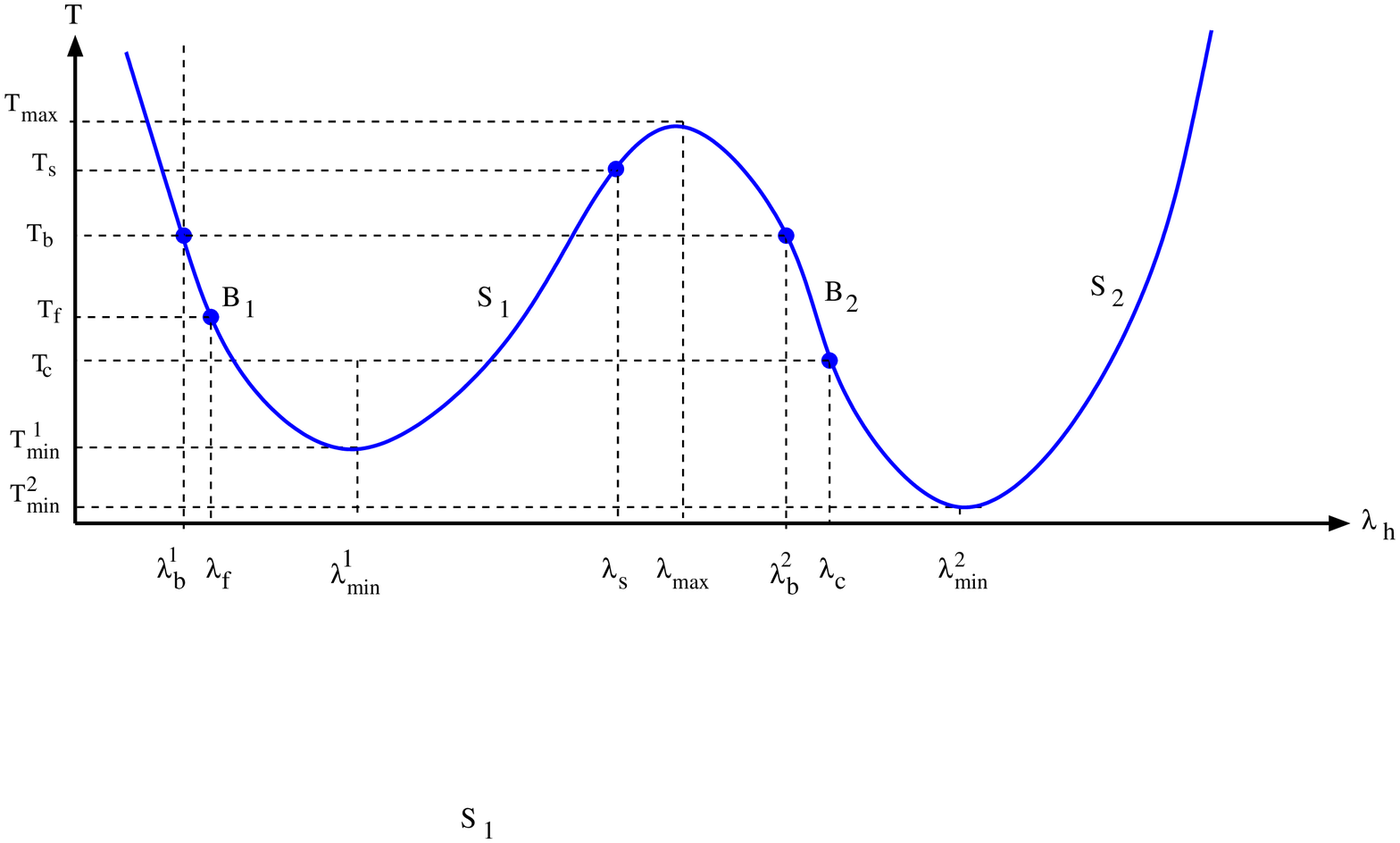}
 \end{center}
 \caption[]{$T$ as a function of $\l_h$ in a case with more than two black hole solutions.}
 \label{f2}\end{figure}

\begin{figure}
 \begin{center}
 \leavevmode \epsfxsize=12cm \epsffile{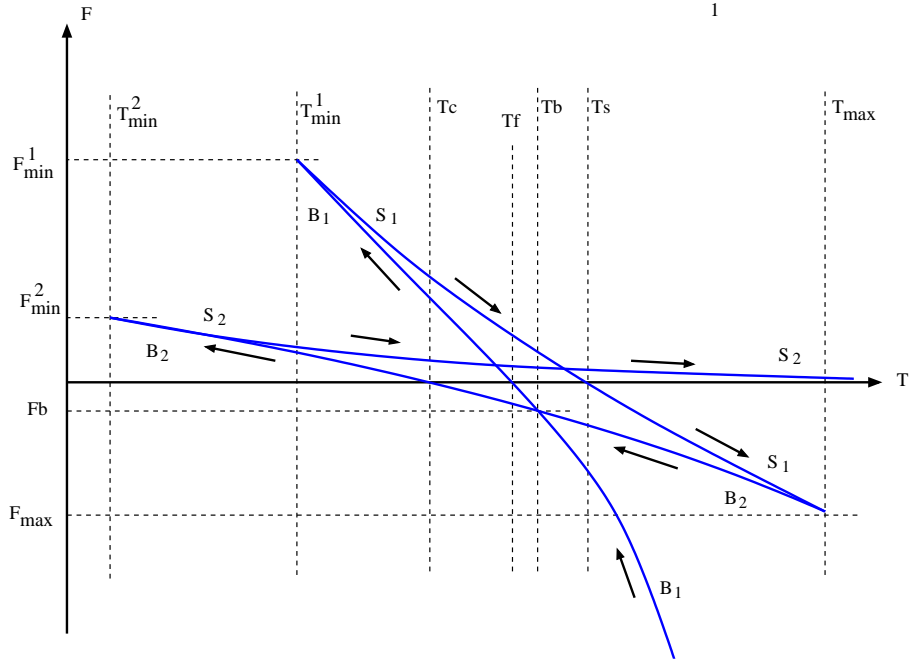}
 \end{center}
 \caption[]{The free energy  $F(T)$ for the case of \Figref{f2} with multiple extrema. $S_1$ and $S_2$ denote small BHs whereas $B_1$ and $B_2$ denote big BHs.
 The arrows represent direction of increasing $\l_h$. There are two first order phase transitions in this case.}
 \label{f3}\end{figure}

In \cite{gkmn2} it was argued that in non-generic cases more than two black hole branches can appear. A case is plotted in \Figref{f2} where there are two local minima in the relation between black hole temperature and horizon position. In such cases multiple first order phase transitions can appear, as indicated in \Figref{f3} where the free energy is plotted as a function of temperature. A concrete example of a potential leading to such a case was given recently in \cite{kajantie}.

\section{The dynamics at finite charge density}

At finite (charge) density the gauge field is non-trivial. We will study solutions where the backreaction of the finite density on the geometry is fully taken into account. In this section we will assume that the gauge field is emerging from the closed string (adjoint) sector and we will therefore use a Maxwell-type of action. This however will also be relevant for fundamental-type conserved charges in the limit in which the DBI action linearizes (weak self-interaction of fundamental charges).

We consider the Einstein-Maxwell Dilaton action (\ref{ActionLiouville}).
Since we are interested in the IR behavior we will take the scalar potential and gauge coupling constant to have their IR asymptotics that we will parametrize as
\be
Z(\phi)=e^{\ga\phi}\,,\qquad V(\phi)=-2\Lambda e^{-\da\phi}\,.
\label{al1}\ee
We have used the freedom to rescale the gauge fields in order to set the overall constant coefficient in $Z$ to one. This however implies that there is an overall unknown (dimensionful)  constant multiplying quantities like the conductivity. This constant can be fixed once the EHT is embedded in a string theory.

This class of EHTs are parameterized by the two exponents $\gamma,\d$ that we take to be real numbers. There is an extra parameter, $\Lambda$ , which sets the length scale of the solutions. It will always appear combined with the asymptotic value of the scalar field $\phi_0$, in the following linear combination, $\Lambda e^{-\d \phi_0}$. Of course, once the asymptotic UV asymptotics are corrected to AdS, this scale is an independent (IR) parameter.

Because the scalar potential in (\ref{al1}) has a runaway minimum when the value of the scalar field is infinite, it will not admit solutions with regular Anti-de Sitter asymptotics.
We will require however from our solutions that they are correctable to asymptotically AdS$_{p+1}$ solutions. This will be done by requiring that when evaluated in the solution,  $V(\phi)\to 0$ in the UV boundary. In this way, simply by adding a constant to the exponential potential the solution will be transformed to an asymptotically AdS$_{p+1}$. More complicated completions are also possible.
For example we can replace the potential by $V_{RN}=-\Lambda \cosh(\da\phi)$ with a similar effect, although in this case the dimension of the perturbing scalar operator is different. Moreover, in this case the RN solution in $p+1$ dimensions is also allowed, as was explored recently in \cite{cadoni}.

We will also be interested in toroidal metrics for the transverse $p-1$ dimensions. We will keep $p$ arbitrary in parts of this paper while we will take $p=3$ when discussing the $\gamma=\delta$ and $\gamma\delta=1$ solutions, as this seems to be a value mostly relevant for applications in CM.

The equations of motion stemming from (\ref{ActionLiouville}) are the following
\bsea
	R_{\mu\nu}-{R+V\over 2}g_{\mu\nu}&=& \half\partial_\mu\phi\partial_\nu\phi-\frac{g_{\mu\nu}}4\le(\partial\phi\ri)^2+{Z(\phi)\over 2}\le[F^{\;\rho}_\mu\ F_{\nu\rho}-\frac{g_{\mu\nu}}4 F^2\ri], \slabel{Einstein_eq} \\
	0&=&\nabla_\mu\le(\sqrt{-g}Z(\phi)F^{\mu\nu}\ri), \slabel{Maxwell_eq} \\
	\square{\phi}&=&\frac{Z'(\phi)}4F^2+V'(\phi)=\frac{\ga}4 e^{\ga \phi}F^2+2\da\Lambda e^{-\da\phi}\,. \slabel{Dilaton_eq} 	
	\label{EOMLiouville}
\esea
It is then easy to deduce from \eqref{EOMLiouville} the value of the Ricci scalar \emph{on-shell} :
\be
	R = \half\le(\partial\phi\ri)^2-{p+1\over p-1}V+{(p-3)Z\over 4(p-1)}F^2\,.
	\label{RicciScalarOnShell}
\ee
The relevant Einstein equations in the general coordinate system \eqref{ToroidalMetrics} and in  the domain-wall coordinate system of \eqref{ToroidalMetrics2} are depicted in Appendix \ref{equo}.

In domain wall coordinates \eqref{ToroidalMetrics2} the equation of motion for the gauge field yields
\be
A_t'={q\over Z(\phi)e^{(p-2)A}}\,,
\label{Einstein1}
\ee
where we have adopted Minkowski signature and primes stand for derivatives with respect to the radial coordinate.
There is a conserved Noether charge, \cite{gubser}, not to be confused with the U(1) charge of the black hole solution
\be
{\cal Q}= e^{pA}g'e^g-qA_t\,,\qquad {d{\cal Q}\over dr}=0\,.
\label{noe}\ee
It is conserved in view of equation  (\ref{Einstein2}).
Evaluating it at the horizon (where $A_t=0$) we obtain that ${\cal Q}\sim TS$. Therefore when it vanishes, it implies extremality.

The finite charge density solutions we will examine have been obtained for the first time in \cite{char}, see \eqref{Sol1},  or are generalisations of solutions found in previous literature \cite{Cai:1997ii,taylor,kachru}, see \eqref{Sol2} and \eqref{Sol3}.

For the sake of completeness and to give an overview of previous literature in four dimensions, let us mention that:
\begin{itemize}
 \item Solutions for zero potential have been obtained by \cite{Gibbons:1987ps} for general gauge coupling and by \cite{gar} for the string case $\ga=1$, while \cite{Monni:1995vu} shows solutions for particular values of $\ga$ with a hyperbolic cosine gauge function. Near-extremal black holes for these theories are studied in \cite{Clement:2002mb,Clement:2004yr,Clement:2004ii}, and in particular are shown to be non-asymptotically flat or AdS.
 \item In the case of Liouville potential, the general uncharged planar solutions were reported in \cite{Charmousis:2001nq} ; charged spherical black holes for the generic action \eqref{ActionLiouville} were obtained in \cite{Chan:1995fr}, while topological solutions were found in \cite{Cai:1997ii} for hyperbolic and toroidal black holes. Wiltshire \emph{et al.} have studied and classified their asymptotics in \cite{Poletti:1994ff,Poletti:1994ww}. All the above dilatonic black hole solutions with a Liouville potential are included in  \cite{char}.
\item Cases of black holes with AdS asymptotics are known analytically for particular potentials in hyperbolic sines or cosines and were studied in \cite{Gao:2004tu,Mignemi:2009ui,cadoni,Sheykhi:2009pf}.
\end{itemize}
The definitions of the various thermodynamic ensembles considered are given in appendix \ref{thermo}.

\section{The AC conductivity}\label{sec:ACconductivity}

In this section we will describe the calculation of the AC conductivity in the spatially homogeneous case both for the Maxwell theory and the DBI case, in view of applications for solutions with string charge interactions.

 \subsection{ The AC conductivity in the Maxwell case}

To compute the appropriate correlator, as we are working in the fully back-reacting case, it will be necessary to also perturb appropriate off-diagonal components for the metric.
We will do our calculations for the general coordinate system where the metric is given in (\ref{ToroidalMetrics}) and $\phi$ and $A_t$ are functions of $r$ only.
$A_t$ is given by
\be
 A_t'(r) = \frac{q\sqrt{DB} }{Z{C^{\frac{p-1}{2}}}}\,.
 \label{27} \ee
We consider the following (small) perturbation to this background\footnote{We can also perturb $g_{ri}$ instead of $g_{ti}$ with similar results.}
\be
A_i=a_i(r)e^{i\omega t}\,,\qquad g_{ti}=z_i(r)e^{i\omega t}\,.
 \label{28}\ee
The first variation of the $rx$ Einstein equation gives
\be
g_{ti}'-{C'\over C}g_{ti} = - Z A_i A_t'\,,
 \label{29}\ee
 while the gauge field fluctuation equation gives
 \be
\partial_r\left(ZC^{p-3\over 2}\sqrt{D\over B} A_i'\right) +
 ZC^{p-3\over 2}\sqrt{B\over D} \omega^2 A_i{+}{q\over C}\left(g_{ti}'-{C'\over C}g_{ti}\right)=0\,.
 \label{30}\ee
 Substituting (\ref{29}) into (\ref{30}) we finally obtain
 a second order equation for $a_i$
\be
 \partial_r\left(ZC^{p-3\over 2}\sqrt{D\over B} a_i'\right) +{ZC^{p-3\over 2}\left[
 \sqrt{B\over D} \omega^2 - {q^2\sqrt{DB}\over {Z C^{p-1}}}\right]}a_i=0\,,
 \label{30a}\ee
where as usual primes stand for derivatives with respect to $r$.

This equation can be eventually mapped to a Schr\"odinger-like equation
by changing the radial coordinate to $z$ and redefining the radial wave-function as
\be
\frac{dz}{dr} = \sqrt{\frac{B(r)}{D(r)}}\,,\qquad %
a_i = {\Psi\over \sqrt{\bar Z}}\,,\qquad \bar Z=ZC^{p-3\over 2}\,,
 \label{31}\ee
 to obtain
\be
-{d^2 \Psi\over dz^2}+V\Psi=\omega^2\Psi\,,\qquad V = {\frac{q^2D}{Z C^{p-1}}} + {1\over 4}{(\partial_z\bar Z)^2\over \bar Z^2} + {1\over 2}\partial_z\left({\partial_z \bar Z\over \bar Z}\ri),
 \label{32}\ee
In many cases, for plotting the potential it is useful to express it in the original $r$ coordinate, reading
\be\label{eq:VSofr}
V(r) = \frac{q^2 D}{Z C^{p-1}} - \frac{D}{B}\left[ \frac{1}{4} ( \partial_r \log\bar Z )^2 - \frac{1}{2} \partial_r^2 \log \bar Z - \frac{1}{2} (\partial_r \log\bar Z) \partial_r \log \left( \bar Z \sqrt{\frac{D}{B}} \right)\right]\,.
\ee

Near an AdS boundary, $z$ is the standard radial coordinate with $z=0$ being the boundary, $D\to {\ell^2\over z^2}$, $B\to {\ell^2\over z^2}$, $C\to {\ell^2\over z^2}$, and $Z$ asymptotes to a constant $Z\to Z_b$.
Then, near the AdS boundary
\be
V(z)\simeq {q^2\over  Z_b}\left({z\over \ell}\right)^{2(p-2)}+{(p-1)(p-3)\over 4z^2}+\cdots
\label{34} \ee

 For $p>3$ the potential diverges quadratically at the boundary and if there is a similar behavior in the IR the spectrum will be gapped and discrete.
For $p=3$ however we must calculate the subleading contributions as the potential  above vanishes in the UV. These come from the scalar perturbation and are
\be
V_{\perp,p=3}=-{k\over 2}{\D(2\D-1)}r^{2\D-2}+\cdots
\label{s12c}\ee
where the parameter $k$ is defined from the UV asymptotics of the gauge function $Z$ as described in (\ref{e1}).

Unitarity implies that $2\D-1\geq 0$.
When $\D<1$, the potential diverges in the UV.
 Regularity then implies that $k<0$, and the potential asymptotes to $V_{\perp,p=3}\to +\infty$.
If the perturbation is softer $\D\geq 1$ the potential vanishes in the UV.
In that case the spectrum is gapless and therefore the system is a  conductor.
We conclude that strongly coupled doped systems in p=3 are generically conducting.

If the UV fixed point is a Lifshitz point instead with Lifshitz exponent $w>1$, then the UV asymptotics of the schr\"odinger potential are modified to
 \be
V(z)\simeq {q^2\over  Z_b}{(|w-2|z)^{2(p-2)\over (2-w)}\over \ell^{2(p-2)}}+{(p-2w+1)(p-3)\over 4z^2(w-2)^2}+\cdots
\label{34a} \ee
The relation above is valid when $w\not=2$. It implies that UV Lifshitz asymptotics with $w=3/2$ make the second term vanish for $p=2$, and $w=5/2$ make it vanish for $p=4$. Therefore Lifshitz behavior affects the presence of a gap in conductivity in the IR.

Finally when $w=2$, the schr\"odinger coordinate is $z=log r$, and the second term in the potential (\ref{34a}) becomes a constant proportional to $(p-3)^2$ and therefore vanishes at $p=3$. The first term as usual asymptotes to zero at the boundary.

Near a regular horizon at $r=r_0$ we have the following expansions for the various functions
\be
D=c_D(r-r_0)+{\cal O}[(r-r_0)^2]\,,\qquad B={c_B\over (r-r_0)}+{\rm regular}\,,
\label{35} \ee
while $C$ and $\bar Z',\bar Z$ asymptote to constants.
{}From (\ref{31}) the horizon is at $z=-\infty$ with
$r-r_0\simeq e^{\sqrt{c_D\over c_B}~z}$, and the potential vanishes exponentially  there,
\be
V(z)\simeq \frac{q^2c_D}{\bar Z_0C^{p-1}_0}e^{\sqrt{c_A\over c_B}~z}+\cdots
\label{36} \ee

Finally near an extremal horizon with finite entropy, we have the following asymptotics
\be
D=c_D(r-r_0)^2+\cdots\,,\qquad B={c_B\over (r-r_0)^2}+{\rm regular}\,,
\label{37} \ee
while $C$ and $\bar Z', \bar Z$ asymptote to constants. Then $z\simeq \sqrt{c_B\over c_D}{1\over r_0-r}$ and the potential vanishes as
\be
V(z)\simeq {q^2c_B\over  Z_0C_0^{p-1}}{1\over z^2}+\cdots
\label{38} \ee

In the case where $Z$ is a constant a general formula for the AC conductivity was given in \cite{roberts},
\be
\sigma(\omega)={1-{\cal R}(\omega)\over 1+{\cal R}(\omega)}\,,
\label{39} \ee
where ${\cal R}(\omega)$ is the (energy-dependent) reflection coefficient of the Schr\"odinger problem above. In the case of a non-trivial gauge coupling constant $Z$, the associated conductivity was derived in \cite{kachru} with the result
\be
\sigma(\omega)={1-{\cal R}(\omega)\over 1+{\cal R}(\omega)}-{i\over 2\omega}{\dot Z\over Z}\Big |_b\,,
\label{40} \ee
where the dot on the extra factor implies derivative with respect to $z$ and this is  evaluated at the AdS boundary.
For regular boundary behavior of $Z$, this factor vanishes, and the  conductivity is given by (\ref{39}).

Finally, in \cite{roberts,kachru}, the zero-temperature asymptotics of the AC conductivity were calculated for a Schr\"odinger potential that asymptotes to
\be
V(z)={\nu^2-{1\over 4}\over z^2}\,,
\label{41} \ee
near the extremal horizon, with the result
\be
\sigma(\omega)\sim \omega^{2|\nu|-1}\,.
\label{42} \ee

We may now use equations above to evaluate the extremal conductivity in the chargeless background solution  (\ref{9}).The Schr\"odinger coordinate in the IR is
\be
z={r^{1-{2\over (p-1)\d^2} }\over {1-{2\over (p-1)\d^2}}}
\ee
and the effective potential (\ref{eq:VSofr}) in the IR becomes
\be
V_{eff}={c\over z^2}\sp c=-{(p-1)(\d^2-1-\d\g)((p-1)\g\d+p-3) \over ((p-1)\d^2-2)^2}
\ee
We observe that $c\geq -{1\over 4}$ always.
When $\d^2 <{2\over p-1}$, $z\to\infty$ in the IR and the potential vanishes there. This signals a continuous spectrum and we obtain the scaling of the AC conductivity as
\be
\s\sim \omega^n\sp n=\sqrt{4c+1}-1
\ee
There is a whole range of parameters where the exponent  becomes negative.
It becomes $n=-{2\over 3}$ when
\be
\g=-{1-\d^2\over 3\d}~~~~{\rm or}~~~~~ \g=-{2(1-\d^2)\over 3\d}
\ee

In the opposite case, $\d^2 >{2\over p-1}$, $z\to 0$ in the IR.
In this case the potential diverges. When it diverges to positive infinity it indicates a discrete spectrum. In $p=3$ an extra conditions is needed, $\D<1$.

As we will see in the rest of the paper, neither the extremal AC conductivity nor the DC conductivity are continuous as the charge density $q\to 0$.

\subsection{The AC conductivity in the DBI case}

We will now consider the DBI action in the Einstein frame and turn on
 a $A_x(r,t)$ as well as a $g_{tx}(r,t)$ perturbation.

The relevant effective action expanded to quadratic order in the perturbation is
\bea
S_{DBI}&=&-\int d^{p+1}x \left[e^{2k\phi}Z(\phi)C^{p-1\over 2}\sqrt{DB-e^{-2k\phi}F_{rt}^2}+\right.\nn\\
	&&\left. + \frac{1}{2} Z C^{(p-3)\over 2}{DF_{rx}^2-BF_{tx}^2+2g_{tx}F_{rx}F_{rt}+Be^{2k\phi}g_{tx}^2\over \sqrt{DB-e^{-2k\phi}F_{rt}^2}}+\cdots\right].
	\label{65}
\eea
Considering $A_x(r,t)=A_x(r)e^{i\omega t}$ we obtain the fluctuation equation

\bea
-\partial_r\left({q\over C}g_{tx}\right)&=&\partial_r\left(e^{-2k\phi}\sqrt{D\over BC^2}\sqrt{e^{2k\phi} q^2+e^{4k\phi}Z^2C^{p-1}} A_x'\right)+\nn\\
&&+\omega^2 e^{-{2k}\phi}\sqrt{B\over DC^2}\sqrt{ q^2e^{2k\phi}+e^{4k\phi}Z^2C^{p-1}} A_x\,,
\label{67}
\eea
while from Einstein's equations we obtain
\be
\partial_r{g_{tx}\over C}+q{\sqrt{DB}\over C^{p-1}} A_x=0\,.
\label{106}\ee
Combining the two we obtain for the gauge field the equation

\bea
0&=&\partial_r\left(e^{-{2k}\phi}\sqrt{D\over BC^2}\sqrt{e^{2k\phi} q^2+e^{4k\phi}Z^2C^{p-1}} A_x'\right)+\nn\\
	&&+\left[\omega^2 \sqrt{B\over DC^2}{\sqrt{ q^2e^{2k\phi}+e^{4k\phi}Z^2C^{p-1}}\over e^{{2k}\phi}}-{q^2\sqrt{DB}\over C^{p-1}}\right]A_x\,	.
\label{68}
\eea
In the strong coupling limit $ q\gg e^{{k}\phi}ZC^{p-1\over 2}$ the fluctuation equation becomes
\be
\partial_r\left(e^{-{k}\phi}\sqrt{D\over BC^2} A_x'\right)+\left[\omega^2 e^{-{k}\phi} \sqrt{B\over DC^2}-{q^2\sqrt{DB}\over C^{p-1}}\right] A_x=0\,.
\label{69}\ee
The Schr\"odinger problem relevant for conductivity calculations that emerges from (\ref{69}) then is
\be
{dz\over dr}=\sqrt{B\over D} \,,\qquad A_x=\sqrt{C}e^{{k\over 2}\phi}\psi\,,\qquad  -\partial_z^2\psi+V\psi=\omega^2\psi\,,
\label{177}\ee
with the Schr\"odinger potential
\be
 V={3\over 4}{C'^2\over C^2}+{k^2\over 4}\phi'^2+{k\over 2}{C'\over C}\phi'-{C''\over 2C}-{k\over 2}\phi''+{q^2 D e^{k\phi}\over C^{p-2}}
\label{178}\,.
\ee
Near an AdS boundary  the DBI action linearizes
and the fluctuation equation becomes
\be
\partial_r\left(\sqrt{D\over B}ZC^{p-3\over 2} A_x'\right)+\left[\omega^2  \sqrt{B\over D}ZC^{p-3\over 2} -  {q^2\sqrt{DB}\over C^{p-1}}\right]A_x=0\,,
\label{70}\ee
matching the equation derived in (\ref{30a}).

\section{The DC conductivity}

For the DC conductivity we will explore several possibilities. In the Maxwell case we will do a drag calculation of the DC conductivity for massive carriers. In the DBI case without  backreaction  the calculation was first done in \cite{karch}. We will also amend this calculation below and apply it to the chargeless solutions of section  \ref{zero}.

\subsection{The DC conductivity in the probe DBI case\label{DCDBI}}

We consider again the DBI action  (\ref{11a}) where we will drop the second piece as it will not be relevant for the conductivity calculation.
Keeping a general coordinate system, working    in $p+1$ dimensions  and considering the ansatz for the gauge field $A_t(r)$, $A_x(r,t)$ we obtain the action
\bea
S_{DBI}&=&-\int d^{p+1}x e^{2k\phi}Z\sqrt{-\det(g+e^{-k\phi}F)}\nn\\
			 &=&-\int d^4x e^{2k\phi}Z~C^{p-2\over 2}\sqrt{DBC-e^{-2k\phi}\left(CA_t'^2+B\dot A_x^2-DA_x'^2\right)}\,,
\label{71}
\eea
and the gauge field equations of motion
\be
{ZC^{p\over 2}A_t'\over \sqrt{DBC-e^{-2k\phi}\left(CA_t'^2+B\dot A_x^2-DA_x'^2\right)}}=q\,,
\label{72}
\ee
\bea
&&\partial_t\left({Z~C^{p-2\over 2}B\dot A_x\over
\sqrt{DBC-e^{-2k\phi}\left(CA_t'^2+B\dot A_x^2-DA_x'^2\right)}}\right)-\nn\\
&&\qquad\qquad-\partial_r\left({Z~C^{p-2\over 2}D A_x'\over
\sqrt{DBC-e^{-2k\phi}\left(CA_t'^2+B\dot A_x^2-DA_x'^2\right)}}\right)=0\,.
\label{73}
\eea
Making the further ansatz $A_x=-Et+a(r)$ we obtain from (\ref{73}) that the first term is identically zero and
\be
{Z~C^{p-2\over 2}D a'\over
\sqrt{DBC-e^{-2k\phi}\left(CA_t'^2+BE^2-Da'^2\right)}}=w\,,
\label{74}\ee
with $w,q$ constants.
We may solve
\be
A_t'=\pm q\sqrt{DB\over C}{\sqrt{DCe^{2k\phi}-E^2}\over \sqrt{(Dq^2-Cw^2)+DC^{p-1}e^{2k\phi}Z^2}}\,,
\label{74a}\ee
\be
a'=\pm w\sqrt{BC\over D}{\sqrt{DCe^{2k\phi}-E^2}\over \sqrt{(Dq^2-Cw^2)+DC^{p-1}e^{2k\phi}Z^2}}\,,
\label{75}\ee
and obtain the on-shell action as
\be
S_{DBI}^{\rm on-shell}=-\int d^4x ~Z^2C\sqrt{DBC}e^{2k\phi}{\sqrt{DCe^{2k\phi}-E^2}\over \sqrt{(Dq^2-Cw^2)+e^{2k\phi}DC^{p-1}Z^2}}\,.
\label{76}\ee

As first observed in \cite{karch} the square roots in numerator and denominator change sign between the horizon and the boundary, and this must happen at the same radial point $r=r_*$,
\be
D(r_*)C(r_*)e^{2k\phi(r_*)}=E^2\,,\qquad C(r_*)w^2-D(r_*)q^2=D(r_*)C(r_*)^{p-1}Z_*^2e^{2k\phi(r_*)}\,,
\label{77}\ee
from where we obtain
\be
w^2=\left({q^2\over C_*^2}+C_*^{p-3}Z_*^2 e^{2k\phi_*}\right)E^2 e^{-2k\phi_*}\,.
\label{78}\ee
Near the AdS boundary, $D=B=C\simeq {\ell^2\over r^2}$ and
\be
A_t=\mu+{q\ell \over (p-2)Z_b}\left({r\over \ell}\right)^{p-2}+\cdots
\,,\qquad
a=a_0+{w\ell \over (p-2)Z_b}\left({r\over \ell}\right)^{p-2} +\cdots
\label{79}\ee
We also identify
\be
\langle J^t\rangle =q\,,\qquad
\langle J^x\rangle =w\,,
\label{80}\ee
so that the (non-linear) conductivity is given by
\be
\sigma={\langle J^x\rangle\over E}={e^{-k\phi_*}\over C_*}\sqrt{q^2+C_*^{p-1}Z_*^2e^{2k\phi_*}}\,.
\label{81}\ee

As first pointed out in \cite{karch} the first term under the square root is proportional to the charge density $q$ and is therefore the contribution to the conductivity from the charge carriers in the system.
 The second term is independent of the charge density and is due to pair creation of charges in the medium. As  pair production is expected to be Boltzmann suppressed at high carrier masses, this term is expected to be dominant for almost massless carriers,
 \be
 \s_{pair}\simeq C_{*}^{p-3\over 2}Z_*
 \label{n81}\ee
For heavier carriers the first term is dominant
\be
\s_{\rm drag} \simeq {q\over C_*e^{k\phi_*}}
\label{ndrag}\ee
and can also be calculated from a drag calculation as we will show in the next section.

For the solution in (\ref{9}), equation (\ref{77}) becomes
\be
r_*^{{8\over (p-1)\d^2}+{4k\over \d}}\left[1-\left({r_0\over r_*}\right)^{{2p\over (p-1)\d^2}-1}\right]=e^{-2k\phi_0}~E^2\,,\qquad e^{\phi_0}={V_0\d^4\over 2\left({2p\over p-1}-\d^2\right)}\,,
\label{82}\ee
and the conductivity is given by
\be
\sigma=e^{-k\phi_0} r_*^{-{2k\over \d}-{4\over (p-1)\d^2}}
\sqrt{{\langle J^t\rangle^2}+e^{2(\gamma+k)\phi_0}r_*^{{4\over \d^2}+{4(\gamma+k)\over \d}}}\,,
\label{83}\ee
where we have taken $Z=e^{\gamma\phi}$. Consistency requires that the electric field is not very strong so that $r_*$ is fully into the IR region.

When the electric field vanishes $r_*\to r_0$ and the conductivity becomes
\be
\sigma=e^{-k\phi_0} r_0^{-{2k\over \d}-{4\over (p-1)\d^2}}
\sqrt{{\langle J^t\rangle^2}+e^{2(\gamma+k)\phi_0}r_0^{{4\over \d^2}+{4(\gamma+k)\over \d}}}\,,
\label{84}\ee
and using
\be
r_0=(\kappa T)^a\,,\qquad \kappa={4\pi\over \left({2p\over (p-1)\d^2}-1\right)}\,,\qquad a={1\over {2\over (p-1)\d^2}-1}\,,
\label{85}\ee
we finally obtain
\be
\sigma=e^{-k\phi_0} (\kappa T)^{2k(p-1)\d+4\over (p-1)\d^2-2}
\sqrt{{\langle J^t\rangle^2}+e^{2(\gamma+k)\phi_0} (\kappa T)^{4(p-1)\left[1+(\gamma+k)\d\right]\over 2-(p-1)\d^2}}\,,
\label{86}\ee
For the resistivity at low density it is the second term in the square root that dominates
\be
\rho={1\over \sigma}\sim T^{{2(p-1)\gamma\d+2(p-3)\over (p-1)\d^2-2}}\,,
\label{87}\ee
while at higher densities it is the first
\be
\rho={1\over \sigma}\sim T^{2k(p-1)\d+4\over 2-(p-1)\d^2}\langle J^t\rangle\,.
\label{88}\ee
This result agrees with the drag-force computation in the next section as was first shown in \cite{karch}.

In the second case, taking into account the bound, $\d^2<{2(p+2)\over 3(p-1)}$ derived in section \ref{spectra} it can be shown that  the resistivity can never be linear.
In the first case we can attain linear resistivity when
\be
\gamma=\gamma_{\rm linear}\equiv {\d^2-{2(p-2)\over (p-1)}\over 2\d}\,.
\label{89}\ee
Note that in this dilute limit the resistivity scaling is independent of the nature of the scalar field, i.e. whether it is dilatonic in nature or not, which was parametrised by the number $k$.

It should be stressed that a DC resistivity formula of the form (\ref{86}) offer the possibility of interpolating between two different powers of the temperature as has been observed in cuprate materials where the resistivity is linear in the strange metal region and quadratic in the overdoped region.
This will happen here if $\delta=-{k\over 4}$.
We will also see in section \ref{dcex} that this can also arise in that case as well.

\subsection{The drag calculation of the DC conductivity}

 Massive fundamental carriers can be thought of as endpoint of strings, as they end on flavor branes. For massive carriers the associated flavor branes are located away from the IR bottom of the bulk space. Their motion can be modeled as the motion of a fundamental string in the semiclassical background described by the dilatonic charged black hole solutions.

We will consider here a massive charged particle moving in the appropriate  background with a world-sheet action given in the Nambu-Goto form
\be
S_{NG}=T_f\int d^2\xi\sqrt{\hat g}+\int d\tau A_{\m}\dot x^{\m}\,,\qquad \hat g_{\a\b}=g_{\m\n}\pa_{\a}x^{\m}\pa_{\b}x^{\n}\,,
\label{43}\ee
where $T_f$ is the fundamental string tension.  The string has two end-points that each couples to the bulk $A_{\mu}$. For the drag calculation we follow one of the endpoints, as the second is very far away. $\xi^{\a}$ are the world-sheet coordinates while $x^{\mu}$ are the space-time coordinates. We have also taken the Chan-Paton charge of the endpoint to be unity.

If the background is invariant under translations we have the following world-sheet Poincar\'e currents
\be
\pi_{\m}^{\a}=\bar\pi^{\a}_{\m}+A_{\m}\eta^{\a\tau}= {T_f}\sqrt{\hat g}\hat g^{\b\a}g_{\nu\m}\pa_{\b}x^{\n}+A_{\m}\eta^{\a\tau}\,,
\label{44}\ee
with conserved momenta
\be
P_{\m}=\int d\s ~\bar \pi^{\tau}_{\m}+A_{\m}={T_f}\int d\s \sqrt{\hat g}\hat g^{\b\tau}g_{\nu\m}\pa_{\b}x^{\n}+A_{\m}\,,
\label{45}\ee
where $\eta^{\a\b}$ is the flat two-dimensional Minkowski metric.

The bulk equations of motion stemming from the effective action for the coordinates for which the background is translational invariant are
\be
\pa_{\a}\bar\pi^{\a}_{\m}=0\,,
\label{46}\ee
while at the open string boundary we obtain
\be
T_f\sqrt{\hat g}\hat g^{\s \b}g_{\m\n}\pa_{\b}x^{\n}+qF_{\m\n}\dot x^{\n}=0\,.
\label{47}\ee

We now consider a space-time metric in a generic coordinate system
\be
ds^2=-g_{tt}(r)dt^2+g_{rr}(r)dr^2+g_{xx}(r)dx^idx^i\,,
\label{48}\ee
while the bulk gauge fields are $ A_{t}(r)$, $A_{x^1}=-Et+h(r)$.
We choose a static gauge with $\sigma=r$ and $\tau=t$ and make the ansatz
\be
x^1=X=vt+\xi(r)\,,
\label{49}
\ee
which is motivated by the expectation that the motion of the string will make it have a profile that is dragging on one side as it lowers inside the bulk space.

We may now calculate the components of the induced metric
\be
\hat g_{\tau\tau}=-g_{tt}+g_{xx}\dot X^2=-g_{tt}+g_{xx}v^2
\,,\qquad
\hat g_{\tau\s}=g_{xx}X'\dot X=g_{xx}v\xi'\,,
\label{50}
\ee
\be
\hat g_{\s\s}=g_{rr}+g_{xx}X'^2=g_{rr}+g_{xx}\xi'^2
\,,\qquad -\det\hat g=g_{tt}g_{rr}+g_{xx}(g_{tt}\xi'^2-g_{rr}v^2)\,,
\label{51}
\ee
and
\be
\bar\pi^{\tau}_{t}=T_f{\hat g_{\s\s}\over \sqrt{-\hat g}}g_{tt}
\,,\qquad
\bar\pi^{\s}_{t}=-T_f{\hat g_{\tau\s}\over \sqrt{-\hat g}}g_{tt}\,,
\label{52}
\ee
\be
\bar\pi^{\tau}_{x}=T_f(\hat g_{\s\s}\dot X-\hat g_{\s\tau}X'){g_{xx}\over \sqrt{-\hat g}}\,,\qquad
\bar\pi^{\s}_{x}=T_f(\hat g_{\tau\tau} X'-\hat g_{\s\tau}\dot X){g_{xx}\over \sqrt{-\hat g}}\,.
\label{53}
\ee
As all $\bar\pi$ are time independent, the only non-trivial bulk equations are $\pa_{\s}\bar\pi^{\s}_{\mu}=0$.
It implies
\be
\bar\pi_{x}=-{T_f}{g_{tt}g_{xx}\xi'\over \sqrt{-\hat g}}={\rm constant}\,.
\label{dr1}\ee
The boundary equation for $\mu=t$ and $\mu=x$ are equivalent and  become
\be
T_f{\hat g_{\s\tau}\over \sqrt{-\hat g}}g_{tt}+Ev=0~~\to~~\bar\pi_{x}=E\,.
\label{dr2}\ee

We can solve (\ref{dr1}) to obtain
\be
\xi'=\sqrt{g_{rr}\over g_{tt}g_{xx}}{\sqrt{g_{tt}-g_{xx}v^2}\over \sqrt{T_f^2g_{tt}g_{xx}-\bar\pi_x^2}}\bar\pi_x\,.
\label{54}\ee
To ensure we have a  real solution,  there must be a turning point at $r=r_s$ where
the numerator vanishes and the denominator should also  vanish at the same point
giving
\be
v^2={g_{tt}(r_s)\over g_{xx}(r_s)}\,,\qquad \bar\pi_x=-T_f\sqrt{g_{tt}(r_s) g_{xx}(r_s)}\,.
\label{dr3}\ee

Finally,  equation (\ref{dr2}) becomes
\be
T_f\sqrt{g_{tt}(r_s) g_{xx}(r_s)}=-E\,,
\label{56}\ee
which gives the electrostatic force on the endpoint.
As v is constant, the equation of motion in the $x$-direction is
\be
{dp\over dt}=-\bar\pi_x+qE\,,
\label{57}\ee
and the steady state solution is $\bar\pi_x=E$.

 We denote by $J^t$ the density of charge as computed from the asymptotic form of $A_t$. From (\ref{dr3}), $\bar\pi_x$ is an implicit function of $v$.
${g_{tt}(r)\over g_{xx}(r)}$ is essentially the blackness function, and therefore vanishes at the horizon $r=r_h$.
 Therefore, the velocity vanishes when $r_s\to r_h$.
For small velocities we obtain
\be
\bar\pi_x\simeq -T_fg_{xx}(r_h)~v+{\cal O}(v^2)\,.
\label{58}\ee
Writing $J^x=J^tv$ we obtain
\be
J^x=J^t~v\simeq {J^t\bar\pi_x\over T_f g_{xx}(r_h)}\simeq {J^t\over T_f g_{xx}(r_h)}E\,,
\label{59}\ee
and we obtain the DC conductivity and related resistivity as
\be
\sigma\simeq {J^t\over T_f g_{xx}(r_h)}\,,\qquad \rho \simeq{T_f g_{xx}(r_h)\over J^t}\,.
\label{60}\ee
The end result depends on temperature only via the dependence of $g_{xx}(r_h)$ on temperature.
Note that the $g_{xx}$ factor should be evaluated in the string metric.
As discussed earlier in (\ref{ein1})
\be
g_{xx}^s=g_{xx}^E e^{k\phi}\,,\qquad 0\leq k\leq \sqrt{2\over p-1}\,,
\label{61}\ee
and $k$ depends on the relation of $\phi$ to the dilaton.

Therefore the final result for the temperature-dependent  DC resistivity is given by
\be
\rho \simeq{T_f g^E_{xx}(r_h)e^{k\phi(r_h)}\over J^t}\,,
\label{55}\ee
for  massive charge careers and is in agreement with (\ref{ndrag}).

In the case that $k=0$ (the scalar field does not involve the dilaton)
we can directly relate the temperature dependence of the resistivity to the entropy.
The reason is that from general principles the entropy is proportional to the area of the horizon
\be
S\sim g^E_{xx}(r_h)^{(p-1)\over 2}\,.
\label{63}\ee
Relation (\ref{63}) implies
\be
{\rho(T)\over S(T)^{2\over p-1}}={\rm constant}\,.
\label{64}\ee
In particular for planar systems this indicates a resistivity linear in the entropy. This is a very interesting correlation that is emerging from holography.


\section{The $\gamma\delta=1$ solutions\label{gd-1}}

For this relation of parameters  the general solution is obtained integrating the equations of motion (\ref{EOMLiouville})\footnote{We refer the reader to \cite{char} for particulars on the method of resolution.}. This range of parameters is most interesting since it encompasses both the string theory case\footnote{This corresponds to 4d string theory with the vector arising from the NS sector.} , $\ga=\da=1$, and the Kaluza-Klein case, $\ga=\sqrt3$, $\da=\frac1{\sqrt3}\,$.
The solution reads,
\bsea
	\ud s^2 &=& - \frac{V(r)\ud t^2 }{\le[1-\le(\frac{r_-}{r}\ri)^{3-\da^2}\ri]^{\frac{4(1-\da^2)}{(3-\da^2)(1+\da^2)}}}+ e^{\da\phi}\frac{\ud r^2}{V(r)}+ \nn\\
			&&+ r^2\le[1-\le(\frac{r_-}{r}\ri)^{3-\da^2}\ri]^{\frac{2(\da^2-1)^2}{(3-\da^2)(1+\da^2)}}\Big(\ud x^2+\ud y^2\Big)\,, \slabel{Metric1} \\
	V(r) &=& \le(\frac r\ell\ri)^2-2\frac{m\ell^{-\da^2}}{r^{1-\da^2}} +\frac{(1+\da^2)q^2\ell^{2-2\da^2}}{4\da^2(3-\da^2)^2r^{4-2\da^2}}\,, \slabel{Pot1} \\
	\le(r_{\pm}\ri)^{3-\da^2} &=& \ell^{2-\da^2}\le[m\pm\sqrt{m^2-\frac{(1+\da^2)q^2}{4\da^2(3-\da^2)^2}}\ri],\\
	e^{\phi}&=& \le(\frac r\ell\ri)^{2\da}\le[1-\le(\frac{r_-}{r}\ri)^{3-\da^2}\ri]^{\frac{4\da(\da^2-1)}{(3-\da^2)(1+\da^2)}}\,, \slabel{Phi1}\\
	\mathcal A &=& \le(\Phi -\frac{q\ell^{2-\da^2}}{(3-\da^2)r^{3-\da^2}}\ri)\ud t\,, \slabel{A1}\\
	\Phi&=&\frac{q\ell^{2-\da^2}}{(3-\da^2)r_+^{3-\da^2}} \slabel{ElectricPotential1}\,,\\
	\ga\da&=&1\,,\nn
	\label{Sol1}
\esea
where the parameters $m$ and $q$ are integration constants linked to the gravitational mass and the electric charge (note that the electric potential \eqref{Sol1} has been fixed accordingly).
 An initial condition for the scalar, $\phi_0$, can aleays be absorbed in a redefinition of $\Lambda\to \Lambda e^{\delta\phi_0}$.
 There is an overall scale $\ell$ which can be fixed freely as the metric is scale-invariant (up to redefinitions of $m$ and $q$, of course). This scale can be identified as an IR scale via
\be
	\ell^2=\frac{3-\da^2}{-\Lambda}\,.
\label{107}\ee
 We could also consider ``de Sitter''-like solutions by setting
\be
	a^2=\frac{3-\da^2}{\Lambda}\,,
\label{108}\ee
in the usual fashion, and these solutions can be obtained from \eqref{Sol1} by making the change $\ell\to ia$. However, this forces to Wick-rotate both the time coordinate $t$ and the radial coordinate $r$, yielding time-dependent solutions rather than black hole space-times. We will thus restrict our analysis to cases where $\ell^2$ is real and positive. It is worth emphasizing that what controls the scalar curvature of the solutions, ie., de Sitter or Anti-de Sitter like, is the sign of the product $(3-\da^2)\Lambda$. We will now focus on the static case, where the radial coordinate $r$ is spacelike:
\begin{enumerate}
 \item $\Lambda<0$ and $\da^2<3$, which will yield black hole solutions;
 \item $\Lambda>0$ and $\da^2>3$, which will yield singular solutions.
\end{enumerate}
The Ricci scalar, computed with \eqref{RicciScalarOnShell} is:
\be
	\mathcal R = 4\Lambda e^{-\da\phi} -\frac{2\da^2\Lambda}{3-\da^2}e^{-\da\phi}\frac{\le[1-\le(\frac{r_+}{r}\ri)^{3-\da^2}\ri]}{\le[1-
\le(\frac{r_-}{r}\ri)^{3-\da^2}\ri]}\le[1-\frac{3-\da^2}{1+\da^2}\le(\frac{r_-}{r}\ri)^{3-\da^2}\ri],
	\label{RicciScalar1}
\ee
and it is straightforward using \eqref{Sol1} to observe that for $\da^2<3$, both $r=0$ and $r_-$ are singular. This confirms that in the generic case, the only event horizon is at $r=r_+$, and space-time only extends up to $r=r_-$. Quite interestingly, when $\da=1$, $r_-$ ceases to be a singular point and turns into an inner horizon, and space-time extends all the way to $r=0$.

The solution has an extremal limit when
\be	
m_e\ell^{2-\da^2}=\frac{\sqrt{1+\da^2}q\ell^{2-\da^2}}{2\da(3-\da^2)}\equiv r_e^{3-\da^2}\,.
\label{109}\ee
The two points $r^\pm$ then merge and become singular, except again in the limit $\da=1$ where the extremal black hole is regular.

{}From \eqref{RicciScalar1}, we observe that in the case where $\da^2>3$ (which is perfectly admissible from previous considerations, provided we take a positive $\Lambda$), the Ricci scalar blows up when $r\to+\infty$ and vanishes  when $r\to0$. This is a hint that our space-like coordinate is ill-chosen, and that we should change $r\to\frac1r$. However, upon doing this, we find that the new radial coordinate is actually spacelike \emph{inside} the horizon radius, so in the end we obtain a space-time describing a cosmological horizon covering a naked singularity. Although this is reminiscent of de Sitter space, the inner singularity spoils the regularity of space-time. A more appropriate picture would rather be that this depicts the interior of the black hole.

The `string' solution $\gamma=\da=1$ can be shown to be the limit of the equivalent solution in \cite{Chan:1995fr,char} with a horizon of non-zero curvature $\ka=\pm1$ by taking the limit $\ka\to0$, $\da\to1$ while keeping the ratio of the two fixed. The question raised in this previous work  \cite{Chan:1995fr} that the limit $\da\to1$ was ill-defined in the string case is then resolved here, a regular limit exists but the horizon's topology is severely constrained and becomes toroidal.

For $\da$ taken in the range $\da^2<3$, we have Anti-de Sitter like solutions. The solution \eqref{Sol1} is then truly a black hole solution with an event horizon situated at $r_+$ and a curvature singularity at $r_-$. The space-time is static outside the horizon and the singularity is timelike. The black hole space-time has to be cut off at $r_-$ and does not extend all the way to $r=0$.  The background space-time obtained either by taking $m=q=0$ or the asymptotic limit $r\to+\infty$ of \eqref{Sol1} goes all the way to $r=0$, where a curvature singularity sits. This does not come as a surprise, as the background solution still contains a non-trivial scalar field which is singular asymptotically and thus cannot be treated as a perturbation. It is also worth pointing out that the two-torii collapse at $r=r_-$, that is for finite ``radius''. This kind of feature seems to be characteristic of general dilatonic solutions, see \cite{gar} for the flat potential case in string theory\footnote{This solution, or more precisely its generalisation to free $\ga$, was proven in \cite{Gibbons:1987ps,char} to be the general solution of the system of equations of motion without a scalar potential.} or references \cite{Henneaux:2004zi,Henneaux:2006hk} for related work.

Finally, the limit $\da\to0$ allows to recover the planar Schwarzschild-Anti-de Sitter solution. Indeed, from \eqref{Sol1}, it is seen that the only consistent way to take the $\da\to0$ limit is also to set $q\to0$, which effectively makes  the Maxwell field vanish. Therefore, the Anti-de Sitter-Reissner-Nordstr\"om solution does not seem to be a limit of \eqref{Sol1}.

The temperature is given by
\be
	 T=\frac{3-\da^2}{4\pi\ell}\le(\frac{r_+}\ell\ri)^{1-\da^2}\le[1-\le(\frac{r_-}{r_+}\ri)^{3-\da^2}\ri]^{1-2\frac{(\da^2-1)^2}{(3-\da^2)(1+\da^2)}}\,.
	\label{Temperature1}
\ee

Let us note already that the behaviour of the temperature in the extremal limit is not uniform in the $\da^2<3$ range. Indeed, for $\da^2<1+\frac2{\sqrt3}$, the temperature vanishes in the extremal limit, whereas it diverges for $\da^2>1+\frac2{\sqrt3}$ and is finite and non-zero for $\da^2=1+\frac2{\sqrt3}$. Thus, the extremal black hole seems ill-defined in the range $1+\frac2{\sqrt3}\leq\da^2<3$, as it does not seem that they can be end states of the evaporation through thermal radiation of the black hole. This behaviour has been noted for a long time in string theory black holes and is commented upon in \cite{Preskill:1991tb,Holzhey:1991bx}. From similar cases of uncharged black holes investigated in \cite{gkmn2}, this behaviour corresponds to the small black hole solution that is unstable, and behaves like the flat space Schwarzschild black holes. In such cases we expect that the AdS asymptotics will generate at least one branch, the large black hole branch, that will dominate eventually the ensemble at larger temperatures.

In order to determine the gravitational energy of the solution, one can integrate the contribution at infinity of the boundary term
\eqref{GravitationalBoundaryVariation},
\be
	M_g = \frac{\omega_2}{8\pi}\ell^{\da^2-2}\le[(r_+)^{3-\da^2}+\frac{2\da^6-15\da^4+20\da^2-3}{(3-\da^2)(1+\da^2)}(r_-)^{3-\da^2}\ri],
	\label{GravitationalMass1}
\ee
where $\omega_2$ is defined as the volume of the compact two-dimensional horizon. This reduces in particular for $\da^2=0,1$ to $M_g=\frac{\omega_2}{4\pi}m$.
However, the boundary term at infinity stemming from the scalar field \eqref{ScalarBoundaryVariation} gives a non-zero contribution,
\be
	M_\phi = \frac{\omega_2}{8\pi}\ell^{\da^2-2}\frac{4\da^2(\da^2-1)}{(3-\da^2)(1+\da^2)}r_-^{3-\da^2}\,,
	\label{ScalarMass1}
\ee
which must be added to find the total energy,
\be
	E = \frac{\omega_2}{8\pi}\ell^{\da^2-2}\le[(r_+)^{3-\da^2}-\frac{2\da^4-5\da^2+1}{(1+\da^2)}(r_-)^{3-\da^2}\ri].
	\label{Energy1}
\ee
The conserved electric charge \eqref{ElectricCharge} is
\be
	Q =  \frac{\omega_2}{16\pi}q\,.
	\label{ElectricCharge1}
\ee

We will make the following rescalings in order to  absorb volume factors in the thermodynamic quantities, throughout the rest of this paper:
\bear
	W[T,Q],\,G[T,\Phi] &\to& \frac{16\pi}{\omega_2}W[T,Q],\,\frac{16\pi}{\omega_2}G[T,\Phi]\nn\,,\\
	E &\to& \frac{16\pi}{\omega_2}E\,,\\
	T &\to& 4\pi T, \quad S\to\frac{4}{\omega_2}S\nn\,,\\
	Q &\to& \frac{4\pi}{\omega_2} Q, \quad \Phi\to4\Phi\nn\,,
\label{rescale}\eear
which will not modify the expression of the first law.

\subsection{Regions of validity}\label{gdregions}

The UV region of the solutions is $r\to \infty$ where the scale factor increases without bound.
For the solutions to be IR regions of asymptotically AdS solutions the scalar potential must become small in the UV region,
\be
V\sim e^{-\delta\phi}\sim \le(\frac r\ell\ri)^{-2\da^2}\le[1-\le(\frac{r_-}{r}\ri)^{3-\da^2}\ri]^{-\frac{4\da^2(\da^2-1)}{(3-\da^2)(1+\da^2)}}\underset{r\to +\infty}{\longrightarrow} 0\,.
\label{110}\ee
Therefore these solutions can  indeed be IR regions of asymptotically AdS solutions.

In the case where the conserved charge is originating on flavor branes with a DBI action, according to (\ref{19})
we can linearize the action when
\be
Z Ce^{k\phi}=\ell^2\left( {r\over \ell}\right)^{4+2k\d}\le[1-\le(\frac{r_-}{r}\ri)^{3-\da^2}\ri]^{\frac{2(\da^2-1)(\d^2+2k\d+1)}{(3-\da^2)(1+\da^2)}}
\gg |q|\,.
\label{aaa}\ee
As $\d<\sqrt{3}$ this approximation is always good in the UV.
In the IR we must distinguish the following cases
\begin{itemize}
\item For $\d^2\leq 1$ the right hand side of (\ref{aaa})  is a function that becomes smaller and smaller in the IR.
 A necessary condition is that at the horizon it is small
and this can be adjusted by adjusting the UV (boundary) value of the scalar $\phi$ or effectively the IR scale $\ell$.
The condition of validity of (\ref{aaa}) at the horizon is
\be
Y_h=\left( {r_+\over \ell}\right)^{4+2k\d}\le[1-\le(\frac{r_-}{r_+}\ri)^{3-\da^2}\ri]^{\frac{2(\da^2-1)(\d^2+2k\d+1)}{(3-\da^2)(1+\da^2)}}>>{| q|\over \ell^2}\,.
\label{111u}\ee
$Y_h$ as a function of $r_+$ is a function that diverges at extremality. As it also diverges at the boundary, it is enough to require
\be
\left( {r_-\over \ell}\right)^{4+2k\d}\gg {q\over \ell^2}\,,
\label{112}\ee
for the Maxwell approximation to be everywhere valid in this case.
\item For $3>\d^2> 1$, the factor $Y_h$ vanishes at extremality
and therefore the Maxwell approximation breaks down there. In this case the charge carriers are in a dense regime.
To test whether in this regime the DBI action can be treated like a probe, we must have that $V\gg e^{2k\phi}Z$ or
$e^{\left(\d+{1\over \d}+2k\right)\phi}\ll 1$.
Indeed, in this case as $\d^2>1$
\be
e^{\left(\d+{1\over \d}+2k\right)\phi}= \le(\frac r\ell\ri)^{2(\da^2+2k\d+1)}\le[1-\le(\frac{r_-}{r}\ri)^{3-\da^2}\ri]^{\frac{4(\d^2+2k\d+1)(\da^2-1)}{(3-\da^2)(1+\d^2)}}\longrightarrow 0\,.
\label{113} \ee
\end{itemize}
We can therefore use the probe techniques to calculate the conductivities  with the result derived already in section \ref{DCDBI}.

\subsection{Thermodynamics}

Throughout this section and both for the grand-canonical and canonical case, several cases will be distinguished depending on the value taken by $\da^2$. However, we will only plot one curve per range, representative of the behaviour. Where useful, we will also display the behaviour at the limiting values for these ranges.

\subsubsection{Grand-canonical ensemble}

The grand-canonical ensemble is defined by keeping the temperature and the electric chemical potential fixed: this means that we can take as the thermal background the space-time with the black hole switched off ($m=q=0$), which also coincides with the asymptotic limit of the solution if $\da^2<3$:
\bsea
	\ud s^2_o &=& r^2\le(-\ud t^2 +\ud x^2 + \ud y^2 \ri) + \le(\frac r\ell \ri)^{2\da^2-2}\ud r^2\,, \slabel{BackgroundMetric1} \\
	\mathcal A_o &=& \Phi\ud t\,,   \slabel{BackgroundA1}  \\
	e^{\phi_o} &=& \le(\frac r\ell \ri)^{2\da}\,,  \slabel{BackgroundPhi1}
\label{LinearDilaton}
\esea
For a flat potential $\da=0$, one recognizes $AdS_4$ as expected.

\begin{figure}[t]
\begin{center}
\begin{tabular}{cc}
	 \includegraphics[width=0.45\textwidth]{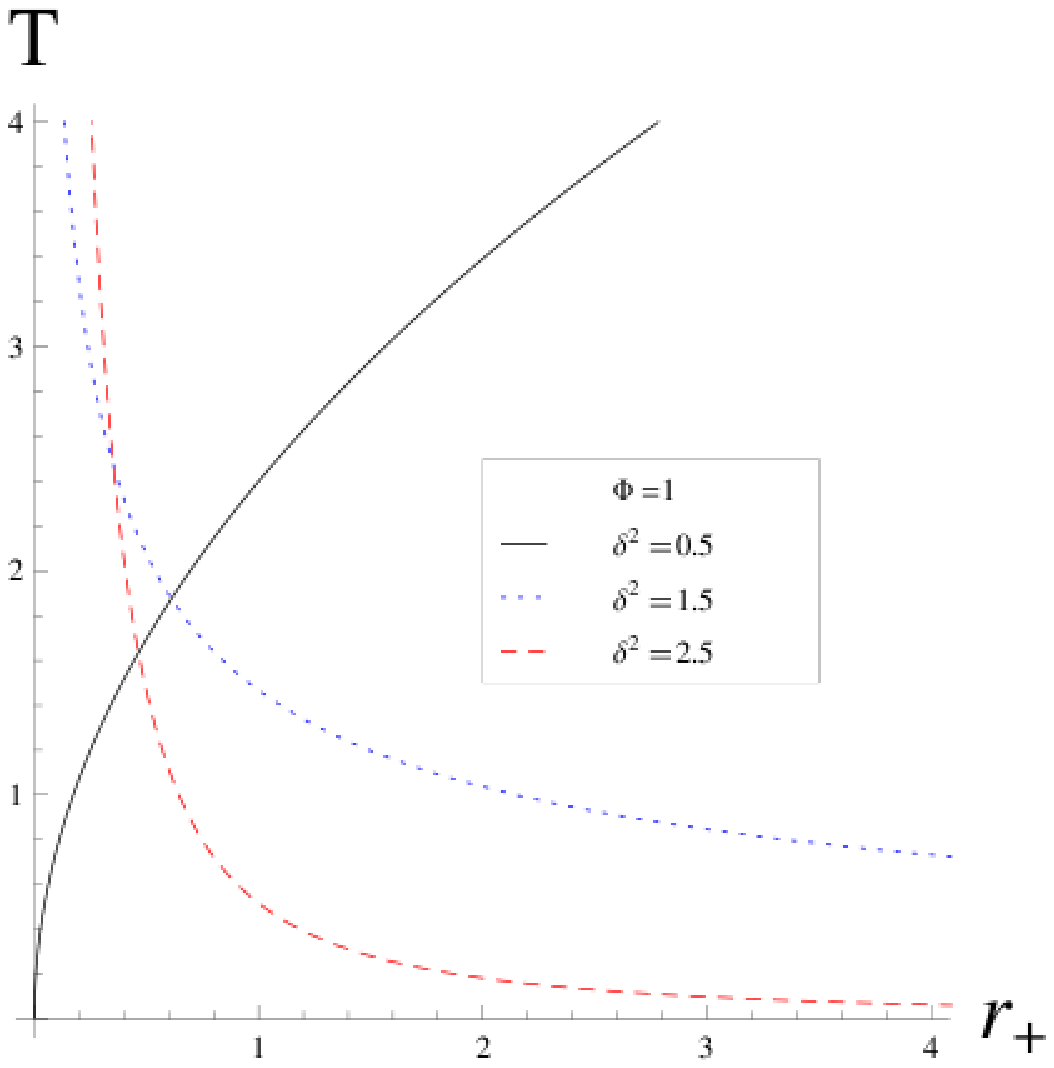}&	
	 \includegraphics[width=0.45\textwidth]{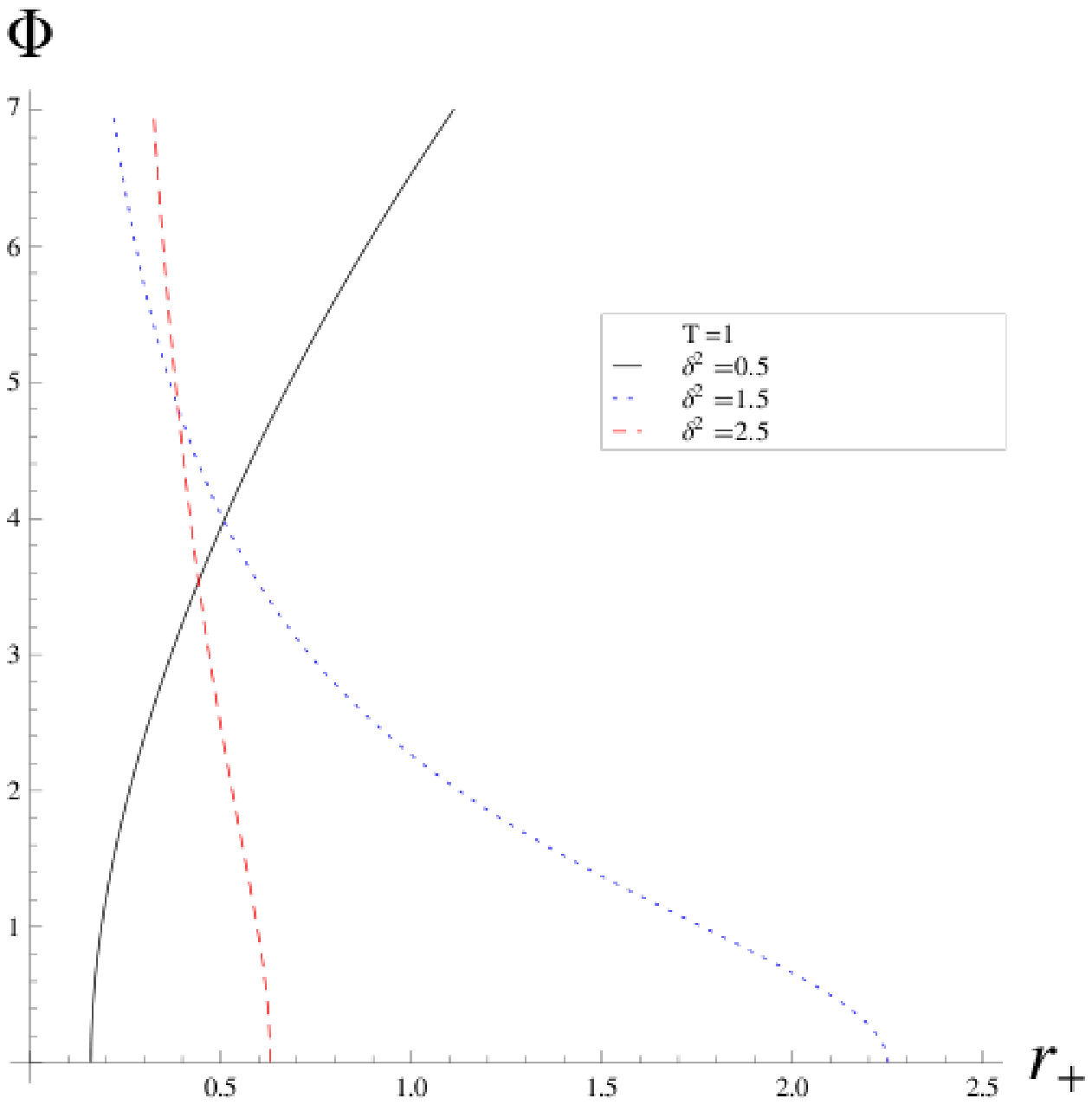}
 \end{tabular}
\caption{Slices of the temperature and the electric potential versus the horizon radius $r_+$.}
\label{Fig:HorizonSizeGaDa1GrandCanonical}
\end{center}
\end{figure}

\begin{figure}[t]
\begin{center}
\begin{tabular}{ccc}
	 \includegraphics[width=0.3\textwidth]{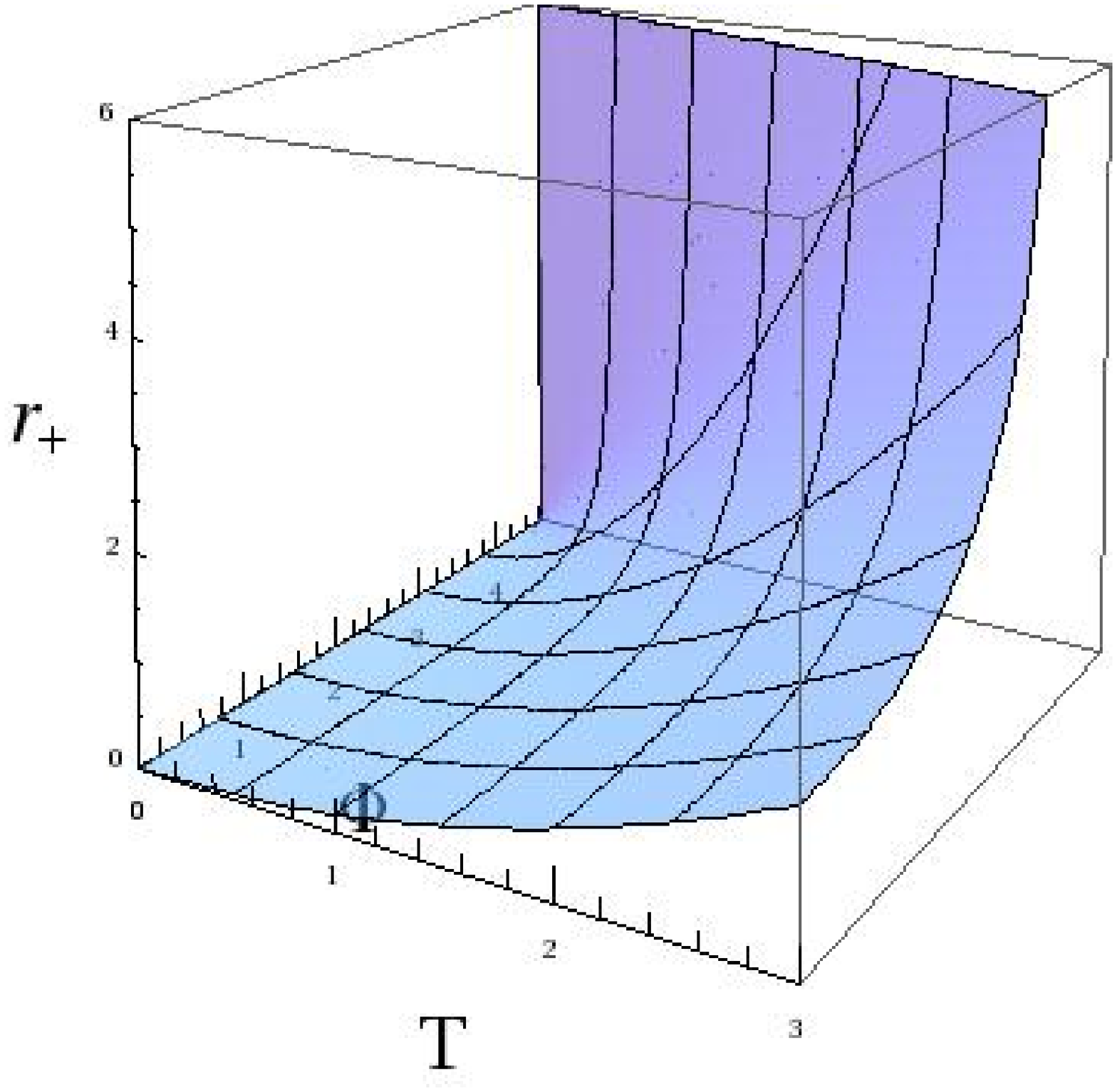} &
	 \includegraphics[width=0.3\textwidth]{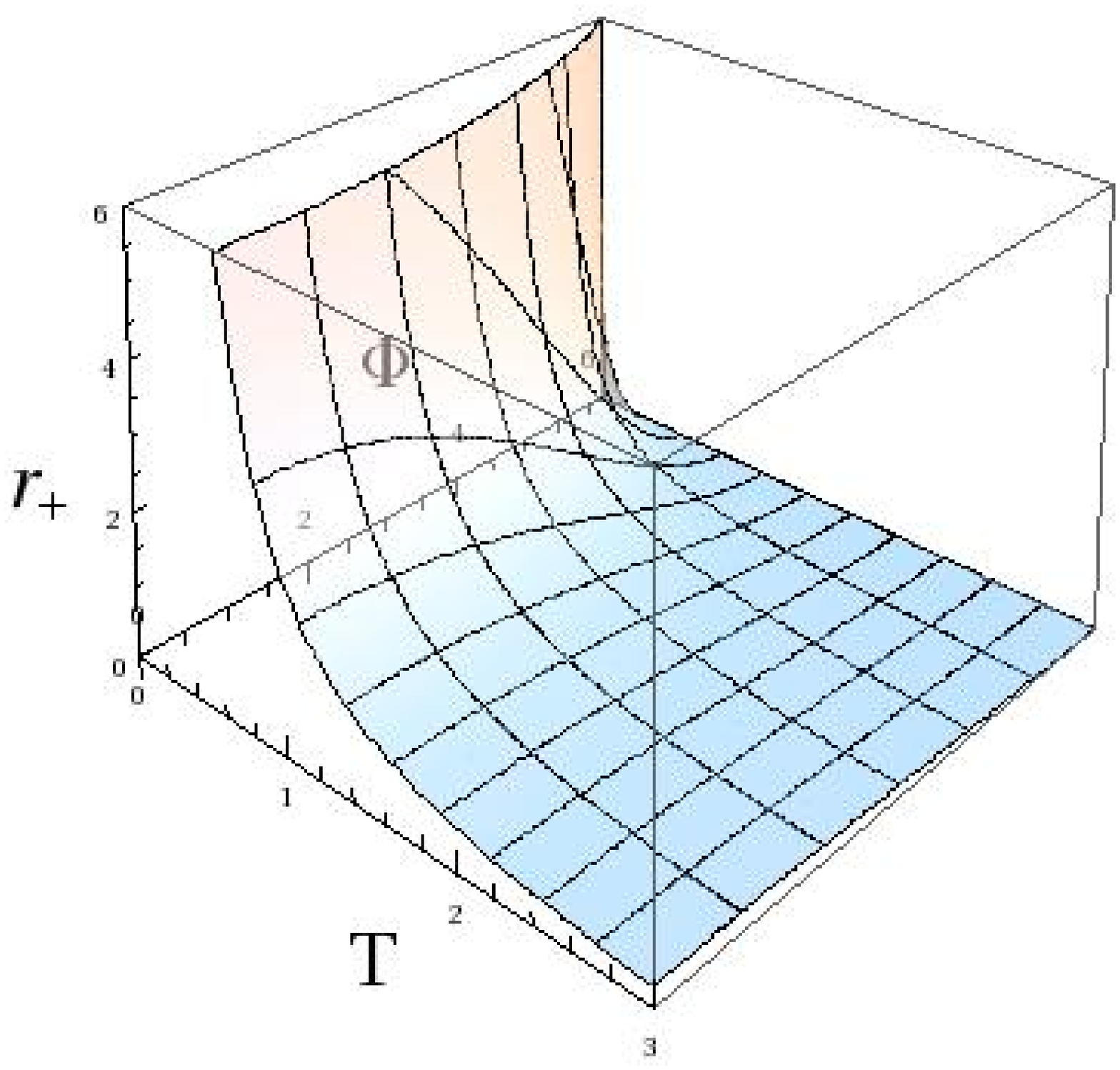} &
	 \includegraphics[width=0.3\textwidth]{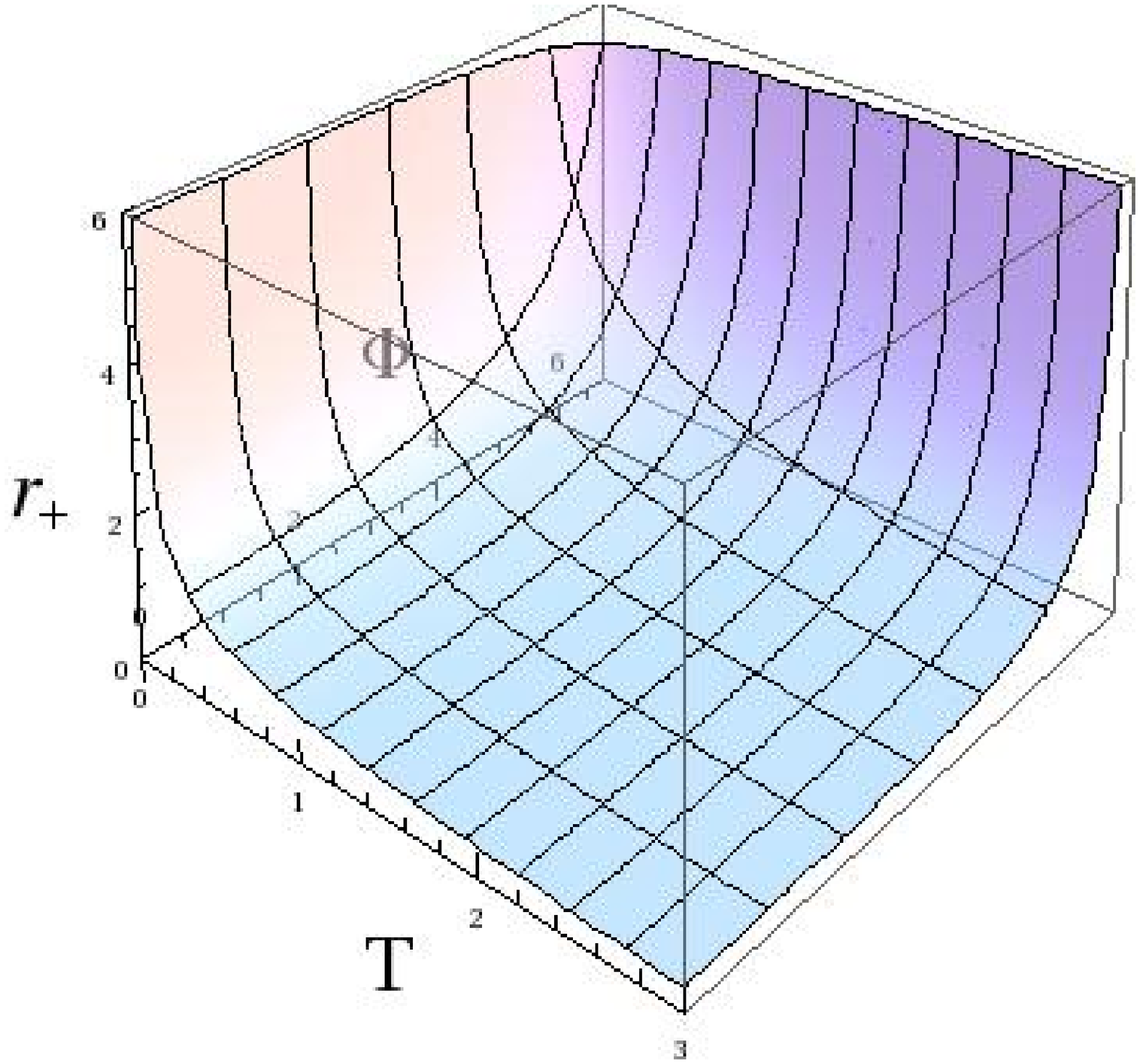}
 \end{tabular}
\caption{Three-dimensional equation of state $r_+(T,\Phi)$ for the lower ($0\leq\da^2<1$), intermediate ($1<\da^2<1+\frac2{\sqrt{3}}$) and upper ($1+\frac2{\sqrt{3}}<\da^2<3$) range from left to right.}
\label{Fig:EquationOfState3DGaDa1GrandCanonical}
\end{center}
\end{figure}

The equation of state \eqref{Temperature1}
\be
	r_+^{1-\da^2}= \ell^{2-\da^2}\frac{T}{3-\da^2}\le[1-\frac{(1+\da^2)\Phi^2}{64\da^2}\ri]^{\frac{2(\da^2-1)^2}{(1+\da^2)(3-\da^2)}-1}\,,
	\label{EqOfStateGaDa1GrandCanonical}
\ee
gives the horizon radius in terms of the grand-canonical thermodynamic variables $(T,\Phi)$. From the previous expression, it becomes apparent why in the string limit the Gibbs potential is identically zero for all values of the thermodynamic variables: this is a signal that the grand-canonical ensemble breaks down in this limit, as the temperature and the chemical potential cannot be considered as independent variables. Thus, in the $\da^2\to1$ limit, the grand-canonical ensemble is ill-defined and we shall have to restrict the discussion to the canonical ensemble.

For all temperatures, there is a maximum value to the chemical potential, corresponding to
\be
	\Phi_e^2 = \frac{64\da^2}{1+\da^2}\,,
\ee
and the region $\Phi\geq\Phi_{e}$ is forbidden because of the non-trivial exponent in \eqref{EqOfStateGaDa1GrandCanonical},
contrary to what happens for AdS-Reissner-Nordstr\"om black holes \cite{RN1}.
The extremal black holes $r_-=r_+=r_e$ are attained only for this particular value of the chemical potential,
and it is straightforward to observe that the equation of state \eqref{EqOfStateGaDa1GrandCanonical} becomes ill-defined in this limit,
as one cannot know both the temperature and the black hole radius.
Thus, one has to conclude that the extremal black holes do not exist in the grand-canonical ensemble.
However, for values of $\Phi\geq\Phi_e$, the neutral black holes can still exist and compete with the dilatonic background. Indeed,
there can be a non-zero chemical potential for a neutral black hole while keeping a zero charge, as only the derivative of the
vector potential has physical meaning. In this case, the equation of state \eqref{EqOfStateGaDa1GrandCanonical} reduces to
\be
	r_+^{1-\da^2}= \ell^{2-\da^2}\frac{T}{3-\da^2}\,,
	\label{EqOfStateNeutralGaDa1GrandCanonical}
\ee
and gives the horizon radius as a function of the temperature only.

We plot slices of the equation of state \eqref{EqOfStateGaDa1GrandCanonical} at fixed chemical potential and temperature in \Figref{Fig:HorizonSizeGaDa1GrandCanonical}, while the full three-dimensional plots are in \Figref{Fig:EquationOfState3DGaDa1GrandCanonical}. Three ranges can be distinguished,
\begin{itemize}
 \item Lower range $0<\da^2\leq1\,$:  The black holes resemble toroidal AdS-Reissner-Nordstr\"om and behave according to the standard physical intuition. Indeed, for finite $\Phi$, the black hole (charged or neutral) disappears (zero horizon radius) as the temperature goes to zero, and grows to cover the whole of space-time as the temperature grows to infinity. This is to be compared with Fig.2 in \cite{RN1}: the term responsible for the vanishing of the radius at non-zero finite temperature is absent because the spatial curvature of the horizon is zero. The extremal limit $\Phi=\Phi_e$ can never be attained on the left plot of \Figref{Fig:HorizonSizeGaDa1GrandCanonical} since $\Phi\neq\Phi_e$ there, so it comes as no surprise that the zero temperature limit switches off the black hole, instead of going to the extremal limit. The right plot of \Figref{Fig:HorizonSizeGaDa1GrandCanonical} is at finite temperature and so does not display extremal black holes. Instead, the extremal limit $\Phi\to\Phi_e^-$ leads to a black hole the engulfs the full  space-time as $r_+\to+\infty$. The horizontal axis $\Phi=0$ shows the uncharged dilatonic black holes.
 \item Intermediate range $1<\da^2<1+\frac2{\sqrt{3}}\,$: This range seems quite different from the usual behaviour as the black hole radius decreases with the temperature! The background space-time, given by $r_+=0$ is attained for infinite temperature and in the zero temperature limit the black hole covers the whole of space-time. The right plot of \Figref{Fig:HorizonSizeGaDa1GrandCanonical} shows a finite radius for the uncharged dilatonic black holes and no black hole for the value $\Phi=\Phi_e$, all of this expected as the temperature is finite and does not allow extremal black holes to be reached.
 \item Upper range $1+\frac2{\sqrt{3}}\leq\da^2<3\,$:  The same analysis of the left plot of \Figref{Fig:HorizonSizeGaDa1GrandCanonical} as in the intermediate range applies, while the right plot corresponds to the same analysis as in the lower range.
\end{itemize}

\begin{figure}[t]
\begin{center}
\begin{tabular}{cc}
	 \includegraphics[width=0.45\textwidth]{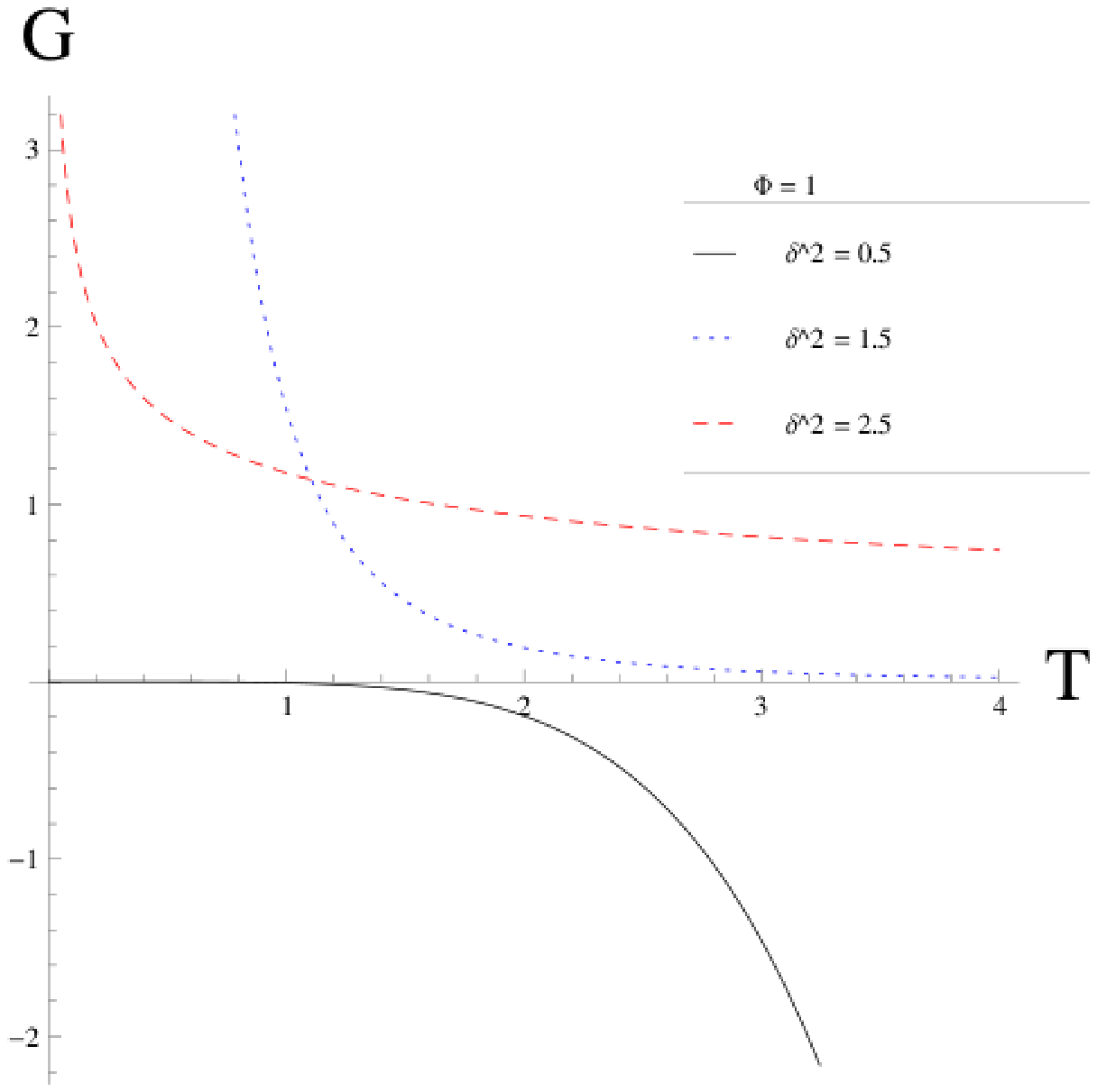}&	
	 \includegraphics[width=0.45\textwidth]{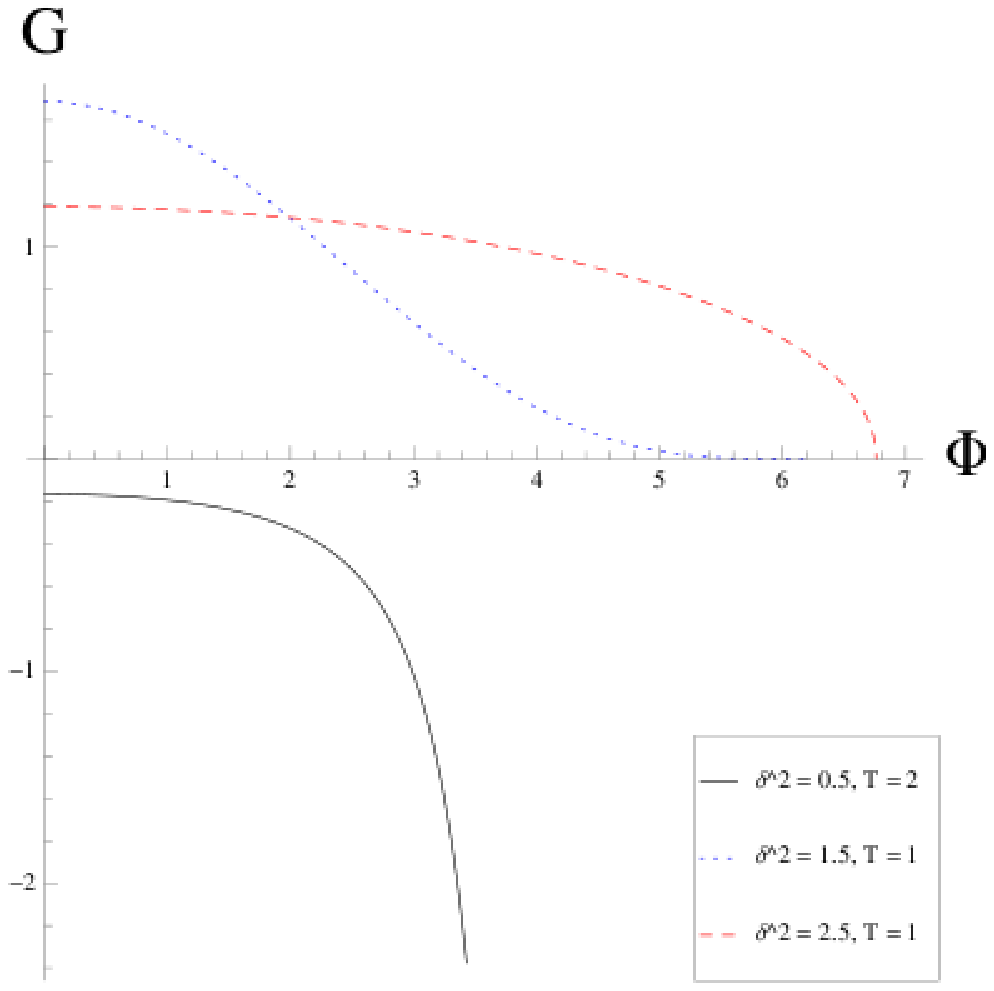}
 \end{tabular}
\caption{Slices at constant chemical potential and constant temperature of the Gibbs potential. The $T=2$ value for $\da^2=0.5$ allows to see graphically the non-zero limit as $\Phi\to0$}
\label{Fig:GibbsGaDa1GrandCanonical}
\end{center}
\end{figure}

\begin{figure}[t]
\begin{center}
\begin{tabular}{ccc}
	 \includegraphics[width=0.3\textwidth]{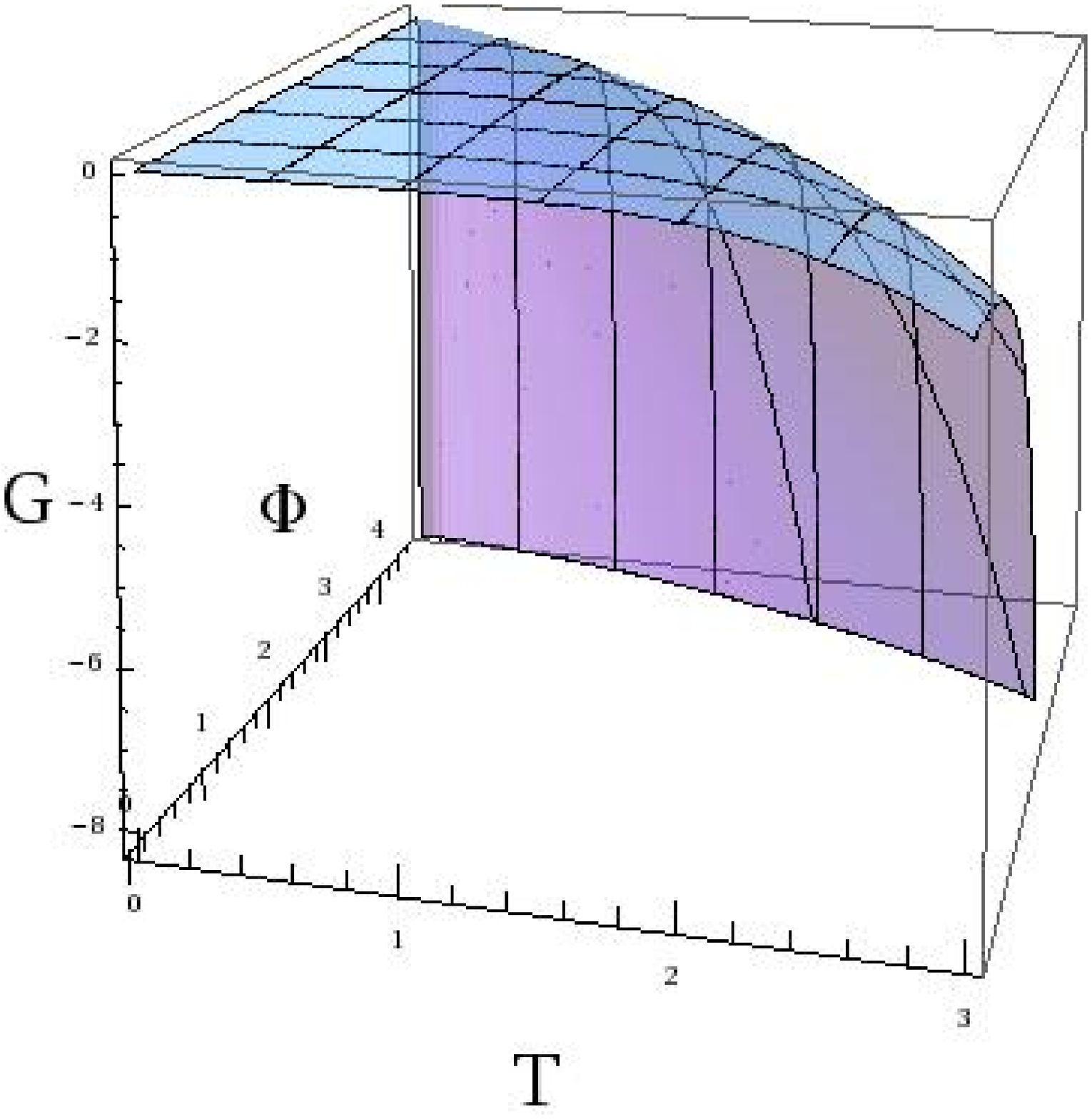} &
	 \includegraphics[width=0.3\textwidth]{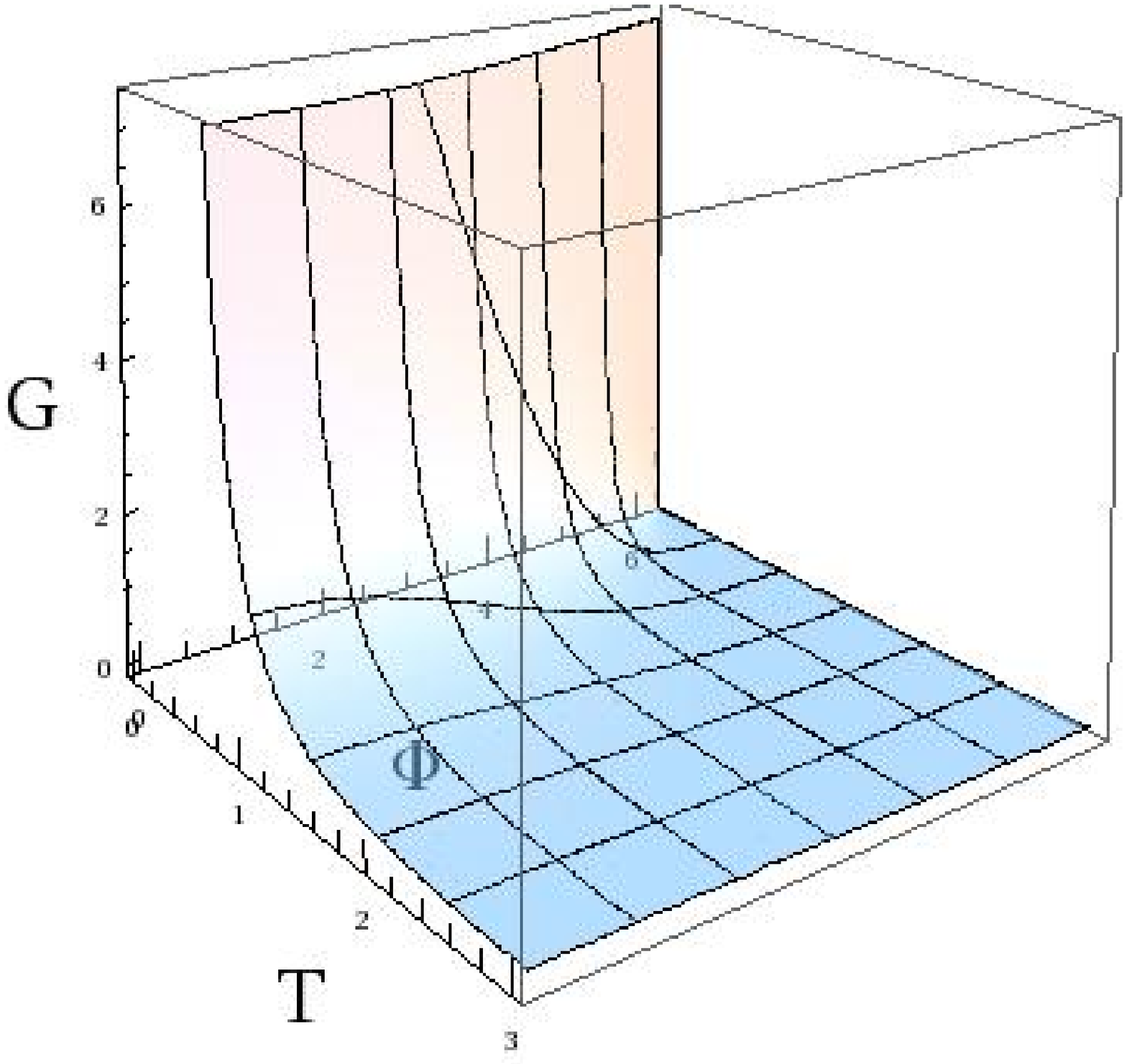} &
	 \includegraphics[width=0.3\textwidth]{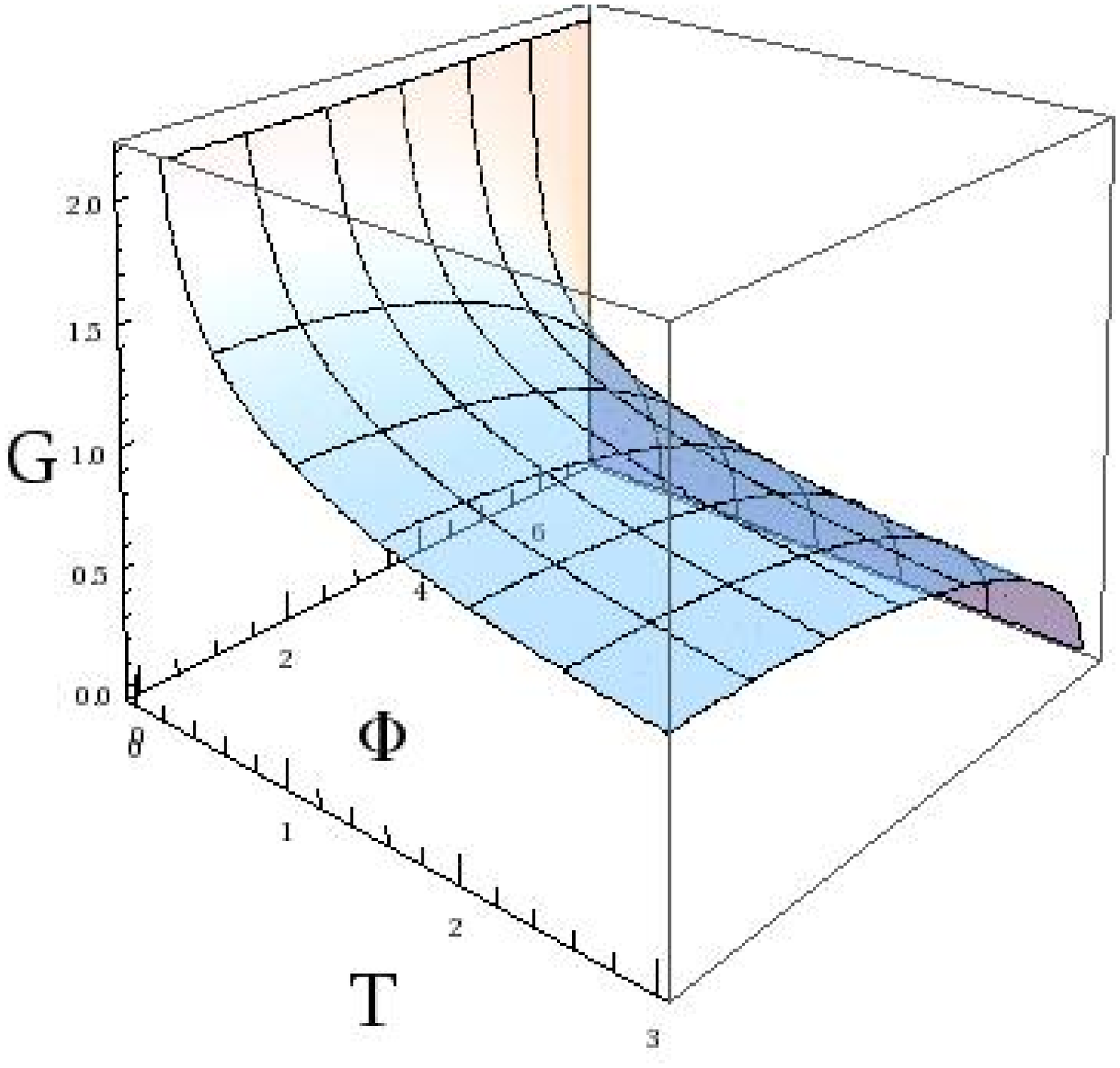}
 \end{tabular}
\caption{Three-dimensional representation of the Gibbs potential for the lower ($0\leq\da^2<1$), intermediate ($1<\da^2<1+\frac2{\sqrt{3}}$) and upper ($1+\frac2{\sqrt{3}}<\da^2<3$) range from left to right.}
\label{Fig:Gibbs3DGaDa1GrandCanonical}
\end{center}
\end{figure}

Then, calculating the value of the Euclidean action \eqref{ActionGrandCanonical} and subtracting the background contribution, we find
\be
	-\beta G = \tilde I - \tilde I_0 = -\beta \ell^{\da^2-2}(\da^2-1)\le[r_+^{3-\da^2}-r_-^{3-\da^2} \ri],
	\label{EuclideanActionGC1}
\ee
where we have identified the temperatures of the black hole and the thermal background on the outer boundary in order to do the subtraction. The Gibbs potential of the black hole in the thermal background \eqref{LinearDilaton} is
\be
	G[T,\Phi] = \ell^{\da^2-2}(\da^2-1)(3-\da^2)^{\frac{3-\da^2}{\da^2-1}}T^{\frac{3-\da^2}{1-\da^2}}\le[1-\frac{\Phi^2}{\Phi_e^2} \ri]^{2\da^2\frac{3-\da^2}{\da^4-1}}\,,
	\label{Gibbs1}
\ee
and here we have to be careful when evaluating the Gibbs potential for extremal black holes. Inspecting the limit $\Phi\to\Phi_e^-$ (e.g. approached from below) in \eqref{Gibbs1}, we find that it yields zero in the intermediate and upper range, but in the lower is minus infinity: in this range we cannot define the extremal limit from \eqref{Gibbs1} as we have an undetermined expression.
 Thus, the proper, unambiguous way to calculate the Gibbs potential of the extremal black holes is by evaluating \eqref{EuclideanActionGC1}, from which we find it is always identically zero. We show slices at constant chemical potential and temperature of the Gibbs potential in \Figref{Fig:GibbsGaDa1GrandCanonical}, while the full three-dimensional representation is in \Figref{Fig:Gibbs3DGaDa1GrandCanonical}.

We also note that the Gibbs potential is always zero if $\da^2=1$, that is in the string case. The explanation is simple: in this limit, the equation of state \eqref{EqOfStateGaDa1GrandCanonical} becomes a relation between $T$ and $\Phi$ and thus they cannot be taken as independent variables; the grand-canonical ensemble is ill-defined in this case.

As it was done in \cite{gkmn2} in the $\delta^2=1$ case a finer parametrization of the potential asymptotics is necessary, $V\sim e^{\pm \phi}\phi^p$ in order to resolve the IR behavior of this regime. We will not pursue this here.

It is also easy to observe, either from \eqref{EuclideanActionGC1} or \eqref{Gibbs1}, that the neutral black holes always have greater Gibbs potential than the charged black holes, and so will not be favoured globally as long as the charged black holes exist.

We may now calculate the entropy \eqref{EntropyGrandCanonical},
\be
	 S=\ell^{\frac{4-2\da^2}{1-\da^2}}\le(\frac{T}{3-\da^2}\ri)^{\frac2{1-\da^2}}\le[1-\frac{(1+\da^2)\Phi^2}{64\da^2}\ri]^{2\frac{(3-\da^2)}{\da^4-1}}\,,
	\label{EntropyGrandCanonical1}
\ee
which is the quarter of the area of the horizon as expected, the electric charge \eqref{ElectricChargeGrandCanonical},
\be
	Q=\frac{(3-\da^2)}{16}\ell^{\da^2-2}\Phi r_+^{3-\da^2}\,,
\label{114}\ee
which is equal to its usual value \eqref{ElectricCharge1}, and  the energy \eqref{EnergyGrandCanonical},
\be
	E = 2\ell^{\da^2-2}r_+^{3-\da^2}\le[1-(2\da^4-5\da^2+1)\frac{\Phi^2}{64\da^2}\ri],
	\label{EnergyGrandCanonical1}
\ee
which coincides with the previous expression \eqref{Energy1}

\begin{figure}[t]
\begin{center}
\begin{tabular}{cc}
	 \includegraphics[width=0.45\textwidth]{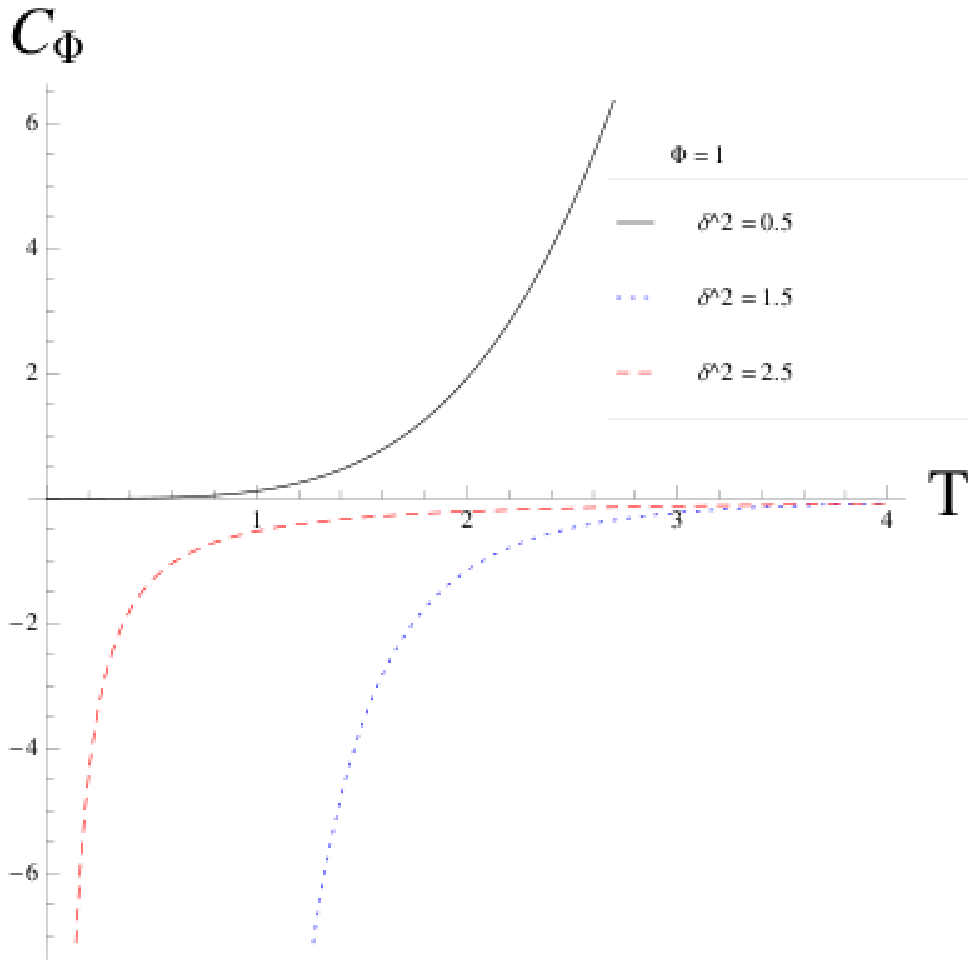}&	
	 \includegraphics[width=0.45\textwidth]{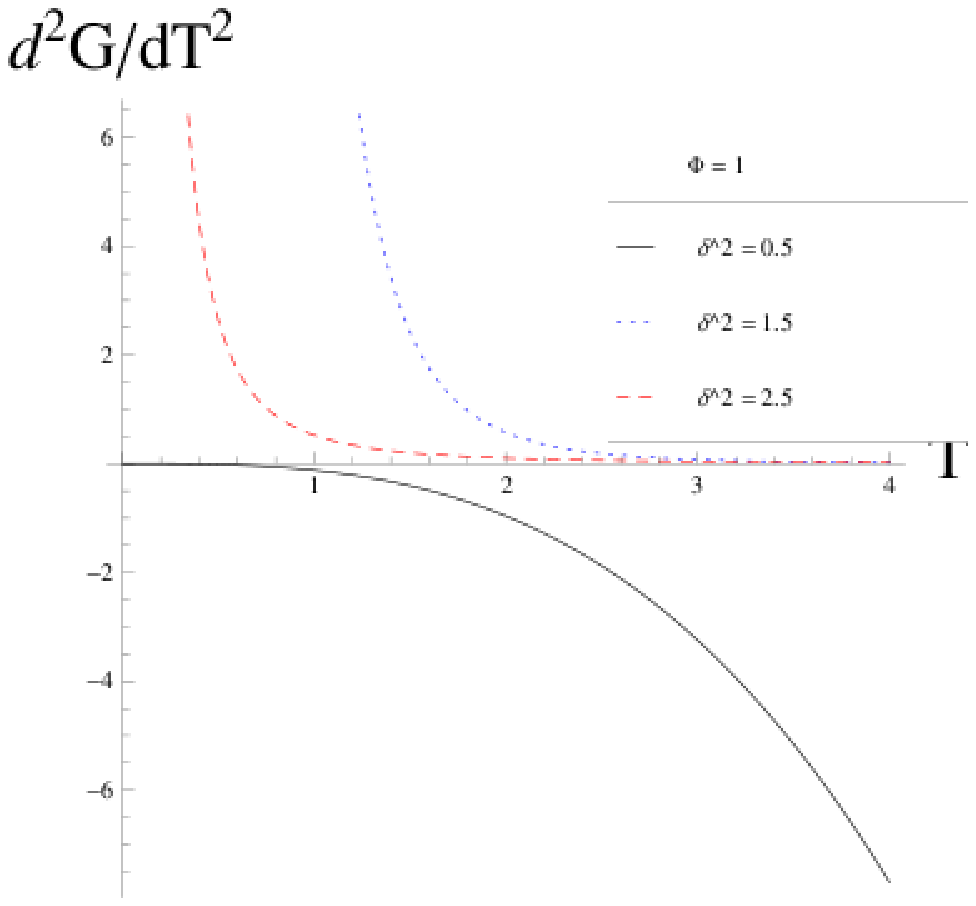}
 \end{tabular}
\caption{Heat capacity and second derivative of the Gibbs potential at constant chemical potential.}
\label{Fig:ThermalStabilityGaDa1GrandCanonical}
\end{center}
\end{figure}

Turning to the thermodynamic analysis, we find that the same three ranges as above must be distinguished.
\begin{itemize}
 \item Lower range: Similarly to the AdS-Reissner-Nordstr\"om black holes, the dilatonic black holes dominate at all values of the temperature, since their Gibbs potential is negative, see \Figref{Fig:GibbsGaDa1GrandCanonical}. The black holes are stable, as can be assessed both from the positivity of the heat capacity \eqref{HeatCapacityCanonical} and negativity of the second derivative of the Gibbs potential at constant chemical potential, see  \Figref{Fig:ThermalStabilityGaDa1GrandCanonical}. They are also stable to electric fluctuations: the slope of the Gibbs potential, as the chemical potential varies, is always negative and decreasing, which corresponds to positive electric permittivity. However, the charged black holes dominate the ensemble only for $\Phi<\Phi_e$, whereas it is the neutral black holes which dominate for greater values of the chemical potential.

The remarkable feature that at zero temperature and for large enough $\Phi$,  the extremal black holes were dominating \cite{RN1} is no longer true, as the extremal black holes are ill-defined except if $\Phi=\Phi_e$. As the temperature approaches zero, the black hole shrinks until it disappears completely and only the dilatonic background is left. This is not a proper phase transition as there are no competing solutions: only the background exists. If both $T=0$ and $\Phi=\Phi_e$, then the dilatonic background and the extremal black holes are competing, but none of them dominate the other.
 \item Intermediate and upper range: The black holes, charged or uncharged, are always globally and locally unstable and decay to the dilatonic background at all values of the temperature and chemical potential.
\end{itemize}

The heat capacity at constant chemical potential is given below,
\be
  C_\Phi = \frac{2}{1-\da^2}\le(\frac{T}{3-\da^2}\ri)^{\frac{2}{1-\da^2}}\le(1-\frac{\Phi^2}{\Phi_e^2}\ri)^{-2\da^2\frac{3-\da^2}{1-\da^4}}\,,
  \label{HeatCapacityGrandCanonical1}
\ee
and is positive for $\da^2<1$ and negative otherwise.


\subsubsection{Canonical ensemble}

The canonical ensemble is defined by keeping the temperature and the electric charge fixed.Then \eqref{Temperature1} can be considered as an implicit equation of state giving $r_+$ in terms of $T$ and $Q$:
\be
	 T=\frac{3-\da^2}{\ell}\le(\frac{r_+}{\ell}\ri)^{1-\da^2}\le[1-\frac{4(1+\da^2)Q^2}{\da^2(3-\da^2)^2r_+^{6-2\da^2}} \ri]^{1-2\frac{(\da^2-1)^2}{(1+\da^2)(3-\da^2)}}\,.
	\label{EquOfStateCanonical1}
\ee

\begin{figure}[t]
\begin{center}
\begin{tabular}{cc}
	 \includegraphics[width=0.45\textwidth]{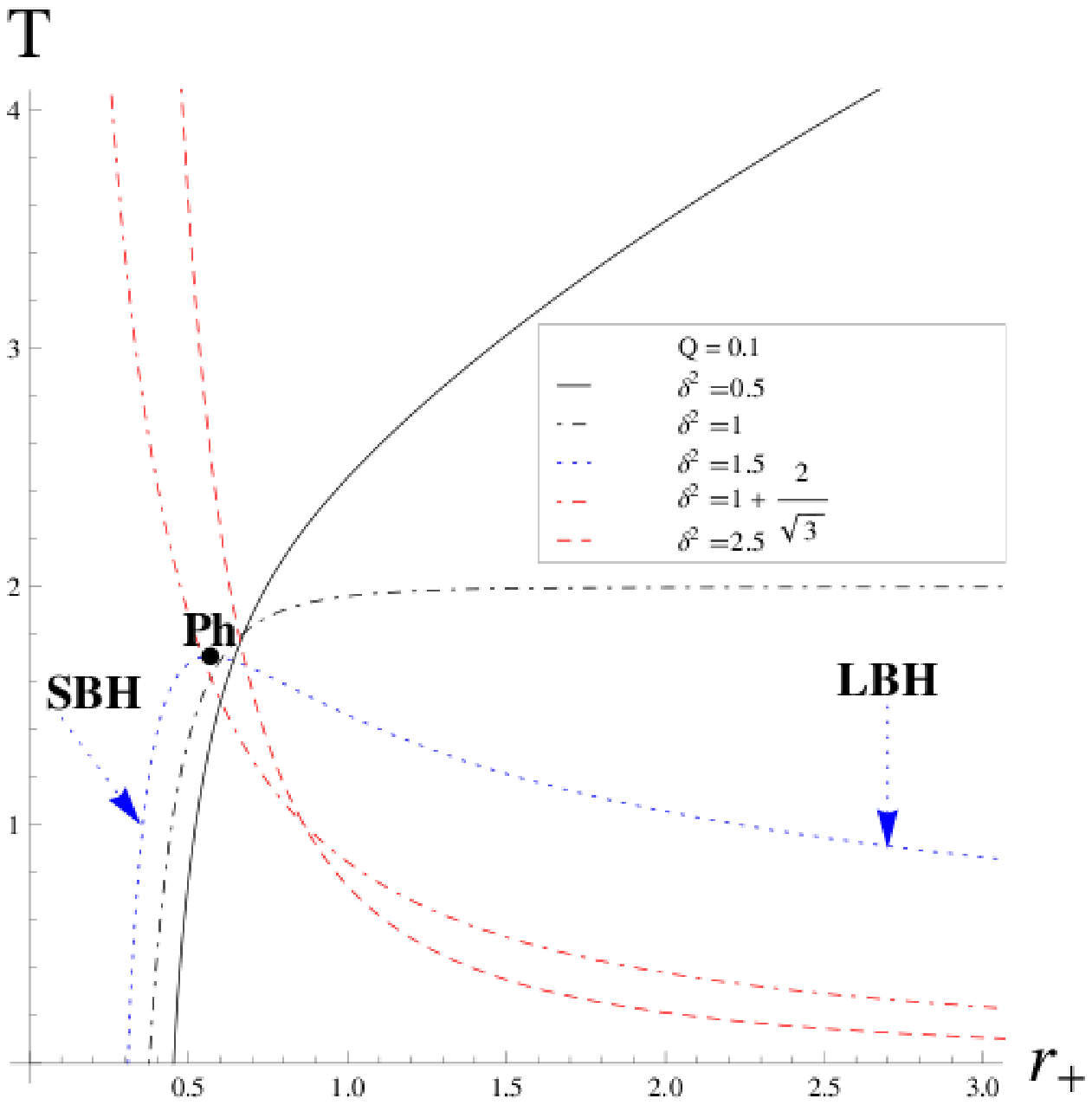}&	
	 \includegraphics[width=0.45\textwidth]{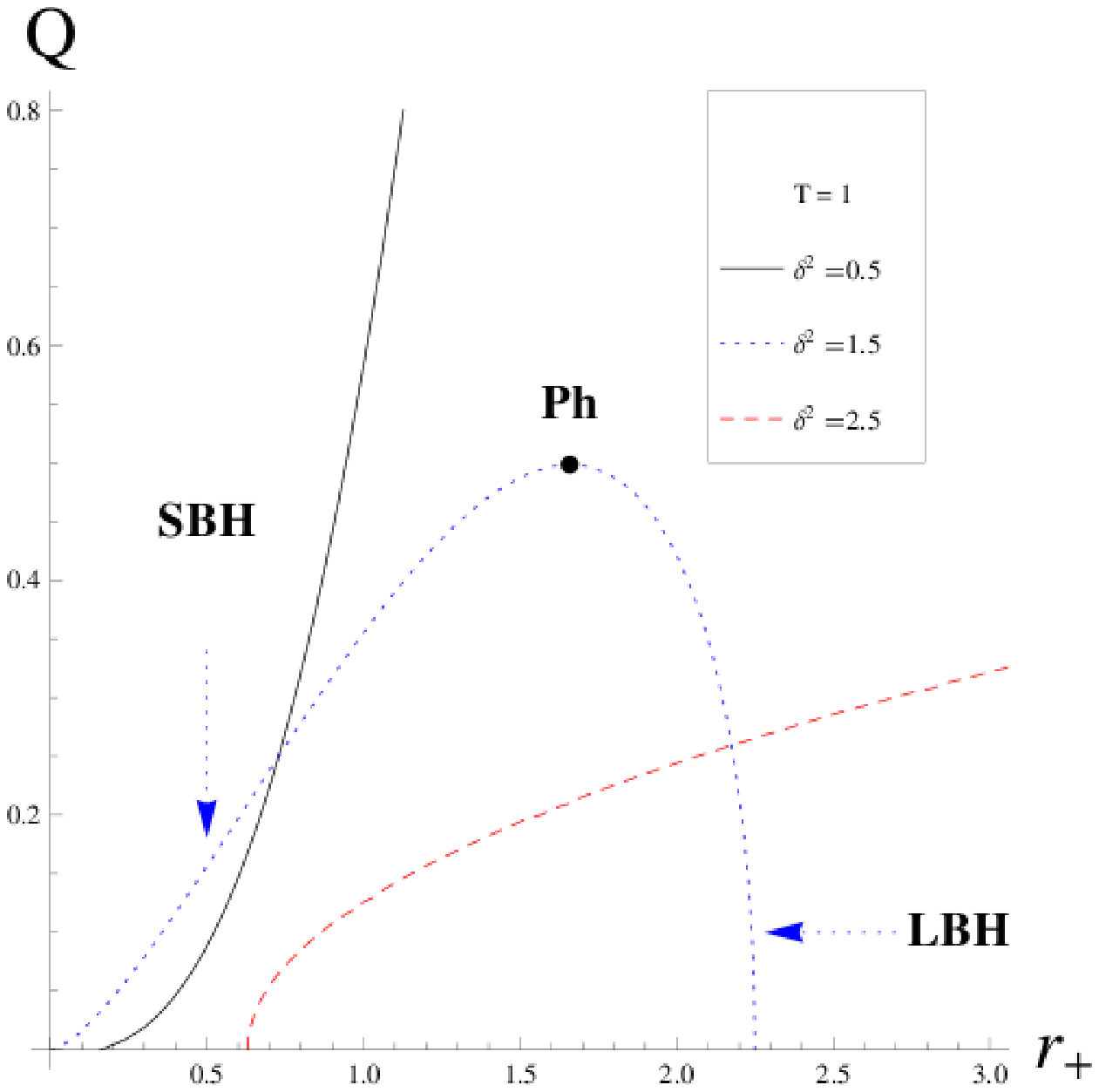}
 \end{tabular}
\caption{Temperature and charge versus horizon radius $r_+$.}
\label{Fig:HorizonSizeGaDa1Canonical}
\end{center}
\end{figure}

\begin{figure}[t]
\begin{center}
\begin{tabular}{cc}
	 \includegraphics[width=0.45\textwidth]{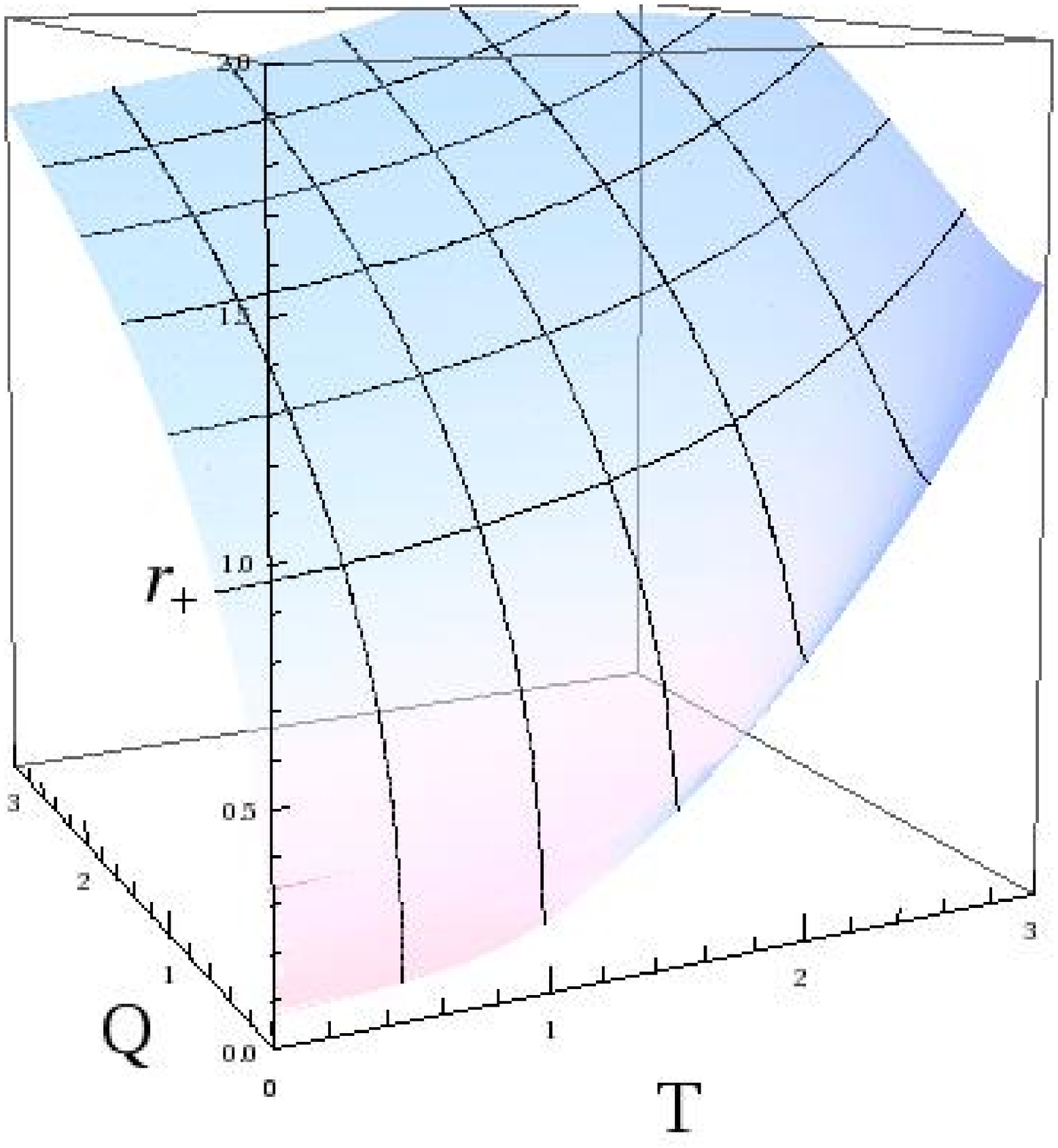}&
	 \includegraphics[width=0.45\textwidth]{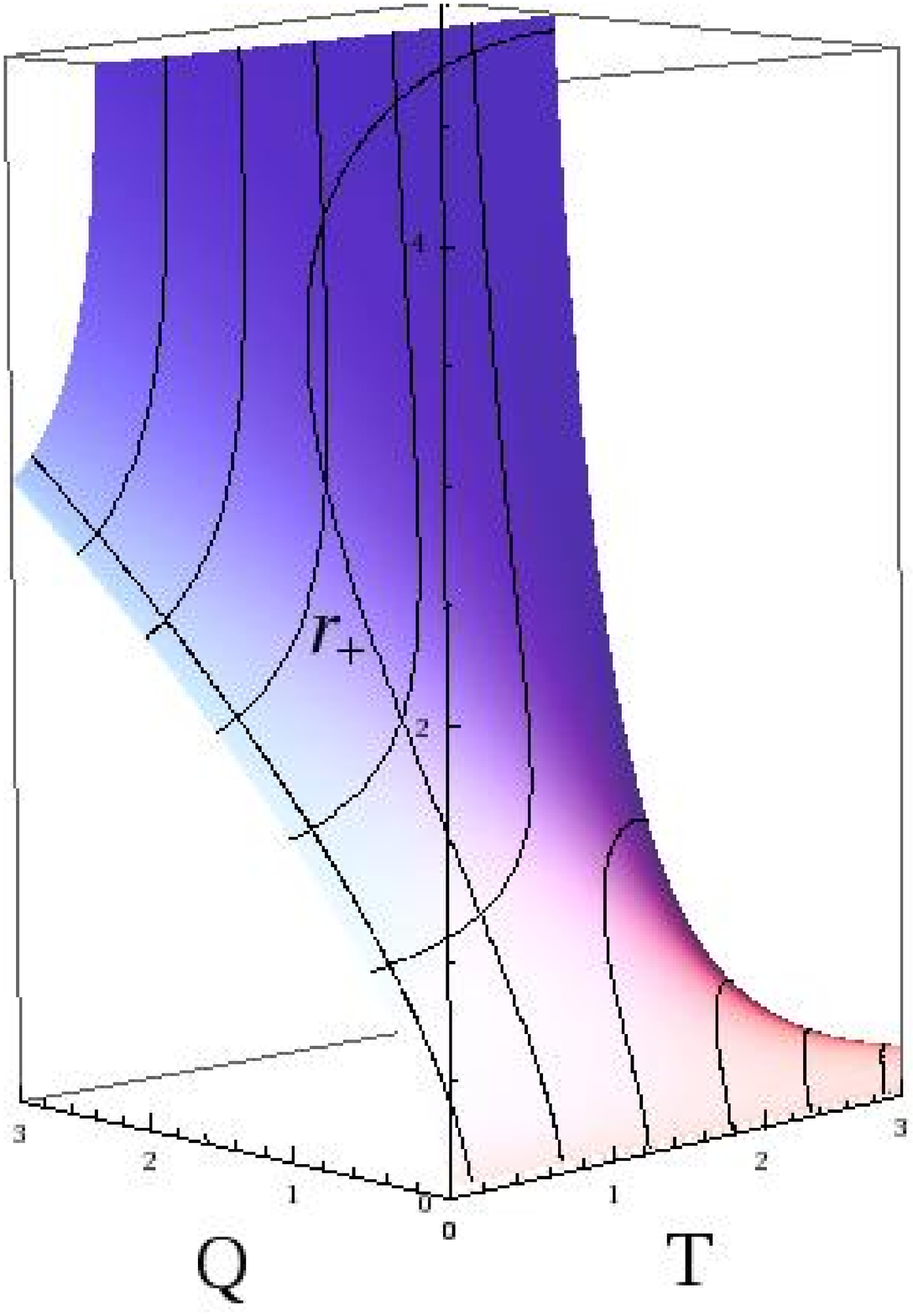}
 \end{tabular}
\caption{Three-dimensional equation of state $r_+(T,Q)$ for the lower ($0\leq\da^2<1$) and intermediate ($1<\da^2<1+\frac2{\sqrt{3}}$) range from left to right.}
\label{Fig:EquationOfState3DGaDa1Canonical}
\end{center}
\end{figure}

We plot it  for various values of the coupling constant $\da$ in \Figref{Fig:HorizonSizeGaDa1Canonical} and \Figref{Fig:EquationOfState3DGaDa1Canonical}. We notice three different types of behaviour:
\begin{itemize}
 \item  In the lower range, $\da^2\leq1$, there is a single black hole branch for each doublet $(T,Q)$.
This is similar to the $\da=0$ AdS-Reissner-Nordstr\"om black holes for $q<q_\mathrm{crit}$, see Fig.3 in \cite{RN1}.
The extremal limit $r_-\to r_+$ has zero temperature, and the radius of the black hole grows with the temperature.
At $Q=0$, the endpoint of the curve on the right in \Figref{Fig:HorizonSizeGaDa1Canonical} is the neutral black hole.
For the limiting case $\da^2=1$, there is a maximal temperature for large black holes, prefiguring the behaviour in the next range.
 \item In the intermediate range, $1<\da^2<1+\frac2{\sqrt3}$, there are two branches, small black holes (SBH) and large black holes (LBH),
which merge at a transition point,
\be
	 r_\mathrm{Ph}^{6-2\da^2}=\frac{\le(-5\da^4+12\da^2+1\ri)}{(\da^4-1)}r_e^{6-2\da^2}\,.
	\label{TransitionRadiusCanonical1}
\ee
This radius corresponds either to a maximal temperature at fixed electric charge, or to a maximum charge at fixed temperature,
see \Figref{Fig:HorizonSizeGaDa1Canonical}. The critical point exists only for $1<\da^2<1+\frac2{\sqrt3}$, and
the critical temperature is given by:
\be
	 T_\mathrm{Ph}=(3-\da^2)\ell^{\da^2-2}r_\mathrm{Ph}^{1-\da^2}\le(\frac{-6\da^4+12\da^2+2}{-5\da^4+12\da^2+1}\ri)^{\frac{-3\da^4+6\da^2+1}{(3-\da^2)(1+\da^2)}}\,.
	\label{TransitionTemperatureCanonical1}
\ee
Again, this relates to Fig.3 in \cite{RN1}, except that the third branch is gone due to the modified UV asymptotics of the solutions. The small black holes branch end on the other side to the extremal black holes at $T=0$ and $r_+=r_e$, while the radius of the large black holes diverges in the zero temperature limit. This alone would cast doubt on their physical relevance, see the energetic analysis below. At $Q=0$, the endpoint of the curve on the right in \Figref{Fig:HorizonSizeGaDa1Canonical} is either the neutral black hole (non-zero radius) or the background extremal black hole, which in this case has zero radius and so coincides with the dilatonic background.
 \item In the upper range $1+\frac2{\sqrt3}\leq\da^2<3$, there is again a single branch. The temperature diverges in the extremal limit $r_-\to r_+$, which consitutes a lower bound for the black hole size. This is however expected for dilatonic black holes \cite{Preskill:1991tb,Holzhey:1991bx} and signals the breakdown of their statistical description. On the other side, the large black holes have arbitrarily small temperature. For the limiting case $\da^2=1+\frac2{\sqrt3}$, the horizon size is actually independent of the electric charge and therefore has no minimum : the temperature diverges for zero radius.
\end{itemize}

\begin{figure}[t]
\begin{center}
\begin{tabular}{cc}
	 \includegraphics[height=3in]{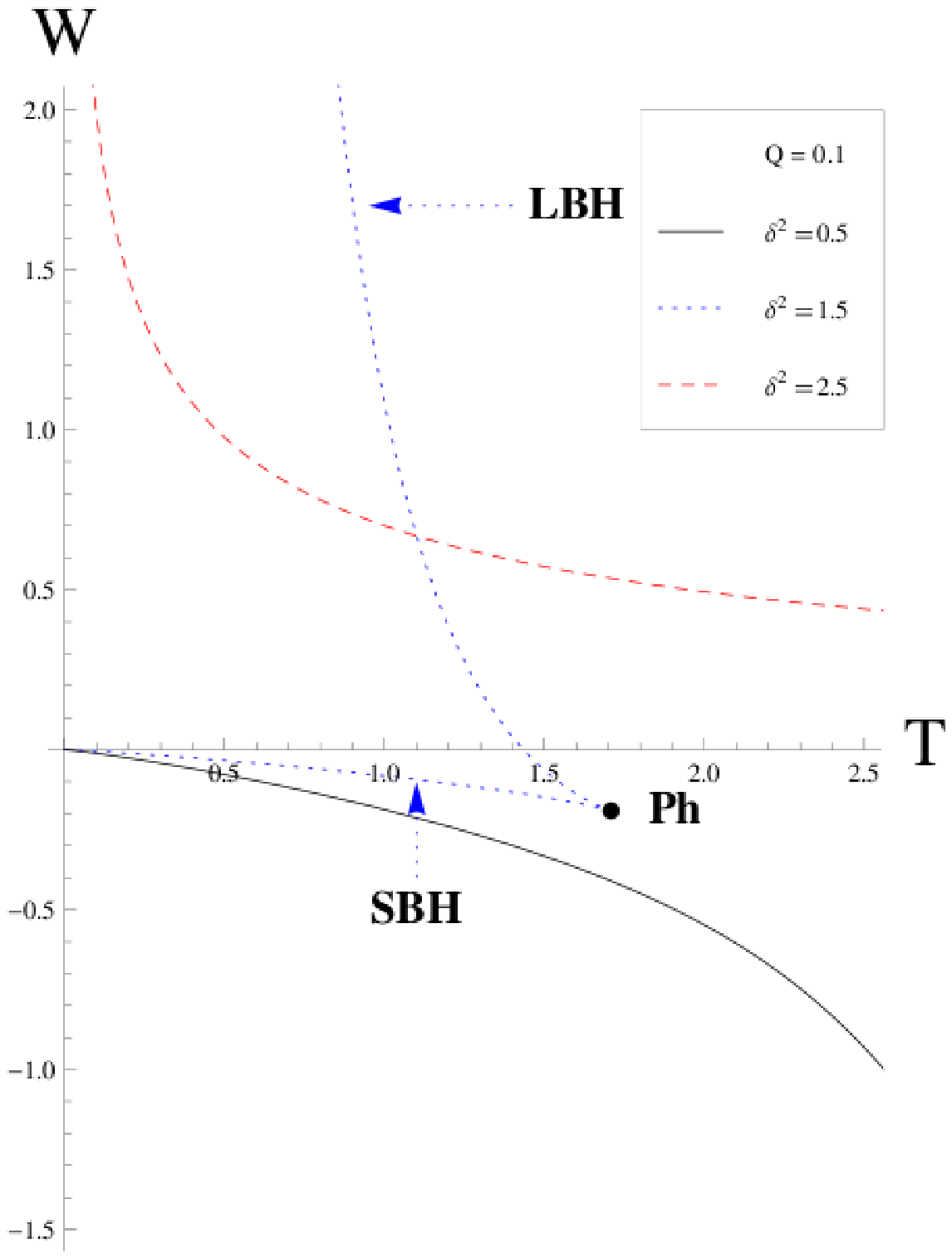}
	 &\includegraphics[height=3in]{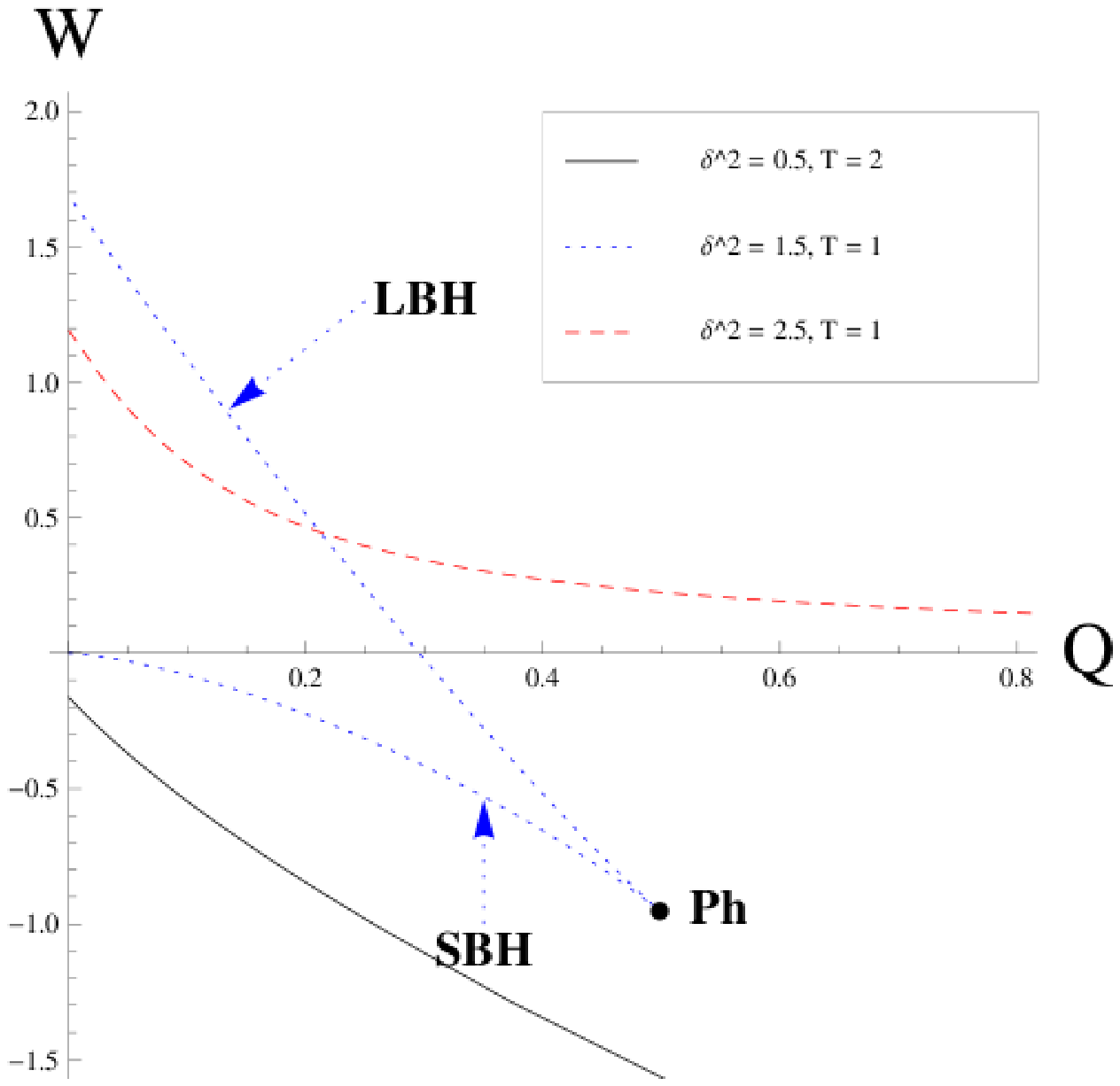}
 \end{tabular}
\caption{Slices of the Helmholtz potential at fixed charge and temperature.}
\label{Fig:FreeEnergyGaDa1Canonical}
\end{center}
\end{figure}

Calculating the value of the action \eqref{ActionCanonical} in the canonical ensemble (where we have subtracted the value of the Euclidean action for the extremal black hole, which we use as the thermal background \cite{RN1}),
\be
	 \tilde I - \tilde I_e = -\beta \ell^{\da^2-2}\le[(\da^2-1)r_+^{3-\da^2}-\frac{5\da^4-12\da^2-1}{1+\da^2}r_-^{3-\da^2}-\frac{4\da^2(3-\da^2)}{1+\da^2}(r_e)^{3-\da^2} \ri],
\ee
yields the Helmholtz potential of the black hole,
\be
	W[T,Q] =\ell^{\da^2-2} \le[(\da^2-1)r_+^{3-\da^2}-\frac{4(5\da^4-12\da^2-1)Q^2}{\da^2(3-\da^2)^2r_+^{3-\da^2}} -8\sqrt{\frac{\da^2}{1+\da^2}}Q\ri],
	\label{Helmholtz1}
\ee
which we plot at fixed charge density  or at fixed temperature in \Figref{Fig:FreeEnergyGaDa1Canonical} for the two ranges.

The energy of the solution is, \eqref{EnergyCanonical},
\be
	E_W = E-E_e = 2\ell^{\da^2-2}\le[(r_+)^{3-\da^2}-\frac{4(2\da^4-5\da^2+1)Q^2}{\da^2(3-\da^2)^2r_+^{3-\da^2}}\ri] - 8Q\sqrt{\frac{\da^2}{1+\da^2}}\,,
	\label{EnergyCanonical1}
\ee
the chemical potential is, \eqref{ElectricPotentialCanonical},
\be
	\Phi_W= \Phi-\Phi_e = \frac{16Q\ell^{2-\da^2}}{(3-\da^2)r_+^{3-\da^2}}-8\sqrt{\frac{\da^2}{1+\da^2}}\,,
	\label{ElectricPotentialCanonical1}
\ee
the entropy \eqref{EntropyCanonical} is equal to one quarter of the area of the horizon,
\be
	S_W = S-S_e =r_+^2\le[1-\frac{4(1+\da^2)Q^2}{\da^2(3-\da^2)^2r_+^{6-2\da^2}} \ri]^{2\frac{(\da^2-1)^2}{(3-\da^2)(1+\da^2)}}\,.
	\label{EntropyCanonical1}
\ee
since the entropy of the extremal background is zero, and goes to zero at extremality, indicating the presence of a non-degenerate ground state, \cite{Preskill:1991tb,Holzhey:1991bx}.
We derive the scaling of the entropy at very low temperatures and find,
\bear
  S_W &=& r_e^2\le[(6-2\da^2)\frac{T}{T_0}\ri]^{\frac{2(\da^2-1)^2}{-3\da^4+6\da^2+1}}\le[1+\frac{2\da^2(3-\da^2)^2}{-3\da^4+6\da^2+1}\le(\frac{T}{T_0}\ri)^{\frac{(3-\da^2)(1+\da^2)}{(-3\da^4+6\da^2+1)}}+\cdots\ri],\nn\\
  T_0&=& \le[2(3-\da^2)\ri]^{2-2\frac{(\da^2-1)^2}{(3-\da^2)(1+\da^2)}}\frac{r_e^{1-\da^2}}{2\ell^{2-\da^2}}\,.
  \label{EntropyScalingCanonical1}
\eear
We observe that for small temperatures, the entropy scales linearly with the temperature for the special value $\da^2=1+\frac25\sqrt5$. The validity of this approximation is for $T\ll T_0$\footnote{the numerical factor in front of the subleading term is of order one for $\da^2=1+\frac25\sqrt5$.}.

Moreover, the entropy will vanish at extremality (zero temperature), for any value of $\da^2$ belonging to the lower and intermediate ranges where the black holes are well-behaved. This indicates the presence of a non-degenerate ground state and is of direct interest for condensed matter systems. We can compare with the neutral black hole entropy, which scales like the temperature as,
\be
  S_N = \ell^2\le[\frac{\ell T}{(3-\da^2)}\ri]^{\frac2{1-\da^2}}\,.
  \label{EntropyNeutralBH}
\ee
The neutral black holes are well-behaved for $\da^2<1$, so it makes sense only to compare them with the charged black holes in this range. The difference between the two leading powers in the temperature is,
\be
  \frac{2(\da^2-1)^2}{-3\da^4+6\da^2+1}-\frac2{1-\da^2}=-\frac{2\da^2(3-\da^2)^2}{(1-\da^2)(-3\da^4+6\da^2+1)}<0\,,\qquad\forall\,\da^2<1\,,
  \label{115}\ee
and this shows that the neutral black holes contribution is always subdominant in the ranges of interest. This implies that the dominant part  of the entropy at very low temperatures  is due to the charge density. As we will see also in the sequel this seems to be a generic property of the theories we are analyzing.

\begin{figure}[t]
\begin{center}
\begin{tabular}{cc}
	 \includegraphics[width=0.45\textwidth]{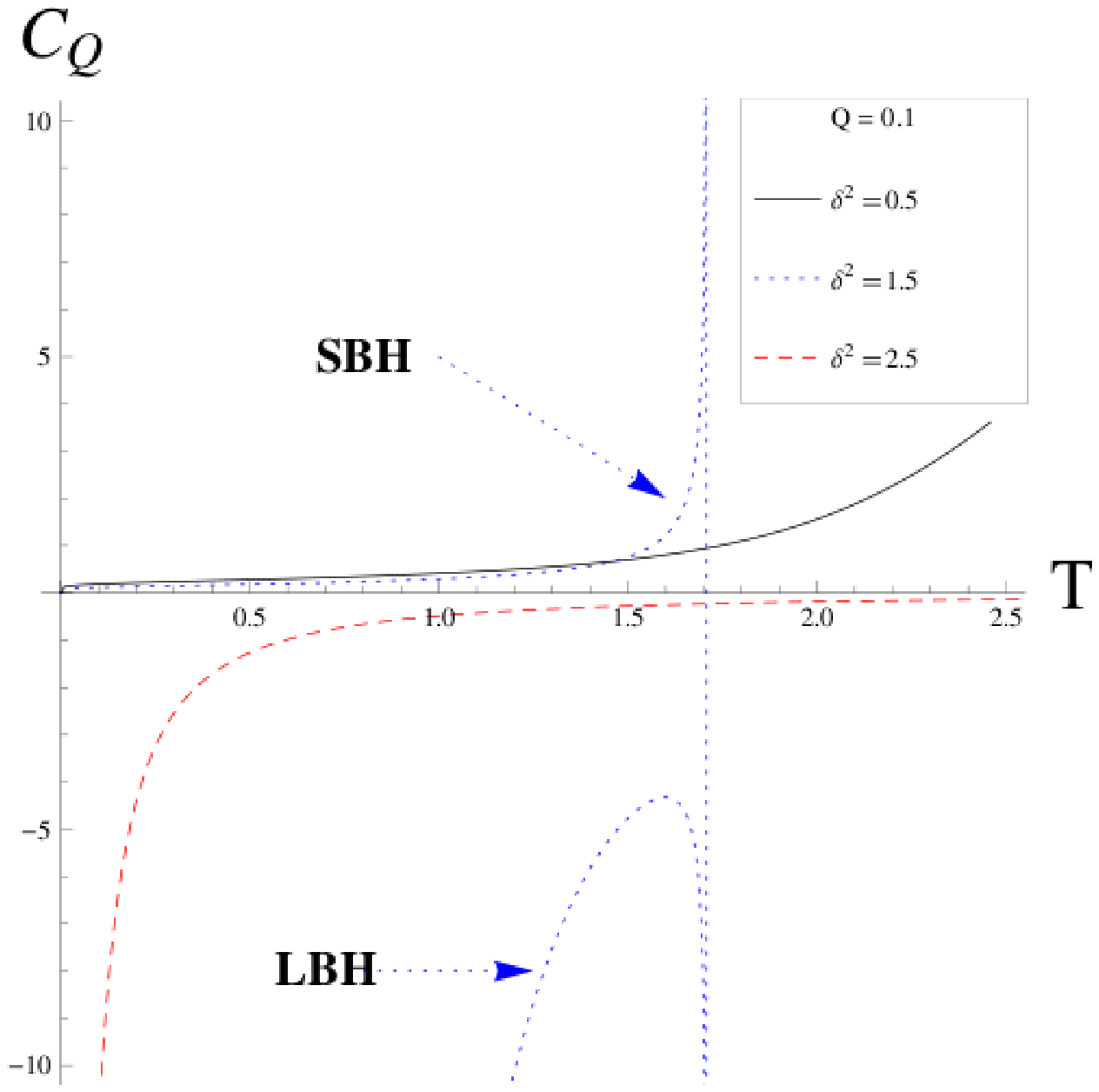}
	 &\includegraphics[width=0.45\textwidth]{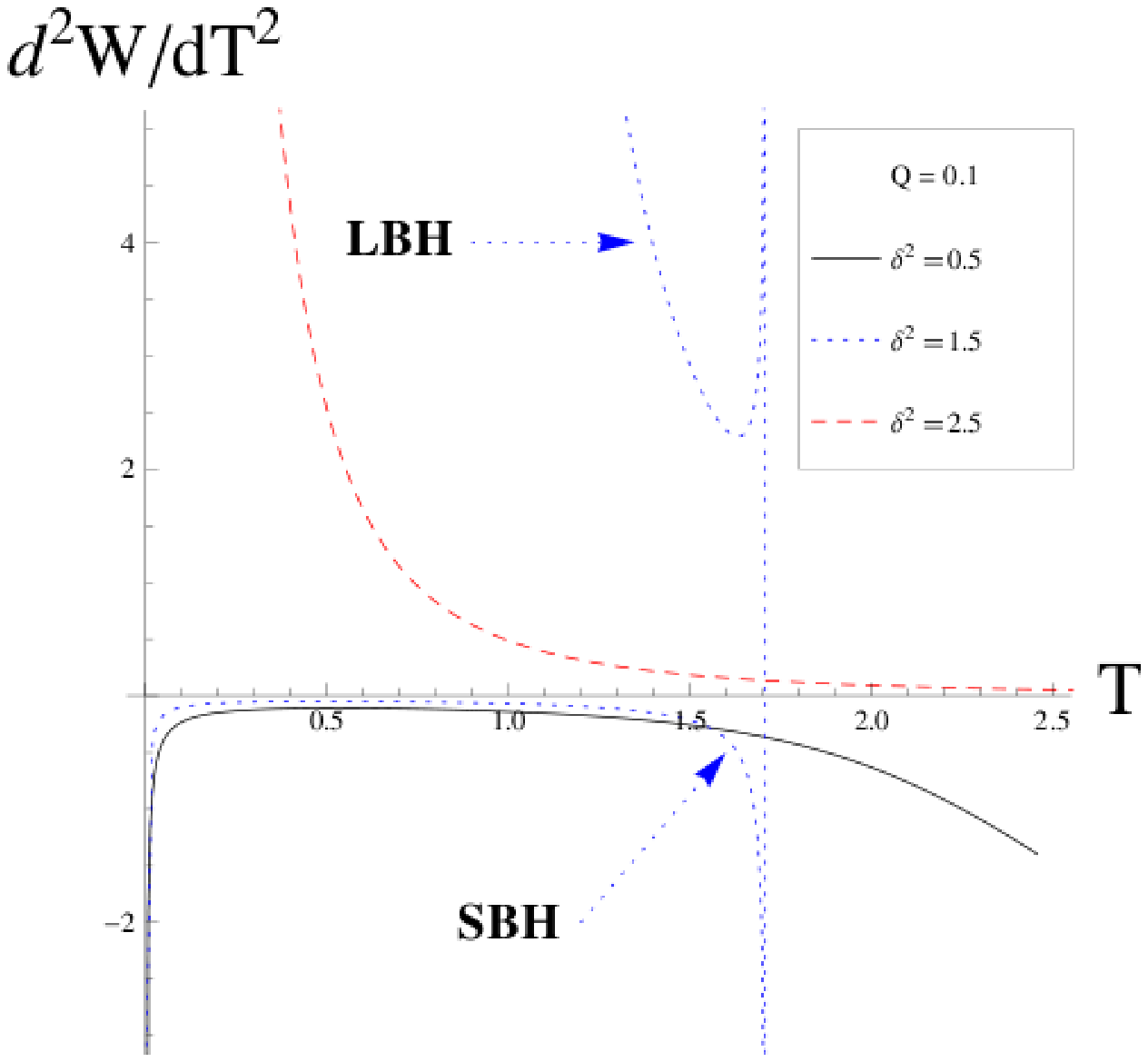}
 \end{tabular}
\caption{Heat capacity and double derivative of the Helmholtz potential against the temperature as a function of the temperature.}
\label{Fig:HeatCapacityGaDa1Canonical}
\end{center}
\end{figure}

We also derive the heat capacity at constant electric charge, \Figref{Fig:HeatCapacityGaDa1Canonical},
\be
	C_{Q} = \frac{2S}{1-\da^2}\le[1-\frac{4(2\da^4-5\da^2-1)Q^2}{\da^2(3-\da^2)^2r_+^{6-2\da^2}}\ri]\le[1-\frac{4(-5\da^4+12\da^2+1)Q^2}{\da^2(3-\da^2)^2(\da^2-1)r_+^{6-2\da^2}}\ri]^{-1}\,.
	\label{HeatCapacity1}
\ee
which scales linearly with the temperature at first-order for the same special value as the entropy, $\da^2=1+\frac25\sqrt5$, and for $T\ll T_0$\footnote{the numerical factor in front of the subleading term is of order one for $\da^2=1+\frac25\sqrt5$.},
\bear
  C_Q &=& r_e^2\frac{(-\da^4+3\da^2+1)}{(-3\da^4+6\da^2+1)}\le[(6-\da^2)\frac{T}{T_0}\ri]^{\frac{2(\da^2-1)^2}{(-3\da^4+3\da^2+1)}}\times \nn\\
      &&\quad\times\le[1+\frac{\da^2(3-\da^2)(\da^6-12\da^4+21\da^2+10)}{(-\da^4+3\da^2+1)(-3\da^4+6\da^2+1)}\le(\frac{T}{T_0}\ri)^{\frac{(3-\da^2)(1+\da^2)}{-3\da^4+6\da^2+1}}+\cdots\ri].
      \label{LowTHeatCapacityCanonical1}
\eear
The linear approximation has the same validity as for the entropy, when $T<<T_0$.

\begin{figure}[t]
\begin{center}
\begin{tabular}{cc}
	 \includegraphics[width=0.45\textwidth]{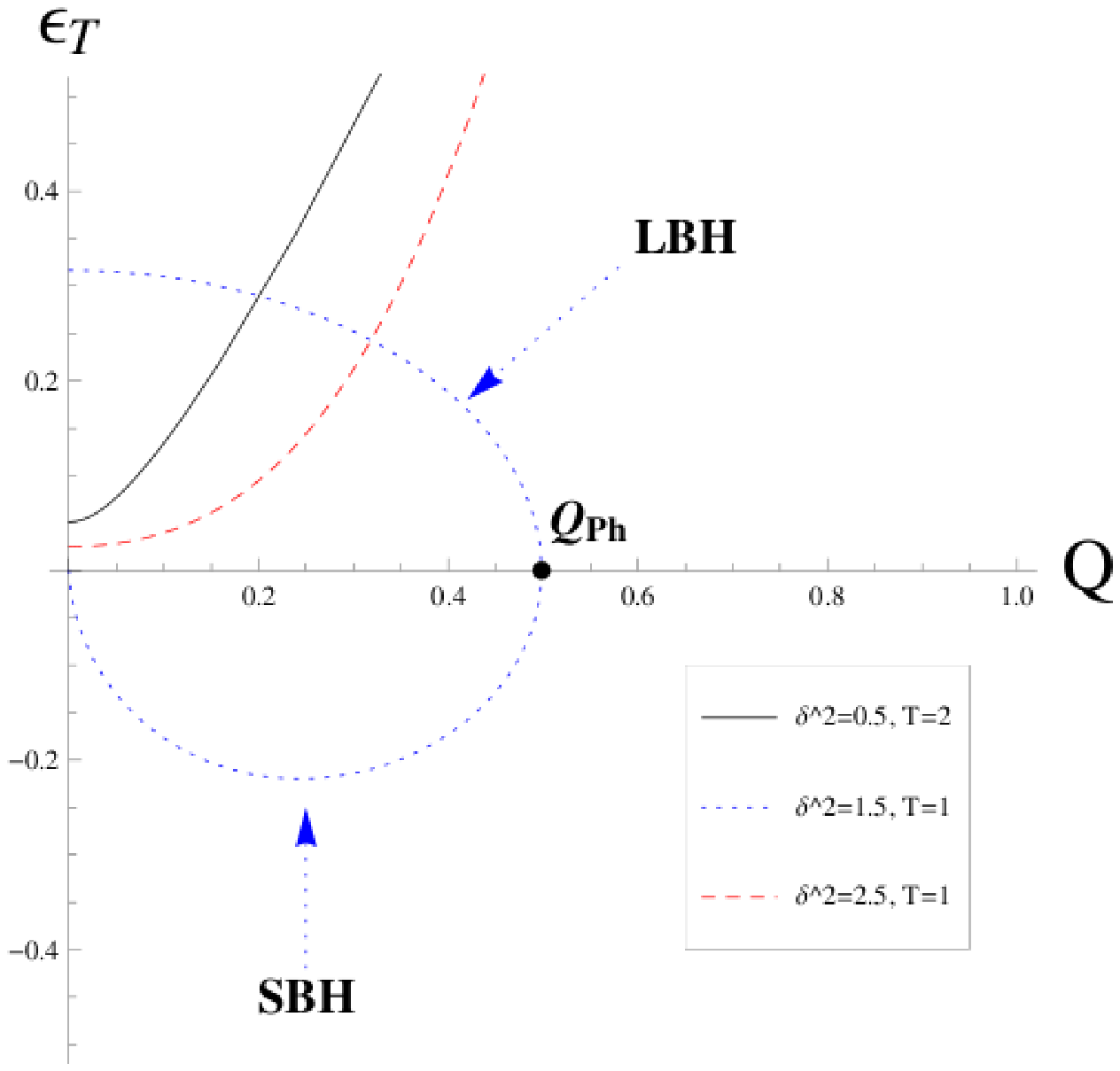}&	
	 \includegraphics[width=0.45\textwidth]{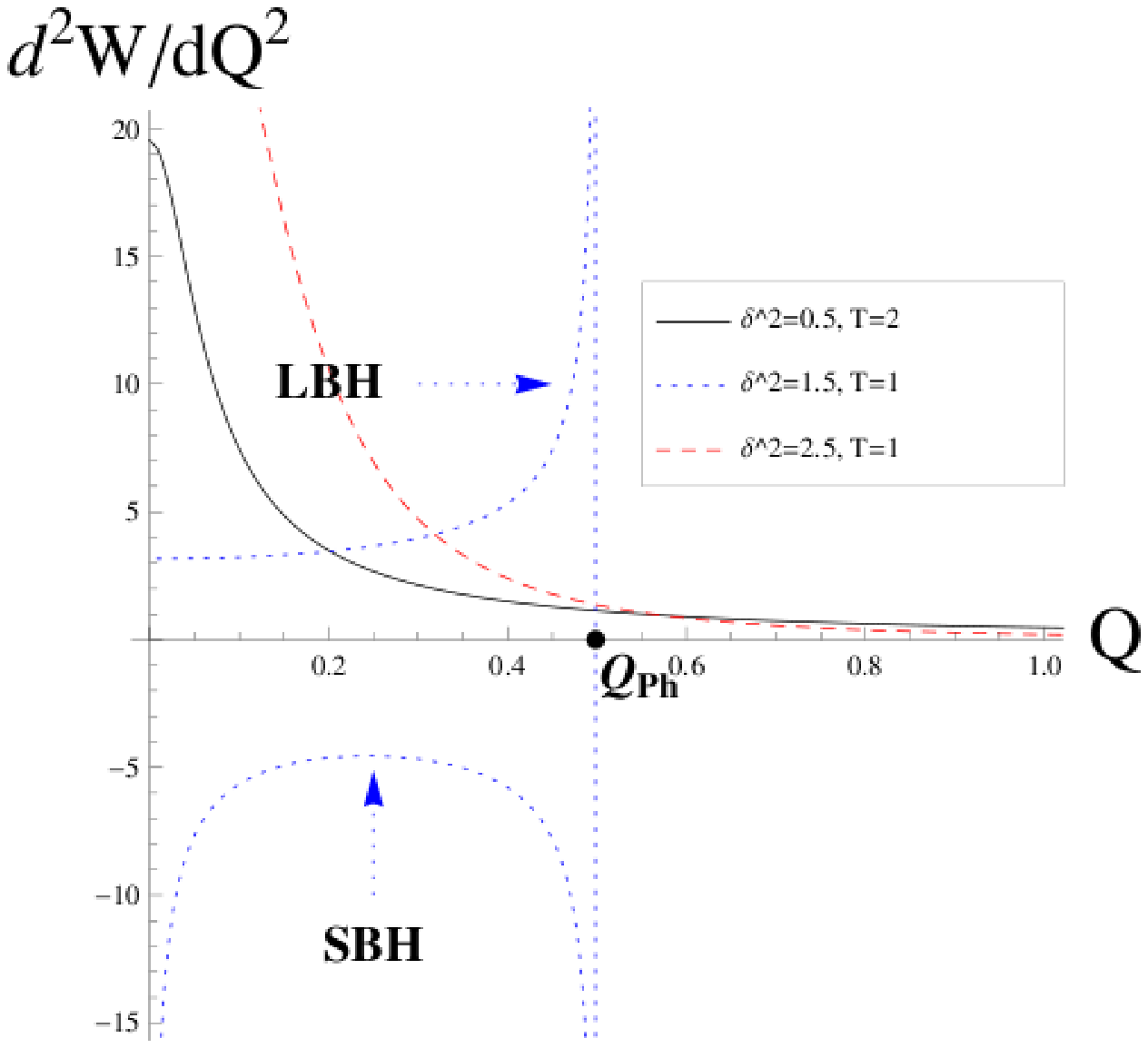}
 \end{tabular}
\caption{Electric permittivity and double derivative of the Helmholtz potential against the charge.}
\label{Fig:ElectricStabilityGaDa1Canonical}
\end{center}
\end{figure}

The electric permittivity at constant temperature is, \Figref{Fig:ElectricStabilityGaDa1Canonical},
\be
	\epsilon_T = \frac{(3-\da^2)r_+^{3-\da^2}}{16\ell^{2-\da^2}}\frac{1-\frac{4(-5\da^4+12\da^2+1)Q^2}{\da^2(3-\da^2)^2(\da^2-1)r_+^{6-2\da^2}}}{1-\frac{4(1+\da^2)l^{4-2\da^2}Q^2}{\da^2(3-\da^2)^2r_+^{6-2\da^2}}}\,.
	\label{ElectricPermittivity1}
\ee

This ensures that the following first law is verified,
\bea
	\ud(E - E_e) = T\ud S + (\Phi-\Phi_e)\ud Q &\Longleftrightarrow& \ud E_W = T\ud S_W +\Phi_W\ud Q\,,
\eea
as expected since we use the extremal black hole as a thermal background.

Let us study the energetic competition between the black holes and the thermal background, e.g. the extremal limit.
\begin{itemize}
 \item Lower range: the single black hole branch is energetically favoured over the thermal background as the Helmholtz potential is always negative, all the way to the $Q=0$ axis for fixed temperature, where the neutral black holes are seen to dominate instead of the extremal background. Moreover, the black holes are stable both to thermal  \Figref{Fig:HeatCapacityGaDa1Canonical} and electric fluctuations \Figref{Fig:ElectricStabilityGaDa1Canonical}.

 \item Intermediate range : the Helmholtz potential is always more negative for the small black holes branch than for the large black holes branch. The latter crosses the $F=0$ plane at some $(T,Q)$ but this is irrelevant to the thermodynamics. Both branches exist only in a certain region of the $(T,Q)$ plane limited by the line $(T_\mathrm{Ph},Q_\mathrm{Ph})$ and beyond which only the thermal background exists.

     At $Q=0$ and for finite temperature, the neutral black holes have positive free energy (they are the endpoint of the large black holes branch) and so it is the extremal background that dominates (which is the endpoint of the small black holes branch). We know from the analyses of \cite{gkmn2} that a large black-hole appears in that case above a certain temperature, $T_{min}$. For lower temperatures it is the extremal background (with discrete spectrum and a mass gap that dominates). At a higher temperature $T_c>T_{min}$ there is a first order phase transition to the large black-hole phase. The transition becomes continuous if $T_{min}=T_c$.

     The small black holes always dominate the phase space, and are stable to thermal fluctuations \Figref{Fig:HeatCapacityGaDa1Canonical} but not to electric fluctuations  \Figref{Fig:ElectricStabilityGaDa1Canonical}. The large black holes display the opposite behaviour.

 \item Upper range: Since the Helmholtz potential is positive for all values $(T,Q)$, the regular black holes are always disfavoured compared to the thermal background. The black-holes are unstable in this range, and the situation is very
\end{itemize}
In the case of the lower and intermediate range, these plots are to be compared with Fig.5 in \cite{RN1} and Figs.4, 6 in \cite{RN2} in which the so-called branch 3 does not exist as argued above.

\subsection{The phase structure}

\subsubsection{The Grand-Canonical ensemble}

\begin{figure}[t]
\begin{center}
\begin{tabular}{cc}
	 \includegraphics[width=0.45\textwidth]{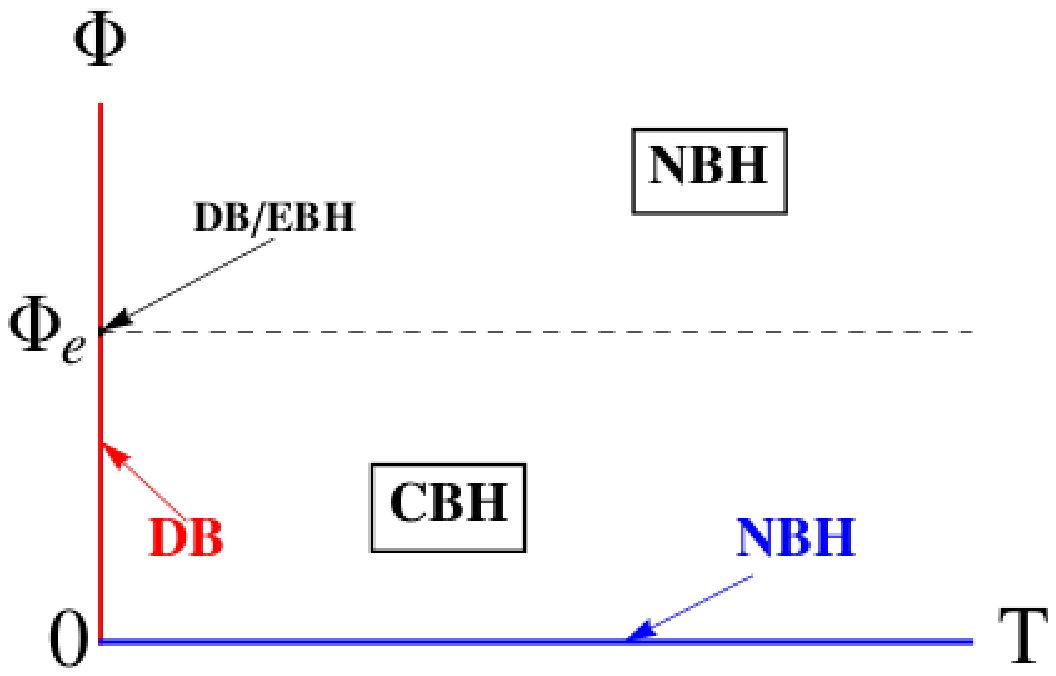}&
	 \includegraphics[height=2in]{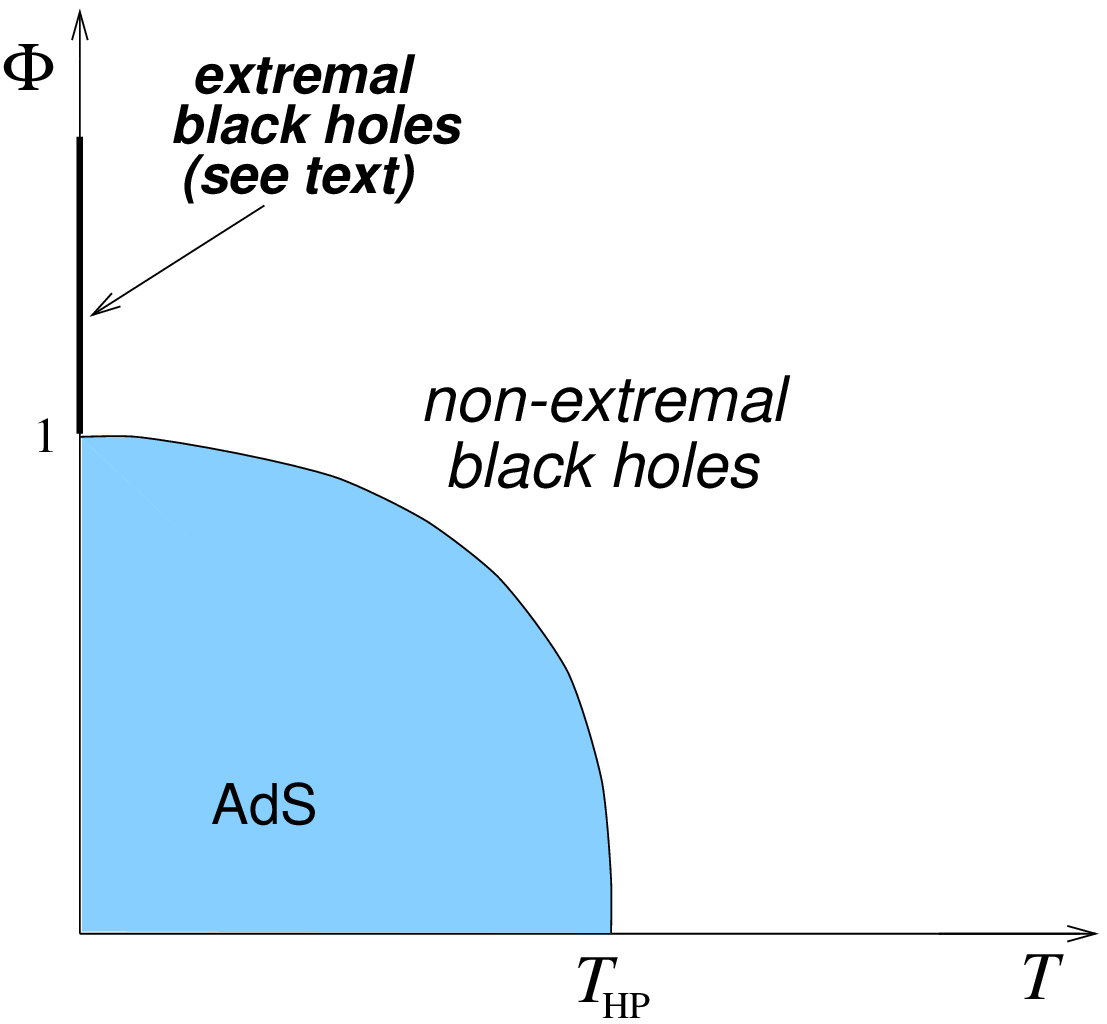}
 \end{tabular}
\caption{On the left, phase diagram in $(T,\Phi)$ space for the lower range, $\da^2 = 0.5$. CBH = Charged Black Holes, NBH = Neutral Black Holes, DB = Dilatonic Background, EBH = Extremal Black Holes. On the right, for comparison, we reproduce Fig.12 of \cite{RN1} displaying the phase space for spherical AdS-RN black holes}
\label{Fig:PhaseDiagramGaDa1GrandCanonical}
\end{center}
\end{figure}

We present the phase diagram for the lower range in \Figref{Fig:PhaseDiagramGaDa1GrandCanonical}, which is the only one that is non-trivial.

In the region $\Phi<\Phi_e$, the charged black holes dominate at finite temperature, and so do the neutral black holes on the vertical axis $\Phi=0$. {The transition to the neutral black holes is smooth, in particular no phase transition happens, as is obvious from \eqref{Gibbs1}.}

On the horizontal axis  $T=0$, only the dilatonic background exists (no black holes) as the extremal black holes can only be reached at the point $\Phi=\Phi_e$. The approach to zero temperature exhibits critical behaviour, there is an $n^\textrm{th}$-order phase transition to the dilatonic background for values
\be
\frac{n-4}{n-2} < \delta^2 < \frac{n-3}{n-1}\,,\quad n=4,5,6\dots
\ee
Note that the upper bound goes to $1$ as $n\to+\infty$, allowing for abitrarily high-order phase transitions.

If $\Phi\to\Phi_e^-$ and $T$ is finite, then the Gibbs potential diverges and there is a zeroth-order phase transition to the neutral black holes. For $\Phi\geq\Phi_e$ and finite temperature, the neutral black holes dominate the ensemble over the dilatonic background. It makes sense on physical grounds to expect that the naked singularity of \eqref{LinearDilaton} should preferably be cloaked by an event horizon. At zero temperature and $\Phi=\Phi_e$, the equation of state \eqref{EqOfStateGaDa1GrandCanonical} for $\delta^2<1$
can still be satisfied, but \eqref{EuclideanActionGC1} yields $G=0$ identically, and so we cannot
discriminate at this particular point between the dilatonic background and
the extremal limit.

We can compare with the phase space in the case of AdS-RN spherical black holes, see Fig.6 of \cite{RN1}. The effect of missing the asymptotically AdS region (and therefore the large black hole branch) is to destroy the Hawking-Page transition, and so the black holes dominate the whole phase space. Then, the effect of the scalar field is that charged black holes exist only up to some critical value for the chemical potential, and then there is a transition to the neutral black holes. Furthermore, the extremal black holes now exist only for a specific value of the chemical potential, and do not dominate the phase space anywhere. This can be seen comparing the equation of state \eqref{EqOfStateGaDa1GrandCanonical} with equation (20) of \cite{RN1}, which we reproduce below (notwithstanding some numerical factors),
\be
  \Phi = \sqrt{r_+^2 -2r_+ T +\kappa}\,,
\label{116}\ee
where we have explicitly introduced the normalised spatial curvature of the horizon, $\kappa=0,\pm1$ for planar, spherical or hyperbolic geometry. Specifically, one sees that for zero temperature zero $\kappa$, one can always obtain in the planar RN case an extremal black hole, whereas \eqref{EqOfStateGaDa1GrandCanonical} forces either $\Phi=\Phi_e$ (fixed value) to keep $r_+=r_e$ finite, or $r_+=0$ (no black hole).

\subsubsection{The Canonical Ensemble}

\begin{figure}[t]
\begin{center}
\begin{tabular}{cc}
	 \includegraphics[width=0.45\textwidth]{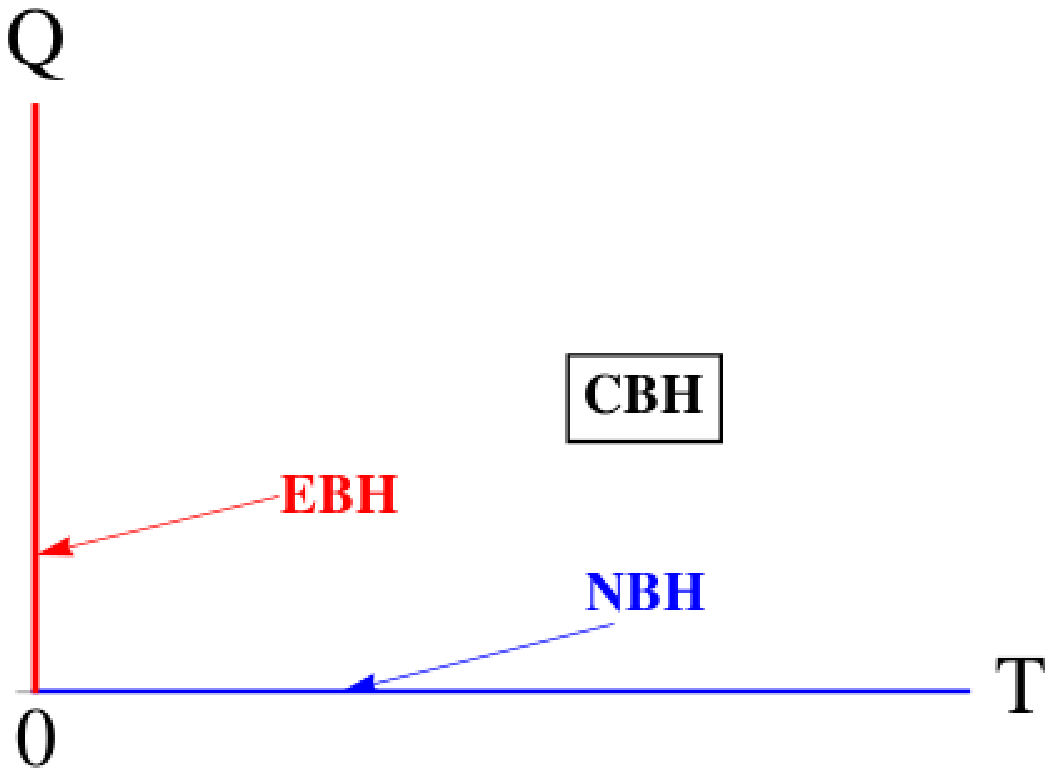}&
	 \includegraphics[width=0.45\textwidth]{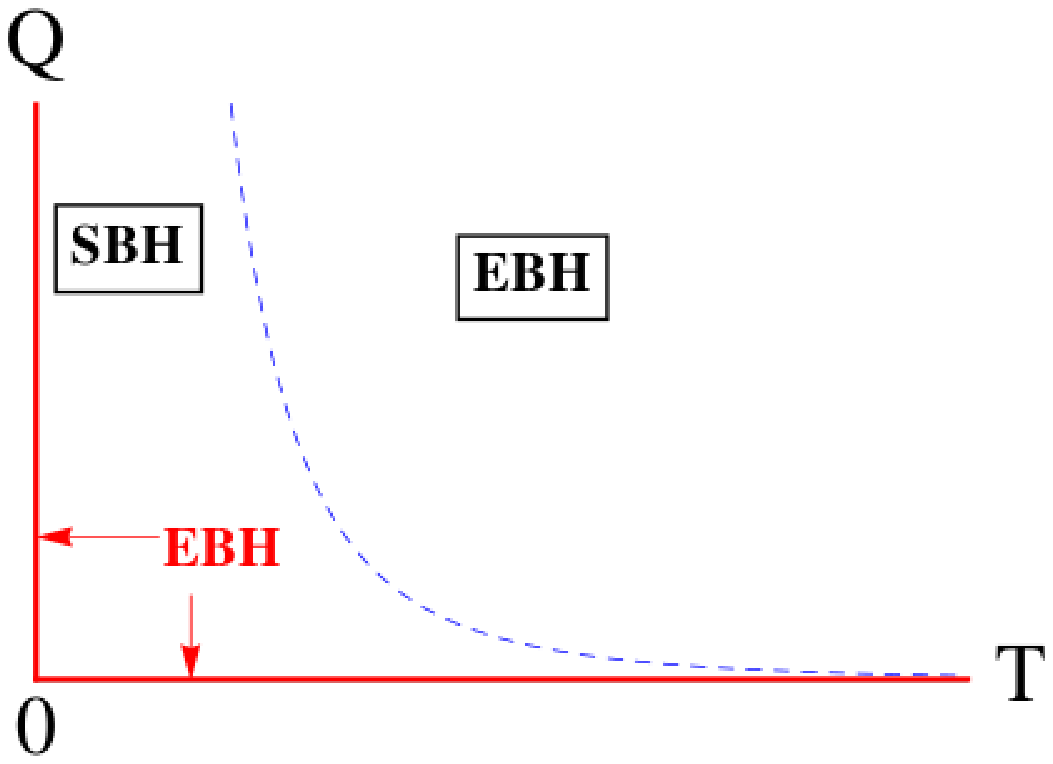}
\end{tabular}
\caption{$(T,Q)$ phase diagram for the lower (left) and the intermediate (right) range. EBH = Extremal Black Holes, CBH = Charged Black Holes, SBH = Small Black Holes.}
\label{Fig:PhaseDiagramGaDa1Canonical}
\end{center}
\end{figure}

\begin{figure}[!ht]
\begin{center}
\begin{tabular}{c}
	 \includegraphics[width=0.45\textwidth]{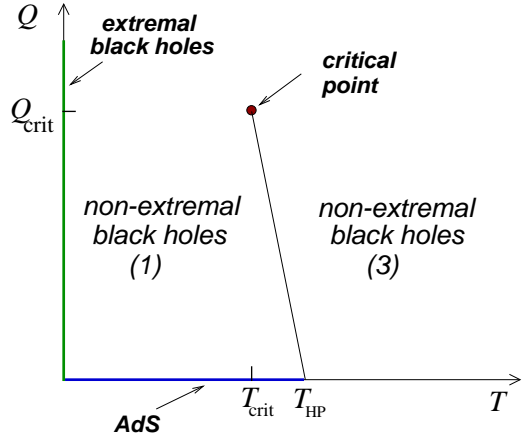}
\end{tabular}
\caption{$(T,Q)$ phase diagram in the case of spherical AdS-RN black holes on the bottom line, taken from \cite{RN1}.}
\label{Fig:PhaseDiagramRNAdSCanonical}
\end{center}
\end{figure}

The order of phase transitions is as follows, inspecting the Helmholtz potential \Figref{Fig:FreeEnergyGaDa1Canonical} and eq.~\eqref{Helmholtz1}.
\begin{itemize}
 \item Lower range: {There is a continuous phase transition to the extremal background at finite charge in the zero-temperature limit. Inspecting \eqref{Helmholtz1}, one finds that the transition is third-order for $0 < \d^2 < 1-\frac{2}{\sqrt{5}}$ and second-order for $1-\frac{2}{\sqrt{5}} < \delta^2 < 1$. There is no discontinuity in any derivative of the free energy at the points $\delta^2=(0,1\pm \frac{2}{\sqrt{5}})$, but subleading corrections to the scalar potential might induce critical behaviour here.

 For zero charge and small temperature, the neutral black holes dominate. This is the range in which (for $p=3$) these black holes have a reasonable thermodynamic limit even without AdS-completion, see sec.~\ref{zero} for details. The phase transition from the charged black holes is at least of fourth-order, and of $n^\textrm{th}$-order for
 \be\label{PTuncharged}
 \frac{n-4}{n-2} < \delta^2 < \frac{n-3}{n-1}\,,\quad n=4,5,6\dots
 \ee
Note that the upper bound goes to $1$ as $n\to+\infty$, thus spanning the whole lower range and admitting phase transitions of arbitrary order. Again, there is no transition at the boundary of these intervals, but transitions can be induced through subleading terms in the scalar potential.

Finally, approaching the zero charge axis of the phase diagram at finite temperature never shows critical behaviour, since the free energy \eqref{Helmholtz1} is a series expansion in integer powers of $Q$. The charged black holes settle continuously and without transition in their stable endpoints, the neutral black holes.
 }
 \item Intermediate range: There is a zeroth-order (discontinuous) phase transition to the thermal background at the point $(T_\mathrm{Ph},Q_\mathrm{Ph})$ since the Helmholtz potential is discontinuous there and jumps to zero\footnote{Once one corrects to full asymptotically AdS solutions then this will be generically becoming a first order phase transition to a RN-like large black hole phase. With special tuning this transition may become a second-  or higher-order transition as in \cite{gkmn2,gursoy1,gursoy2}.}.

{At finite charge, in the zero-temperature limit, the Helmholtz potential and its first derivative are continuous, but higher-derivatives  diverges for the small black holes branch: There is a continuous phase transition to the thermal background of $n^\textrm{th}$-order for the parameter values
\be
1 + 2 \sqrt{\frac{n-2}{3n-4}} < \delta^2 < 1 + 2 \sqrt{\frac{n-1}{3n-1}}\,.
\ee
Note that as $n\to+\infty$, one reaches the endpoint of the Intermediate Range, $\da^2=1+{2\over \sqrt3}$, so phase transitions of any order are possible. At the endpoints of these intervals there are no phase transitions with our choice of potential, but again subleading corrections to the scalar potential might change this. }

At zero charge and finite temperature, the (small) neutral black holes are unstable and the thermal background dominates. Above some finite temperature a first order transition  exists to a large black-hole phase.

At finite temperature, in the zero charge limit, one follows the stable branch (the small black-holes) to its endpoint, which has zero free energy and coincides with the thermal background. This is consistent with the fact that in the neutral case,  the black holes are stable only for $\da^2<1$. There is no phase transition because the free energy is an integer-power expansion in $Q$.

 \item Upper range: The non extremal (small black holes) are always unstable, so the extremal background dominates. There will be large black holes in the higher energy completion of the theory, that will dominate above some finite temperature.

\end{itemize}

We have plotted the phase diagram ($T,Q$) for the lower and intermediate ranges in \Figref{Fig:PhaseDiagramGaDa1Canonical}, and reproduced the phase diagram for spherical RNAdS black holes from Fig.6 in \cite{RN1}. As branch 3 is absent in our case, there is no first-order phase transition with branch 1 in the lower and intermediate ranges : the Hawking-Page transition is absent. In the latter, this also explains the zeroth-order phase transition, which in the vocabulary of \cite{RN1} corresponds to the maximal temperature where branch 1 (small black holes in our case) and branch 2 (large black holes in our case) merge. However, the fact that the extremal black hole background dominates on the $T=0$ axis is of course unchanged.

\subsection{The AC conductivity}

In this section we calculate the optical (AC) conductivity for small frequencies at zero temperature, i.e. for the extremal case of \eqref{Sol1}. We proceed along the lines of section~\ref{sec:ACconductivity}. The understanding of the small frequency, small temperature behaviour is particularly interesting, since the theory of quantum critical behaviour predicts universal scaling laws of the form
\be
\Re  \sigma (\omega) \sim \omega^n\,,
\label{117}\ee
in the strange metallic region around a quantum critical point.

The extremal limit of the solution~\eqref{Sol1} is reached for
\be
\left( \frac{r_\ast}{\ell} \right)^{3-\d^2}= \frac{m}{\ell} = \frac{q}{\ell} \frac{\sqrt{1+\d^2}}{2\d(3-\d^2)} = \tilde q\,.
\label{118}\ee
Since the low-frequency behaviour of the AC conductivity is only sensitive to the near-horizon geometry, it suffices to analyse the fluctuation problem \eqref{28} in the extremal near horizon geometry of \eqref{Sol1}, which reads
\bsea\nonumber
\ud s^2 &=& - \ell^2 \frac{ \rho^{2\frac{1+4\d^2-\d^4}{(3-\d^2)(1+\d^2)}} }{{\tilde q}^{\frac{2\d^2}{1+\d^2}}} \ud \tilde t^2
+ \frac{\ell^2 {\tilde q}^{\frac{2\d^2}{1+\d^2}}}{(3-\d^2)^2}  \rho^{2\frac{3\d^4 - 4\d^2 - 3}{(3-\d^2)(1+\d^2)}} \ud \rho^2 + \\
&&+ \ell^2 {\tilde q}^{\frac{2\d^2}{1+\d^2}} \rho^{\frac{2(\d^2-1)^2}{(3-\d^2)(1+\d^2)}} (\ud x^2 + \ud y^2)\,, \\
e^\Phi &=& {\tilde q}^{\frac{2\d}{1+\d^2}} \rho^{\frac{4\d(\d^2-1)}{(3-\d^2)(1+\d^2)}}\,.
\label{119}\esea
Here the near-horizon coordinate
\be
\rho = \left(\frac{r}{\ell}\right)^{3-\d^2} - \tilde q\,,
\label{120}\ee
has been introduced, and the time coordinate $t = \ell \tilde t$ has been rescaled in order to work with dimensionless coordinates only. Note that the AC conductivity is calculated in the canonical ensemble, in which the temperature and the charge density of the solution are fixed. Following the thermodynamic discussion of \eqref{Sol1}, we may restrict our attention for now to the case $\d^2 < 1 + \frac{2}{\sqrt{3}}$, which admits a sensible extremal limit in which the temperature of the black hole horizon tends to zero. \footnote{Note that the restriction coming from the consideration of boundary conditions at the naked singularity (see section~\ref{spectra}) of the uncharged background, $\d^2\leq \frac{5}{3}$, is even stronger than this bound. From the point of view of the construction of ``natural'' effective holographic theories the parameter range $\d^2> \frac{5}{3}$ is thus ruled out both by the breakdown of the thermodynamic description for $\d^2>1+\frac{2}{\sqrt{3}}$ and by the ``bad'' nature of the IR singularity at $\d^2>\frac{5}{3}$.}
The relevant coordinate change to transform the fluctuation problem \eqref{30a} into the Schr\"odinger problem \eqref{32} then is given by
\be
z(\rho) = \frac{(1+\d^2)  {\tilde q}^{\frac{2\d^2}{1+\d^2}}}{3\d^4-6\d^2-1} \rho^{\frac{3\d^4-6\d^2-1}{(3-\d^2)(1+\d^2)}}\,.
\label{121}\ee
In this coordinate system the extremal near-horizon metric reads
\bea
\ud s^2 &=& D(z) (-\ud t^2 + \ud z^2) + C(z)( \ud x^2 + \ud y^2)\,,\\
D(z) &=& \ell^2 {\tilde q}^{-\frac{2\d^2}{1+\d^2}} \left[ {\tilde q}^{-\frac{2\d^2}{1+\d^2}} \frac{3\d^4-6\d^2-1}{1+\d^2} z \right]^{2\frac{1+4\d^2-\d^4}{3\d^4-6\d^2-1}}\,,\\
C(z) &=& \ell^2 {\tilde q}^{\frac{2\d^2}{1+\d^2}} \left[ {\tilde q}^{-\frac{2\d^2}{1+\d^2}} \frac{3\d^4-6\d^2-1}{1+\d^2} z \right]^{\frac{2(\d^2-1)^2}{3\d^4-6\d^2-1}}\,,\\
e^\phi &=& {\tilde q}^{\frac{2\d}{1+\d^2}} \left[ {\tilde q}^{-\frac{2\d^2}{1+\d^2}} \frac{3\d^4-6\d^2-1}{1+\d^2} z \right]^{\frac{4\d(\d^2-1)}{3\d^4-6\d^2-1}}\,.
\label{122}\eea
The effective Schr\"odinger potential \eqref{32} in the near-horizon region (the horizon is at $z_H=-\infty$ for $\d^2<1+\frac{2}{\sqrt{3}}$) then reads
\be
V = \frac{c}{z^2}\,,\qquad c = \frac{2(\d^2(\d^2-4)-1)(\d^2(2\d^2-5)-1)}{(3\d^4-6\d^2-1)^2}\,.
\label{gd1Veff}
\ee
Note in particular that the extremal near-horizon potential does not depend on the charge density $q$. Since the Schr\"odinger potential has the same form as observed in \cite{kachru}, we would like to apply the matching procedure developed there in order to obtain the scaling behaviour of the AC conductivity. To justify this procedure, the full Schr\"odinger problem needs to be frequency-dominated near the horizon and near the boundary in an AdS completion, and potential-dominated in between.

\begin{figure}[!ht]
\begin{center}
\begin{tabular}{ccc}
	\includegraphics[width=0.3\textwidth]{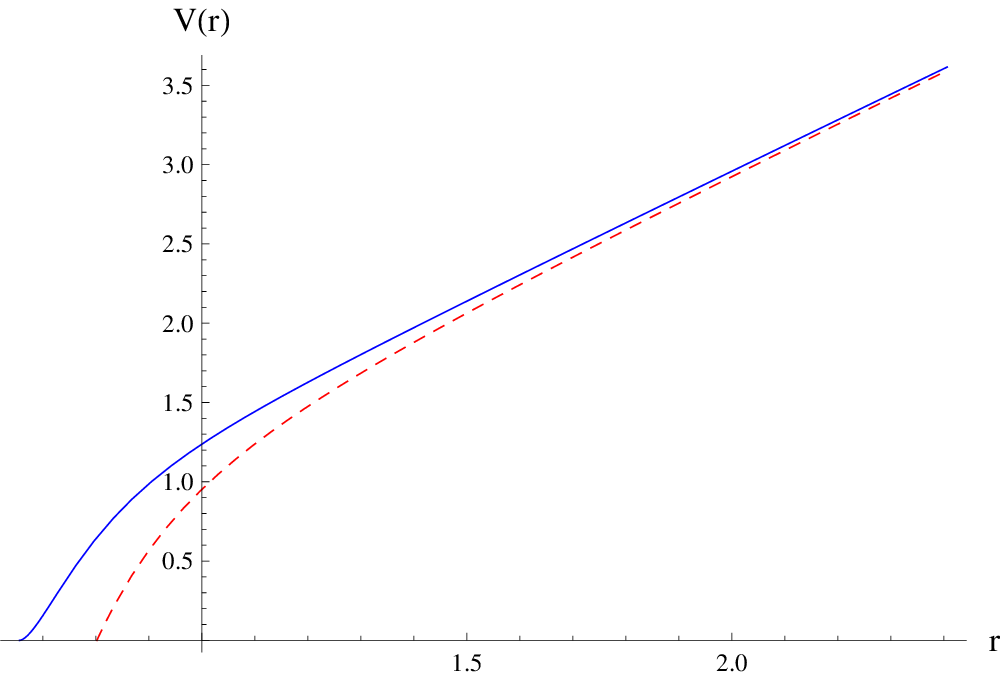}&
	\includegraphics[width=0.3\textwidth]{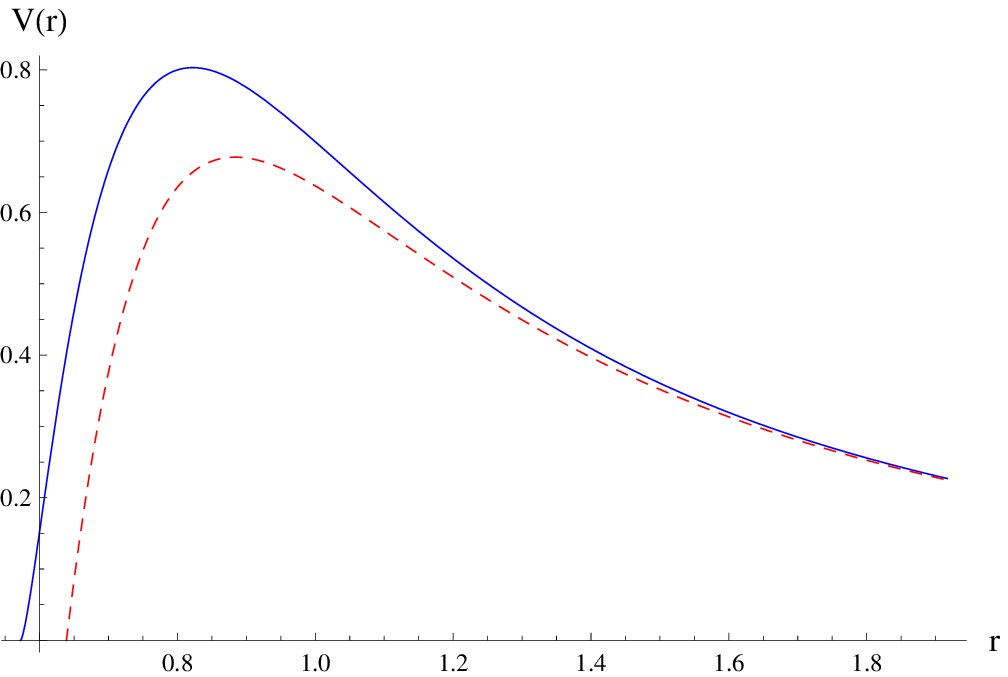}&	 \includegraphics[width=0.3\textwidth]{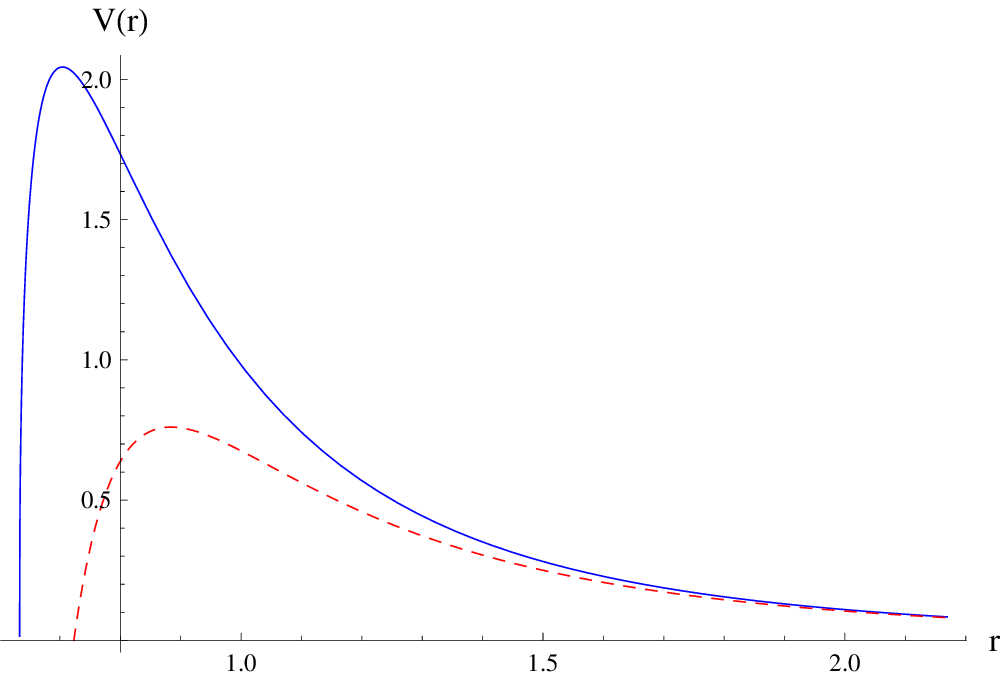}
\end{tabular}
\caption{The Schr\"odinger potential $V(r)$ for the $A_x$ fluctuation around the $\gamma\delta=1$ solution at extremality (blue solid curves) and slightly away from extremality (red dashed curves) in the range $\d^2<1+2/\sqrt{3}$. The non-extremality is parametrized by the non-extremality parameter $\epsilon$ via $m = \frac{q\sqrt{1+\d^2}}{2\d(3-\d^2)} + \frac{\epsilon^2}{2}$. From left to right, the plots show the respective potentials for  $(\d,\epsilon)=(0.7,0.3)$, $(1.3,0.1)$ and $(\sqrt{1+2/\sqrt{3}}-0.03,0.1)$. The horizon is at the left of each plot at the point where the the potential hits zero. The boundary is at the right of each plot. However the behavior of the potential there depends on the UV asymptotics not included here, but which were described in section \protect\ref{current}.}
\label{Fig:gd1VSlowmiddle}
\end{center}
\end{figure}

\begin{figure}[!ht]
\begin{center}
	 \includegraphics[width=0.45\textwidth]{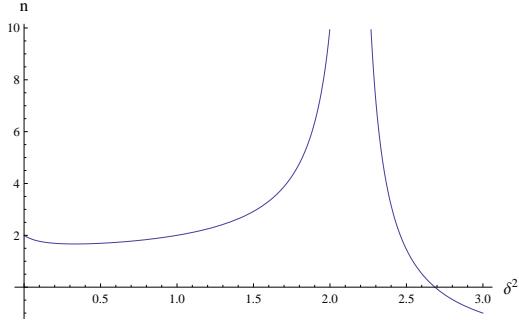}
\caption{The scaling exponent $n$ of the optical conductivity $\sigma (\omega) \sim \omega^n$ as a function of $\d^2$ ($0 \leq \d^2 \leq 3$).}
\label{gd1ACconductivity}
\end{center}
\end{figure}

\Figref{Fig:gd1VSlowmiddle} shows the full Schr\"odinger potential for various values of $\d^2<1+\frac{2}{\sqrt{3}}$, both at extremality (blue solid curves) and slightly away from extremality (red dashed curves). We find that away from extremality the potential changes qualitatively only near the horizon, according to the results in sec.~\ref{sec:ACconductivity}, and in particular that the near-horizon region is separated from the asymptotic region by a potential wall.\footnote{In the case when the full Schr\"odinger potential seems to diverge at large $r$ one needs to keep in mind that in an AdS completed solution the potential will drop to zero for $p=3$ at the boundary according to \eqref{34}.}
Applying the results of \cite{kachru}, we can thus immediately deduce the small frequency scaling of the optical conductivity to be
\be\label{gd1ACscaling}
\sigma(\omega) \simeq \omega^n \,,\qquad n = \sqrt{4c+1}-1 = \frac{(3-\d^2)(5\d^2+1)}{|3\d^4-6\d^2-1|} - 1\,.
\ee
The exponent $n(\d^2)$ is plotted in \Figref{gd1ACconductivity} for $0\leq \d^2 \leq 3$. It diverges at $\d^2 = 1+\frac{2}{\sqrt{3}}$, and otherwise is positive in the region $0 \leq \d^2 < 1+ \frac{2}{\sqrt{3}}$.
At the point $\d^2 = \frac{5+\sqrt{33}}{4}$
where $n$ vanishes (which is also the point at which the coefficient $c$ in the potential vanishes), subleading corrections to the Schr\"odinger potential will become important.
At $\d^2 = 1/3$ the minimum $n = 5/3$ is attained.

We therefore conclude that in the region where the IR black-holes are stable $\d^2<1+\frac{2}{\sqrt{3}}$, the solution \eqref{Sol1} reproduces low-frequency zero temperature scaling laws for the AC conductivity with positive scaling exponents $n$ only. Furthermore, the scaling exponent is bounded from below,
\be
n \ge \frac{5}{3}\,.
\label{124}\ee

The spectrum in this class  is continuous without a mass gap and the system remains a conductor at arbitrarily low temperatures.

\begin{figure}[!ht]
\begin{center}
\begin{tabular}{cc}
	 \includegraphics[width=0.45\textwidth]{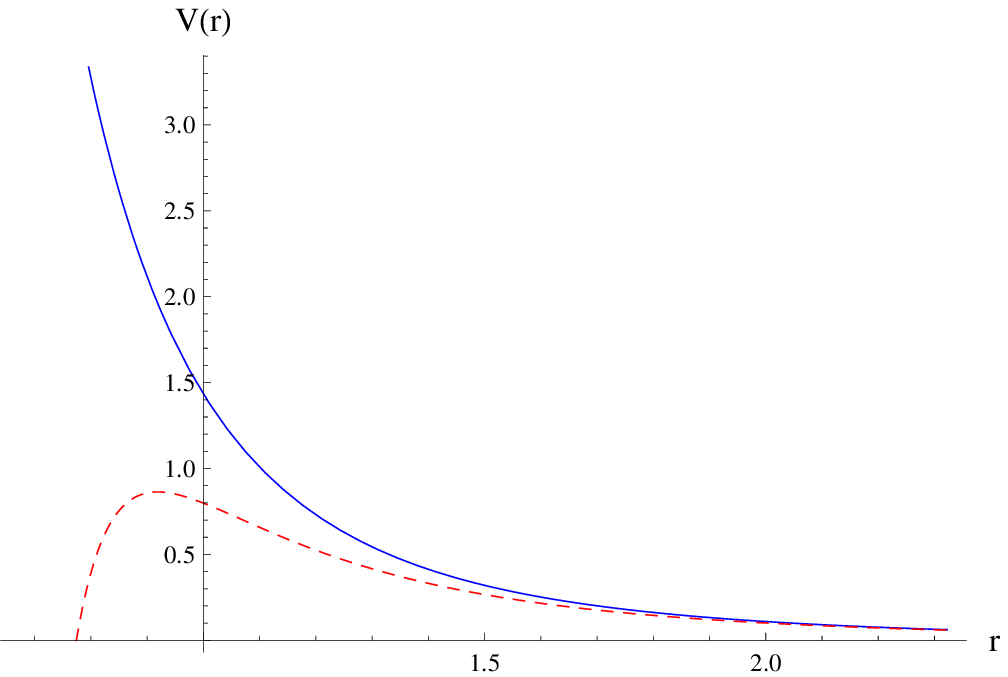}&
	 \includegraphics[width=0.45\textwidth]{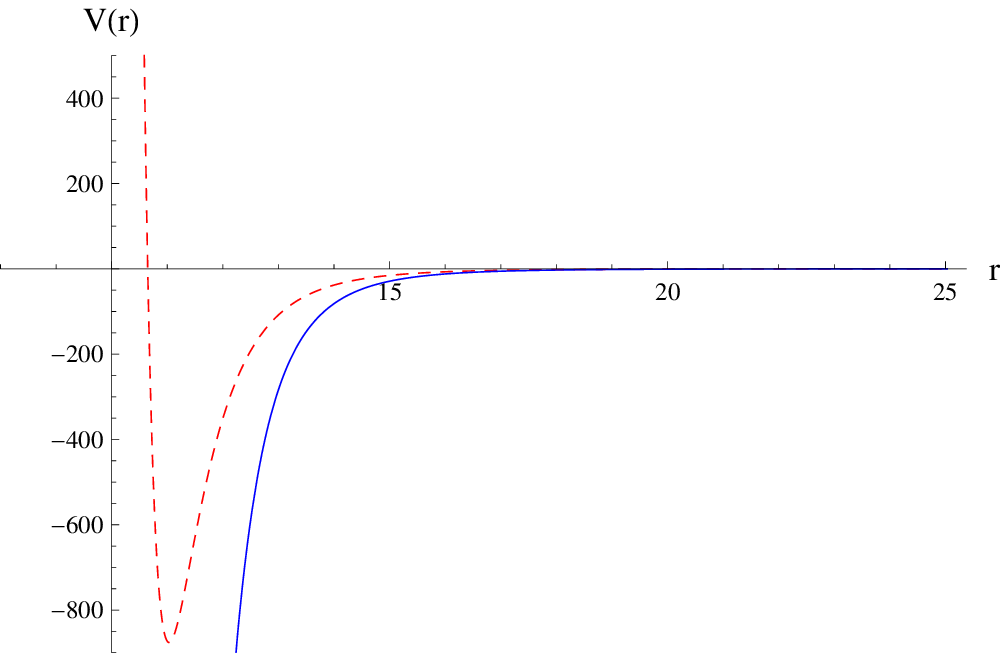}
	 \end{tabular}
\caption{The Schr\"odinger potential $V(r)$  for the $A_x$ fluctuation around the $\gamma\delta=1$ solution  at extremality (blue solid curves) and slightly away from extremality (red dashed curves) in the range $\d^2\ge1+2/\sqrt{3}$. The non-extremality is parametrized by the non-extremality parameter $\epsilon$ via $m = \frac{q\sqrt{1+\d^2}}{2\d(3-\d^2)} + \frac{\epsilon^2}{2}$. From left to right, the plots show the respective potentials for  $(\d,\epsilon)=(\sqrt{1+2/\sqrt{3}},0.1)$ and $\left(\frac{1}{2}\sqrt{5+\sqrt{33}}+0.002,0.01\right)$. The horizon and asymptopia is as in figure \protect\ref{Fig:gd1VSlowmiddle}, with the difference that the non-extremal case here is without thermodynamic significance.}
\label{Fig:gd1VSupper}
\end{center}
\end{figure}

On the other hand, in the region $3\ge \d^2\ge 1+2/\sqrt{3}$, the situation is different, as can be seen from \Figref{Fig:gd1VSupper}. In this region of $\d$, the horizon is at $z=0_+$, and the asymptotically AdS region is also at $z=0_+$, i.e. the $z$-coordinate is multi-valued. There are two qualitatively different regimes, depending on whether the near-horizon  potential at extremality \eqref{gd1Veff} is diverging to $+\infty$ ($c>0$), or to $-\infty$ ($c<0$). In both cases $V\simeq {c\over z^2}$ with $z\to 0$ and $c$ still given by  \eqref{gd1Veff}.

The division line between these two regimes is
$$\d_\ast = \frac{1}{2}\sqrt{5+\sqrt{33}}\,,$$
at which $c(\d_\ast)=0$. The left plot in \Figref{Fig:gd1VSupper} shows the situation for $c>0$: At extremality the potential diverges as $+1/z^2$.
 Taking into account the behavior at the UV, detailed in section \ref{current} we conclude that when the dimension of $\phi$ in the UV is ${1\over 2}\leq \Delta\leq 1$ then this describes a finite density system whose charged excitations have a gap and a discrete spectrum.\footnote{The spectrum is continuous and gapless if  $\Delta\geq 1$.}  It is therefore an insulator.
 It resembles very much to a 2+1-dimensional Mott insulator, where insulating behavior is expected due to strong interactions.\footnote{John McGreevy informed us that they also found a qualitatively similar solution in a theory with $\delta=0$.}

 As the temperature is raised from zero, the IR black-hole we found is not thermodynamically dominant. In analogy with the chargeless case analyzed in a UV complete context in \cite{gkmn2}, we expect that the extremal solution will dominate the thermodynamics up to a temperature $T_c>0$. At some lower temperature $T_{min}<T_c$ a new black hole solution will appear due to the behavior of the potential at higher energies. There will generically be a first-order phase transition from the gapped solution to the large black-hole solution that will be a conductor. This will be an example of a insulator/conductor transition. As $\d^2 \to 1+\frac{2}{\sqrt{3}}$ this transition is expected to become continuous.

 On the other hand for $c<0$ (right plot in \Figref{Fig:gd1VSupper}),
 the coefficient $c\geq -{1\over 4}$ as long as $\d^2\leq 3$.
 Although the potential diverges, the quantum mechanical problem is well defined and we expect not to have any negative energy levels. Therefore in this case the system seems to have  again a continuous spectrum and is probably gapless.

\subsection{The DC conductivity}

We calculate the DC resistivity, for $\gamma\delta  =1$ solution using the drag force formula derived in section (7.2). The DC resistivity is given in (\ref{55}).
For $\gamma\delta  = 1$ case, it reads
\be
	\rho
	\simeq \frac{T_f }{J^t}   \left( \frac{r_+}{\ell} \right)^{2 + 2 \delta k}
	\le[ 1 -\left(\frac{r_e^2}{r_+^2}  \right)^{3-\delta^2} \ri]^{\frac{2(\delta^2 -1)
	(\delta^2 - 1 + 2 k d)}{(3-\delta^2) (1+\delta^2)}}
	 \ ,
\label{125}\ee
where $r_e$ is the horizon radius in the extremal limit and we use the relation $r_+ r_- = r_e^2  $.
For the canonical ensemble, the total charge density  is fixed and thus $J^t$ is constant. It can be readily checked
that the resistivity has dimension $[\rho]= 2 - p$ using the convention of \cite{hpst}, because
$[T_f] = 2,  ~[J^t] = p$ and all the others are in dimensionless combinations. It is therefore dimensionless in the planar case.

\begin{figure}[!ht]
\begin{center}
\begin{tabular}{c}
	 \includegraphics[width=0.45\textwidth]{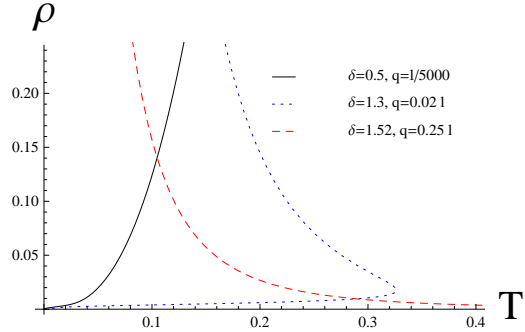}
\end{tabular}
\caption{DC resistivity for the $\gamma\delta  = 1 $ case. Plots are for the three different
ranges of the parameter $\delta$. The lower range $0< \delta^2 <1$ is denoted with the black solid line, the intermediate range
$1< \delta^2 <1+\frac{2}{\sqrt{3}}$ with the blue dotted line and the upper range $1+\frac{2}{\sqrt{3}}< \delta^2 <3$
with the red dashed line. All the plots display  dimensionless quantities, the $x$-axis being $\ell T$,
and the $y$-axis being $\frac{J^t}{T_f} \rho $ with $k=0$. The lower part of the $1< \delta^2 <1+\frac{2}{\sqrt{3}}$ line corresponds to the thermodynamically dominant small black hole.}
\label{Fig:dg=1Resistivity}
\end{center}
\end{figure}

To investigate a temperature-dependent resistivity, we use the temperature expression (\ref{Temperature1})
in a slightly different form using $r_e$
\be
	T=\frac{3-\da^2}{4\pi \ell}\le(\frac{r_+}\ell\ri)^{1-\da^2}\le[1-\le(\frac{r_e^2}{r_+^2}\ri)^{3-\da^2}\ri]^{1-2\frac{(\da^2-1)^2}{(3-\da^2)(1+\da^2)}}\,.
\label{126}\ee
Putting these two equations together we generate three plots in \Figref{Fig:dg=1Resistivity}
for three different ranges of $\delta$, namely, lower range with $0< \delta^2 <1$, intermediate
range with $1< \delta^2  < 1 + \frac{2}{\sqrt{3}}$ and upper range with $1 + \frac{2}{\sqrt{3}}< \delta^2 <3$.

Resistivity reveals different behaviors for different ranges of $\delta$, while
inside a given range  the qualitative features are the same. For example, it increases as we increase
temperature for $0< \delta^2 <1$. It flattens out for decreasing $k$, decreasing $\ell$,
but does not change much when we vary $q$.
In the intermediate range, it has two different branches corresponding to the two black hole phases.
The resistivity of one branch increases and that of the other decreases as the temperature increases.

A special case, $\delta^2 =1$, reveals a limiting behavior of the lower range case,
while another special case, $\delta^2 = 1+ \frac{2}{\sqrt{3}}$, is a limiting case
of the upper range case. Those two cases are special because the temperature expression simplifies,
and we can convert the temperature expression to obtain $r_+$ in terms of temperature $T$.
Thus,  for those cases we obtain the resistivity in a closed form
\be
\rho_{\da = 1} \sim  \frac{T_f }{ J^t}   \left( \frac{r_e}{\ell} \right)^{2 + 2  k}
		\left( 1 -   2 \pi   \ell T \right)^{-\frac{1+k}{2}}   \ ,
\label{126a}\ee
while for $\delta^2=1+ \frac{2}{\sqrt{3}}$ we obtain
\be
\rho \sim  \frac{T_f }{J^t}
  \left( \frac{ 2 \sqrt{3}  \pi }{\sqrt{3} - 1}  \ell T  \right)^{-
\sqrt{3} - \sqrt{3+ 2\sqrt{3}} k}
		\Big[ 1 -   \left( \frac{r_e}{\ell} \right)^{4-\frac{4}{\sqrt{3}}}
\left( \frac{ 2 \sqrt{3}  \pi }{\sqrt{3} - 1} \ell T
\right)^{4(\sqrt{3} -1)} \Big]^{1+\sqrt{3+ 2\sqrt{3}}k}   \ .
\label{127}\ee
For $\delta^2 = 1$ and $k=1$, there exists a temperature where the resistivity diverges,
$ T = \frac{1}{2 \pi \ell} $. This suggests some sort of critical behaviour that deserves further attention.
 We have also checked that the result for $\delta^2 = 1$ is compatible with the one calculated for  $\gamma = \delta $ solution in the  next section.

We will now consider the small temperature limit of the resistivity for the lower and intermediate range of $\d$.
This limit  corresponds to the extremal limit $r_+ \rightarrow r_e$. Therefore, the
leading  behavior of the resistivity is
\begin{eqnarray}
	\rho_{\rm leading} &&\sim
	\frac{T_f}{ J^t}   \left( \frac{r_e}{\ell} \right)^{2 + 2 \delta k} 	 \Big[ \frac{4 \pi }{3-\delta^2} \left(\frac{\ell}{r_e}\right)^{1-\delta^2} (\ell T) \Big]^{\frac{2 (\delta^2 -1) (\delta^2 -1 + 2 k \delta)}{1 + 6 \delta^2 - 3 \delta^4}}  \nonumber \\ 	&&\sim
	\frac{T_f}{ J^t}   \left( \frac{q}{\ell} \right)^{\frac{2 \da  (\da (3-\da^2 )+ (1+\da^2 ) k )}{1+6 \da^2-3 \da^4}}	(\ell T)^{\frac{2 (\delta^2 -1) (\delta^2 -1 + 2 k \delta)}{1 + 6 \delta^2 - 3 \delta^4}}\,. 
\label{128}\end{eqnarray}

\begin{figure}[!ht]
\begin{center}
\begin{tabular}{c}
	 \includegraphics[width=0.45\textwidth]{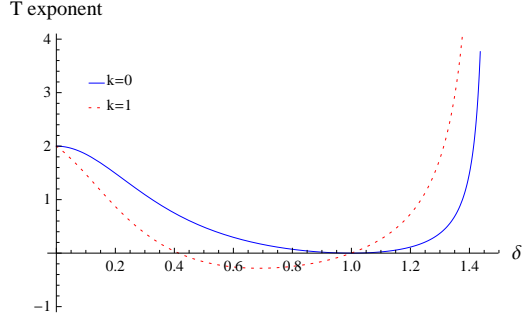}
\end{tabular}
\caption{Exponent of temperature in resistivity, $\rho_{small T}$, for two extreme cases, $k=0, 1$.
There are always two distinct points
for the exponent to be 1, which reveals the linear temperature dependence of resistivity. }
\label{Fig:dg=1-SmallTExponent}
\end{center}
\end{figure}

\begin{figure}[!ht]
\begin{center}
\begin{tabular}{c}
	 \includegraphics[width=0.45\textwidth]{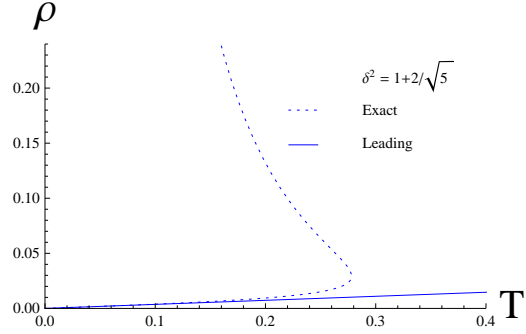}
\end{tabular}
\caption{The exact and leading order resistivity  in the intermediate range with
particular values $\delta^2 = 1+ \frac{2}{\sqrt{5}}$, which gives the linear temperature dependence to  leading order. }
\label{Fig:dg=1-d=1.3763-DC-Exact-Leading}
\end{center}
\end{figure}

 To leading order at low temperatures, there are two different values of $\delta$, one in the lower and the other
in the intermediate range, for which the resistivity has a linear temperature dependence.
This is the case for all  $0< k <1$.
This is seen in \Figref{Fig:dg=1-SmallTExponent}. In particular,
for $k=1$, the two values are $\delta \sim 0.17782 $ and $\delta \sim 1.2359$, while
$\delta =\sqrt{\frac{1}{5} \left(5-2 \sqrt{5}\right)} \sim  0.3249 $ and $\delta = \sqrt{\frac{1}{5} \left(5+2 \sqrt{5}\right)} \sim  1.3763$ for $k=0$. In \Figref{Fig:dg=1-d=1.3763-DC-Exact-Leading}, we compared the exact result and
the leading order approximation with a set of values, $k=0\,,\, \delta = \sqrt{\frac{1}{5} \left(5+2 \sqrt{5}\right)}$,
corresponding to the linear temperature dependence of the resistivity.

 In order to assess the range of validity of the leading approximation we will compare the leading term to the subleading one by varying $r_+$ as $r_+ = r_e + \delta r$ with $\delta r$ given by
\begin{eqnarray}
\delta r	\sim \frac{ r_e }{2(3-\delta^2)} \left(\frac{r_e}{\ell}\right)^{\frac{(-1+\delta^4)(3- \da^2 )}{1+6 \delta^2-3 \delta^4}} \left(\frac{4 \pi }{3-\delta^2}  \ell T \right)^{\frac{(3-\da^2)(1+\da^2)}{1+6 \delta^2-3 \delta^4}} \ .
\label{129}\end{eqnarray}
We obtain
\begin{eqnarray}
{\rho_{\rm next-to-leading}\over
 \rho_{\rm leading} } \simeq C \cdot \left(\frac{r_e}{\ell}\right)^{\frac{(-1+\delta^4)(3- \da^2 )}{1+6 \delta^2-3 \delta^4}} \left(\frac{4 \pi }{3-\delta^2}  \ell T \right)^{\frac{(3-\da^2)(1+\da^2)}{1+6 \delta^2-3 \delta^4}}  \ ,
\label{130}\end{eqnarray}
where the constant factor is $C = \frac{-1+20 \da (\da + k)-14 \da^3 k+\da^4 (-13+2 \da (\da+k))}{2 \left(-3+\da^2\right)^2 \left(1+\da^2\right)}$.

We require
$\rho_{\rm next-to-leading} \ll \rho_{\rm leading}$,
which is equivalent  to
\be
\ell T \ll \left(\frac{r_e}{\ell}\right)^{1-\da^2} \sim  \left(\frac{q}{\ell}\right)^{\frac{1-\da^2}{3-\da^2}} \ .
\label{131}\ee
Therefore  the leading  approximation is valid when the  dimensionless temperature scale is small compared to an appropriate power of the dimensionless charge density.
In particular for $\d<1$, the linear range can be substantial at larger charge densities while for $\d>1$, for small charge densities.


\section{The $\gamma=\delta$ solutions}

Taking $\ga=\da$, one finds the following scaling solution\footnote{We refer to \cite{char} for details on the method of resolution. In the language of this paper, the solution \eqref{Sol2} is the general solution of the equations of motion provided we set the integration constant $h=0$.}:
\bsea
	\ud s^2 &=& - V(r)\ud t^2 + e^{\da\phi}\frac{\ud r^2}{V(r)} + r^2\Big(\ud x^2+\ud y^2\Big)\,, \slabel{Metric2} \\
	V(r) &=& \le(\frac r \ell\ri)^{2}-2m\ell^{-\da^2}r^{\da^2-1} +\frac{q^2}{4(1+\da^2)r^2}\,, \slabel{Pot2} \\
	e^{\phi}&=& \le(\frac r\ell\ri)^{2\da} \slabel{Phi2}\,, \slabel{ElectricPotential2}\\
	\mathcal A &=& \le(\Phi - \frac{\ell^{\da^2}q}{(1+\da^2)r^{1+\da^2}}\ri)\ud t\,, \slabel{A2}\\
	\Phi&=&  \frac{q\ell^{\da^2}}{(1+\da^2)r_+^{1+\da^2}}\,,\\
	\ga&=&\da\,. \nn
	\label{Sol2}
\esea
We observe that this solution reduces when $\da=1$ to the solution \eqref{Sol1} for the same value of $\da$. Let us also note that for $\da=0$ (and so $\ga=0$, too), we recover the planar version of the Anti-de Sitter-Reissner-Nordstr\"om black hole, as expected from the action.

The IR scale $\ell^2=\frac{3-\da^2}{-\Lambda}$ is the same as in the previous solution, and, for the same reasons, we shall restrict our attention to black hole solutions, that is $\Lambda<0$ and $\da^2<3$.

The Ricci scalar on-shell is:
\be
	\mathcal R = 4\Lambda e^{-\da\phi}\le[ 1-\frac{2\da^2\ell^2V(r)}{(3-\da^2)^2r^2}\ri],
	\label{RicciScalar2}
\ee
so there is a curvature singularity at $r=0$ for $\da^2<3$, radial infinity is regular if $\da^2<3$ and there are event horizons at the zeros of the potential \eqref{Sol2}. The size of the torus vanishes now at $r=0$.

 We now analyze the conditions for $V(r)$ to vanish. Indeed, solving simultaneously for $V(r)=0$ and $\frac{\partial V(r)}{\partial r}=0$, we find the extremal value,
\be
	r_e^4=\frac{q^2\ell^2}{4(3-\da^2)}\,,\qquad m_e = \frac{2r_e^{3-\da^2}}{(1+\da^2)\ell^{2-\da^2}}\,.
\label{132}\ee
This implies that for $r_+\geq r_e$, or equivalently $m>m_e$, there are two event horizons, one inner and another outer, which are degenerate in the extremal case (where the bound is saturated). Below the bound, a naked singularity exists. The extremal black hole is defined only for $\da^2<3$.

The temperature of the solution is,
\be T=\frac{(3-\da^2)}{4\pi}\ell^{\da^2-2}r_+^{1-\da^2}\le[1-\le(\frac{r_e}{r_+}\ri)^{4}\ri],
	\label{Temperature2}
\ee
and vanishes in the extremal case. The integrated gravitational mass of the solution from \eqref{GravitationalBoundaryVariation} is
\be
	M_g = \frac{\omega_2}{4\pi}m\,,
	\label{GravitationalMass2}
\ee
and in this case the scalar contribution \eqref{ScalarBoundaryVariation} is zero. This is expected since the scalar field always has its background value, it is not backreacted on by the black hole, contrarily to \eqref{Sol1}. The electric charge \eqref{ElectricCharge} is
\be
	Q = \frac{\omega_2}{16\pi}q\,.
	\label{ElectricCharge2}
\ee

Thus, there are two independent integration constants specifying the solution, as well as an independent overall scale that we have fixed to its `natural' value, e.g. the maximal number allowed by the equations of motion. They are $m$, $q$ and $\ell$ as before and there is no relation between them. $m$ and $q$ can be considered as the `reduced' mass and charge, while $\ell$ is the IR radius. However, it is possible that this is not the most general solution for $\ga=\da$, because of the method by which we obtained \eqref{Sol2}.

For the remainder of this section, we shall use the same rescalings as for the previous $\ga\da=1$ solution, as in (\ref{rescale}).

\subsection{Regions of validity}\label{g=dregions}

The UV region of the solutions is $r\to \infty$ where the scale factor increases without bound.
For the solutions to be IR regions of asymptotically AdS solutions the scalar potential must become small in the UV region,
\be
V\sim e^{-\delta\phi}\sim \le(\frac r\ell\ri)^{-2\da^2}\underset{r\to +\infty}{\longrightarrow} 0\,.
\label{133}
\ee
Therefore these solutions are indeed IR regions of asymptotically AdS solutions.

In the case where the conserved charge is originating on flavor branes with a DBI action, according to (\ref{19})
we can linearize the action when
\be
Z Ce^{k\phi}=\ell^2\left( {r\over \ell}\right)^{2+2\d(\d+k)}
\gg |q|\,.
\label{aaa1}\ee
As $\d<\sqrt{3}$ this approximation is always good in the UV and it always breaks down in the IR at extremality, ($r\to 0$).

To test whether in the IR extremal  regime the DBI action can be treated like a probe, we must have that $V\gg e^{2k\phi}Z$ or
$e^{(\d+k)\phi}\ll 1$.
Indeed, in this case since $k\ge 0$,
\be
e^{(\d+k)\phi}= \le(\frac r\ell\ri)^{2\da(\d+k)}\underset{r\to 0}{\longrightarrow}  0\,.
 \label{134}\ee
We can therefore use the probe techniques to calculate the conductivities in that case with the result derived already in section \ref{DCDBI}.

\subsection{Thermodynamics}

Throughout this section and both for the grand-canonical and canonical case, several cases will be distinguished depending on the value of  $\da^2$. However, we will only plot one curve per range, representative of the behaviour in this range. Where useful, we will also display the behaviour at the limiting values for these ranges.

\subsubsection{Grand-Canonical ensemble}

\begin{figure}[ht]
\begin{center}
\begin{tabular}{cc}
	 \includegraphics[width=0.45\textwidth]{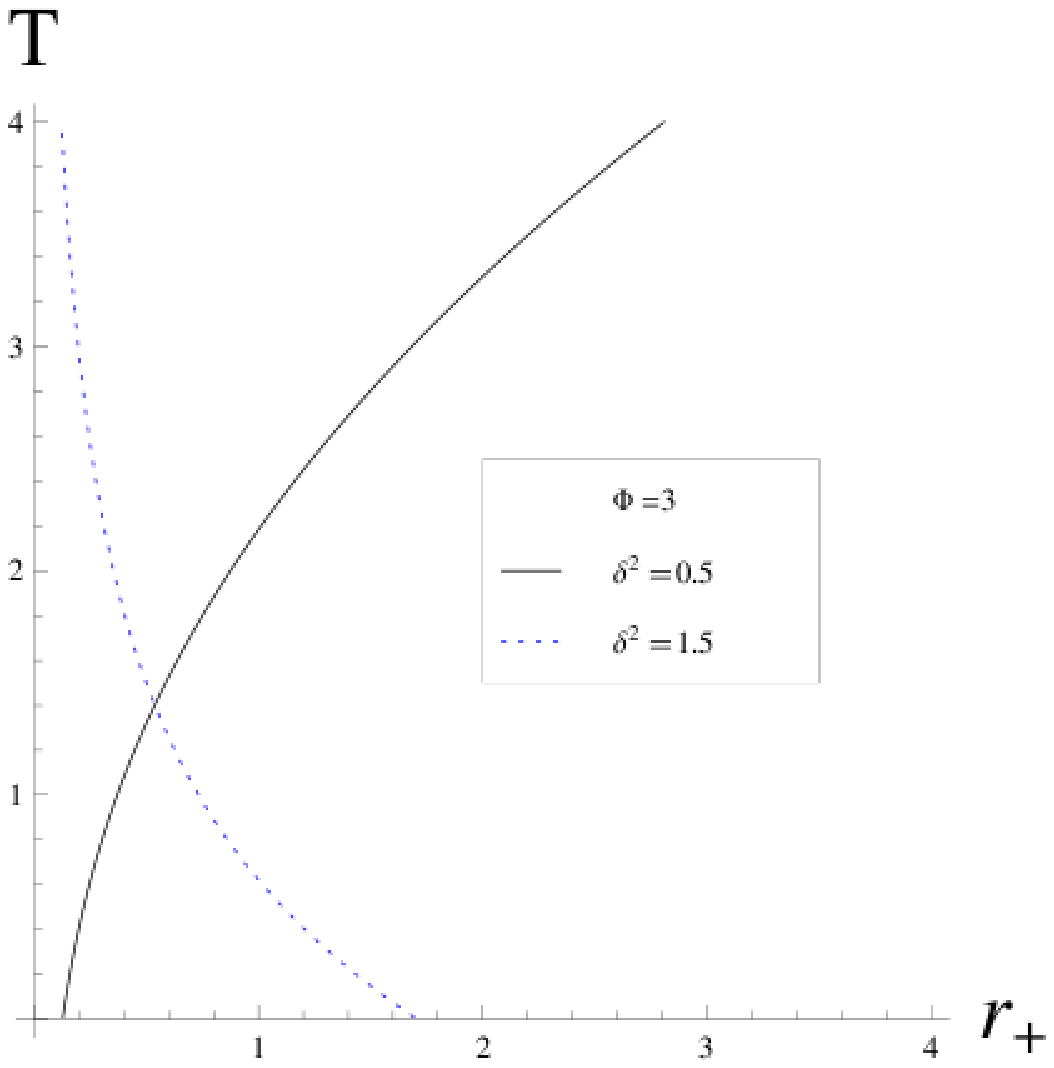}&
	 \includegraphics[width=0.45\textwidth]{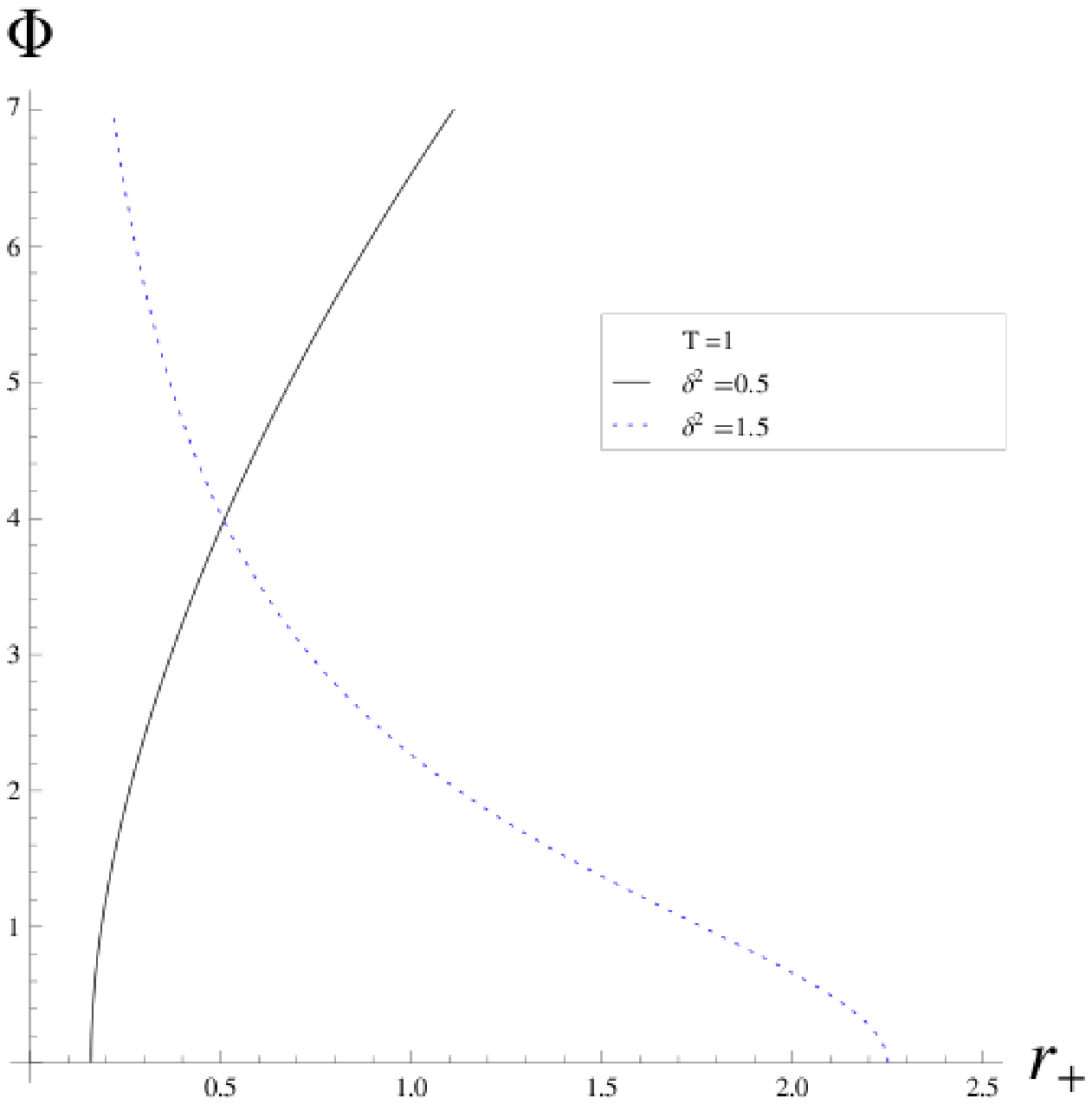}
 \end{tabular}
\caption{Slices of the equation of state $T(r_+,\Phi)$ at fixed potential (left) and $\Phi(r_+,T)$ at fixed temperature (right)}
\label{Fig:EqOfStateGa=DaGrandCanonical}
\end{center}
\end{figure}

\begin{figure}[ht]
\begin{center}
\begin{tabular}{cc}
	 \includegraphics[width=0.45\textwidth]{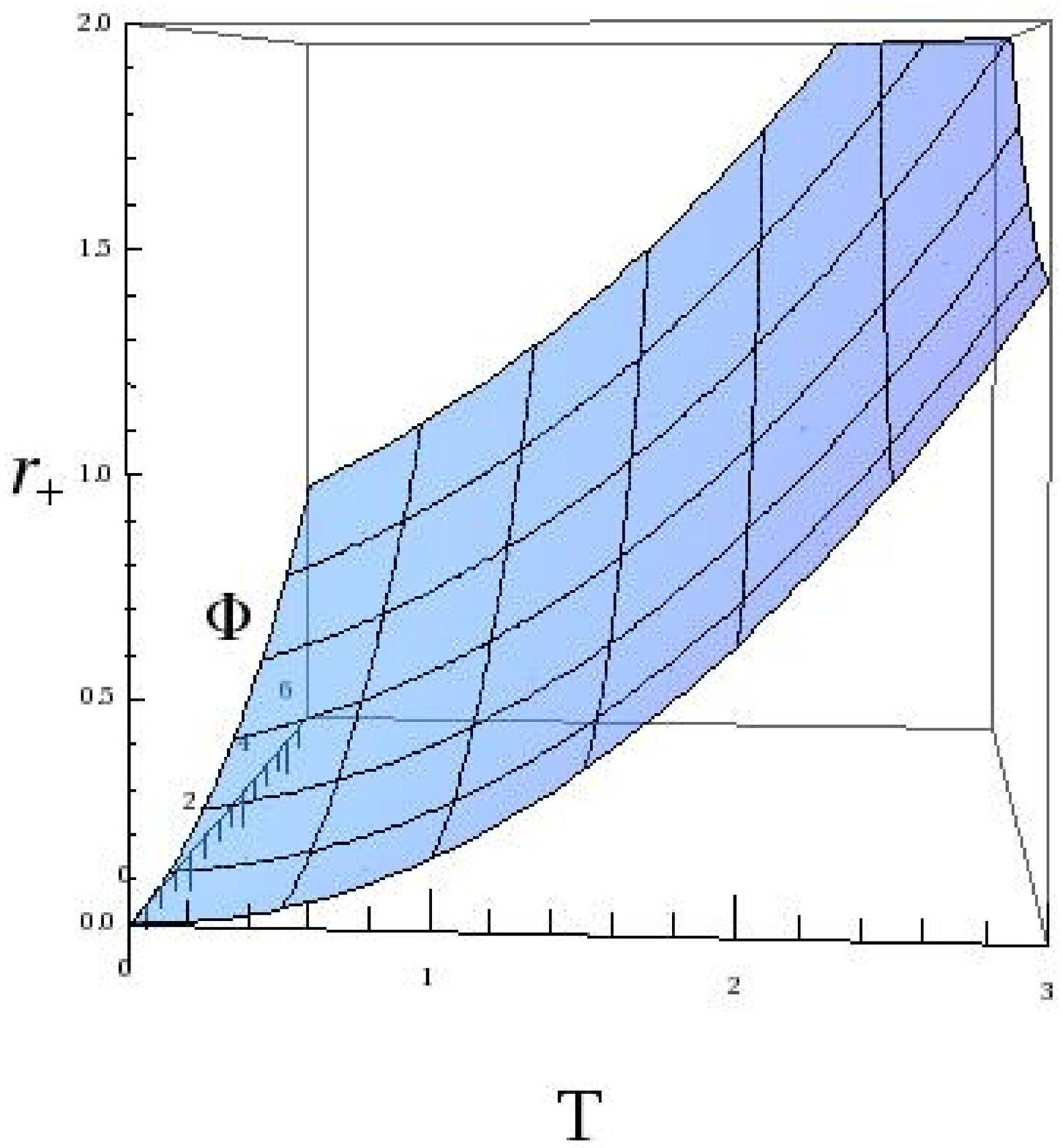}&	
	 \includegraphics[width=0.45\textwidth]{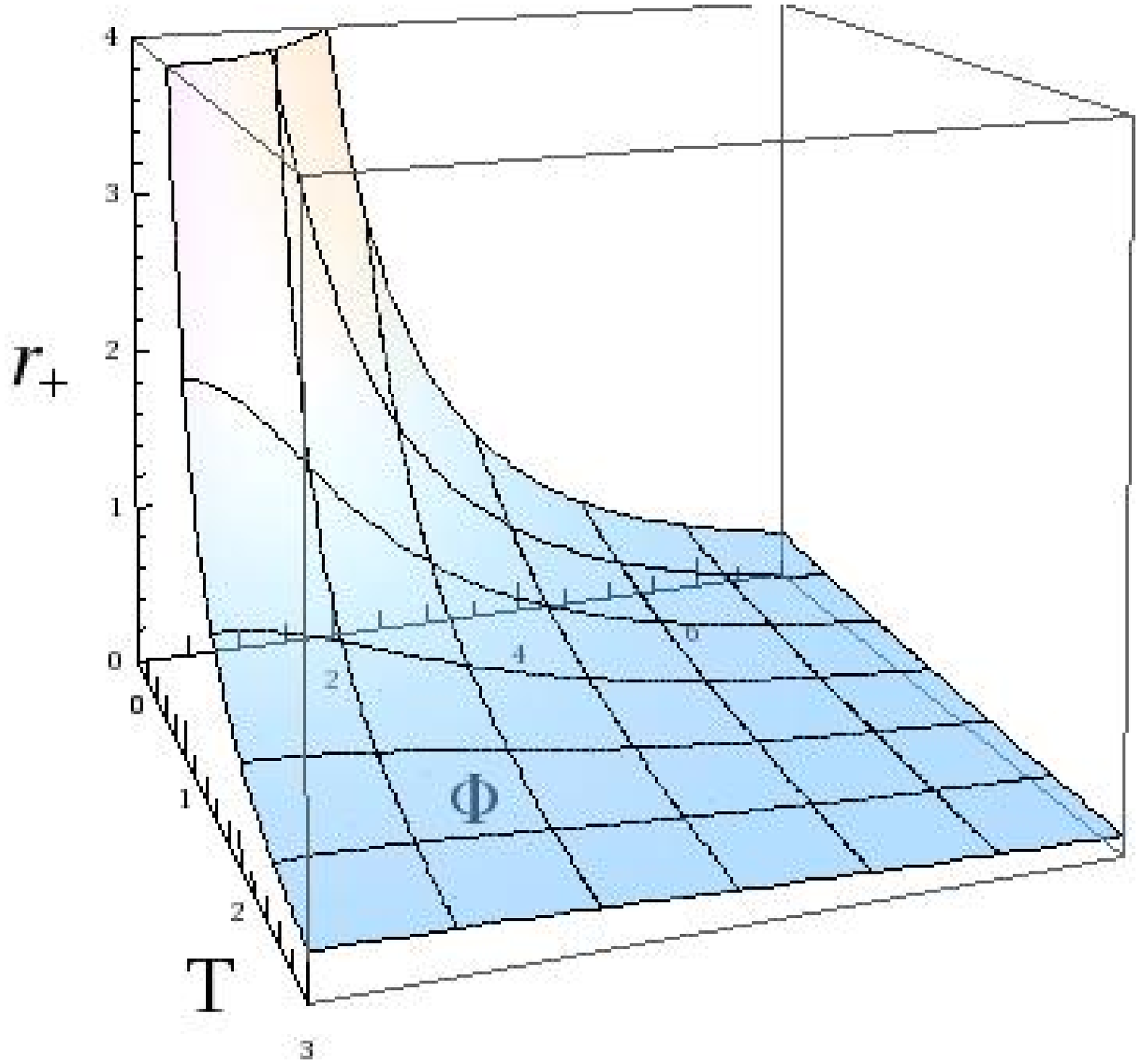}
 \end{tabular}
\caption{Equation of state $r_+(T,\Phi)$in the lower ($0\leq\da^2<1$) and upper ($1<\da^2<3$) range from left to right.}
\label{Fig:EqOfState3DGa=DaGrandCanonical}
\end{center}
\end{figure}

The equation of state \eqref{Temperature2}
\be
	T= (3-\da^2)\ell^{\da^2-2}r_+^{1-\da^2}\le[1-\frac{(1+\da^2)^2\ell^{2-2\da^2}\Phi^2}{64(3-\da^2)r_+^{2-2\da^2}} \ri],
\label{135}\ee
can be rewritten so as to express the horizon radius as a function of the thermodynamic variables $(T,\Phi)$
\be
	r_+^{\da^2-1} = \frac{32(3-\da^2)}{(1+\da^2)^2\Phi^2}\le[-\frac{T}{3-\da^2}+\sqrt{\frac{T^2}{(3-\da^2)^2}+\frac{(1+\da^2)^2\Phi^2}{16(3-\da^2)}} \ri].
	\label{EqOfStateGa=DaGrandCanonical}
\ee
 As in the previous solution, the grand-canonical ensemble breaks down in the string limit ($\d^2= 1$) since the temperature and the chemical potential cease to be independent variables. We plot slices at fixed chemical potential and temperature of the equation of state in \Figref{Fig:EqOfStateGa=DaGrandCanonical} and the full three-dimensional representations in \Figref{Fig:EqOfState3DGa=DaGrandCanonical}.

We have to distinguish between two ranges,
\begin{itemize}
 \item Lower range $\da^2<1\,$: There is a single black hole branch. At fixed potential, the black hole grows with the temperature and the endpoint of the curve is the extremal black hole. Contrary to the previous solution, the extremal black hole can now be reached for any value of the chemical potential. At fixed temperature, the vertical axis $\Phi=0$ corresponds to the neutral black holes.
 \item Upper range $1<\da^2<3\,$: There is a single black hole branch. At fixed chemical potential, the curve starts at zero temperature at the extremal black hole, and then the radius actually decreases for the non-extremal black holes as the temperature increases. At fixed temperature, the vertical axis $\Phi=0$ displays the neutral black holes, and here also the radius diminishes as the chemical potential grows.
\end{itemize}

\begin{figure}[t]
\begin{center}
\begin{tabular}{cc}
	 \includegraphics[width=0.45\textwidth]{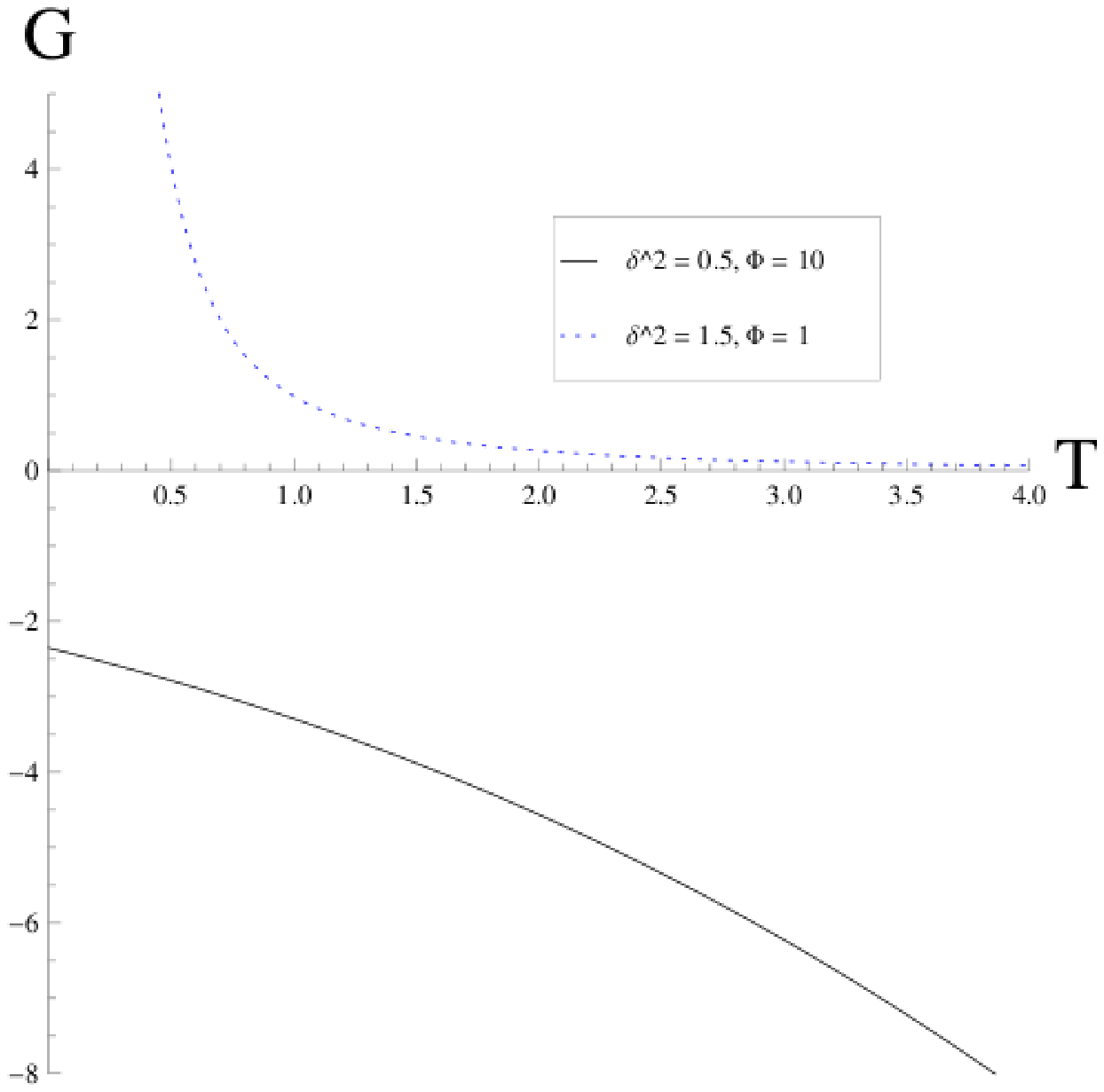}&	 \includegraphics[width=0.45\textwidth]{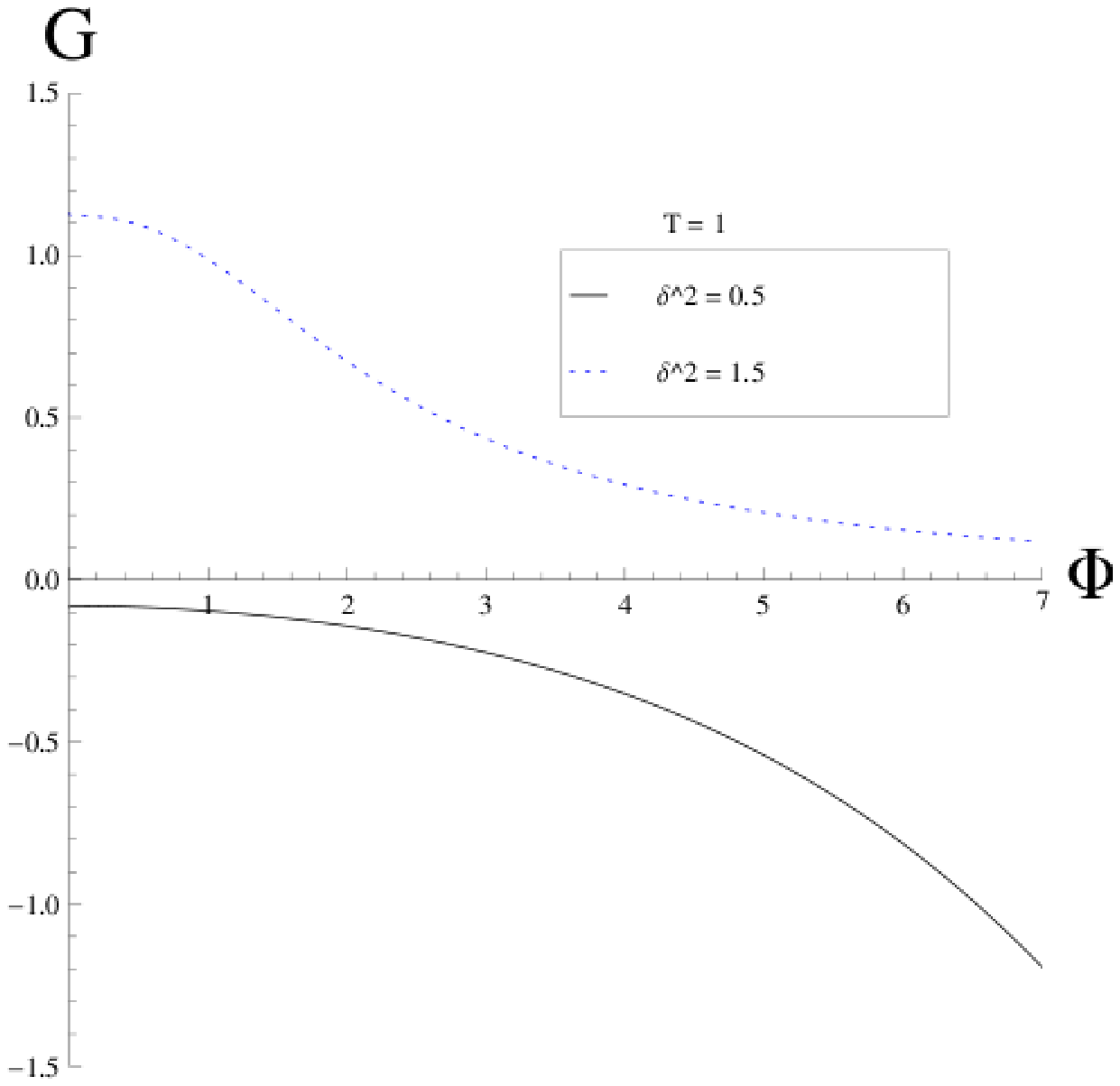}
 \end{tabular}
\caption{Slices of the Gibbs potential at fixed potential (left) and fixed temperature (right)}
\label{Fig:GibbsGa=DaGrandCanonical}
\end{center}
\end{figure}

\begin{figure}[t]
\begin{center}
\begin{tabular}{cc}
	 \includegraphics[width=0.45\textwidth]{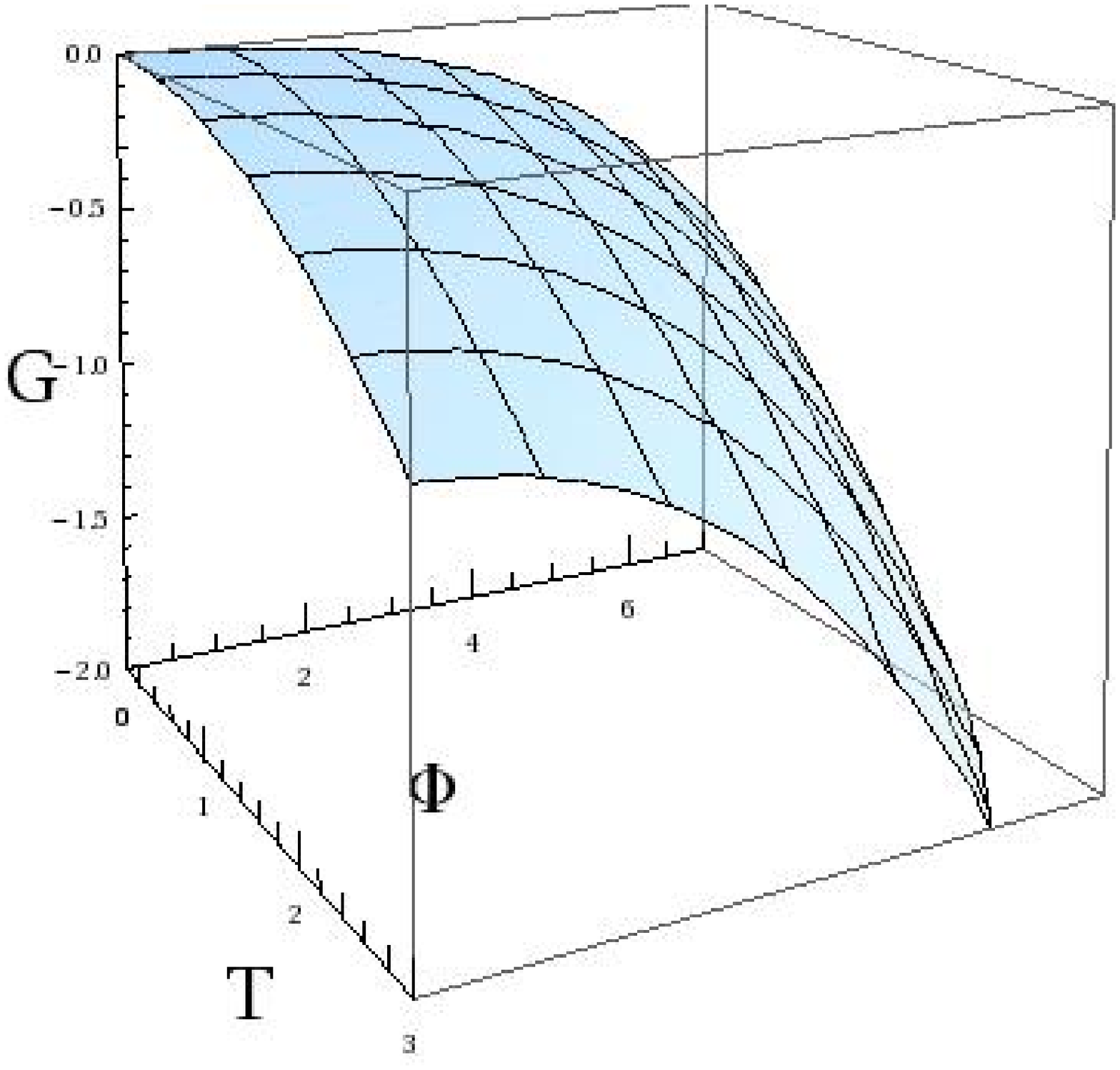}&	
	 \includegraphics[width=0.45\textwidth]{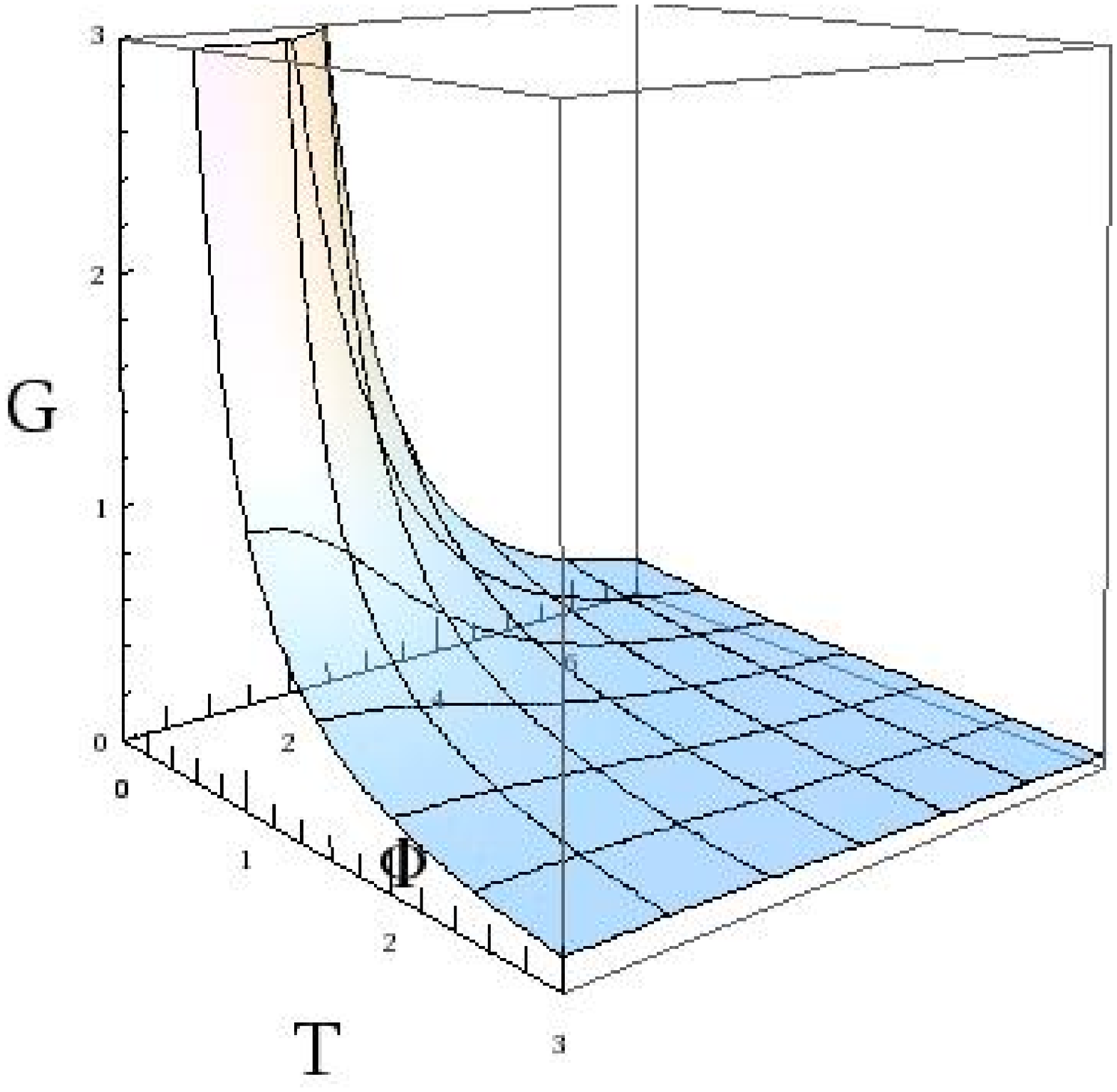}
 \end{tabular}
\caption{Three-dimensional representation of the Gibbs potential in the lower ($0\leq\da^2<1$) and upper ($1<\da^2<3$) range from left to right.}
\label{Fig:Gibbs3DGa=DaGrandCanonical}
\end{center}
\end{figure}

We now turn to the calculation of the Gibbs thermodynamical potential. Evaluating the value of the Euclidean action \eqref{ActionGrandCanonical} and subtracting the background contribution, we find
\be
	-\beta G = \tilde I - \tilde I_0 = -\beta \ell^{\da^2-2}(\da^2-1)\le[r_+^{3-\da^2}+\frac{4Q^2\ell^2}{(1+\da^2)r_+^{1+\da^2}}\ri],
	\label{EuclideanActionGC2}
\ee
where we have identified the temperatures of the black hole and the thermal background on the outer boundary in order to do the subtraction. The Gibbs potential of the black hole in the thermal background \eqref{LinearDilaton} is then
\be
	G[T,\Phi] = \ell^{\da^2-2}(\da^2-1)r_+^{3-\da^2}\le[1+\frac{(1+\da^2)\Phi^2}{64}r_+^{2\da^2-2} \ri],
	\label{Gibbs2}
\ee
slices of which at fixed chemical potential or fixed temperature are plotted in \Figref{Fig:GibbsGa=DaGrandCanonical}, while the three-dimensional representation is in \Figref{Fig:Gibbs3DGa=DaGrandCanonical}.

From this expression, we derive the entropy \eqref{EntropyGrandCanonical},
\be
	S=r_+^2\,,
	\label{EntropyGrandCanonical2}
\ee
which is the quarter of the area of the horizon as expected, the charge density \eqref{ElectricChargeGrandCanonical},
\be
	Q=\frac{(1+\da^2)}{16\ell^{\da^2}}r_+^{1+\da^2}\Phi\,,
\ee
which is equal to its usual value \eqref{ElectricCharge2}, and  the energy \eqref{EnergyGrandCanonical},
\be
	E = 2\ell^{\da^2-2}\le[r_+^{3-\da^2}+\frac{4Q^2\ell^2}{(1+\da^2)r_+^{1+\da^2}}\ri]=4m = M_g\,,
	\label{EnergyGrandCanonical2}
\ee
which is equal to the gravitational mass \eqref{GravitationalMass2} as expected for a scaling dilaton. The quantities calculated by use of \eqref{EnergyGrandCanonical}, \eqref{EntropyGrandCanonical}, \eqref{EntropyGrandCanonical} satisfy the first law \eqref{FirstLawCanonical}.

\begin{figure}[t]
\begin{center}
\begin{tabular}{cc}
	 \includegraphics[width=0.45\textwidth]{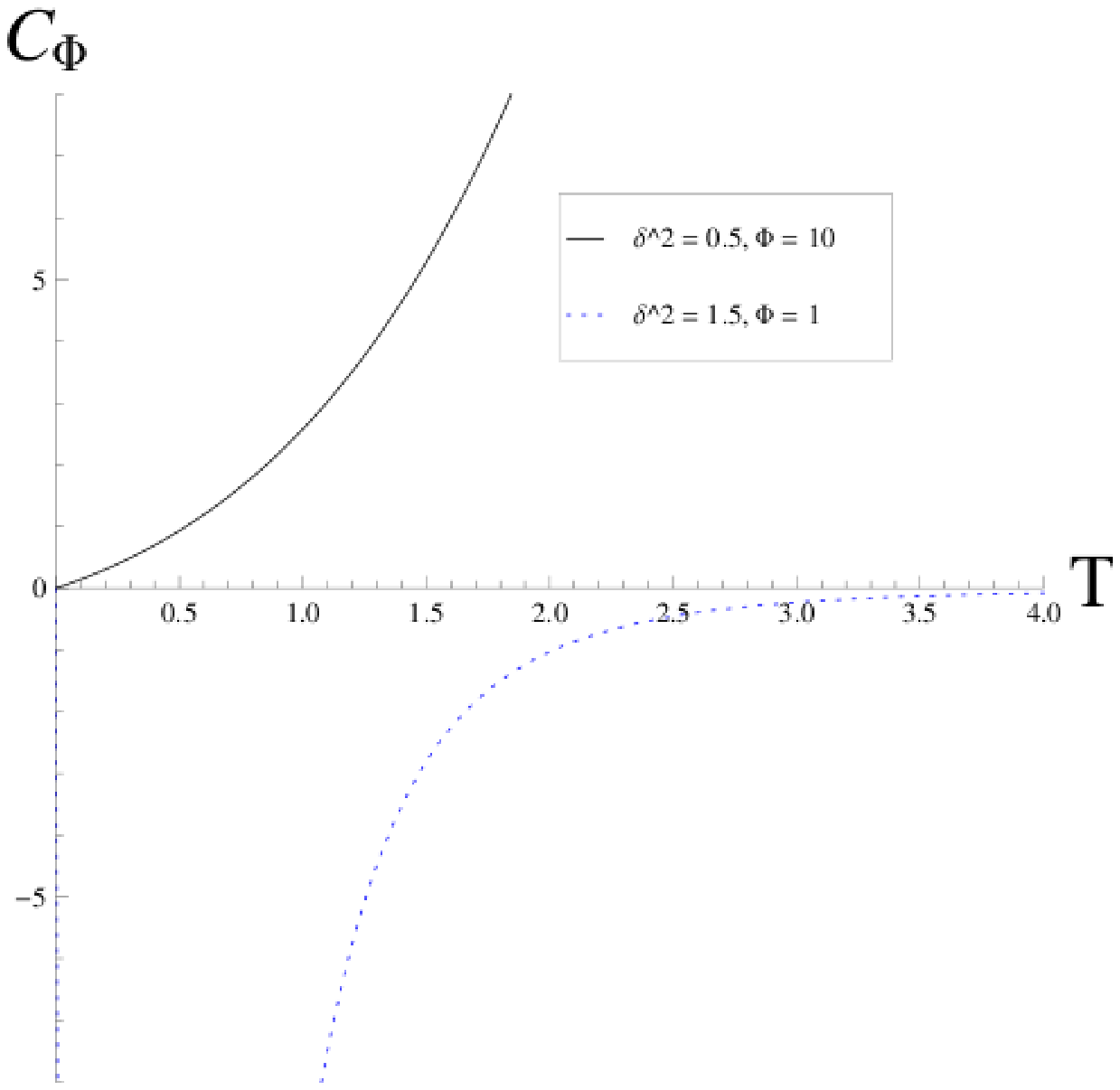}&	 \includegraphics[width=0.45\textwidth]{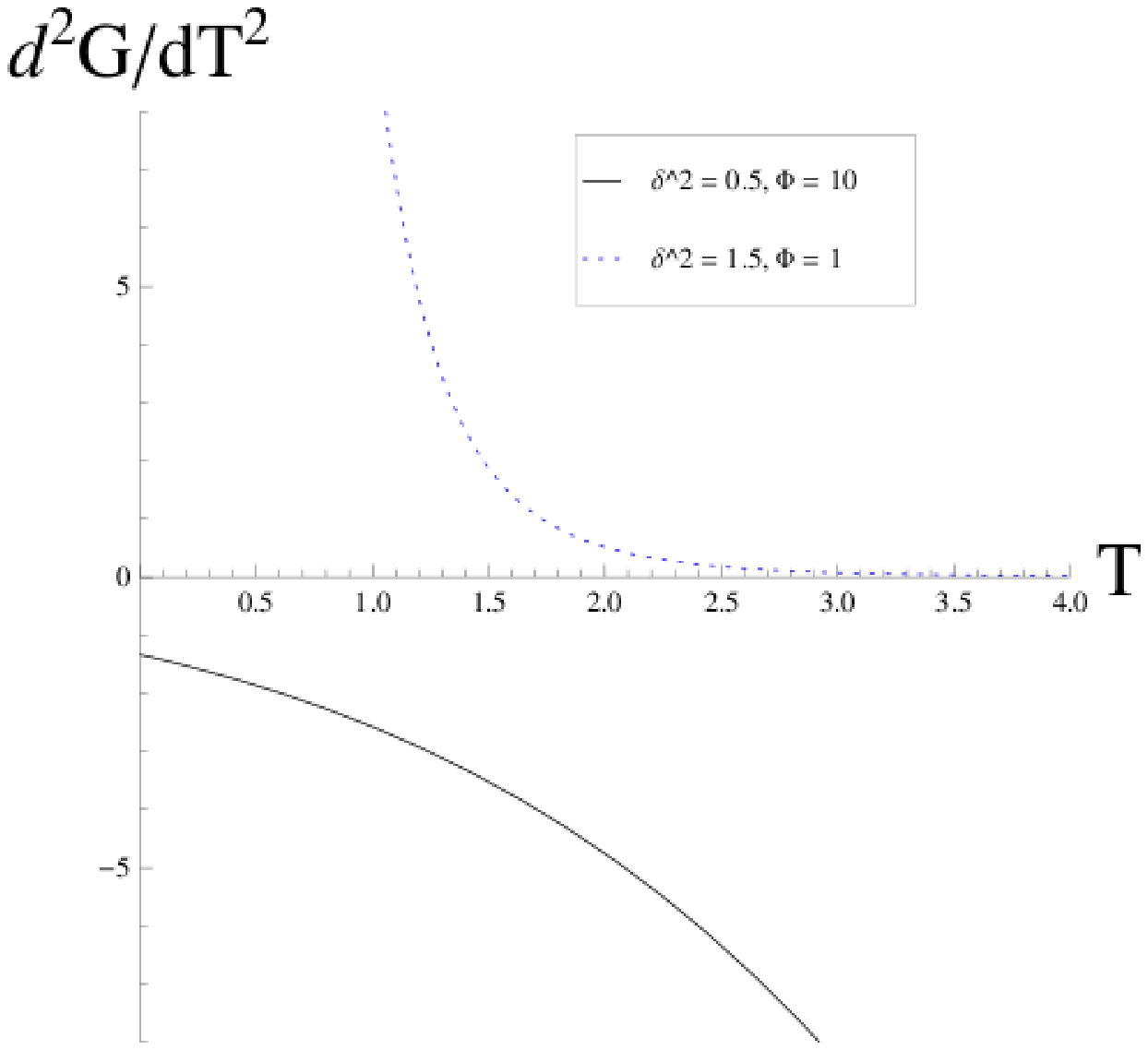}
 \end{tabular}
\caption{Heat capacity (left) and second derivative of the Gibbs potential with respect to the temperature (right) at fixed potential.}
\label{Fig:ThermalStabilityGa=DaGrandCanonical}
\end{center}
\end{figure}

\begin{figure}[t]
\begin{center}
\begin{tabular}{c}
	 \includegraphics[width=0.45\textwidth]{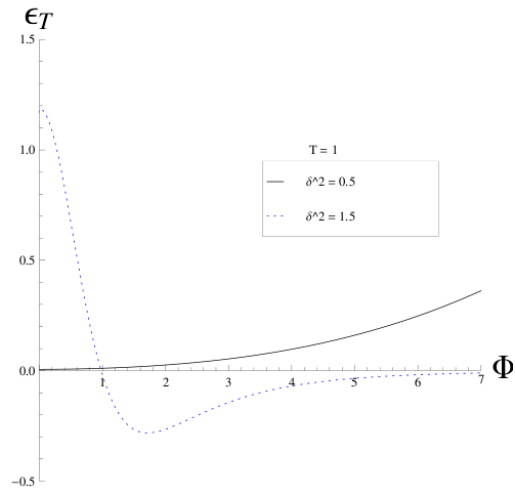}
 \end{tabular}
\caption{Electric permittivity at fixed temperature.}
\label{Fig:ElectricStabilityGa=DaGrandCanonical}
\end{center}
\end{figure}

Studying the thermodynamics of the solution, two behaviours have to be distinguished:
\begin{itemize}
 \item Lower range: At fixed potential, the charged black holes dominate the ensemble (\Figref{Fig:GibbsGa=DaGrandCanonical} and \Figref{Fig:Gibbs3DGa=DaGrandCanonical}) , at all temperatures, even in the extremal limit. At fixed potential, the black holes also dominate at all values of the chemical potential, including the neutral black holes at $\Phi=0$. The black holes are stable thermally and electrically, see \Figref{Fig:ThermalStabilityGa=DaGrandCanonical} and \Figref{Fig:ElectricStabilityGa=DaGrandCanonical}.
 \item Upper range: The ensemble is dominated by the dilatonic background for all $(T,\Phi)$ values.
\end{itemize}

\subsubsection{Canonical ensemble}

\begin{figure}[!ht]
\begin{center}
\begin{tabular}{cc}
	 \includegraphics[width=0.45\textwidth]{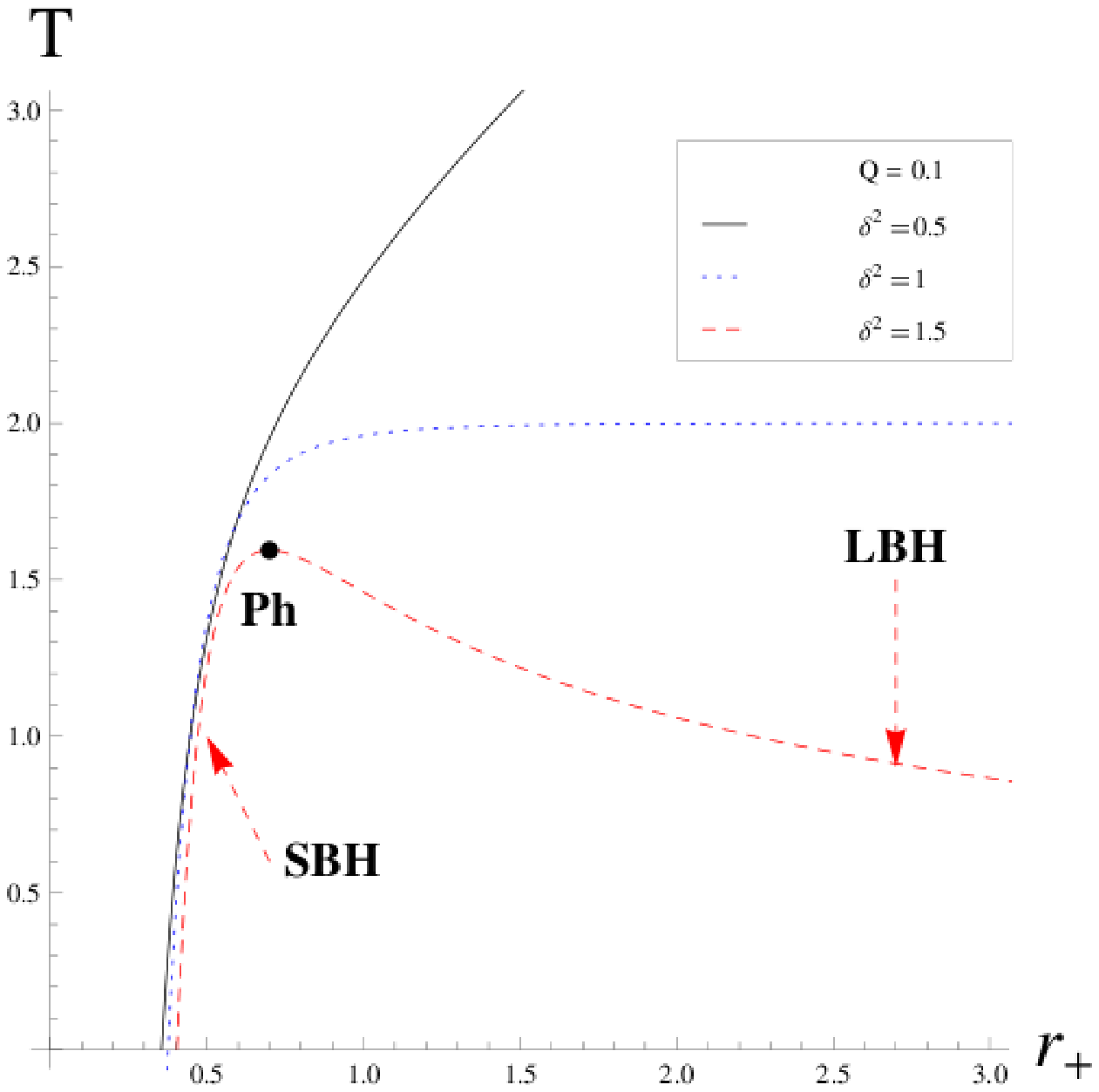}&
	 \includegraphics[width=0.45\textwidth]{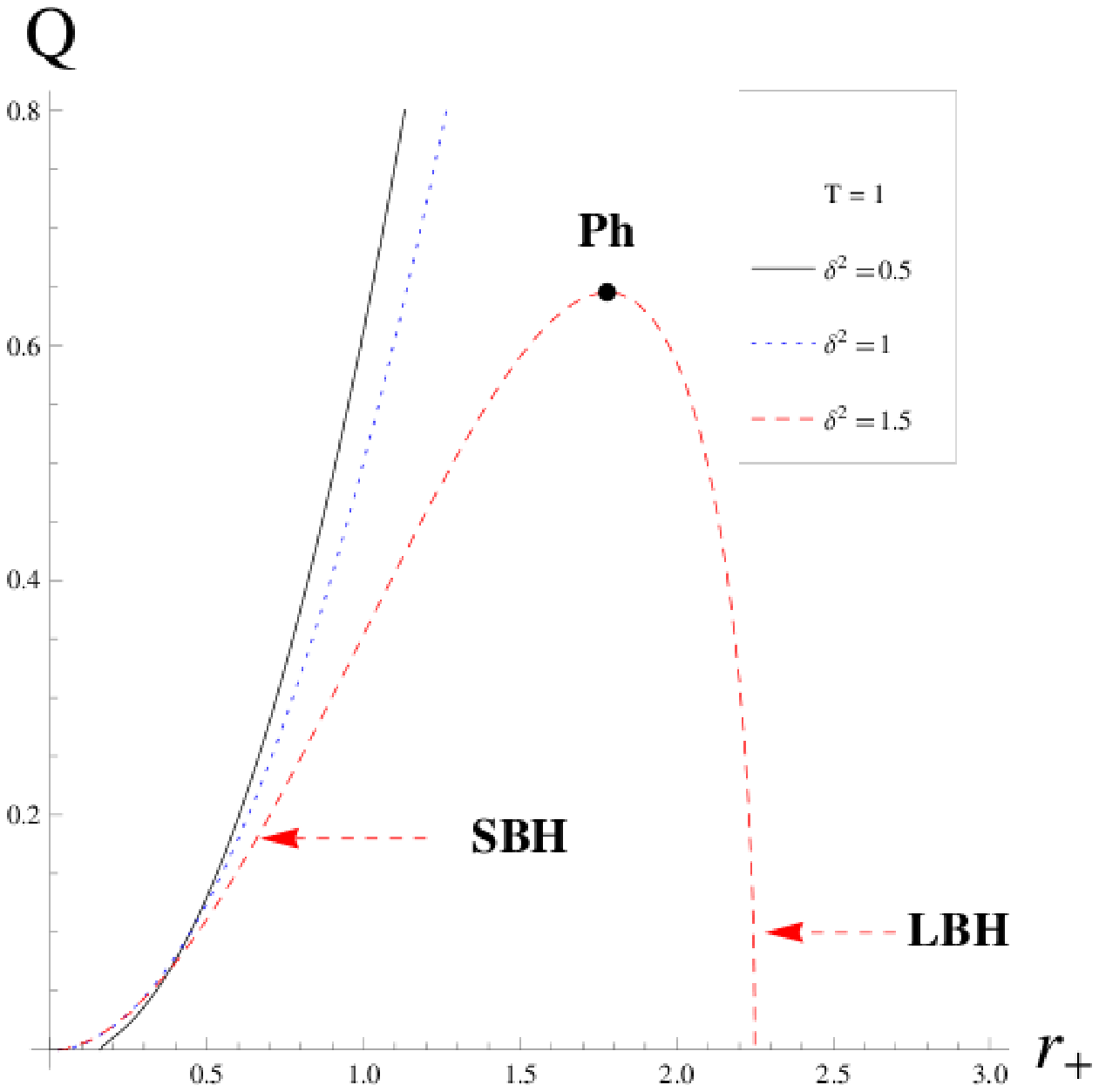}
 \end{tabular}
\caption{Slices of the equation of state $T(r_+,Q)$ at fixed charge (left) and $Q(r_+,T)$ at fixed temperature (right).}
\label{Fig:HorizonSizeCanonicalGa=Da}
\end{center}
\end{figure}

\begin{figure}[t]
\begin{center}
\begin{tabular}{cc}
	 \includegraphics[width=0.45\textwidth]{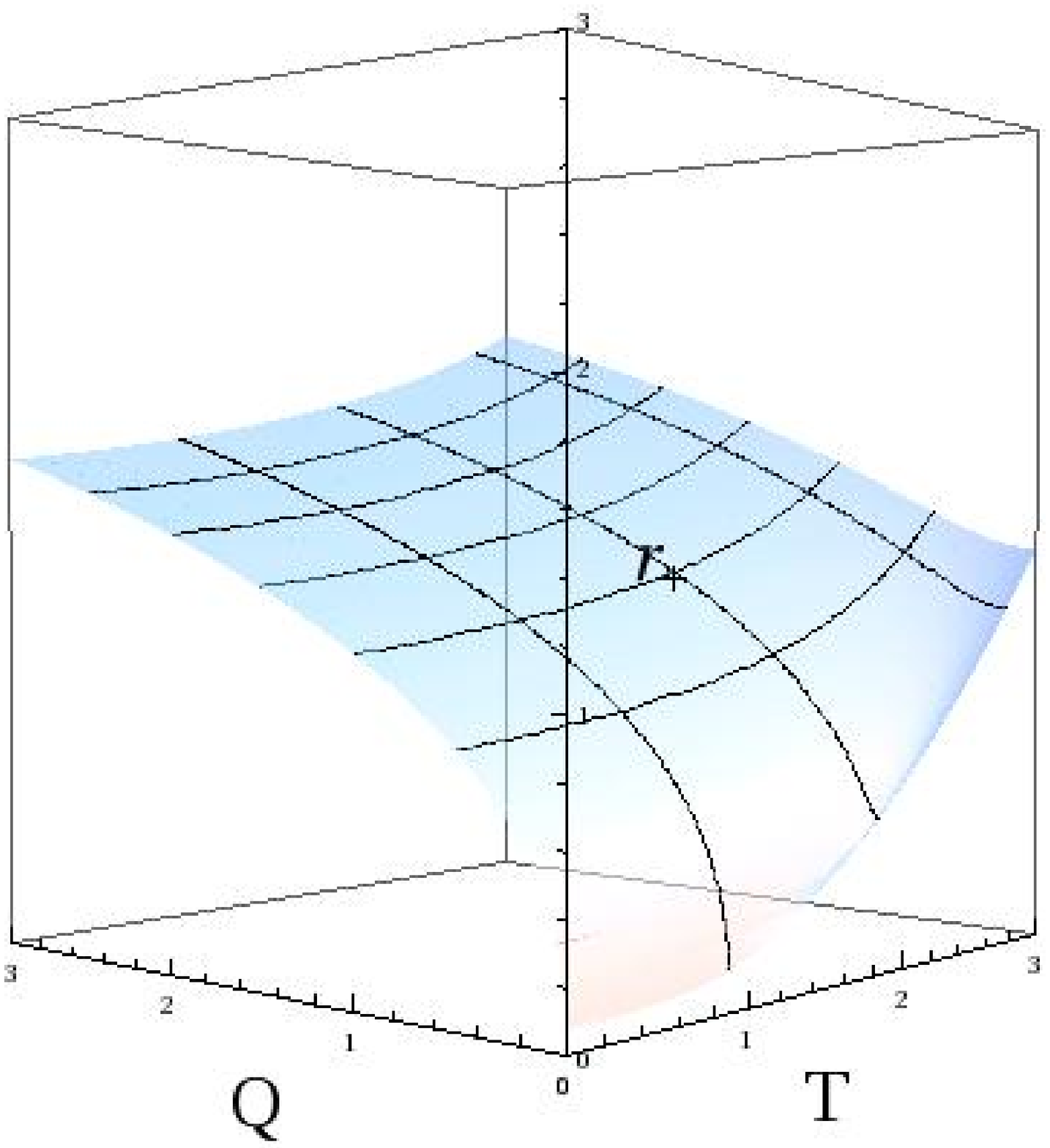}&
	 \includegraphics[width=0.45\textwidth]{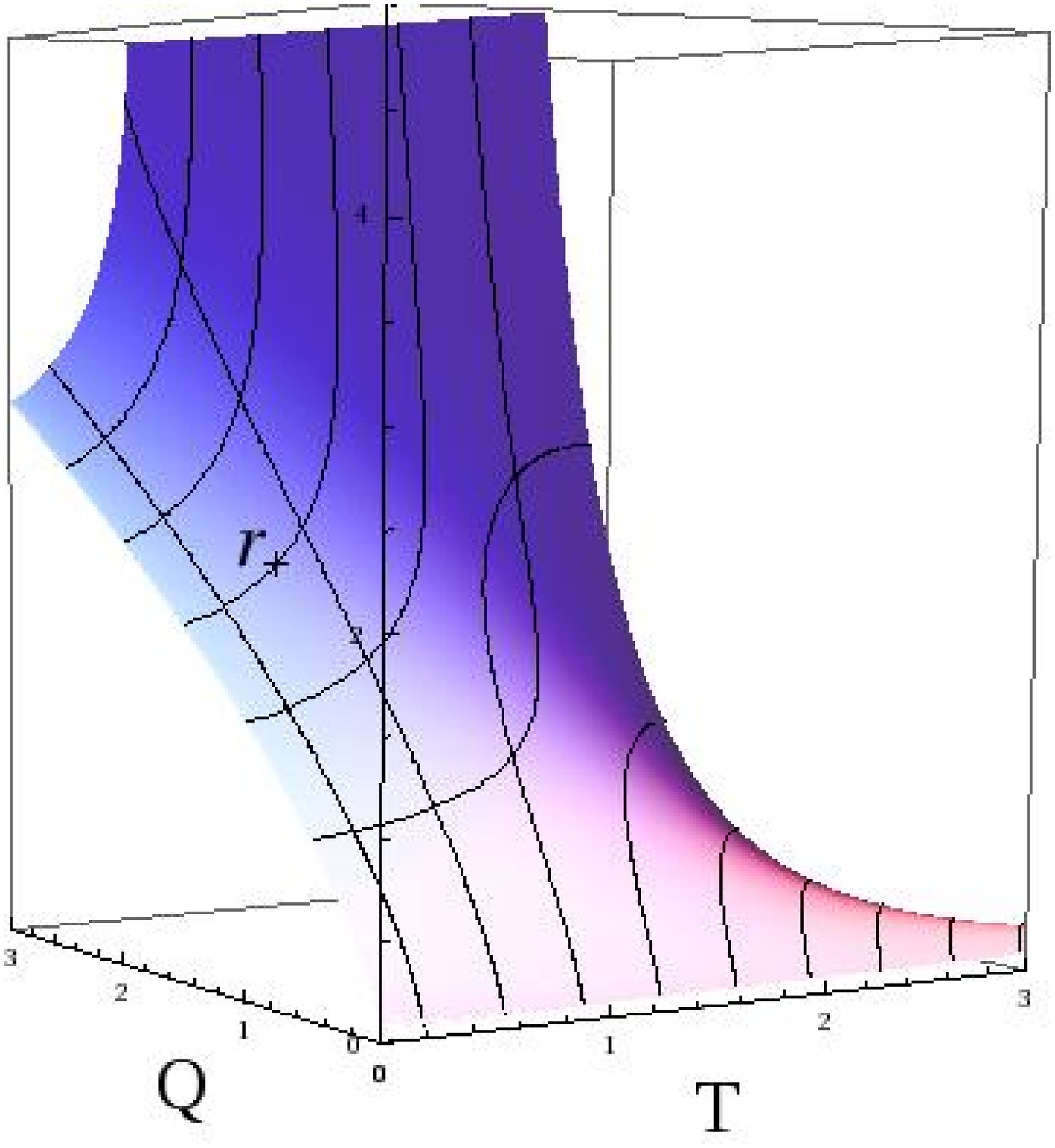}
\end{tabular}
\caption{Equation of state $r_+(T,Q)$ for the lower ($0\leq\da^2<1$) and upper ($1<\da^2<3$) range from left to right.}
\label{Fig:EqOfState3DCanonicalGa=Da}
\end{center}
\end{figure}

In the canonical ensemble, the equation of state is
\be
	 T=(3-\da^2)\ell^{\da^2-2}r_+^{1-\da^2}\le[1-\frac{4\ell^2Q^2}{(3-\da^2)r_+^{4}}\ri],
	\label{EquationOfStateCanonical2}
\ee
and allows to determine implicitly the horizon radius in term of the temperature and the charge density. Slices at fixed charge density or fixed temperature are plotted in \Figref{Fig:HorizonSizeCanonicalGa=Da}, while the three-dimensional version is in \Figref{Fig:EqOfState3DCanonicalGa=Da}. We notice two behaviours depending on the value of $\da^2$:
\begin{itemize}
 \item  In the lower range, $\da^2\leq1\,$, there is a single black hole branch for each doublet $(T,Q)$, and the limiting case $\da^2=1$ has a maximal temperature at large radius, again signalling the change of behaviour in the upper range.
 \item In the upper range, $1<\da^2<3\,$, there are two branches, small black holes and large black holes, which merge at radius,
\be
	r_{Ph}^4=\frac{(3+\da^2)}{(\da^2-1)}r_e^4\,.
	\label{CriticalRadiusCanonical2}
\ee
This corresponds either to a maximal temperature at fixed charge density, or to a maximum charge at fixed temperature, see \Figref{Fig:HorizonSizeCanonicalGa=Da}. The critical point exists only for $\da^2>1$, and the line in the $(T, Q)$ phase space so defined is given by
\be T_{Ph}=4\ell^{\da^2-2}\frac{3-\da^2}{3+\da^2}\le(\frac{4(3+\da^2)\ell^2Q_{Ph}^2}{(\da^2-1)(3-\da^2)}\ri)^{\frac{1-\da^2}{4}}\,.
	\label{CriticalTemperatureCanonical2}
\ee
\end{itemize}

\begin{figure}[t]
\begin{center}
\begin{tabular}{cc}
	 \includegraphics[width=0.45\textwidth]{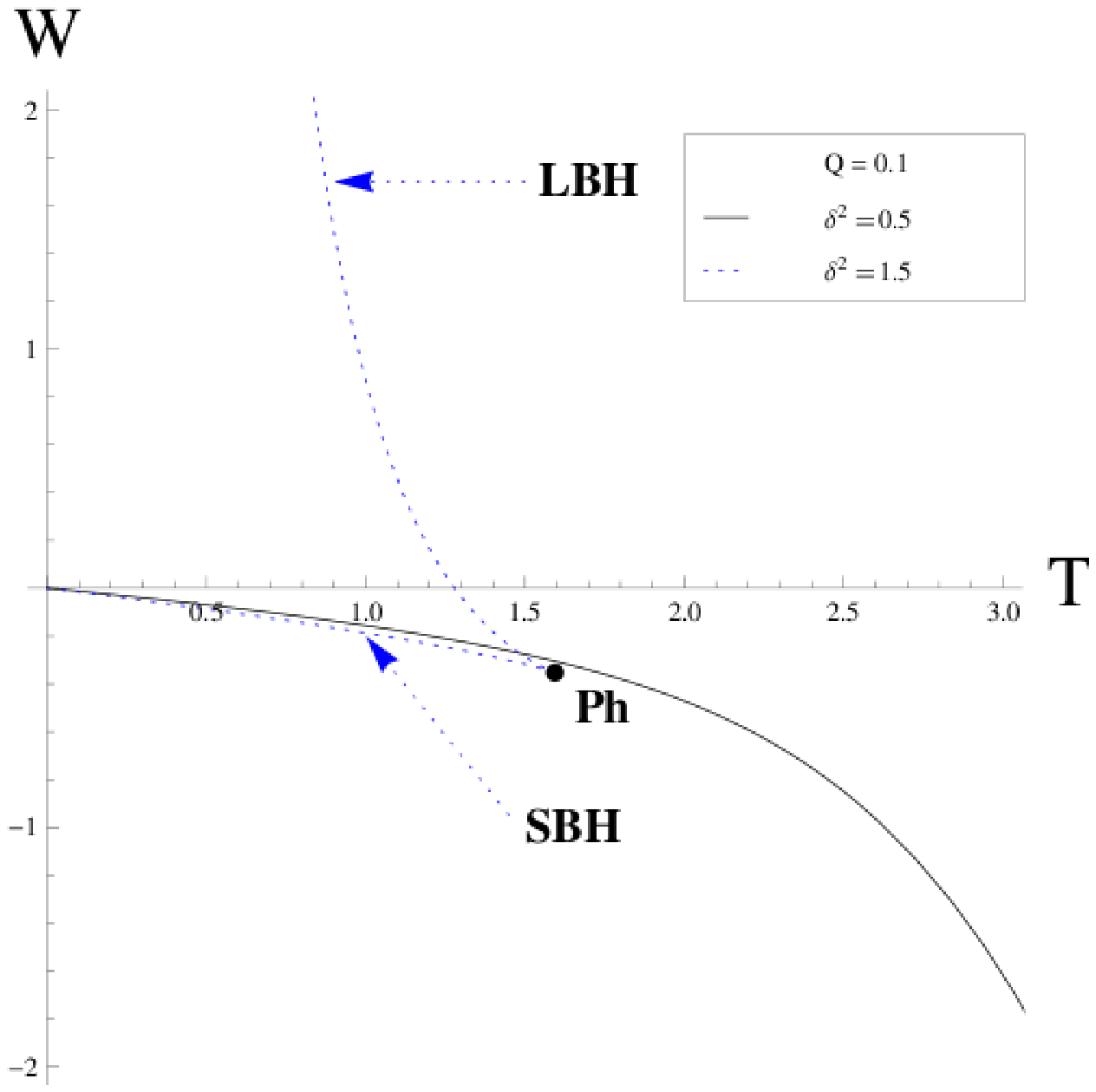}&
	 \includegraphics[width=0.45\textwidth]{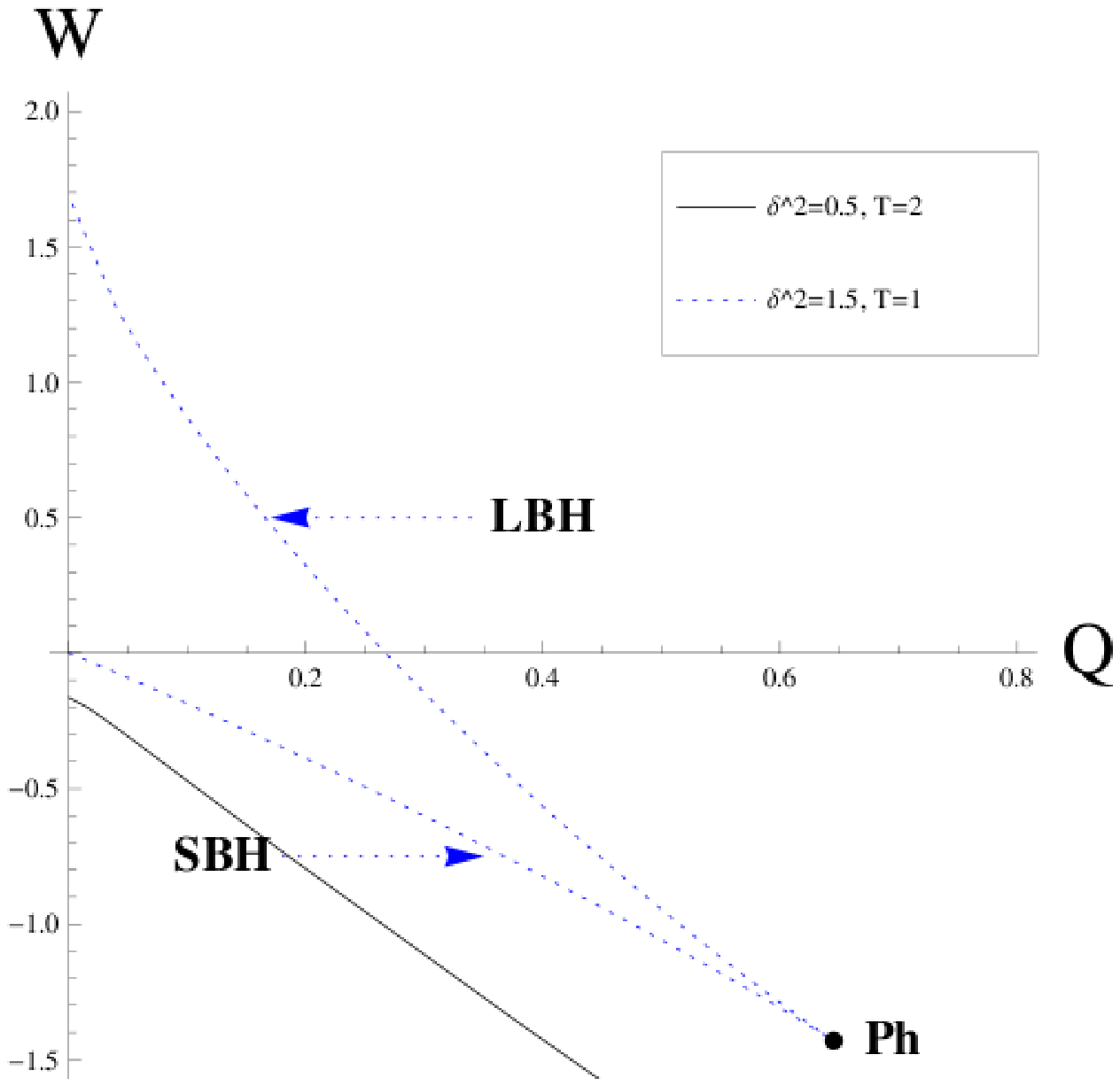}
 \end{tabular}
\caption{Helmholtz potential versus the temperature at fixed charge (left), or versus charge at fixed temperature (right), for $\da^2=0.5$ (solid line), $\da^2=1$ (dotted line)  and $\da^2=1.5$ (dashed line).}
\label{Fig:FreeEnergyGa=DaCanonical}
\end{center}
\end{figure}

We now calculate the value of the Euclidean action \eqref{ActionCanonical}, subtracting the appropriate extremal black hole background,
\be
	\tilde I - \tilde I_e = \beta \ell^{\da^2-2}\le[(1-\da^2)r_+^{3-\da^2}-\frac{4(3+\da^2)\ell^2Q^2}{(1+\da^2)r_+^{1+\da^2}} +\frac{8}{(1+\da^2)}r_e^{3-\da^2}\ri],
\label{136}\ee
from which we can deduce the Helmholtz potential in terms of the thermodynamic variables,
\be
	W[T,Q] = \ell^{\da^2-2}\le[(\da^2-1)r_+^{3-\da^2}+\frac{4(3+\da^2)\ell^2Q^2}{(1+\da^2)r_+^{1+\da^2}} -\frac{8}{(1+\da^2)}\le(\frac{4\ell^2Q^2}{3-\da^2}\ri)^{\frac{3-\da^2}4}\ri].
	\label{Helmholtz2}
\ee
Slices at fixed charge density  or at fixed temperature are plotted in \Figref{Fig:FreeEnergyGa=DaCanonical}.

 We can check that the first law is satisfied using \eqref{EnergyGrandCanonical}, \eqref{EntropyGrandCanonical} and \eqref{EntropyGrandCanonical}:
\bsea
	E_W&=& E - E_e = 4(m-m_e)\,,\\
	\Phi_W&=& \Phi-\Phi_e = \frac{16Q\ell^{\da^2}}{(1+\da^2)}\le(r_+^{-1-\da^2}-r_e^{-1-\da^2}\ri),\\
	S_W&=& S= r_+^2 = \frac{A_h}{4} \,.
\esea
The entropy, similarly to regular AdS-RN black holes, is finite at extremality, suggesting a large degenerate ground state. Its low temperature behaviour is as follows,
\be
  S_W = r_e^2\le[1+2\frac{T}{T_0}+(4\da^2-9)\le(\frac{T}{T_0}\ri)^2+\cdots\ri],\qquad T_0 = \frac{4(3-\da^2)}{\ell^{2-\da^2}}r_e^{1-\da^2}\,,
  \label{LowTEntropyCanonical2}
\ee
valid for $T\ll T_0$\footnote{the numerical coefficient in front of the subleading term will be of order one in the range of interest.}.

 Here again, we can compare with the uncharged black holes, for which the entropy scales with the temperature as
\be
  S_N\sim T^{\frac2{1-\da^2}}\,,
\label{138}\ee
which for $0<\da^2<1$ is always at least quadratic and so is subdominant both with respect to the finite and linear piece in \eqref{LowTEntropyCanonical2}, in the range of interest.
\begin{figure}[t]
\begin{center}
\begin{tabular}{cc}
	 \includegraphics[width=0.45\textwidth]{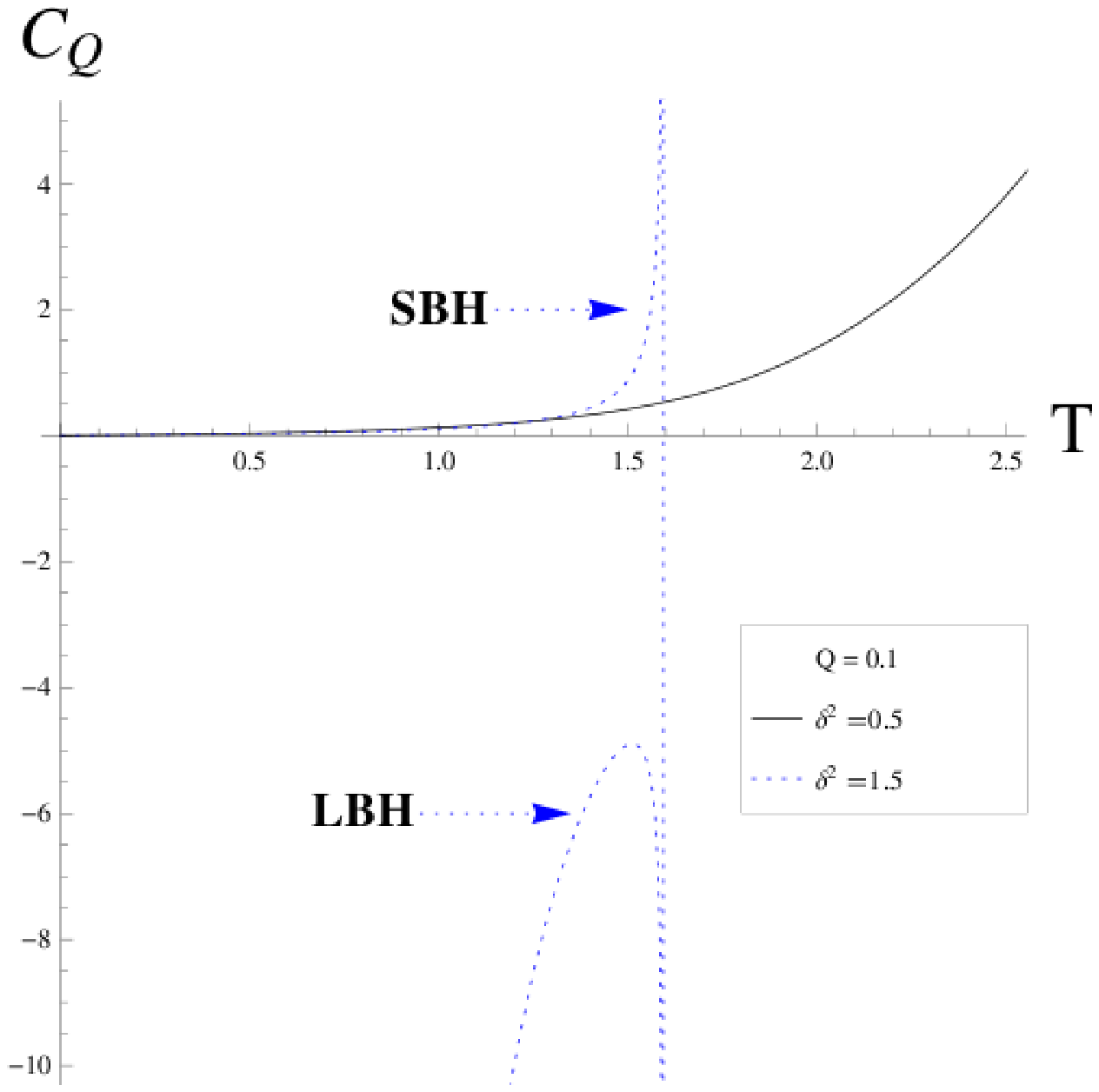}&
	 \includegraphics[width=0.45\textwidth]{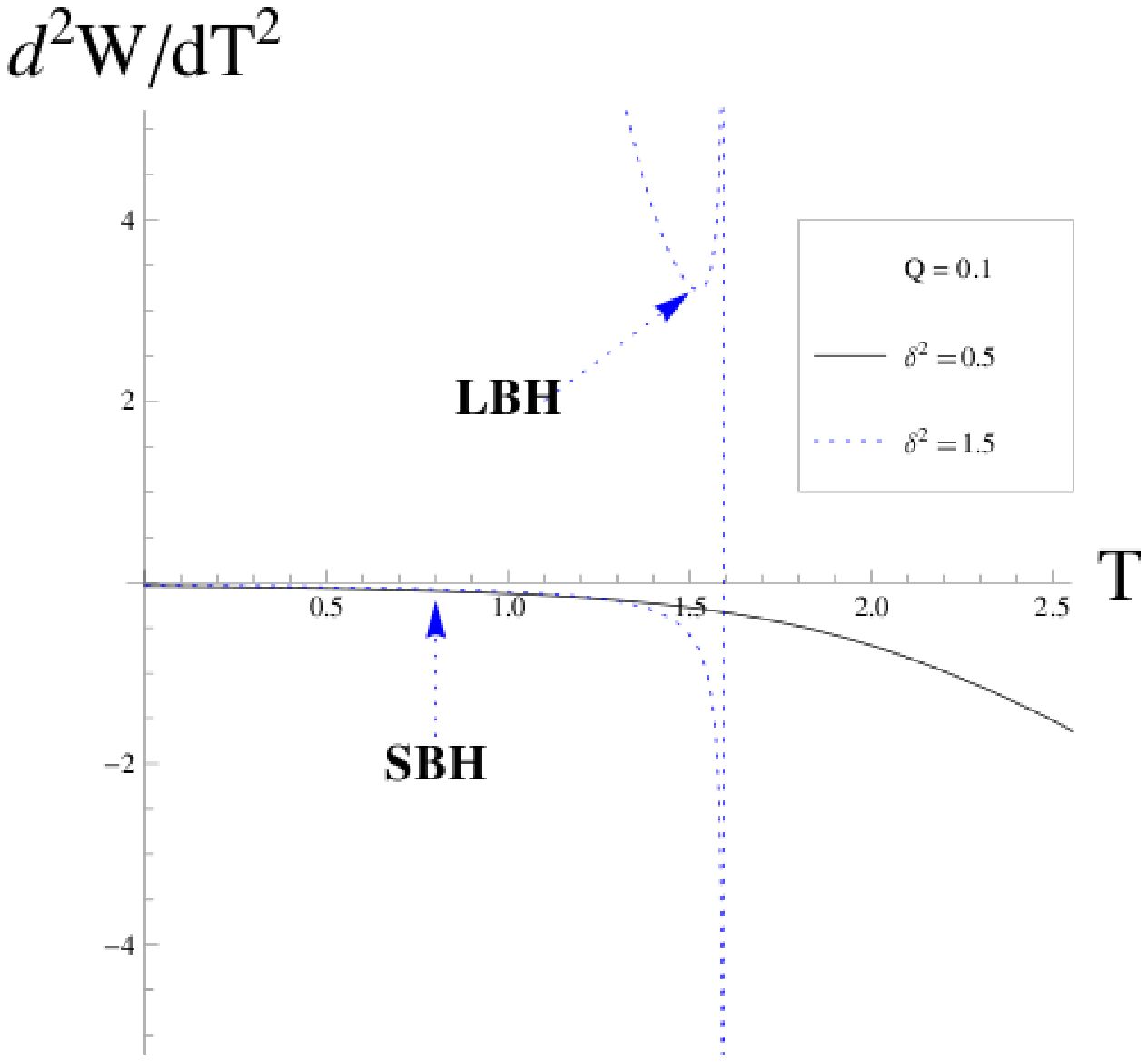}
 \end{tabular}
\caption{These plots show the heat capacity (left) and the second derivative of the Helmholtz potential with respect to the temperature (right) versus the temperature, for $\da^2=0.5$ (solid) and $\da^2=1.5$ (dot).}
\label{Fig:ThermalStabilityGa=DaCanonical}
\end{center}
\end{figure}

The heat capacity is explicitly:
\be
	C_{Q} = \frac{2r_+^2}{1-\da^2}\frac{r_+^4-r_e^4}{r_+^4-r_{Ph}^4}\,,
	\label{HeatCapacity2}
\ee
and is displayed in \Figref{Fig:ThermalStabilityGa=DaCanonical}. It scales linearly with the temperature near extremality,
\be
  C_Q = 2r_e^2\frac{T}{T_0}\le(1+(2\da^2-3)\frac{5T}{2T_0}+\cdots\ri),
  \label{LowTHeatCapacityCanonical2}
\ee
valid for $T\ll T_0$\footnote{the numerical coefficient in front of the subleading term will be of order one in the range of interest.}.

\begin{figure}[t]
\begin{center}
\begin{tabular}{cc}
	 \includegraphics[width=0.45\textwidth]{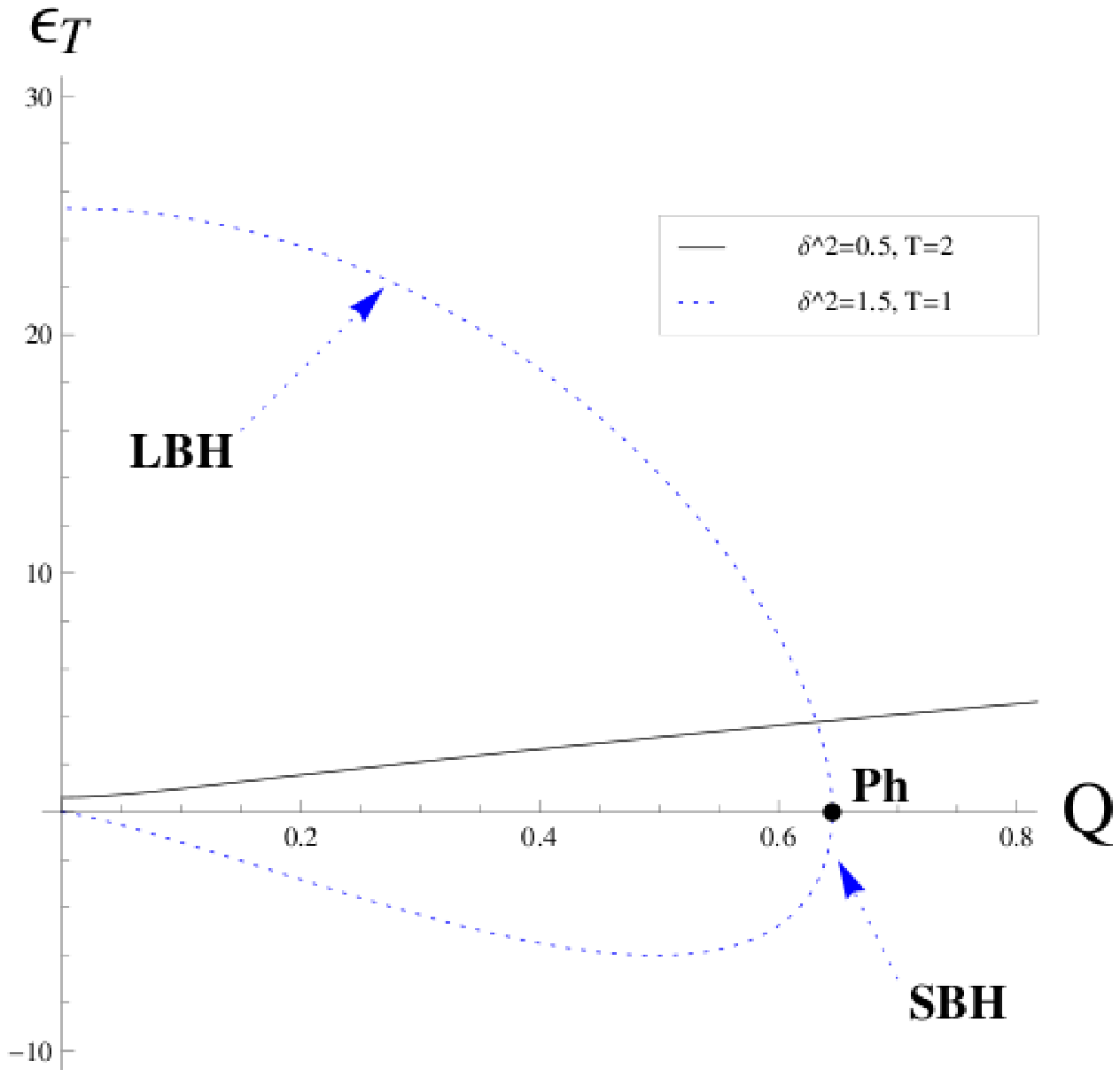}&
	 \includegraphics[width=0.45\textwidth]{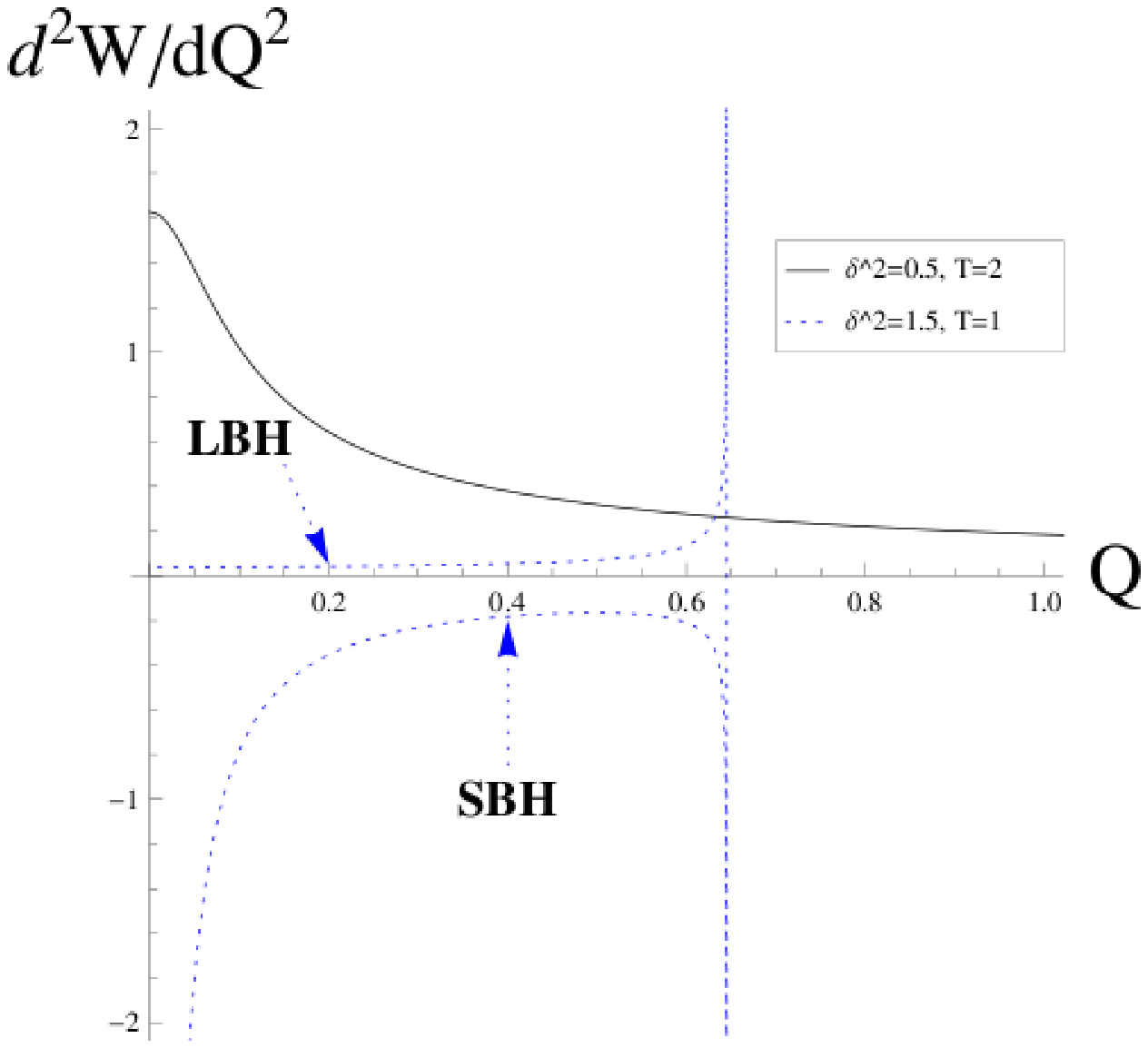}
 \end{tabular}
\caption{Electric permittivity (left) and second derivative of the Helmholtz potential (right) versus charge, for $\da^2=0.5$ (solid line) and $\da^2=1.5$ (dotted line).}
\label{Fig:ElectricPermittivityGa=DaCanonical}
\end{center}
\end{figure}

The electric permittivity is
\be
	 \epsilon_T=2\frac{1+\da^2}{3-\da^2}r_+^{1+\da^2}\frac{1+\frac{4(3+\da^2)Q^2}{(1-\da^2)(3-\da^2)r_+^4}}{1+\frac{4Q^2}{(3-\da^2)r_+^4}}\,,
	\label{ElectricPermittivityCanonical2}
\ee
see \Figref{Fig:ElectricPermittivityGa=DaCanonical}

We will now study the energetic competition between the black holes and the thermal background, e.g. the extremal limit.
\begin{itemize}
 \item Lower range: The single black hole branch is energetically favoured over the thermal background. The $\da^2=1$ displays an interesting maximal temperature, even though the separation into two branches has not yet occurred. The black holes are stable both against thermal and charge fluctuations, as seen in  \Figref{Fig:ThermalStabilityGa=DaCanonical} and  \Figref{Fig:ElectricPermittivityGa=DaCanonical}.
 \item Upper range: The small black holes are energetically favoured both with respect to the background and to the large black holes, up until the critical point where the two branches merge and cease to exist. When they are small enough, the large black holes become energetically favoured compared to the background, but still have a greater free energy than the small black holes. The small black holes are stable against thermal fluctuations, see  \Figref{Fig:ThermalStabilityGa=DaCanonical}, but not against electric fluctuations, see  \Figref{Fig:ElectricPermittivityGa=DaCanonical}. The large black holes have the opposite properties.
\end{itemize}

\subsection{The phase structure}

\subsubsection{Grand-Canonical ensemble}

\begin{figure}[!ht]
\begin{center}
\begin{tabular}{c}
	 \includegraphics[width=0.45\textwidth]{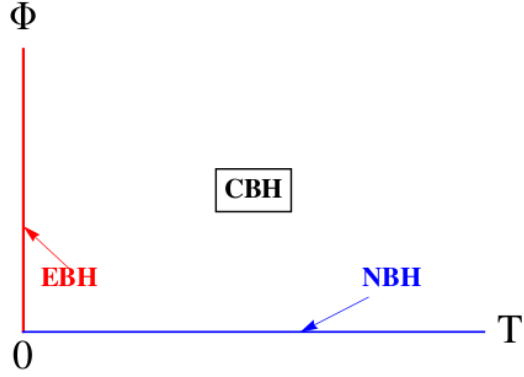}
 \end{tabular}
\caption{Phase diagram in the $(T,\Phi)$ phase space for the lower range.}
\label{Fig:PhaseDiagramGa=DaGrandCanonical}
\end{center}
\end{figure}

The phase space is non-trivial only in the lower range $\da^2<1$, and is depicted in \Figref{Fig:PhaseDiagramGa=DaGrandCanonical}. In that case, the charged black holes dominate for all $(T,\Phi)$, except when $T=0$ (extremal black hole) or $\Phi=0$ (neutral black holes). There are no phase transitions to the dilatonic background in the interior of the phase diagram, the extremal black holes can exist for any values $(T,\Phi)$, contrary to the case $\ga\da=1$. These black holes behave more like AdS-RN black holes. {The only critical behaviour appearing is when approaching zero temperature at $\Phi=0$, or zero chemical potential on the $T=0$ axis. In both cases there are phase transitions of $n^\textrm{th}$-order to the corresponding dominating solution in the parameter range
\be
\frac{n-4}{n-2}< \delta^2 < \frac{n-3}{n-1}\,,\quad n= 4,5,6\dots
\ee
In particular, these transitions are fourth-order or higher.
}

In the upper range, the dilatonic background dominates everywhere and there are no phase transitions.

\subsubsection{Canonical ensemble}

\begin{figure}[!ht]
\begin{center}
\begin{tabular}{cc}
	 \includegraphics[width=0.45\textwidth]{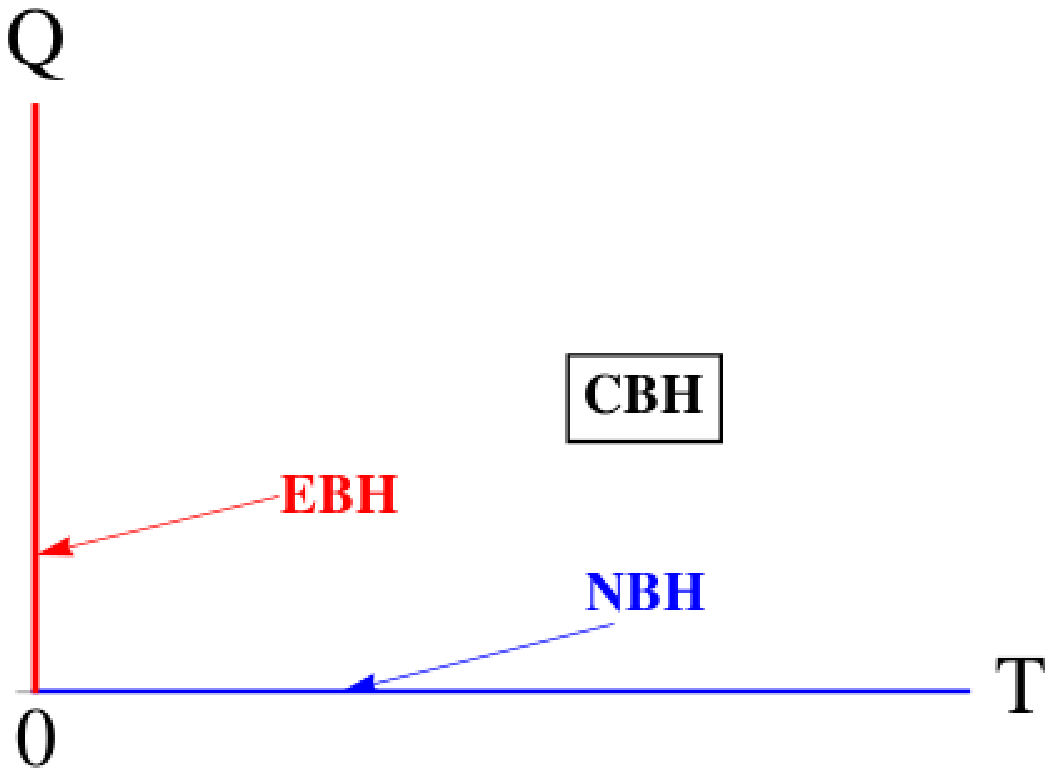}&
	 \includegraphics[width=0.45\textwidth]{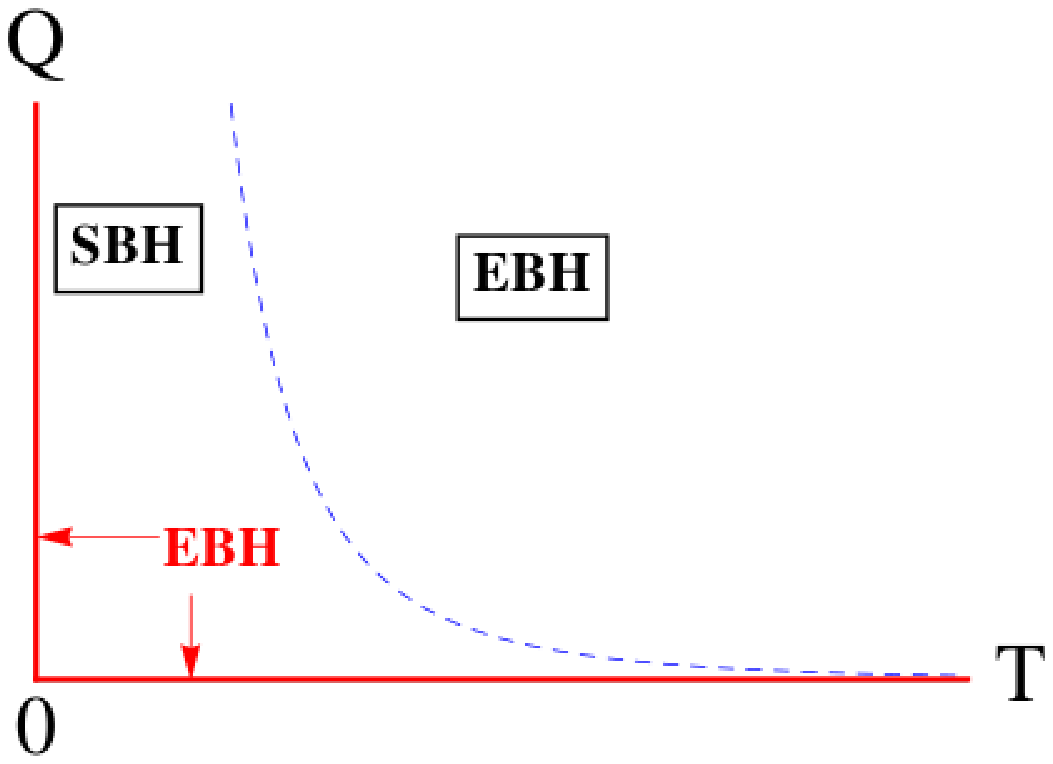}
\end{tabular}
\caption{$(T,Q)$ phase diagram for the lower range (left) and the intermediate range (right). EBH = Extremal Black Holes, CBH = Charged Black Holes, NBH=Neutral Black Holes, SBH = Small Black Holes.}
\label{Fig:PhaseDiagramGa=DaCanonical}
\end{center}
\end{figure}

The order of phase transitions is as follows, inspecting the free energy eq.~\eqref{Helmholtz2}
\begin{itemize}
 \item {Lower range: There is no phase transition to the extremal background at zero temperature for non-zero charge of the black hole. There appear however diverging mixed derivatives of the free energy as one approaches $Q=T=$ simultaneously, which are of at least third-order.

In the zero-charge limit at finite temperature, the neutral black holes dominate and there is a continuous phase transition from the charged black holes in this parameter range.
Finally, higher-order phase transitions appear as $T\rightarrow 0_+$ at exactly zero charge, as discussed around eq.~\eqref{PTuncharged}.

 }
 \item Intermediate range: There is a zeroth-order phase transition to the thermal background at the point $(T_\mathrm{Ph},Q_\mathrm{Ph})$ since the Helmholtz potential is discontinuous there and jumps to zero. {This is again to be remedied by an AdS-completion.

 At zero-temperature and finite charge, the small black holes again settle continuously in their extremal state, as seen by the continuity of the Helmholtz potential and its second derivative.

 Aproaching zero charge at finite temperature, the first derivative of the Helmholtz potential w.r.t. $Q$ diverges, so there is a first-order phase transition to the thermal background (which is the endpoint of the stable small black holes branch). In particular, on the vertical axis $Q=0$, the neutral black holes do not dominate (they are the endpoints of the unstable large black holes branch).
}
\end{itemize}

We have plotted the phase diagram ($T,Q$) for the lower and intermediate ranges in \Figref{Fig:PhaseDiagramGa=DaCanonical}.

\subsection{The AC conductivity}

We now turn to the optical (AC) conductivity for small frequencies at zero temperature, i.e. for the extremal case of \eqref{Sol2}. We proceed along the lines of section~\ref{sec:ACconductivity}. The extremal limit of the solution~\eqref{Sol2} is reached for
\be
 \left( \frac{r_\ast}{\ell} \right)^{3-\d^2}= \frac{m(1+\d^2)}{2\ell} = \left[ \frac{1+\d^2}{3-\d^2} \tilde q^2 \right]^{\frac{3-\d^2}{4}},\qquad \tilde q^2 = \frac{q^2}{4\ell^2(1+\d^2)}\,.
\label{140}\ee
Since the low-frequency behaviour of the AC conductivity is only sensitive to the near-horizon geometry, it suffices to analyse the fluctuation problem \eqref{28} in the extremal near horizon geometry of \eqref{Sol2}, which reads
\bsea\nonumber
\ud s^2 &=& - 2 \ell^2 (3-\d^2) \rho^2 \ud t^2 + \ell^2 \rho_\ast^{2\d^2} \frac{\ud \rho^2}{2(3-\d^2) \rho^2}
 + \ell^2 \rho_\ast^2 (\ud x^2 + \ud y^2)\,,\\
e^\phi &=& \rho_\ast^{2\d}  = \text{const}\,.
\label{141}\esea
Here the near-horizon coordinate
\be
\rho = \frac{r}{\ell} - \rho_\ast\,,\qquad \rho_\ast^4 = \frac{1+\d^2}{3-\d^2} \tilde q^2\,,
\label{142}\ee
has been introduced, and the time coordinate $t \mapsto \ell t$ has been rescaled in order to work with dimensionless coordinates only. Note that the AC conductivity is calculated in the canonical ensemble, in which temperature and charge of the solution is fixed. The relevant coordinate change to transform the fluctuation problem \eqref{30a} into the Schr\"odinger problem \eqref{32} then is given by
\be
z(r) = - \frac{\rho_\ast^{\d^2}}{2(3-\d^2) \rho}\,.
\label{143}\ee
The effective Schr\"odinger potential \eqref{32} in the near-horizon region (the horizon is at $y_H=-\infty$) then reads
\be
V = \frac{2}{z^2}\,.
\label{g=dVeff}
\ee
 Since the Schr\"odinger potential has the same form as observed in \cite{kachru}, we immediately deduce the small frequency scaling of the optical conductivity to be
\be\label{g=dACscaling}
\sigma(\omega) \simeq \omega^n\,,\qquad n=2\,.
\ee
We thus find that the solution~\eqref{Sol2} gives rise to a low-frequency scaling law for the AC conductivity independent of the parameter $\d$. A similar case of ``universality'', i.e. independence of the scaling exponent from the parameters in the action, was observed in \cite{kachru}. As a consistency check we can use the fact that the solution~\eqref{Sol1} coincides with the $\ga=\d$ solution \eqref{Sol2} for $\d=1$, and hence the scaling exponent \eqref{gd1ACscaling} agrees with \eqref{g=dACscaling} for this particular choice of $\d$.

In particular it indicates that in this case the charged carriers are gapless with a continuous spectrum everywhere.

\subsection{The DC conductivity}

Similar to the DC conductivity calculation for $\gamma \delta  = 1$, we consider a DC resistivity
for $\gamma =\delta $ solution. The resistivity reads
\be
	\rho \simeq{T_f g^E_{xx}(r_+)e^{k\phi(r_+)}\over  J^t}
	= \frac{T_f }{J^t}  \left( \frac{ r_+}{\ell} \right)^{2 + 2 \delta k},
\label{144}\ee
which is much simpler than $\gamma \delta = 1$ case and the resistivity is a simple power of $r_+$.

\begin{figure}[!ht]
\begin{center}
\begin{tabular}{c}
	 \includegraphics[width=0.45\textwidth]{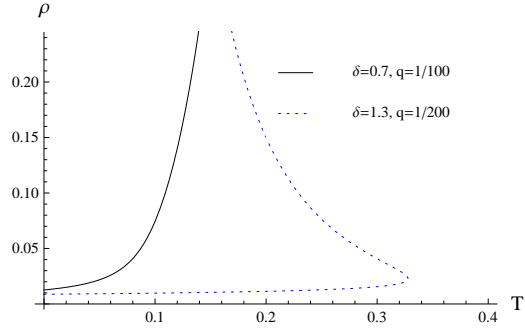}
\end{tabular}
\caption{DC resistivity for $\gamma = \delta$ case. At the left side is a plot for $\gamma =\delta =0.7$ in the lower range
and at the right one for $\gamma = \delta =1.3$ in the upper range, both with $k=0$.
Plots are generated for dimensionless quantities, $x$-axis being $\ell T$,
$y$-axis being $\frac{J^t}{T_f} \rho $.  The lower part of the upper range line corresponds to the thermodynamically dominant small black hole.}
\label{Fig:d=gResistivity}
\end{center}
\end{figure}

To consider a temperature-dependent resistivity, we use the temperature expression (\ref{Temperature2}).
Resistivity in the $ \gamma =\delta$ case has  slightly different features compared to the $\gamma \delta  =1$ case.
The upper range in the $ \gamma \delta =1$ case disappears and merges to the intermediate range. Thus
there are two distinct regimes, which are $0< \delta^2 <1$ and  $1< \delta^2 <3$.
For given ranges of $\delta$, the qualitative behavior of the resistivity does not change as we can observe
in the plots in \Figref{Fig:d=gResistivity}.

\begin{figure}[!ht]
\begin{center}
\begin{tabular}{c}
	 \includegraphics[width=0.45\textwidth]{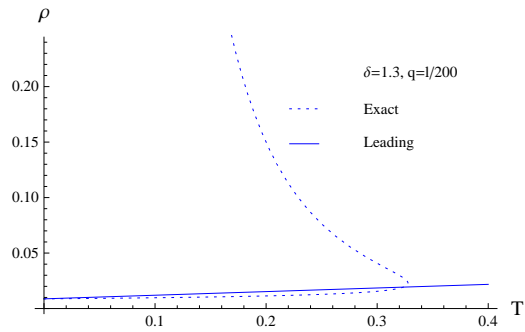}
\end{tabular}
\caption{Resistivity for both leading order (solid line) and exact case (dotted line), which is in the
lower branch of the upper range with $k=0$. Leading order resistivity grows faster than the exact case at first, and then
they meet each other near the turning point. Plots share the same axis conventions as the previous one. }
\label{Fig:d=g=1.2-DC-Exact-Leading}
\end{center}
\end{figure}

For small temperatures, $r_+$ approaches the extremal limit $r_e = \left(\frac{\ell^2 q^2}{4(3-\delta^2)}\right)^{\frac{1}{4}}$.
Therefore we expand $r_+$ around the extremal $r_e$, $r_+ = r_e + \delta r$, to obtain $\delta r$ as a function of $T$.
Then,  we substitute $\delta r (T)$ in the resistivity expression to obtain the small temperature expression for the resistivity,
\be
	\delta r \sim  \frac{ \pi}{3 - \delta^2} (\ell^2 T) \left(\frac{r_e}{\ell}\right)^{\delta^2}.
\label{145}\ee
The low temperature resistivity, including the leading non-trivial order of $T$, is
\begin{eqnarray}
	\rho &&\simeq  \frac{T_f }{ J^t}   \left( \frac{r_e}{\ell} \right)^{2 + 2 \delta k}
	\left( 1 +   \frac{ \pi(2+2 \delta k)}{3 - \delta^2} \left(\frac{r_e}{\ell}\right)^{\delta^2 - 1} ~\ell T  +{\cal O}(T^2)\right).
\label{146}\end{eqnarray}
Thus the first non-trivial temperature dependence is always linear unless we have $\delta k = -1$, which always gives
a constant resistivity. Note that there is a non-zero resistivity, $\rho_{0}$, at zero temperature. This is  different
from $\gamma \delta =1$ case. These features are depicted in the \Figref{Fig:d=g=1.2-DC-Exact-Leading}.

 We estimate the range of the validity of the leading order approximation
by requiring the next-leading order contribution to the resistivity to be small compared to
the leading order term. Note that this condition automatically makes the leading terms much larger than the linear term in the temperature.

We obtain
\be
\ell T \ll \left(\frac{r_e}{\ell}\right)^{1-\da^2} \sim  \left(\frac{q}{\ell}\right)^{\frac{1-\da^2}{2}}.
\label{147}\ee

\section{Near-extremal scaling solutions\label{nearex}}

A solution can be obtained without any constraints on the coupling constants for general values of $\gamma,\d$.
\bsea
	\ud s^2 &=& -V(r)r^{-4\frac{\ga(\ga-\da)}{wu}}\ud t^2 +\frac{e^{\da\phi}\ud r^2}{-w\Lambda V(r)} + r^{2\frac{(\ga-\da)^2}{wu}}\le(\ud x^2+\ud y^2\ri), \slabel{Metric3bis}\\
	e^{\phi} &=& e^{\phi_0}r^{-4\frac{(\ga-\da)}{wu}}\,, \slabel{Phi3bis}\\	
	\mathcal A&=& 2\sqrt{\frac{-v}{wu}}e^{-\frac\ga2\phi_0}\left[r-2m\ri]\ud t\,, \slabel{A3bis}\\
	V(r)&=& r(r-2m)\,, \slabel{Pot3bis} \\
	wu&=&3\ga^2-\da^2-2\ga\da+4\,, \quad u=\ga^2-\ga\da+2\,, \quad v=\da^2-\ga\da-2\,,\nn
	\label{Sol3}
\esea

For future use, we also give the solution in the so-called \emph{domain-wall} coordinates \eqref{ToroidalMetrics2}.
This is achieved with the change of coordinates $r=\rho^{\frac{wu}4}$, $t=4\sqrt{\frac{-\Lambda}{wu^2}}e^{-\frac\da2\phi_0}\bar{t} $, and yields
\bsea
	\ud s^2 &=&\rho^{\frac{(\ga-\da)^2}2}\le[\ud x^2+\ud y^2  - f(\rho)\ud \bar t^2\ri] + \frac{\ud\rho^2}{f(\rho)}\,, \slabel{DWMetric3}\\
	f(\rho) &=& \frac{16\le(-\Lambda\ri)}{wu^2}e^{-\da\phi_0}\rho^{1-\frac34(\ga-\da)^2}\le(\rho^{\frac{wu}4}-2m\ri), \slabel{DWPot3}\\
	e^{\phi} &=& e^{\phi_0}\rho^{-(\ga-\da)}\, \slabel{DWDilaton3}\\
	\mathcal A &=& \frac8{wu}\sqrt{\frac{v\Lambda}{u}}e^{-\frac{(\ga+\da)}2\phi_0}\left[\rho^{\frac{wu}4}-2m\ri]\ud t\,. \slabel{DWA3}
	\label{DomainWall3}
\esea

The solution is well defined only if $uv<0$ ( since we take $\Lambda<0$ to have a potential bounded from below).

The Ricci scalar is
\be
	\mathcal R = 4\Lambda e^{-\da\phi_0}r^{-\frac{3(\ga-\da)^2+4}{wu}}\le\{\le[1-\frac{2(\ga-\da)^2}{wu^2}\ri]r+4m\frac{(\ga-\da)^2}{wu^2}\ri\}\,.
	\label{RicciScalar3}
\ee
There is an event horizon at $r_h=\rho_h^{\frac14(\ga-\da)(3\ga+\da)+1}=2m$ and a curvature singularity at $r=\rho=0$ for all $\da^2<3$.
Indeed, it is not hard to see that if $wu>0$ (which is true in particular for all $\da^2<3$),
\be
	wu=(3-\da^2)(1+\ga^2)+(1-\ga\da)^2>0\,,\qquad\forall \da^2<3\,, \nn
\ee
then the Ricci scalar diverges at $r=0$ and is regular at $r_h$. On the other hand, if $wu<0$, then we need to change variables $r\to\frac1r$ in order for the Ricci scalar not to diverge asymptotically, but then this changes the nature of the solution and we have a cosmological space-time.
This is precisely the {\em Gubser bound} for such solutions.
The region in the ($\g,\d$) plane that gives a cosmological solution violates the Gubser bound and so should be considered as unphysical: the conditions for having an acceptable naked singularity at extremality are
\be
wu>0\sp uv<0
\ee

These respective regions in the upper half $(\ga,\da)$ plane are plotted in the left part of figure \Figref{Fig:NatureSolutionNearExtremal},
Note the symmetry by a rotation of $\pi$ around the origin, reflecting that any simultaneous change of sign in $\ga$ and $\da$ can always be absorbed by a change of sign in $\phi$.

Here again, $m$ is related to the mass of the solution, and $\phi_0$ to the overall scale (and thus can be fixed at will), but we lack a second dimensionful integration constant. As such, the charge density  built through the usual Gaussian integral is finite but universal. This hints that the solution is not the generic one.

 We will now proceed and take both $\ga\da=1$ and $\ga=\da$ in \eqref{Sol3}, and compare with both full solutions we have in our possession in these two limits, \eqref{Sol1} and \eqref{Sol2}, by going to the so-called \emph{near-extremal} limit for both.

Let us determine the near-extremal behaviour of the full solution \eqref{Sol1} that is, set\footnote{We could as well work with $r_+$, this is just a matter of convenience.} $r^{3-\da^2} = r_-^{3-\da^2} + \xi$, with $\xi$ small compared to $r_-$. We obtain:
\bsea
	e^\phi &=& r_-^{2\da\frac{(3-\da^2)}{(1+\da^2)}}\xi^{\frac{4\da(\da^2-1)}{(3-\da^2)(1+\da^2)}}\,, \\
	\mathcal A & = & \le[\Phi + \sqrt{\frac{\da^2}{1+\da^2}}r_-^{-\frac{(3-\da^2)}{(1+\da^2)}}\xi \ri]\ud t\,, \\
	\ud s^2 &=& -\le[\xi(\xi-2\tilde m)\ri]\xi^{\frac{4(\da^2-1)}{(3-\da^2)(1+\da^2)}}  \ud t^2 +\frac{e^{\da\phi}}{(3-\da^2)^2}\frac{\ud\xi^2}{\xi(\xi-2\tilde m)}+ \nn\\
					&&+\xi^{\frac{2(\da^2-1)^2}{(3-\da^2)(1+\da^2)}}  \le(\ud x^2+\ud y^2 \ri),
	\label{NearExtremal1}
\esea
where we have set $r_-\sim r_+$ at zeroth order and $\tilde m = r_+^{3-\da^2}-r_-^{3-\da^2}$. The extremal limit is seen to be exactly $\tilde m=0 \Longleftrightarrow r_+=r_-=r_e$. Then, identifying in \eqref{Sol3}
\bea
	e^{\phi_0} &=&  r_-^{2\da\frac{(3-\da^2)}{(1+\da^2)}}\,, \nn\\
	\frac{-\Lambda}{3-\da^2} &=& \ell^{-2} =1\,, \nn
\eea
the near-extremal limit of the $\ga\da=1$ solution \eqref{Sol1} and the $\ga\da=1$ limit of \eqref{Sol3} agree.

We now turn to the second full solution for $\ga=\da$ \eqref{Sol2}, and set this time $r^{1+\da^2} = r_-^{1+\da^2} + \xi$. Then, being careful to keep all the terms to high-enough order, this yields:
\bsea
	e^\phi &=& r_-^{2\da}\,, \\
	\mathcal A & = & \le[\Phi + \sqrt2r_-^{-\da^2}\xi \ri]\ud t\,, \\
	\ud s^2 &=& -\xi(\xi-2\tilde m) \ud t^2 +\frac{r_-^{2\da^2}}{2(3-\da^2)\xi(\xi-2\tilde m)} + \le(\ud x^2+\ud y^2 \ri).
	\label{NearExtremal2}
\esea
This is just a product of $AdS_2\times\mathbb{R}_2$, with a constant dilaton but with a non-zero gauge field. $\tilde m$ has been redefined:
\be
	\tilde m = \frac{1}2r_-\le(\frac{r_e^4}{r_-^4}-1\ri), \nn
\ee
where here again $\tilde m=0$ corresponds precisely to the extremal limit $r_-=r_+=r_e$. We then compare with \eqref{Sol3}, and find perfect agreement, provided we set in \eqref{Sol3}
\bea
 	e^{\phi_0} & = & r_-^{2\da}\,, \nn \\
	-(3-\da^2)\Lambda &=& \ell^{-2}= 1\,.
\label{148}\eea

Here,  we are able to ascribe a precise interpretation to the generic $\ga$, $\da$ solution \eqref{Sol3}. Indeed, it turns out that, both in the limit $\ga\da=1$ and $\ga=\da$, it coincides exactly with the near-extremal limit of the full solution. Therefore {\em we interpret \eqref{Sol3} as the near-extremal limit of the full generic solution}. Yet, such a general solution remains elusive and we keep a more detailed search for a later work.

\subsection{Extension to arbitrary dimension\label{p+1}}

From the $p+1$-dimensional equations of motion, one can find the $p+1$-dimensional generalisation of the solution \eqref{DomainWall3}, using a scaling Ansatz:
\be
  \ud s^2 = e^{2A}\left[-f(r)\ud t^2 +\ud x_i\ud x^i \right] + \frac{\ud r^2}{f(r)}\,,
  \label{Sol3DW}\ee
  \be
  e^{A} = \left({r\over \ell}\right)^{\frac{(\ga-\da)^2}{2(p-1)}}\sp
  e^\phi = e^{\phi_0}\left({r\over \ell}\right)^{(\da-\ga)}
  \label{dd1}\ee
  \be
  f(r) = \frac{8(p-1)\ell^2\le(-\Lambda\ri) e^{-\da\phi_0}}{u^2 w_p}\left({r\over \ell}
  \right)^{1-\frac{p(\ga-\da)^2}{2(p-1)}}\le[\left({r\over \ell}\right)^{\frac{w_pu}{2(p-1)}} -2m \ri],\label{88e}
  \ee
  \be
  A_t = \frac{4(p-1)}{w_pu}\sqrt{\frac{\ell^2\Lambda v}{u}}e^{-\frac{(\ga+\da)}2\phi_0}
  \left[\left({r\over \ell}\right)^{\frac{w_pu}{2(p-1)}}-2m\right]\,,
  \label{dd2}\ee
  \be
  w_pu = 2(p-1)+p\ga^2-2\ga\da-(p-2)\da^2\sp
  u = \ga^2-\ga\da+2\sp
  v = \da^2-\ga\da -2\,.\label{wpu}
  \ee
We have also reintroduced the IR scale $\ell$.

\begin{figure}[t]
\begin{center}
\begin{tabular}{cc}
	 \includegraphics[width=0.45\textwidth]{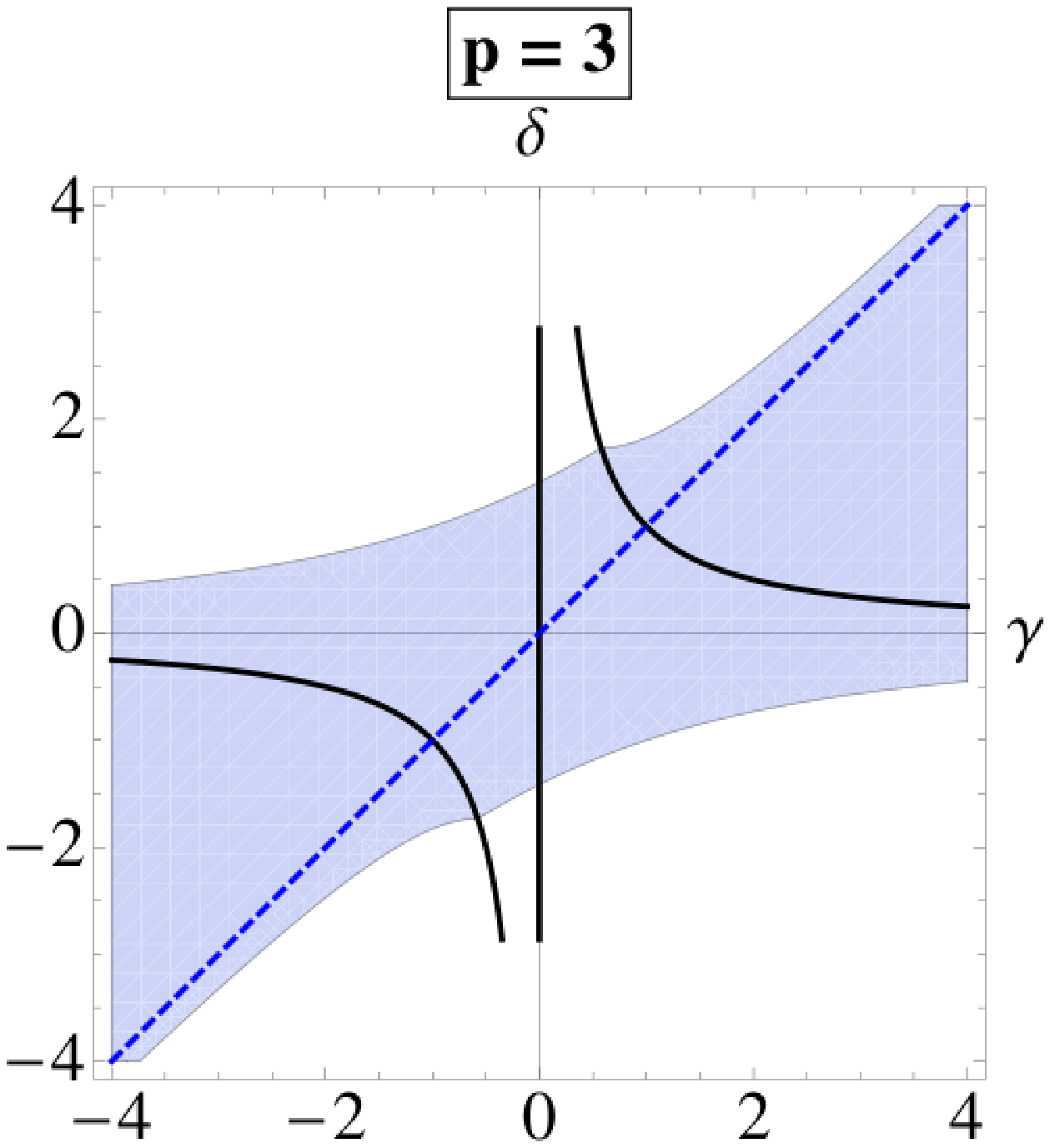}
	 \includegraphics[width=0.45\textwidth]{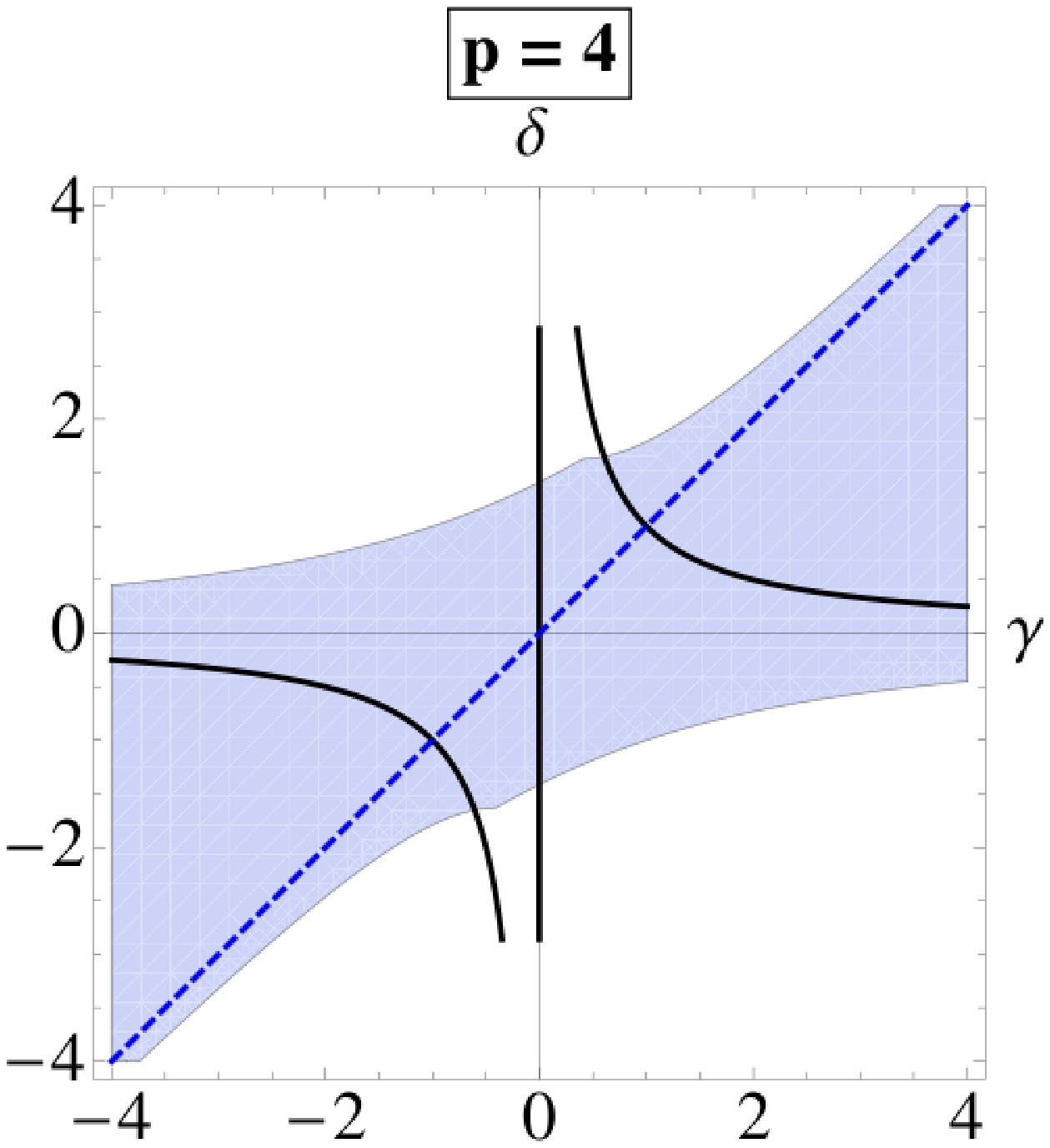}
\end{tabular}
\caption{
This graph shows the Gubser bounds on the near extremal solution on the whole  of the $\le(\ga,\da\ri)$ plane for $p=3$ and $p=4$.
The blue regions are the allowed regions where the near extremal solutions are black-hole like. The white regions are solutions of a cosmological type and therefore fail the Gubser bound.
The constraints for a black-hole type solution are given in (\protect\ref{d3}) The dashed blue line is the $\g=\delta$ solutions while the solid black line corresponds to the $\g\d=1$ solutions.
}
\label{Fig:NatureSolutionNearExtremal}
\end{center}
\end{figure}

The extremality Noether charge (\ref{noe}) is given by
\be
{\cal Q}={8e^{-\delta\phi_0}\ell^2\le(-\Lambda\ri)\over u}m\,,
\label{d1}\ee
the temperature $T\sim m^{1-{(p-1)(\gamma-\d)^2\over w_pu}}$  while the chemical potential as a function of temperature
\be
\Phi={4(p-1)\over w_p}e^{-{\gamma+\d\over 2}\phi_0}\sqrt{{v\over \ell^2\Lambda u}}\left({\pi u T\over \ell \le(-\Lambda\ri) e^{-\d\phi_0}}\right)^{w_p\over w_p-p(\gamma-\d)^2}\,.
\label{d2}\ee
There are constraints on the parameters:
for the solution to exist and have AdS character
\be
u>0\sp v<0\sp w_p>0\,.
\label{d3}\ee
These capture the Gubser criterion for the associated naked singularity at extremality. They are portrayed in figure \ref{Fig:NatureSolutionNearExtremal} for $p=3$ and $p=4$.

\subsection{Regions of validity}

\begin{figure}[!ht]
\begin{center}
\begin{tabular}{cc}
	 \includegraphics[width=0.45\textwidth]{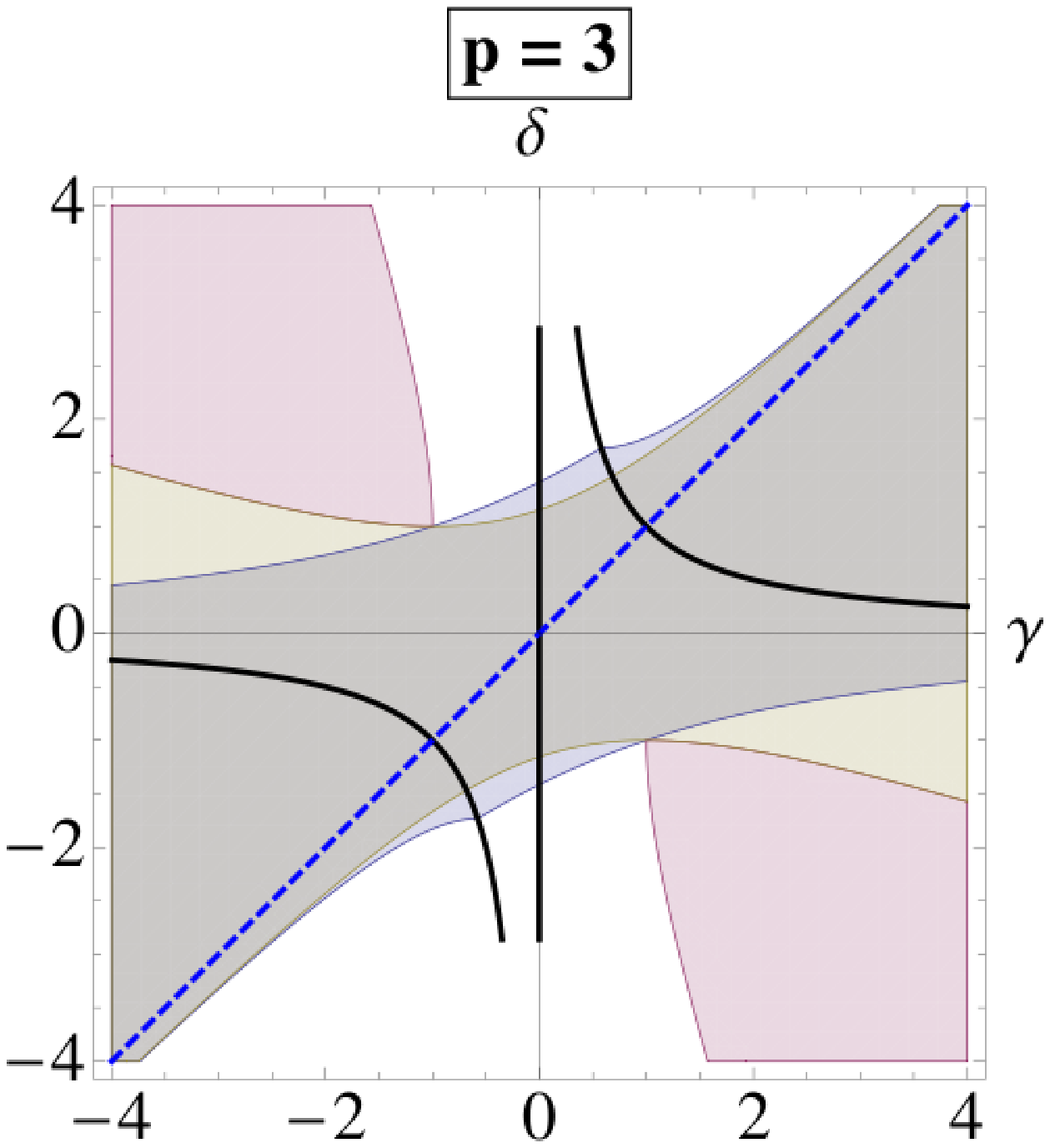}	 \includegraphics[width=0.45\textwidth]{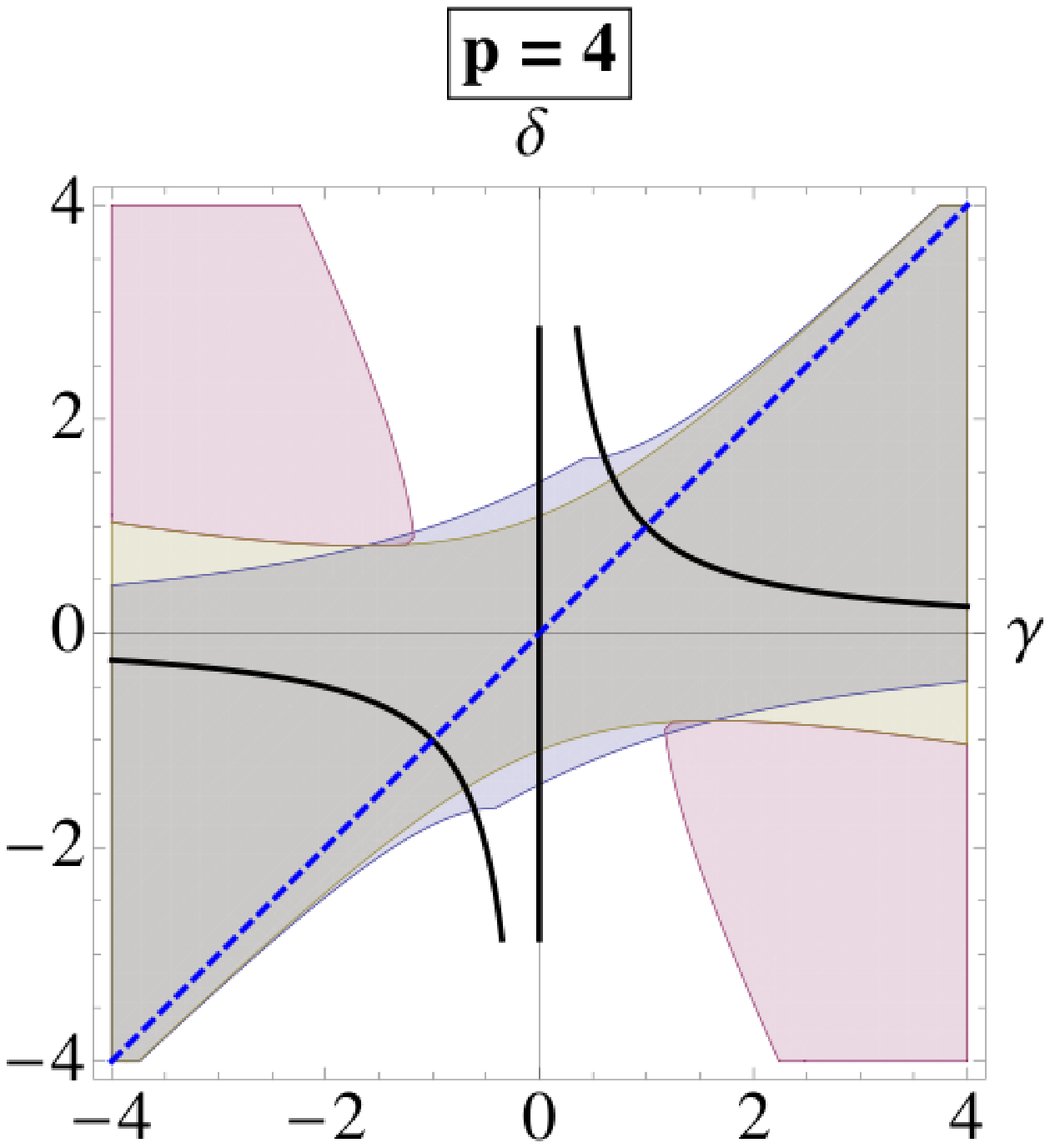}
\end{tabular}
\caption{These figures summarize the reliability constraints on the spin-2 fluctuations for $p=3$ and $p=4$.  The blue region depicts the part of the $(\gamma,\d)$ plane which satisfies the Gubser bound constraints in (\protect\ref{d3}).
The yellow-brown region is determined by $b>1$ and is a reliable region.
The second reliable region is $b<1$ and $a>3$ from  (\protect\ref{d10}) and is portrayed as the purple region. Note that this barely touches the Gubser allowed region for $p=3$ and has a small overlap for $p=4$. The solid black line corresponds to $\g\d=1$ and the dashed blue line to $\g=\d$.  }
\label{Fig41}
\end{center}
\end{figure}
\indent
The UV region of the solutions is $\rho\to \infty$ where the scale factor increases without bound. We have to be careful with the interpretation of this class of solutions in the UV.
Although they represent exact solutions of the system, the question as to whether their UV region can be successfully modified to different asymptotics is subtle.

 For example the solution \eqref{Sol3} has a potential that behaves as
\be
V\sim e^{-\delta\phi}\sim \rho^{\d(\gamma-\d)}\,.
\label{149}\ee
Therefore, for $\d(\gamma-\d)> 0$, the potential diverges in the UV, and implies that there is a direct problem of considering such solutions as IR limits of asymptotically AdS solutions for example.
A hint of what really happens is given by the $\g\d=1$ solutions, which have been shown to asymptote to the extremal scaling solutions in the IR, and to the extremal uncharged solutions in the UV, (see (\ref{NearExtremal1}) and appendix \ref{gd11}).

In appendix \ref{phasev} we give a detailed analysis of the system of equations with the extremal uncharged and charged solutions as fixed points. We then study the stability of these fixed points and show that the uncharged fixed point solution is always unstable, while the extremal charged solution is stable once an initial condition is chosen to avoid a bad singularity. Therefore, with an appropriate tuning of UV conditions, there is a solution that asymptotes to the uncharged solution (\ref{9}) in the UV and the charged solution (\ref{Sol3}) or (\ref{Sol3DW})-(\ref{wpu}) in the IR. This is supported by the stability analysis as well as the explicit $\g\d=1$ and $\g=\d$ solutions\footnote{In reference \cite{perl} that appeared after the present paper, this same statement was argued and shown numerically to be true in several concrete examples. The discussion here has been added after the appearance of \cite{perl}, and in order to justify the UV completability of the solutions.}.

 Because we have shown in section \ref{zero} that the uncharged solutions have vanishing potential in the UV, this implies that all charged scaling solutions are UV completable.

In the case where the conserved charge is originating on flavor branes with a DBI action, according to (\ref{19})
we can linearize the action when
\be
Z e^{(p-1)A+k\phi}\sim \rho^{{1\over 4}(\gamma-\d)\left[-4(k+\gamma)+(\gamma-\d)(p-1)\right]}\gg |q|\,.
\label{150}\ee
When $(\gamma-\d)\left[-4(k+\gamma)+(\gamma-\d)(p-1)\right]>0$, the Maxwell approximation breaks down near the UV.
In the opposite case it breaks down in the deep IR.

Finally the probe approximation is valid
\be
{ZVe^{2(p-1)A}}\sim \rho^{{(p-3)\over 2}(\gamma-\delta)^2}\gg q^2\,,
\label{151}\ee
in the Maxwell limit. The probe approximation thus always breaks down in the IR for $p\ge 3$.
In the strong self-interaction limit this same condition becomes $\rho^{(\delta-\gamma)(\d+\ga+2k)}\ll 1$ according to section \ref{probe}.

\subsection{Reliability of the IR fluctuation problem at extremality}

The extremal black-hole solutions are generically naked singularities (except  when $\g=\d$).  The region in the $(\g,\d)$ plane where these satisfy the Gubser bound is portrayed in figure  \ref{Fig:NatureSolutionNearExtremal}.  To asses the IR reliability of such naked solutions, we must investigate the Sturm-Liouville problems for the fluctuations. This is done in appendix \ref{slc} and we review our findings here.

The spin-2 Sturm-Liouville problem is reliable at extremality when
\be
b>1 ~~~{\rm   or}~~~ \le(b<1 ~~~{\rm and}~~~{a>3}\ri),
\ee
where $a,b$ were defined in (\ref{d5}) and (\ref{d6}).
The regions are portrayed in \Figref{Fig41}.

\begin{figure}[!ht]
\begin{center}
\begin{tabular}{cc}
	 \includegraphics[width=0.45\textwidth]{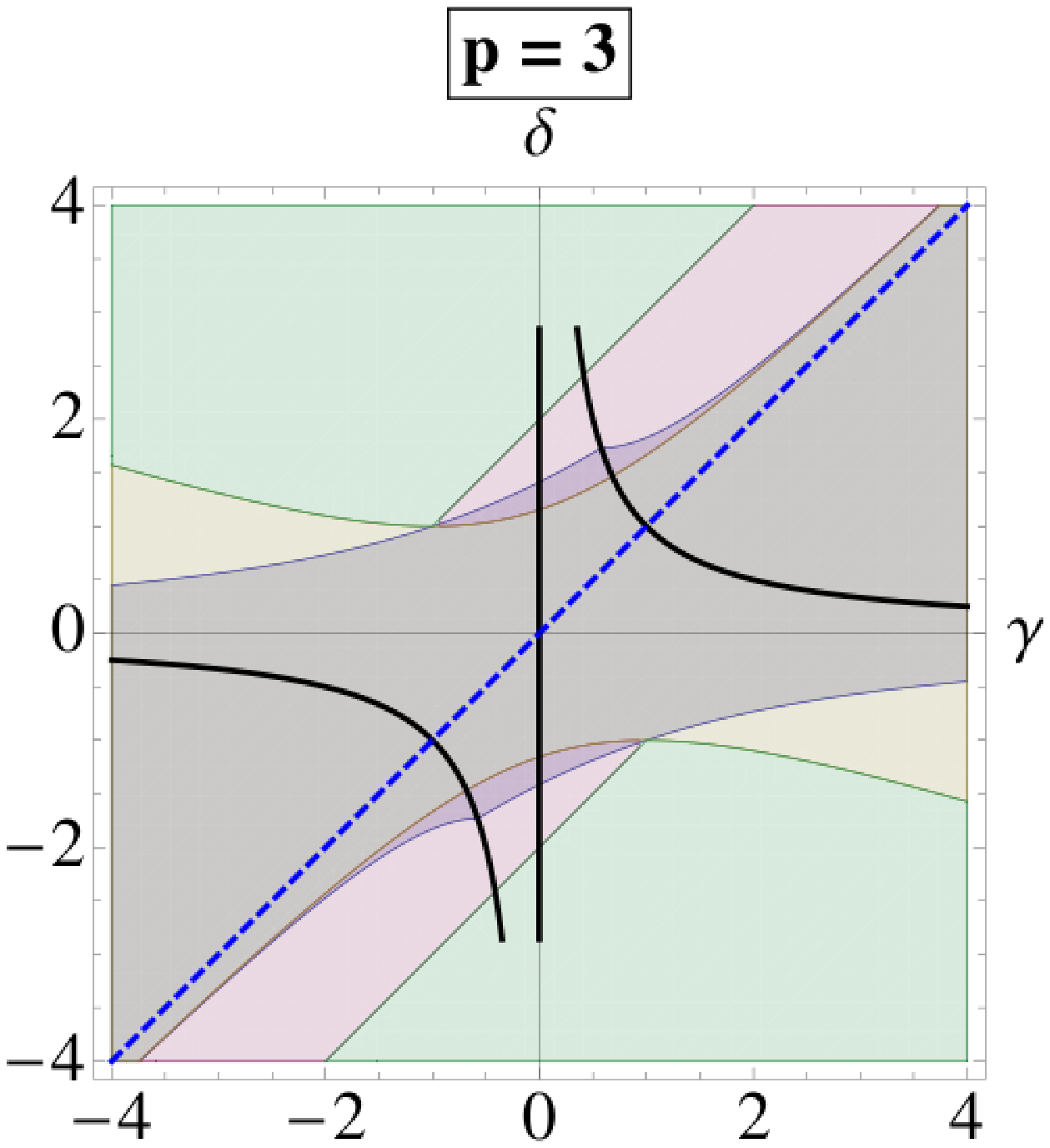}
	 \includegraphics[width=0.45\textwidth]{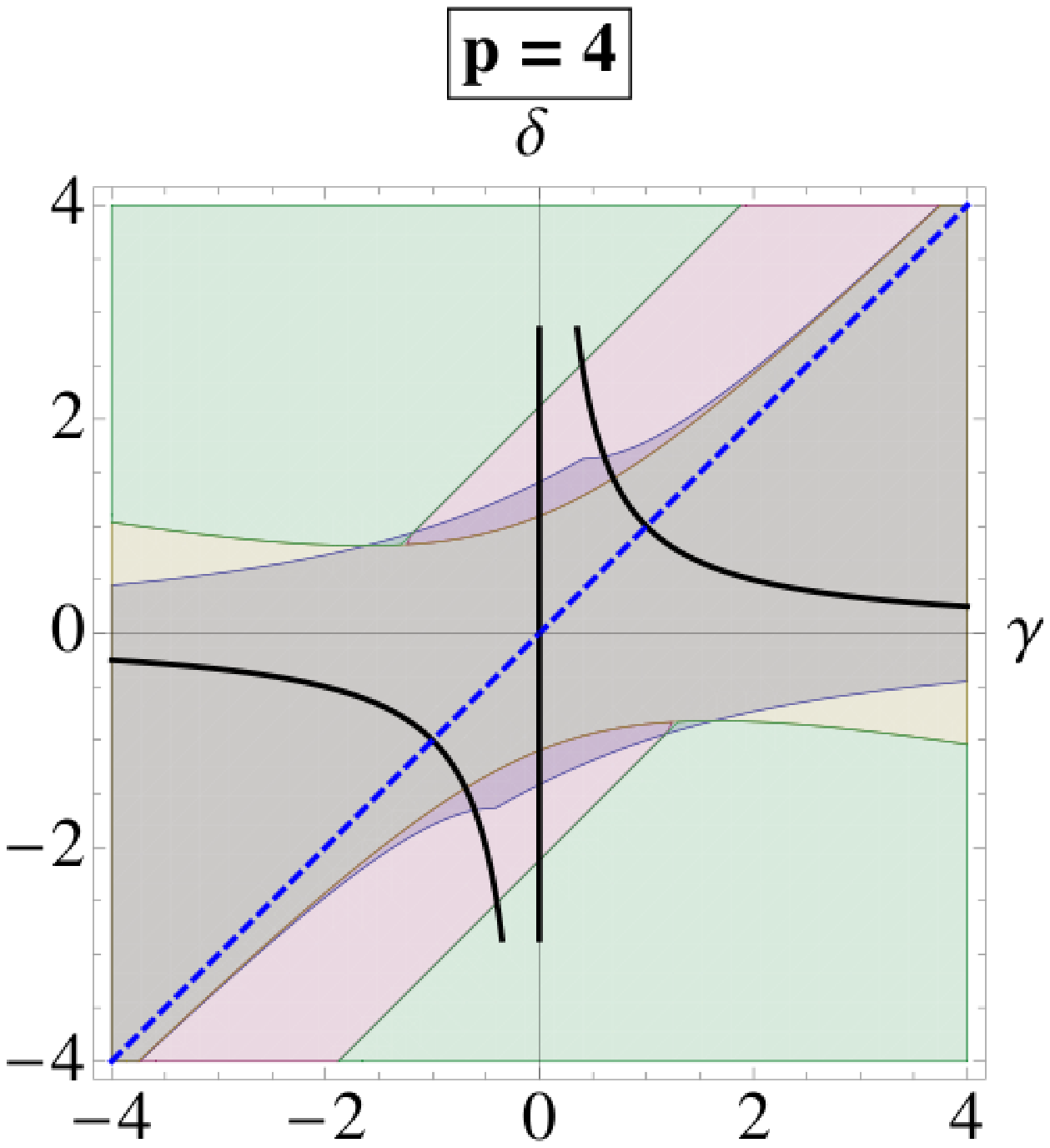}
\end{tabular}
\caption{These figures summarize the reliability constraints on the spin-1 fluctuations for $p=3$ and $p=4$.  The blue region depicts the part of the $(\gamma,\d)$ plane which satisfies the Gubser bound constraints in (\protect\ref{d3}).
The yellow region is determined by $b>1$ and is identical to both spin-2 and thermal stability constraints.
The purple region is determined by $b<1$ and $-1<\tilde a<2$ in   (\protect\ref{d16}), with a lower bound on charge density $q^2>q_0^2$ in (\protect\ref{tan4}), while the green region is $b<1$ and $\tilde a<-1$. The overlap of the latter region with the Gubser one is trivial in $p=3$, and small but non-trivial for $p=4$.
The solid line corresponds to $\g\d=1$ and the dashed blue line to $\g=\d$.  }
\label{Fig42}
\end{center}
\end{figure}

On the other hand, the spin-1 (current) Sturm-Liouville problem is reliable at extremality when
\begin{itemize}
\item $b>1$ or

\item
$b<1$  and $\tilde a<-1$ where $\tilde a$ is defined in (\ref{tan3}) or

\item
 $ b<1$ and $2>\tilde a>-1$ and $q^2>q_0^2$ where $q_0$ is defined in (\ref{tan4}).
 \end{itemize}
Therefore, as is shown in  \Figref{Fig42}, there are no constraints on the $(\ga\,,\,\da)$ plane from the spin-1 fluctuations (though the third constraint imposes that the charge density should be large enough). For $p=3$, only the first and the third constraints have some overlap with the Gubser constraints, while for $p=4$ the second spin-1 constraint has a (quite) small overlap with the Gubser ones, too. Overall, the spin-1 problem is always reliable in the Gubser region for $p=3$, while for $p=4$ there is a small part of the Gubser region where there is a lower bound on the charge density.

\begin{figure}[!ht]
\begin{center}
\begin{tabular}{cc}
	 \includegraphics[width=0.45\textwidth]{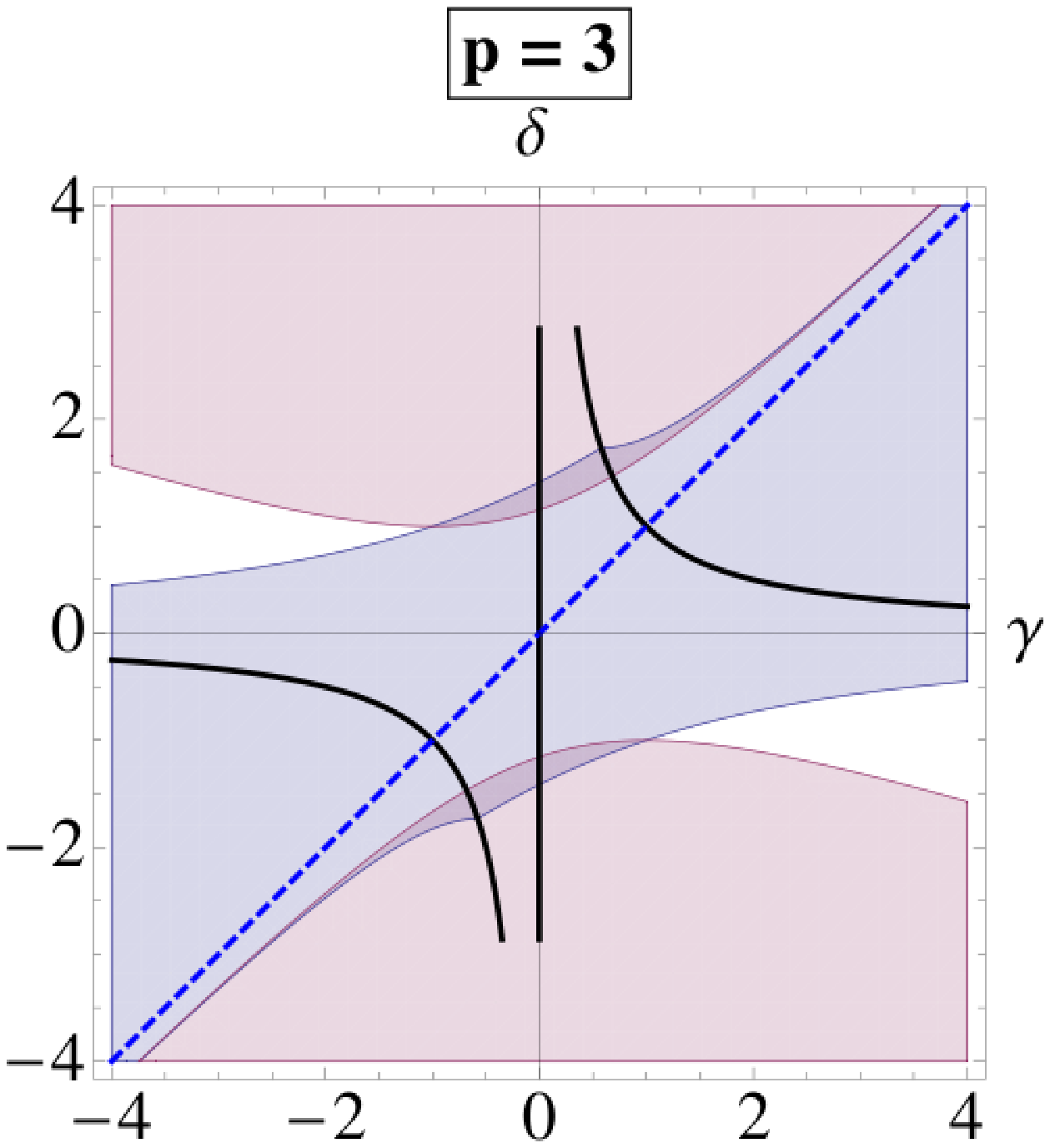}
	 \includegraphics[width=0.45\textwidth]{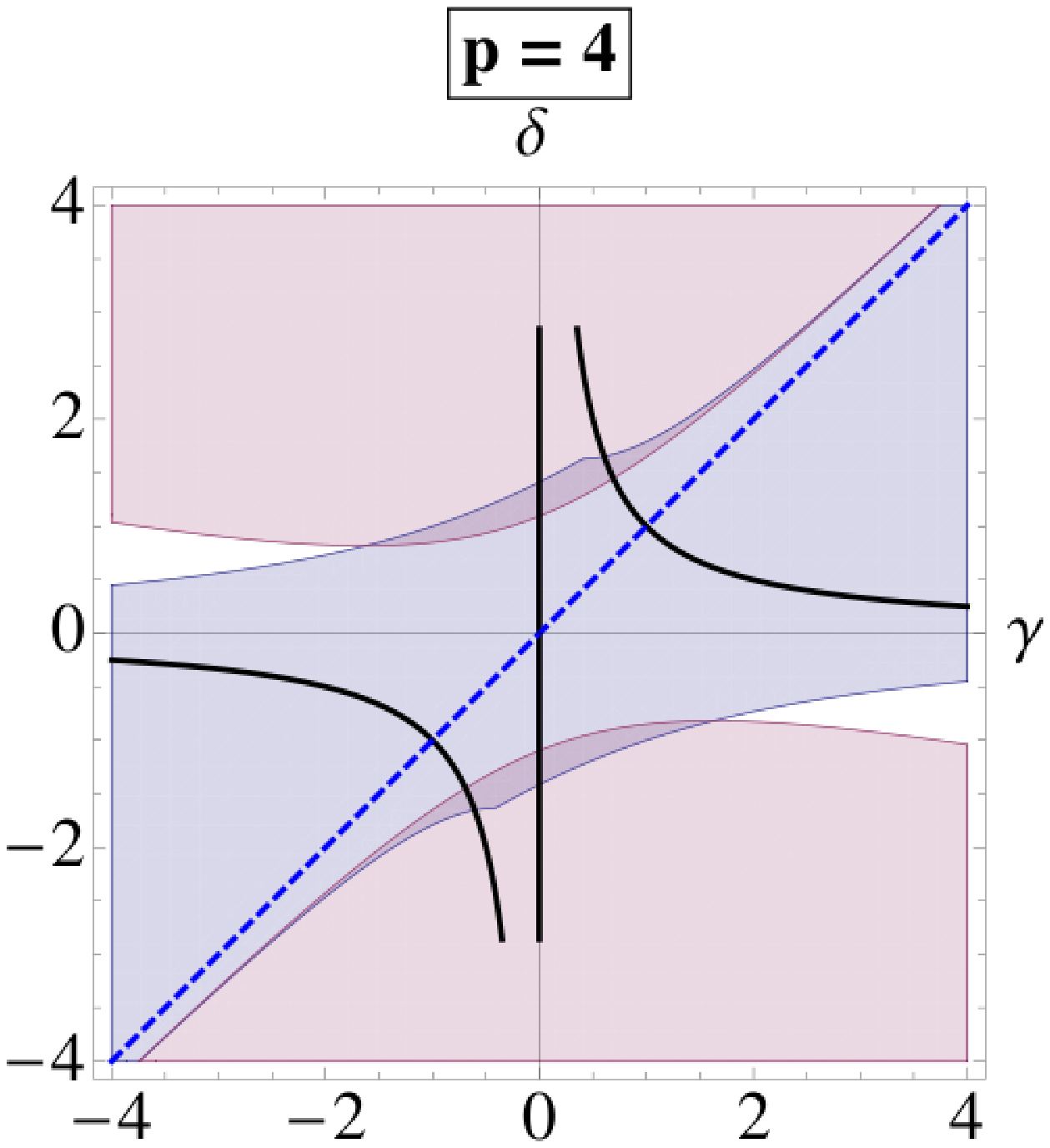}
\end{tabular}
\caption{These figures summarize the region where the (small) near-extremal black holes are thermodynamically unstable.
The blue region depicts the part of the $(\gamma,\d)$ plane which satisfies the Gubser bound constraints in (\protect\ref{d3}).
The purple region is the thermodynamically unstable region determined by $1-2\frac{(\ga-\da)^2}{w_pu}<0$, as indicated in equation (\protect\ref{Temperature3}). The overlap region  is the region where there will be a gap in the black hole temperatures.
The black solid line corresponds to $\g\d=1$ and the dashed blue line to $\g=\d$.
}
\label{Fig43}
\end{center}
\end{figure}

 A further comment concerns the correlation of the constraints above and the behavior of black holes near extremality. As discussed in section   \ref{secthermo}
the near extremal black holes are thermodynamically unstable in the region
\be
{(p-1)(\g-\d)^2\over w_pu}>1
\ee
where the temperature at extremality diverges. In analogy with the uncharged case, in such situations the system is expected to have no stable black hole solutions up to a  temperature $T_{min}$ above which new large black holes will appear due to a modified UV potential. This region is described in \Figref{Fig43}.

This region is exactly the same as the $b<1$ region above.
Therefore, in the thermodynamically unstable region, one also needs $a>3$, so that the spin-2 problem is reliable. For $p=3$, this is never really the case as this region touches the allowed region at two points, $(\g,\d)=(1,-1)$ and   $(\g,\d)=(-1,1)$ where $v=0$.
We conclude, that in the $p=3$ case, for the thermodynamically unstable region the spin-2 problem is not reliable. The spin-1 problem in the same region is on the other hand reliable except the case $b<1,\tilde a<-1$, which has a a non-zero overlap with the allowed region and where the spin-1 problem is reliable only at sufficiently high charge density.

Finally, in the thermodynamically stable region, the spin-1 and -2 problems are always reliable since it corresponds to $b>1$.

\subsection{Lifshitz solutions}

A generalization of AdS solutions involves asymmetric scaling symmetries of the Lifshitz type, \cite{kliu}.
A generalized Lifshitz metric can be written as
\be
\ud s^2=\tilde\ell^2{\ud r^2\over r^2}-{\ud t^2\over r^{2zx}}+{\ud x^i\ud x^i\over r^{2x}}\,.
\label{152}\ee
By changing variables to $u=r^x$, the metric can be rewritten as
\be
\ud s^2={\ell^2}{\ud u^2\over u^2}-{\d t^2\over u^{2z}}+{\ud x^i\ud x^i\over u^{2}}\,,\qquad \ell={\tilde \ell\over x}\,.
\label{153}\ee
To bring this metric to the domain wall frame we must identify
\be
e^{A(r)}={1\over u}\,,\qquad e^{g(r)}={1\over u^{2(z-1)}}\,,\qquad \ud r=-\ell{\ud u\over u^z}\to {r\over \ell}={u^{1-z}\over z-1}\,,
\label{154}\ee
so that
\be
e^{A(u)}=\left((z-1){r\over \ell}\right)^{1\over z-1}\,,\qquad e^{g(u)}=(z-1)^2{r^2\over \ell^2}\,.
\label{li1}\ee

Comparing with the scaling solutions (\ref{DomainWall3}), we find that they are of the Lifshitz form only when $\delta=0$ and $\gamma<0$.
In that case the non-extremal solutions read
\be
e^{A(r)}=\left((z-1){r\over \ell}\right)^{1\over z-1}\,,\qquad e^{g(r)}=(z-1)^2{r^2\over \ell^2}\left[1-\left({r_0\over r}\right)^{p+z-1\over z-1}\right]
\,,\qquad e^{\phi}=e^{\phi_0}~r^{\sqrt{2(p-1)\over (z-1)}}\,,
\label{155}\ee
with
\be
e^{-\gamma\phi_0}={2(z+p-1)(z-1)\over q^2\ell^2}~\left({z-1\over \ell}\right)^{2(p-1)\over z-1}
\label{156}\ee
\be
 {1\over \ell^2}={2\Lambda\over (p+z-1)(p+z-2)}\,,\qquad \gamma=\pm\sqrt{2(p-1)\over (z-1)}\,.
\label{157}\ee
Specializing to the planar case $p=3$ the parameters are
\be
e^{-\gamma\phi_0}={2(z-1)^{1+{4\over z-1}}(-2\Lambda)^{1+{2\over z-1}}\over q^2(z+2)^{2\over z-1}(z+1)^{1+{2\over z-1}}}\,,\qquad
 {1\over \ell^2}={2\Lambda\over (z+2)(z+1)}\,,\qquad \gamma=\pm\sqrt{4\over (z-1)}\,,
\label{158}\ee
Such solutions were interpreted as Lifshitz solutions  before in \cite{taylor}.

The non-extremal version provides explicit Lifshitz black-holes
\be
ds^2=-\left(1-{u^{p+1-z}\over u_0^{p+1-z}}\right){dt^2\over u^{2z}}+{dx_idx^i\over u^2}+{1\over (-2\Lambda)(z+p-1)(z+p-2)}{du^2\over u^2\left(1-{u^{p+1-z}\over u_0^{p+1-z}}\right)}
\ee
\be
e^{\phi}=e^{\phi_0}~u^{-\sqrt{2(p-1)(z-1)}}\sp A_t={2\ell\over z+p-1}\sqrt{-\Lambda(z-1)\over (z+p-2)}{u^{1-z-p}\over (z-1)^{p\over z-1}}\left(1-{u^{p+1-z}\over u_0^{p+1-z}}\right)
\ee
The fluctuation spectra for such solutions pose no constraints.

We conclude that such solutions are generated by a scalar along a flat direction of a potential, which drives the coupling constant of the U(1) to weak coupling.

\subsection{Thermodynamics\label{secthermo}}

We will discuss here the thermodynamics for $p=3$. Other dimensions show similar behavior, and the equivalent expressions are straightforward to compute.

The temperature is,
\be
	 T=\frac1{4\pi}\sqrt{-w\Lambda}e^{-\frac\da2\phi_0}(2m)^{1-2\frac{(\ga-\da)^2}{wu}}\,,
	\label{Temperature3}
\ee
and we observe that it vanishes in the $m=0$ (extremal) limit if and only if the exponent is positive, that is
\be
wu-2(\ga-\da)^2=(3-\da^2)(1+\ga^2)+(1-\ga\da)^2-2(\ga-\da)^2>0.
\label{159} \ee
 This precisely reduces in the limit we already observed in the $\ga\da=1$ solution, $\da^2<1+\frac2{\sqrt3}$.  We recover the two standard dilatonic cases,
\begin{itemize}
 \item If  $1-2\frac{(\ga-\da)^2}{wu}>0$ (lower range), the extremal temperature is zero.
 \item If  $1-2\frac{(\ga-\da)^2}{wu}<0$ (upper range), the temperature diverges in the extremal limit.
\end{itemize}
The gravitational mass is
\be
	M_g = \frac{\omega_2}{4\pi}\sqrt{\frac{-\Lambda}{wu^2}}(\ga-\da)^2m\,,
	\label{Mass3}
\ee
and reduces to the corresponding near-extremal limits of the $\ga\da=1$ and $\ga=\da$ solutions. Notice that if $\ga=\da$ the mass is identically zero, which is expected since in that case the geometry is not a black hole but simply the direct product $AdS_2\times S_2$.
\noindent
The electric charge
\be
	Q = \frac{\omega_2}{8\pi}\sqrt{\frac{v\Lambda}{u}}e^{\frac{\ga-\da}2\phi_0}\,,
	\label{ElectricCharge3}
\ee
is universal and does not contain an independent integration constant. Finally, the entropy is
\be
	S = \frac{\omega_2}{4}(2m)^{2\frac{(\ga-\da)^2}{wu}}\,,
	\label{Entropy3}
\ee
and we observe that it vanishes  at extremality provided that $\ga\neq\da$ (otherwise it is finite) and $1-2\frac{(\ga-\da)^2}{wu}>0$. This is consistent with the behaviour of the two previous solutions for $\ga\da=1$ and $\ga=\da$.

\begin{figure}[t]
\begin{center}
\begin{tabular}{cc}
	 \includegraphics[width=0.45\textwidth]{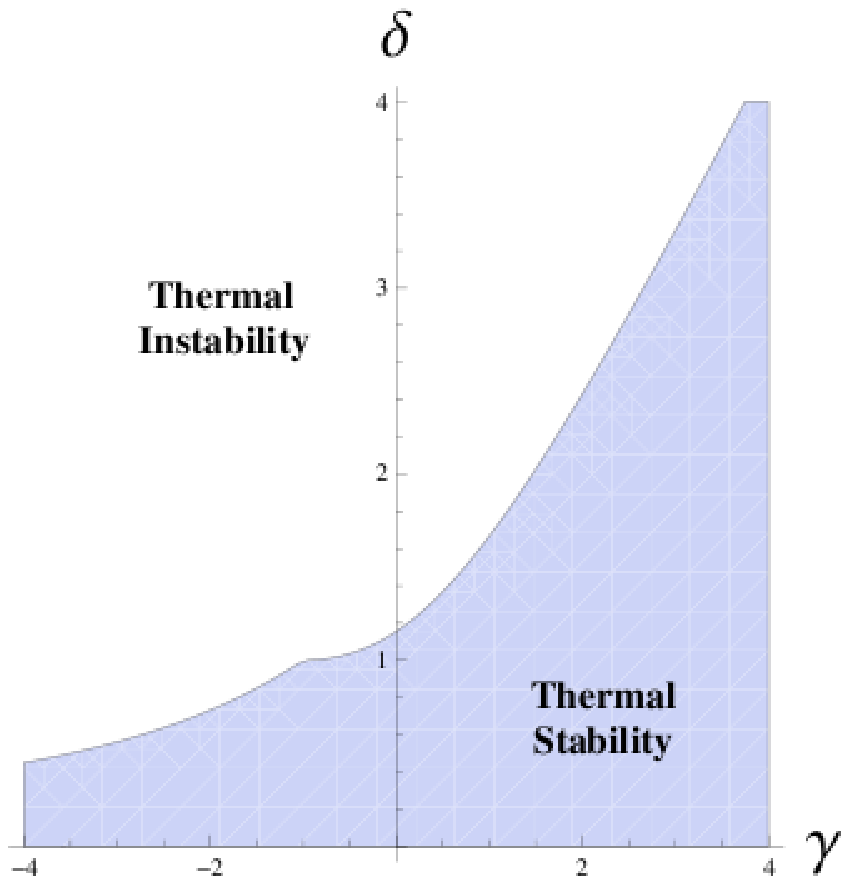} &
	 \includegraphics[width=0.45\textwidth]{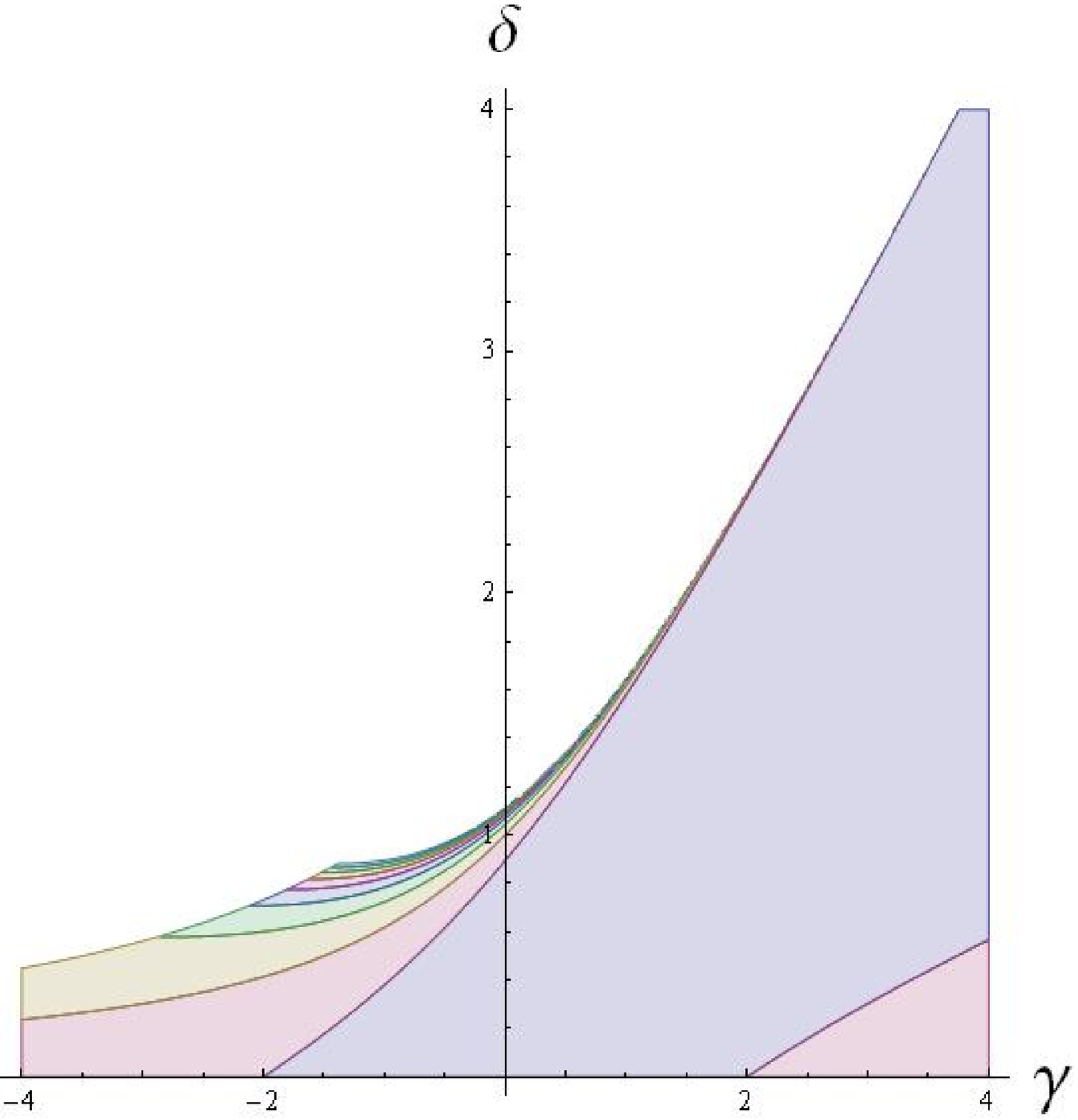}
\end{tabular}
\caption{{On the left panel, we plot the region of local stability of the near extremal black hole. The right panel shows a variety of phase transitions of the near extremal black hole to the background at zero temperature. In the blue region continuous transitions occur, in the purple region adjacent to the blue one the transitions are of third-order. The stripes starting with yellow to the left of the blue and purple regions depicts transitions of fourth-(yellow) up to tenth-order. Above them all higher-order transitions also occur.}}
\label{Fig:PhaseTransitionsNearExtremal}
\end{center}
\end{figure}

Given all of this, we work in the canonical ensemble, where only the temperature is allowed to vary (and the  charge density  is fixed to its value \eqref{ElectricCharge3}. Then the Helmholtz potential is
\be
	W = -\frac{\omega_2}{8\pi}\sqrt{-w\Lambda}e^{-\frac\da2\phi_0}\le[1-2\frac{(\ga-\da)^2}{wu} \ri]m\,,
	\label{Helmholtz3}
\ee
and again reduces both to \eqref{Helmholtz1} and \eqref{Helmholtz2} in the appropriate limits. Unsurprisingly, inspection of its sign reveals that the black holes are globally stable in the lower range but not in the upper range, similarly to the $\ga\da=1$ solution. Moreover, computing the heat capacity, we find
\be
	C_Q = \frac{\omega_2}{4}\frac{(\ga-\da)^2}{wu}\le[1-2\frac{(\ga-\da)^2}{wu}\ri]^{-1}\le(e^{\frac\da2\phi_0}\frac{4\pi T}{\sqrt{w\Lambda}} \ri)^{\frac{2(\ga-\da)^2}{wu-2(\ga-\da)^2}}\,,
	\label{HeatCapacity3}
\ee
which will be positive in a range a little bit smaller than the lower range and negative in the upper range, thus rendering the black holes locally stable in the former and locally unstable in the latter, as displayed in the left panel of \Figref{Fig:PhaseTransitionsNearExtremal}.

{Finally, second- or higher-order phase transitions will occur in the zero temperature limit if the appropriate derivative of the Helmholtz potential \eqref{Helmholtz3} diverges. In particular an $n^\textrm{th}$-order transition occurs if
\be
	-1 < \frac{wu}{wu-2(\gamma-\delta)^2} - n < 0\,.
\label{160}\ee
This is depicted in the right panel of \Figref{Fig:PhaseTransitionsNearExtremal}, showing the regions in which transitions of second  to tenth order occur. Two remarkable features are that, in most of the parameter space that is thermodynamically stable (which amounts to requiring a well-defined extremal limit and $vu<0$), second-order transitions appear, and there are no first-order transitions. We note that it does not make sense to analyse the thermodynamically unstable region in this way as long as the AdS-completion has not been taken into account, since our experience with the solutions for $\gamma \delta=1$ show us that this is crucial for the correct structure of the phase transitions.
}

Finally we would like to compare the entropy at finite charge density to that for zero charge density. We have already done this for the $\g=\d$ and the $\g\d=1$ solutions and found that in both cases, the finite density entropy is dominant compared to that of zero density for low temperatures.
For the extremal solutions we are discussing (at finite charge density) the entropy behaves at low $T$ as  \be
	S \sim T^{2\frac{(\ga-\da)^2}{wu-(\g-\d)^2}}\,,
	\label{e4}
\ee
while at zero density from   (\ref{EntropyNeutralBH}) $S\sim T^{2\over 1-\d^2}$.
When $\d^2\leq 1$, both powers in (\ref{e4}) and  (\ref{EntropyNeutralBH}) are positive and  the entropy at finite charge density dominates the one for zero density. When $\d^2\ge 1$, there are two regions. In one, both  (\ref{e4}) and  (\ref{EntropyNeutralBH}) are negative. In this region the theory has a gap and the theory remains in the extremal solution as $T\to 0$. In the rest, the neutral extremal solution dominates at low temperature with a temperature of ${\cal O}(1)$, while at finite density the entropy is non-zero and ${\cal O}(N^2)$. Therefore in this last case again  the entropy at finite charge density dominates the one for zero density. This is summarized in figure \ref{CEntropy}.

\FIGURE[!ht]{
\begin{tabular}{c}
	 \includegraphics[width=0.55\textwidth]{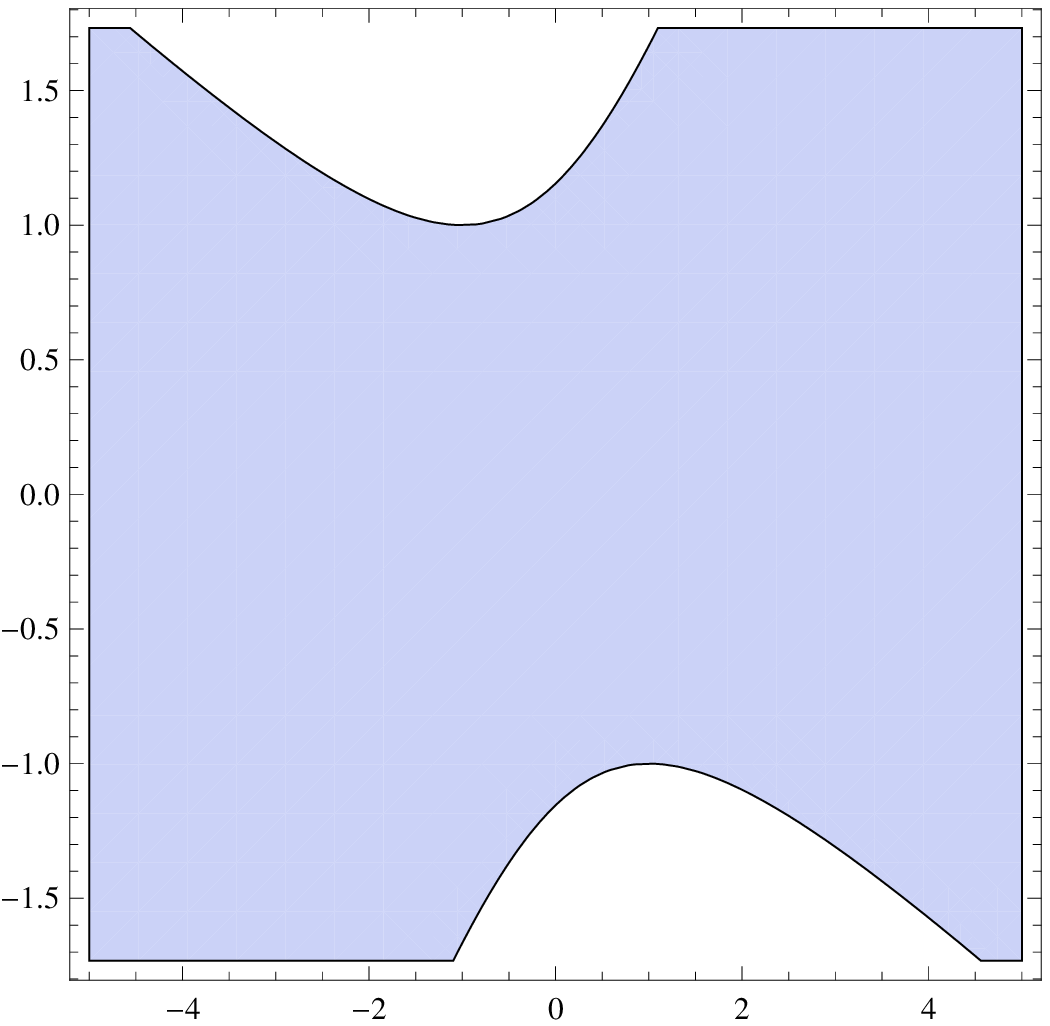}
\end{tabular}
\caption{In the shaded region,  the entropy at finite charge density dominates the one for zero density at very low temperatures. In the rest of the ($\g,\d$) diagram, the comparison cannot be made as both entropies are expected to be of  ${\cal O}(1)$ instead of ${\cal O}(N^2)$. The vertical axis represents the value of $\da$, while the horizontal axis the value of $\gamma$.}
\label{CEntropy}
}

We stress again that all these considerations should be taken as valid in the near-zero temperature regime: as the temperature grows, one will need the (yet unknown) full solution and not its near-extremal approximation in order to study finite temperature thermodynamics.

\subsection{The AC conductivity}\label{sec:NearExtremalACConductivity}

We now turn to the calculation of the AC conductivity at zero temperature, i.e. in the extremal limit, and low frequencies. We will first work in general dimension $p+1$, and then analyse the special cases of planar systems ($p=3$) and bulk systems ($p=4$). The scaling solutions at arbitrary dimension are given in subsection \ref{p+1}.

The low-frequency scaling of the AC conductivity for the solutions \eqref{Sol3} is again calculated from the general formulae \eqref{32}, \eqref{41} and \eqref{42}, wherever these are applicable. It is convenient to work with the solution written in domain wall coordinates \eqref{Sol3DW}-(\ref{wpu}).
From \eqref{DomainWall3} it is evident that the extremal limit is $m\rightarrow 0$, after which the coordinate $\rho$ (or equivalently $\xi$ in \eqref{Sol3DW}) is already the near-horizon coordinate - the horizon is at $\rho=0$. Requiring the temperature of the solution \eqref{Sol3DW} to go to zero in the extremal limit yields a constraint
\be\label{extremalcond}
(wu)_p > (p-1)(\gamma-\d)^2\,.
\ee
The relevant coordinate change to arrive at the Schr\"odinger problem is given by
\be\label{Sol3coordchange}
\frac{\ud z}{\ud \rho} = \frac{u(wu)_p e^{\d\phi_0}}{8\Lambda (p-1)} \rho^{\frac{(p-1)(\gamma-\d)^2-(wu)_p}{2(p-1)}-1}\,.
\ee
Thus if the condition for a extremal zero-temperature limit \eqref{extremalcond} is fulfilled, the power in \eqref{Sol3coordchange} is always smaller than minus one, and hence the coordinate change is polynomial. The Schr\"odinger potential found from \eqref{32} is again
\be\label{Sol3Veff}
V(z) = \frac{c}{z^2}\,,\qquad c = \frac{(p-1)( \d^2 - \gamma^2 - 4) (8(1 -
    p) + (\d - \gamma) ( \gamma(1+p) + \d(5  p - 7)))}{4 (2(1 -
    p) + (\d - \gamma) (\gamma + \d (2 p - 3)))^2}\,.
\ee
The AC conductivity at zero temperature scales then for small frequencies as
\be\label{Sol3ACsigma}
\sigma \sim \omega^n\,,\qquad n = \left|\frac{6(1-p)+(\d-\gamma)(p\gamma + (3p-4)\d)}{2(1 -
    p) + (\d - \gamma) (\gamma + (2 p - 3)\d )}\right| - 1\,.
\ee
As a consistency check we observe that both \eqref{Sol3Veff} and \eqref{Sol3ACsigma} reduce to corresponding results for the special cases $\ga=\d$ (eqs.~\eqref{g=dVeff}~and~\eqref{g=dACscaling}) and $\ga\d=1$ (eqs.~\eqref{gd1Veff}~and~\eqref{gd1ACscaling}), and reproduce the results of \cite{kachru}. For the Lifshitz solutions ($\d=0$, $\ga=-\sqrt{\frac{2(p-1)}{z-1}}$)
 we find
\be
n = \sqrt{\frac{(3(z-1)+p)^2}{z^2}} - 1\,,
\label{161}\ee
which reduces to $\sigma \sim \omega^2$ for $p=3$. This is different than the $\omega^{-2/z}$ result of \cite{hpst}. The difference can be traced to including the full backreaction as well as the non-trivial scalar factor in front of the gauge field action.

We now turn to a more detailed analysis of the special cases of planar systems, $p=3$, and bulk systems, $p=4$. The above calculation again made use of the matching procedure of \cite{kachru} which, from experience with the $\ga\d=1$ solution \eqref{Sol1} and the fact that \eqref{Sol3} reduces to \eqref{Sol1} for $\ga=1/\d$, is applicable in the region of the $(\ga,\d)$ plane given by \eqref{extremalcond}, i.e. whenever the solution has an extremal limit in which the temperature drops to zero.

\FIGURE[!ht]{
\begin{tabular}{cc}
	 \includegraphics[width=0.45\textwidth]{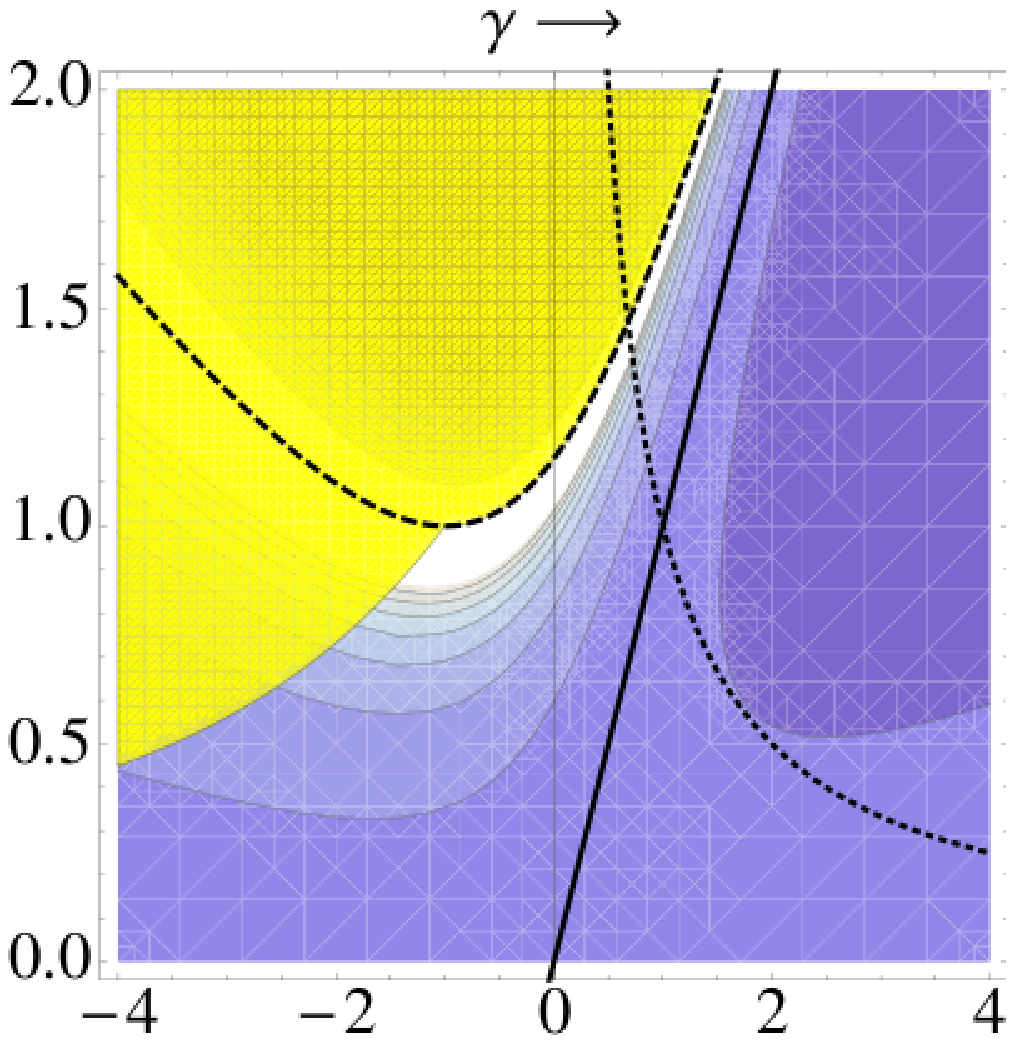}&
	 \includegraphics[width=0.45\textwidth]{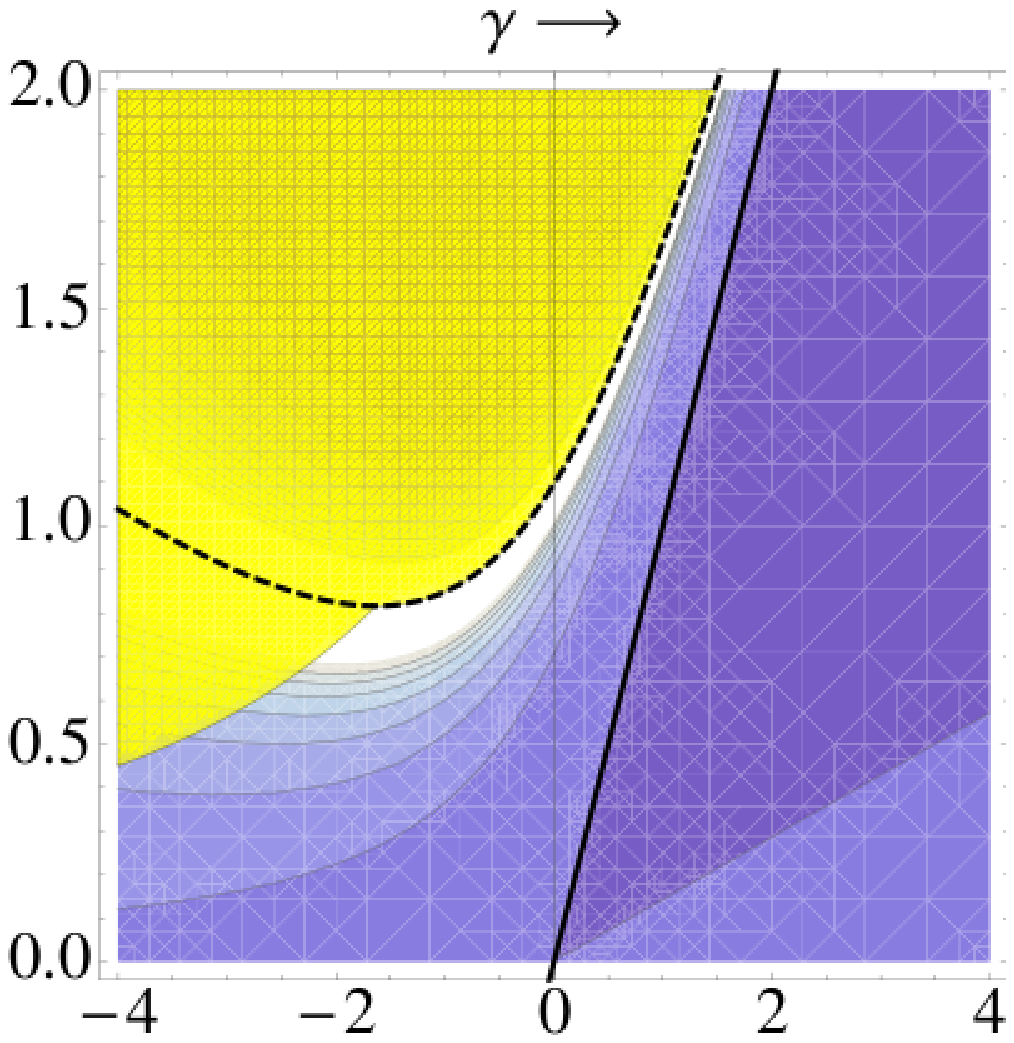}
\end{tabular}
\caption{Contour plot of the scaling exponent $n$ in the $(\gamma,\d)$ upper half plane for $p=3$ (left figure $0\leq \d \leq \sqrt{\frac{5}{3}}$) and $p=4$ (right figure, $0 \leq \d\leq \sqrt{\frac{4}{3}}$). Left figure: Contours correspond to $n=1.52,\dots,8.36$, starting with $n=1.52$ in the upper right corner and increasing in steps of $0.76$. The black solid line $\ga=\d$ is $n=2$, and brighter colors correspond to larger $n$. Right figure: Contours correspond to $n=2.2,\dots,12.1$, starting with $n=2.2$ in the lower right corner and increasing in steps of $1.1$. The black solid line $\ga=\d$ is again at $n=2$. In the yellow regions the computation of $n$ cannot be trusted, since an explicit AdS completion of the space-time is needed to render the thermodynamics well-defined. The scaling exponent diverges to $+\infty$ along the dashed black line in both cases.}
\label{Fig:ACNegExpSol3}
}

\FIGURE[!ht]{
\begin{tabular}{cc}
	 \includegraphics[width=0.45\textwidth]{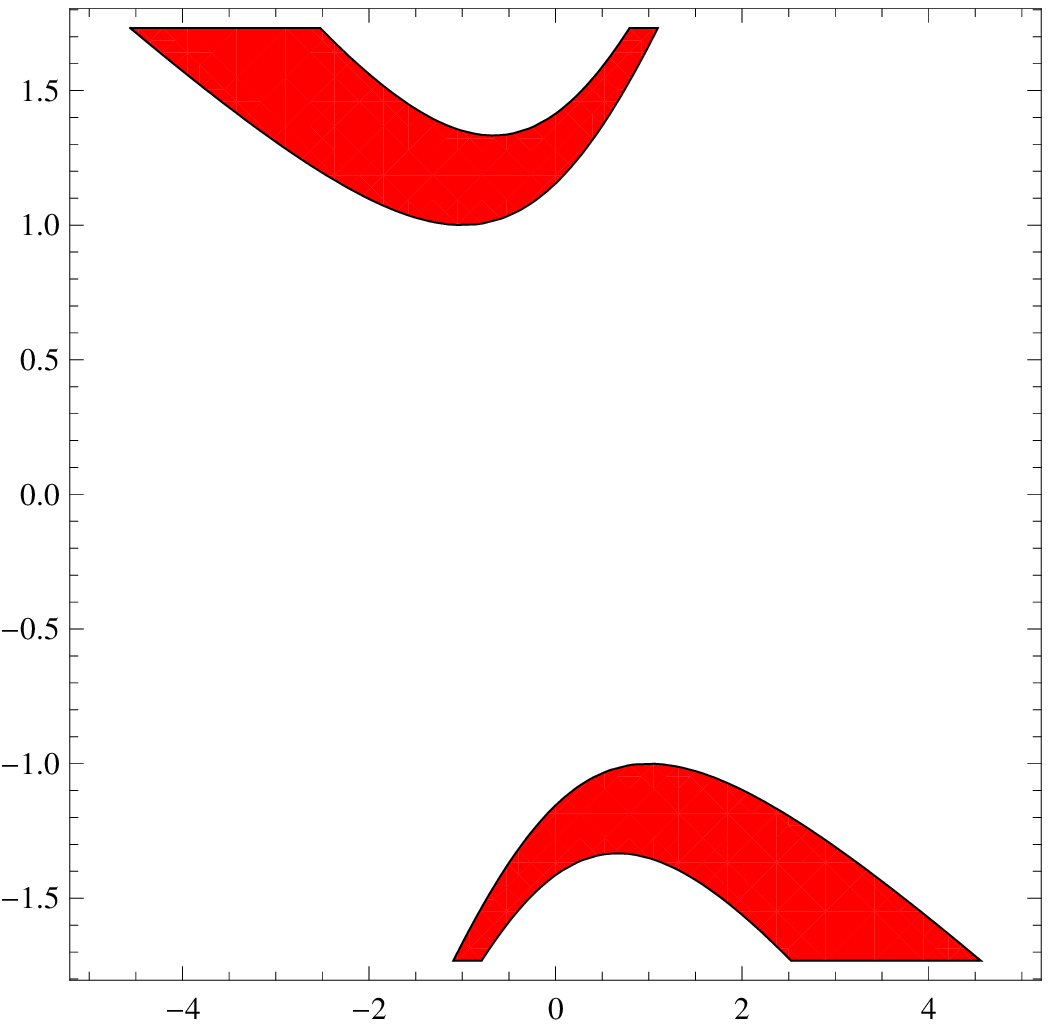}&
	 \includegraphics[width=0.45\textwidth]{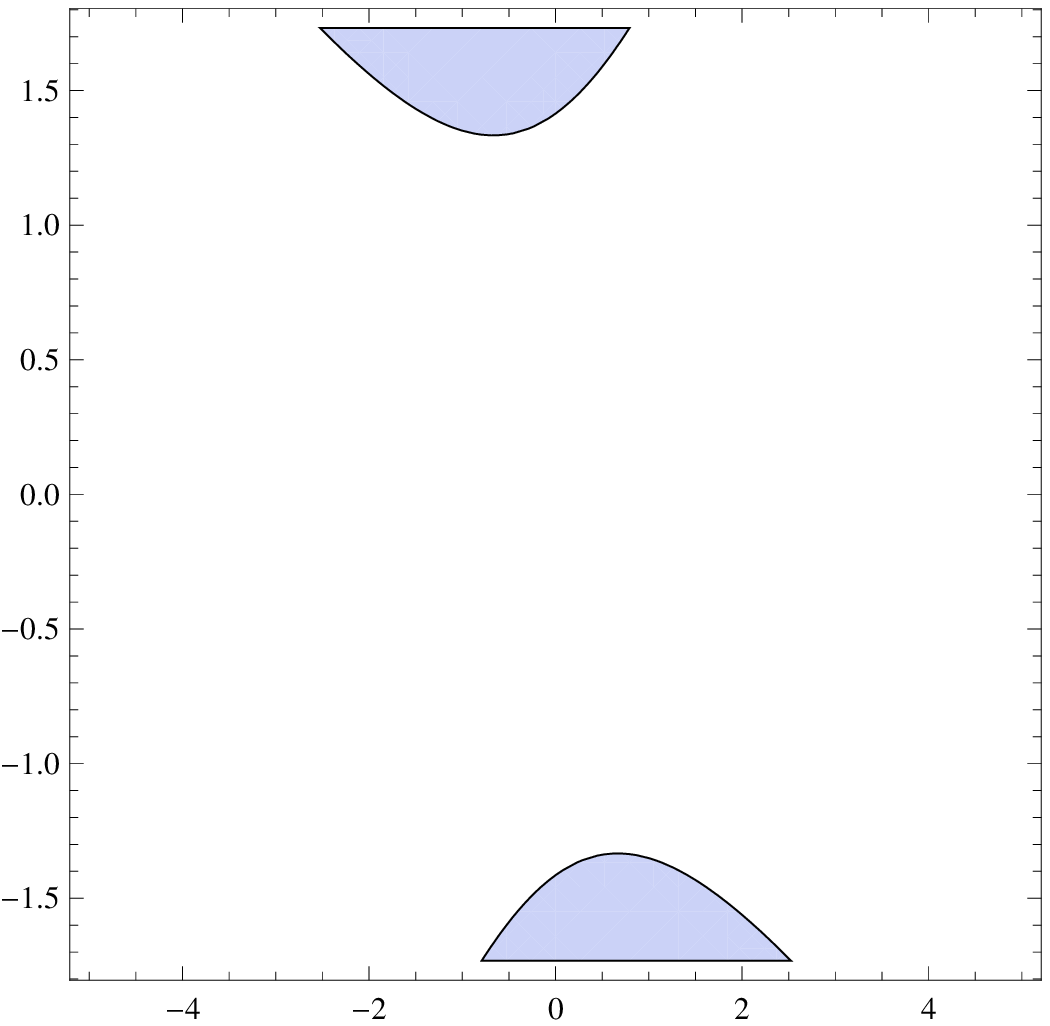}
\end{tabular}
\caption{p=3. Left figure: The region on the $(\gamma\d)$ plane where the IR black holes are unstable, and the coefficient $c$ is positive. In this region the extremal finite density system has a mass gap and a discrete spectrum of charged excitations, when the UV dimension of $\phi$ is $\Delta<1$. This resembles a Mott insulator and the figure provides the Mott insulator ``islands" in the $(\gamma,\d)$ plane. Right figure: The region on the $(\gamma\d)$ plane where the IR black holes are unstable, and the coefficient $c$ is negative. In this region the extremal finite density system has a gapless continuous spectrum at zero temperature. In both figures the horizontal axis parametrizes $\gamma$, whereas the vertical axis $\d$.}
\label{mott}
}

\paragraph{Planar systems ($p=3$)} The left plot in \Figref{Fig:ACNegExpSol3} shows the scaling exponent $n$ in the upper half $(\ga,\d\leq \sqrt{5/3})$ plane for $p=3$. The maximum value of $\d$ is set by the spin 2 fluctuation constraint, see sec.~\ref{spectra}.  The yellow shaded region is either forbidden by thermal instability (above the black dashed line, see \Figref{Fig:PhaseTransitionsNearExtremal}),\footnote{Note that the thermal instability constraint is not to be taken too serious since, as discussed before, an AdS completed solution with these parameters will have proper thermodynamics and hence a meaningful holographic interpretation.} by the spin 1 fluctuation constraint \eqref{s13a} (the wedge-shaped region touching $\ga=\d=0$), or by requiring that the r-coordinate is space-like (yielding $u>0$, $v<0$, which forbids the yellow region on the left below the black dashed line). See figure~\ref{Fig:NatureSolutionNearExtremal} for a discussion of all the constraints on $(\ga,\d)$. In the allowed region the scaling exponent is always positive, $n>0$. We find $n=2$ along the solid black $\ga=\d$ curve, with $n$ increasing until it diverges along the dashed black curve
\be
\d(\ga) = \frac{1}{3}\left( \ga + 2 \sqrt{3+\ga^2} \right)\,.
\ee
For comparison, the $\ga\d=1$ line is also depicted as the dotted black curve. We thus conclude that whenever the underlying scaling solution is thermodynamically stable and completable to AdS space, our system describes a AC conductivity scaling law at zero temperature and small frequencies with positive scaling exponent, $\sigma \sim \omega^{n>0}$.

In particular this implies that in this regime the spectrum of charged fluctuations is gapless and continuous and therefore describes a conductor.

The region for which
\be\label{extremalcond1}
(wu)_p < (p-1)(\gamma-\d)^2\,.
\ee
is the one where the ``small" black holes found in the IR EHT are thermodynamically unstable and the region where the extremal background is dominant thermodynamically.
In this case we have two possible behaviors for $c$ defined in (\ref{Sol3Veff}):

(a) $c>0$. This region is depicted at the left of figure \ref{mott}. Here the potential for charged excitations at T=0 diverges to infinity in the IR, and also in the UV if $\D<1$. These systems have a gap and a discrete spectrum of charged excitations at zero temperature and therefore resemble Mott insulators.

(b) $c<0$. This region is depicted at the right of figure \ref{mott}. In this case the system is gapless with continuous spectrum at zero temperature. It is therefore a conductor.

The behavior above generalizes what was found in the $\gamma\d=1$ and $\gamma=\d$ solutions.

\paragraph{Bulk systems ($p=4$)} This situation is shown in the right plot in \Figref{Fig:ACNegExpSol3}. The conclusions are similar to the planar case, in particular the positivity of the scaling exponent in the physically reasonable region, and the divergence along the dashed black curve, in this case given by
\be
\d(\ga) = \frac{1}{5} \left(2 \ga + \sqrt{3} \sqrt{10 + 3 \ga^2}\right)\,.
\ee
The spin 1 fluctuation constraint is now violated in the horn-shaped region, and the $r$-coordinate is timelike in the yellow region on the left below the black dashed line.
One should also note that for general dimension $p$, the scaling exponent always behaves as
\be
n_p(\ga,\ga) = 2
\ee
along the line $\ga=\d$.

Concerning the low energy spectrum some general properties are different when $p\geq 4$. The Schr\"odinger potential always diverges near the AdS boundary, and therefore when it also diverges positively in the IR the spectrum is gapped and discrete. This indeed happens on ``islands" in the $(\gamma, \d)$ plane analogous to those of the $p=3$ case. Moreover one again expects here the insulating phase to transform via a finite-temperature  phase transition to a conducting phase.

All the rest of the allowed regions have a continuous and gapless spectrum at extremality.

\subsection{The DC conductivity\label{dcex}}

From (\ref{55})  we evaluate the DC conductivity for massive carriers using
\be
\rho_{\rm massive} \sim g_{xx}^s(\rho_0)\sim e^{2A(\rho_0)}e^{k\phi(\rho_0)}\sim T^{m_p}\,,\qquad m_p\equiv {2k(p-1)
(\d-\gamma)+2(\d-\gamma)^2\over 2(p-1)(1-\d(\d-\gamma))+(\d-\gamma)^2}\,,
\label{162}\ee
where we have extended the calculation to $p+1$ dimensions.
This contribution is inversely proportional to the charge density.

In the planar case $p=3$ the relevant exponent reads
\be
m_3= {4k
(\d-\gamma)+2(\d-\gamma)^2\over 4(1-\d(\d-\gamma))+(\d-\gamma)^2}\,.
\label{163}\ee
The exponent  becomes unity for two values of $\gamma$
\be
\gamma_{\pm}=3\d+2k\pm 2\sqrt{1+(\d+k)^2}\,.
\label{164}\ee
 For a non-dilatonic scalar, $k=0$ and the temperature dependence of the entropy and the resistivity are the same.
 As the entropy scales with the same factor as the heat capacity we find that linearity of the first implies linearity of the second.

For the Lifshitz solutions, we must take $\delta=0$ and $\gamma=-\sqrt{2(p-1)\over (z-1)}$. In this case we obtain that
\be
m_p={2+k\sqrt{2(p-1)(z-1)}\over z}\,,
\label{166}\ee
in agreement with \cite{hpst} when $k=0$.

We can also estimate the conductivity from pair production from (\ref{n81}),
\be
\rho_{pair}\sim g_{xx}^s(\rho_0)^{3-p\over 2}Z_0^{-1}\sim T^{n_p}\sp n_p=
{(\g-\d)\left[2(p-1)\g+(p-3)(\g-\d-2k(p-1))\right]\over 2(p-1)(1-\d(\d-\gamma))+(\d-\gamma)^2}
\label{n166}\ee
For the Lifshitz solutions the exponent becomes
\be
n_p={3p-5\over z}+k{(p-3)\sqrt{2(p-1)(z-1)}\over z}
\ee

In the planar case $p=3$ and
\be
n_3={4\g(\g-\d)\over 4(1-\d(\d-\gamma))+(\d-\gamma)^2}
\label{n167}\ee

This exponent will become 2, if $k=-{\d\over 2}$ together with (\ref{164}).
Therefore when
\be
k=-{\delta\over 2}\sp \gamma_{\pm}=2\d\pm \sqrt{4+\d^2}
\label{n168}\ee
the DC conductivity interpolates between a $~T$ behavior for massive carriers and a $~T^2$ behavior for light carriers.
As the linear behavior is inversely proportional to the charge density we expect the $T^2$ behavior to dominate eventually at higher doping as indeed happens in cuprates.

There are two special values in (\ref{n168}), for $k=0$, $\delta=0$ and $\gamma=\pm 2$. One of them corresponds to the $z=2$ Lifshitz geometry.
The other is $k=1$, $\delta=-2$ and $\gamma=-4\pm 2\sqrt{2}$.

\section{Strong self-interaction solutions}

In section \ref{strong} we have derived the equations for solutions that describe strong self-interactions. We reproduce them here in the domain wall frame
\be
(p-1)A^{\prime\prime}+{1 \over 2}\phi^{\prime 2}=0\,,\qquad
pA^{\prime}+g^{\prime}+{g^{\prime\prime} \over g^{\prime}} -
{|q| e^{-g+{k}\phi} \over  g^{\prime}e^{(p-1)A}} =0\,,
\label{22a}\ee

\be
(p-1)A'(g'+pA')-{1\over 2}\phi^{\prime 2}-e^{-g}V
+|q|e^{-g-(p-1)A+{k}\phi}=0\,,
\label{23a}\ee
with the gauge field being in this limit
\be
A_t'=e^{A+k\phi}\,.
\label{167}\ee

We have found the following three charged solutions of the equations above:

\subsection{Strong coupling solution I}

\be
e^A=r^{c_p^2\over 2(p-1)}\,,\qquad e^{\phi}=e^{\phi_0}r^{c_p}\,,\qquad e^g=e^{g_0} r^{c_1}\left[1-\left({r_0\over r}\right)^{c_3}\right],
\label{168}\ee
with
\be
c_p=0\,,\qquad c_1=2\,,\qquad c_3=1,2\,,\qquad e^{g_0}={q\over 2}e^{k\phi_0}\,,\qquad q=-2\Lambda e^{-(\d+k)\phi_0}\,.
\label{1681}\ee
For this solution
\be
 T\sim r_0\,,\qquad S={\rm constant}\,.
\label{169}\ee
These are  AdS$_2$ black holes driven by strong charge self-interactions.

\subsection{Strong coupling solution II}

\be
e^A=r^{c_p^2\over 2(p-1)}\,,\qquad e^{\phi}=e^{\phi_0}r^{c_p}\,,\qquad e^g=e^{g_0} r^{c_1}\,,
\label{170}\ee
with
\be
c_p=2(\d+k)\,,\qquad c_1=1-{2p\over p-1}(\d+k)^2\,,\qquad q=-2\Lambda e^{-(\d+k)\phi_0}\,.
\label{171}\ee
This is an extremal solution.

\subsection{Strong coupling solution III}

For $\d=\pm 1$ and $p=4$, there is another solution
\be
e^{A}=r^{1\over 4}\,,\qquad e^{\phi}=r^{\mp{3\over 4}}\,,\qquad f={16|q|\over 25}(r^{5\over 4}-r_0^{5\over 4})\,,\qquad |q|={5\over 8}V_0
\,,\qquad A_t={4\over 5}(r^{5\over 4}-r_0^{5\over 4})\,.
\label{172}\ee
We can obtain the temperature and entropy as
\be
4\pi T=\Lambda{\sqrt{r_0}}
\,,\qquad S\sim e^{3A}=r_0^{3\over 4}\,,
\label{173}\ee
giving
\be
Q={4\over 5}\,,\qquad \Phi=-{4\over 5}\left({4\pi T\over \Lambda}\right)^{5\over 2}\,,\qquad S\sim  T^{3\over 2}\,.
\label{174}\ee
Consistency with the strong coupling limit implies that
\be
q>> e^{(\gamma+k)\phi+3A}\,.
\label{175}\ee
This can be valid in the IR depending on the sign of $\gamma+k$.

The DC resistivity for massive carriers is
\be
\rho\sim T^{1\pm {3\over 2}k}\,.
\label{176}\ee
It is linear if the scalar is not the dilaton.

\section{Acknowledgements}\label{ACKNOWL}

We would like to thank E. Fradkin, U. Gursoy, S. Hartnoll, C. Herzog, S. Kachru, A. Kitaev, M. Lippert, H. Liu, J. McGreevy, F. Nitti, A. O'Bannon, H. Ooguri, N. Papanicolaou, N. Read, S. Sadchdev, K. Skenderis, T. Takayanagi, M. Taylor and X.G. Wen for useful conversations and correspondence. We would like to especially thank C. Panagopoulos, whose invaluable input on the physics of strange metals was a strong guiding principle for this effort.

B.G. would like to thank Prof. Marc Henneaux from the Service de Physique Th\'eorique et Math\'ematique, Universit\'e Libre de Bruxelles, for precious comments on Hamiltonian formalism and energy definitions as well as Prof. Renaud Parentani from LPT, Univ. Paris-Sud 11 for interesting discussions on thermodynamic ensembles and Hamiltonian formalism. Finally, he would like to thank the Crete Center for Theoretical Physics at University of Crete for its kind hospitality at the later stages of this work.

R.M. would like to thank the members of the Center for Quantum Spacetime (CQUeST), Sogang University, Seoul, as well as Prof. Deog-Ki  Hong at Pusan National University for their kind hospitality during his visits in the late stages of the project.

B.S.K.  would like to thank the members of CQUeST and KIAS, Seoul, for their warm hospitality during his visits.

This work was  partially supported by  a European Union grant FP7-REGPOT-2008-1-CreteHEPCosmo-228644, Excellence grant MEXT-CT-2006-039047
and by ANR grant  STR-COSMO, ANR-09-BLAN-0157.

\vskip 2cm
\section*{Note added}
\addcontentsline{toc}{section}{Note added\label{nadd}}

After the appearance of this paper, \cite{perl} appeared which independently analyzed the scaling solutions of section \ref{nearex}. Moreover it has connected them with the uncharged extremal solutions in the UV, and defined the physics by using the non-conformal holography. Our appendix \ref{phasev}, added later to respond to questions posed by the referee is in agreement with \cite{perl} although the techniques used here are somewhat different. We thank E. Perlmutter for discussions on his work.

\newpage
\appendix
\renewcommand{\theequation}{\thesection.\arabic{equation}}
\addcontentsline{toc}{section}{APPENDIX\label{app}}
\section*{APPENDIX}

\section{Equations of motion for the EMD system\label{equo}}

The relevant Einstein equations stemming from (\ref{ActionLiouville}) in the general coordinate system \eqref{ToroidalMetrics} are
\be
\phi'^2+(p-1){C''\over C}-{p-1\over 2}\left({D'\over D}+{B'\over B}+{C'\over C}\right){C'\over C}
=0\,,
\label{16bb}\ee
\be
{D''\over D}-{C''\over C}+{1\over 2}\left({C'\over C}-{D'\over D}\right)\left({B'\over B}+{D'\over D}+(3-p){C'\over C}\right)-{q^2 B\over ZC^{{p-1}}
}=0\,,
\label{16bbb}\ee
\be
-{1\over 2}\phi'^2-BV+{p-1\over 2}{C'\over C}\left({p-2\over 2}{C'\over C}+{D'\over D}\right)+{q^2B\over 2ZC^{p-1}}=0\,,
\label{17b}\ee
along with the solution for the gauge field
\be
F_{rt}=A_t'={q \sqrt{DB}\over ZC^{(p-1)\over 2}}\,.
 \label{14bb}\ee
In the domain-wall coordinate system of \eqref{ToroidalMetrics2} the equations of motion read
\begin{eqnarray}
 \label{Einstein2}
0 &=& (p-1)A'' + {\frac{1}{2}} \phi'^2\,,\\
\label{Einstein3}
0 &=& \frac{g''}{g'} + g' + pA' - \frac{q^2}{e^{2(p-1)A+g} Z(\phi) g'}\,,\\
\label{Einstein4}
0 &=& (p-1)A'(g'+pA') - e^{-g} V(\phi) - {\frac{1}{2}} \phi'^2 + \frac{q^2}{2e^{2(p-1)A+g} Z(\phi)}\,.
\end{eqnarray}

\section{Definition of thermodynamic ensembles\label{thermo}}

\subsection{Grand-canonical ensemble}

In this appendix  we will identify the various thermodynamical ensembles we will work with. To do this, we consider the Euclidean version of action \eqref{ActionLiouville} in $p+1$ dimensions and supplement it with appropriate p-dimensional boundary terms, in order to make the variational problem well-defined. Although these boundary terms might be neglected in order to obtain classical solutions of the equations of motion, they are crucial once one tries to calculate the Hamiltonian associated to these solutions \cite{Regge:1974zd}. We will start by adding only the Gibbons-Hawking boundary term \cite{Gibbons:1976ue} to the usual Lorentzian action:
\be
	I_E =\frac1{16\pi} \int_{\mathcal M} \ud^{p+1}x\sqrt{g_E}\le[R-\frac12(\partial \phi)^2-\frac14 e^{\ga\phi}F^2+2\Lambda e^{-\da\phi}\ri]+\frac1{8\pi}\int_{\partial\mathcal M}\sqrt{h_E^p} K_p\,,
	\label{ActionGrandCanonical}
\ee
where $\partial\mathcal M$ is the boundary of our space at radial infinity, $h^p_{\mu\nu}$ the induced metric on the $p$-boundary and $K_p=h_p^{\mu\nu}\nabla_{\mu}n_\nu$ its extrinsic curvature. $n^\mu$ is a vector orthogonal to the boundary and pointing outwards.  A convenient way to calculate the boundary term is to use the following relationship:
\be
	\int_{\partial\mathcal M}\sqrt{h_E^p} K_p = \frac{\partial}{\partial n}\int_{\partial\mathcal M}\sqrt{h_E^p}\,,
\ee
see \cite{Gibbons:1976ue}. We then  consider the variation of this action \eqref{ActionGrandCanonical} \cite{Hawking:1995ap}:
\bea
	\delta I_E &=& (\textrm{Terms giving the equations of motion})[\delta g^{\mu\nu},\,\delta A_\mu,\,\delta \phi]+ \nn\\
			&&+ (\textrm{Gravitational boundary term})\delta g^{\mu\nu}* \nn\\
			&&- \frac1{16\pi}\int_{\partial\mathcal M} \sqrt{h_E^p}e^{\ga\phi} F^{\mu\nu}n_\mu\delta\le(A_\nu\ri)- \nn\\
			&&-	 \frac1{16\pi}\int_{\partial\mathcal M} \sqrt{h_E^p} \da\le(\phi\ri) n^\mu\partial_\mu\phi \,.
\eea
The Gibbons-Hawking boundary term allows to fix only the metric components on the boundary and leave their first derivatives free. Variation of \eqref{ActionGrandCanonical} will yield the equations of motion if the variation is at fixed chemical potential on the boundary, which we will denote $\Phi$, as well as at zero variation of the scalar field at the boundary (we will see however that this variation is not independent of the other quantities varied).

This will define the grand-canonical ensemble as we use the gauge potential at infinity and the temperature as parameters and they are intensive variables. Again, it will not be necessary to define a thermodynamic variable related to the scalar field. The other extensive variables $S$ and $Q$ are determined as functions of them. The Gibbs thermodynamical potential in the grand-canonical ensemble is
\be
	G[T,\Phi] =  E - \Phi Q - TS\,,
	\label{ThermPotentialGrandCanonical}
\ee
and is related for a particular solution to the Euclidean continuation of the Lorentzian action by $I_{E}-I_E^0=-\beta G$ where $I^0$ is the background contribution. The first law is expressed as:
\be
	\ud E = \Phi\ud Q+ T\ud S\,,
	\label{FirstLawGrandCanonical}
\ee
or equivalently in terms of the Gibbs potential:
\be
	\ud G = \ud E -T\ud S -S\ud T - \Phi\ud Q - Q\ud\Phi= -S\ud T - Q\ud\Phi\,.
\ee
We can then deduce the other thermodynamic quantities of the solution by taking the appropriate derivatives of the Gibbs potential with respect to the thermodynamic variables,
\bea
	E&=&G-T\frac{\partial G}{\partial T}\Big|_\Phi -\Phi\frac{\partial G}{\partial \Phi}\Big|_T\,, \label{EnergyGrandCanonical}\\
	S&=&-\frac{\partial G}{\partial T}\Big|_\Phi \label{EntropyGrandCanonical}\,,\\
	Q&=&\frac{\partial G}{\partial\Phi}\Big|_T \label{ElectricChargeGrandCanonical}\,.
\eea

Specialising to $p=3$ (four-dimensional case), we can also determine the various thermodynamic quantities as conserved charges stemming from the symmetries realised asymptotically
on the boundary, following the Regge-Teitelboim-Henneaux procedure, \cite{Regge:1974zd,Henneaux:1985tv}.
 We detail this procedure in the case of the conserved charge associated to translations in time on the boundary, that is, the energy.
Indeed, it has been shown in other contexts that a scalar field could backreact strongly on the metric in such a way as to modify the asymptotic
boundary conditions, and thus generating extra contributions to the variation of the conserved charge associated to the energy,\cite{Martinez:2004nb,Henneaux:2004zi,Henneaux:2006hk}.

Although our setup generically breaks the $SO(4,2)$ symmetry of the Anti-de Sitter space-time on the asymptotic boundary, see \eqref{LiouvilleBackground},
considering the timelike Killing vector $\frac{\partial}{\partial t}$ generating the time isometry associated to the energy will be enough for our purposes
 and allows to follow the procedure defined in \cite{Regge:1974zd,Henneaux:1985tv,Martinez:2004nb,Henneaux:2004zi,Henneaux:2006hk}.
Referring to these papers for details of the computation, one uses the following minisuperspace metric
\be
  \ud s^2 = -N^2f^2(r)\ud t^2 + \frac{\ud r^2}{f^2(r)} + R^2(r)\left(\ud x^2 + \ud y^2\right),
  \label{MetricMinisuperspace}
\ee
which yields the following boundary terms (the simplified analysis of \cite{Martinez:2004nb} is still valid in our case, with appropriate
boundary conditions)
\be
  \delta B_g = \frac{\beta\omega_2}{8\pi G}\left[N\left(RR'\delta f^2-(f^2)'R\delta R \right)+2f^2\left(N\delta R'-N'\delta R \right)\right]^\infty_{r_+},
  \label{GravitationalBoundaryVariation}
\ee
with primes denoting derivatives with respect to $r$, $\beta$ the inverse Euclidean temperature, $\omega_2$ the volume of the compact planar horizon.
 $r_+$ means the inner boundary of the space-time, which for a black hole is the outer event horizon.

To this gravitational contribution, one must add a scalar term, \cite{Martinez:2004nb,Henneaux:2004zi,Henneaux:2006hk},
\be
  \delta B_\phi = \frac{\beta\omega_2}{16\pi G}\left[NR^2f^2\phi'\delta\phi \right]^\infty_{r_+}.
\label{ScalarBoundaryVariation}
\ee
The contribution on the asymptotic boundary of these two terms will yield the variation of the energy, that is
\be
 \delta E = -\delta B_g^\infty -\delta B_\phi^\infty \,,
\ee
which can then be integrated. One should also take into account the boundary term stemming from the integration by parts of the Maxwell Lagrangian,
but this does not present any particular difficulty. Summing all these boundary terms, both at the horizon and at infinity, will give the Gibbs potential.

The electric energy is
\be
	-\Phi Q =  \frac1{16\pi}\int_{S_t^\infty} \sqrt\sigma N \le[e^{\ga\phi} F^{\mu0}n_\mu A_0\ri],
	\label{ElectricEnergy}
\ee
which allows to identify the conserved electric charge, the $A_0$ term on the boundary being the chemical potential at infinity. The chemical potential
for a black hole is usually fixed such that the Maxwell vector $A_\mu$ is always regular on the horizon of the black hole. This implies fixing $\Phi$
 so that $A_0(r_+)=0$.

This yields finally,
\bea
	E &=& M_g + M_\phi \,,\\
	G &=& M_g+M_\phi-\Phi Q - TS\,,
\eea
for the energy and the Gibbs potential.

\subsection{Canonical ensemble}

On the other hand, consider adding to the action the following Maxwell boundary term
\be
	\tilde I_E = I_E +\frac1{16\pi}\int_{\mathcal M} \sqrt{h_E^p}e^{\ga\phi} F^{\mu\nu}n_\mu A_\nu\,,
	\label{ActionCanonical}
\ee
which will combine with the previous electric two-dimensional boundary term when the action is varied and yield:
\bear
	\delta \tilde I_E &=& (\textrm{Terms giving the equations of motion})[\delta g^{\mu\nu},\,\delta A_\mu,\,\delta \phi] \nn\\
			&&+ \textrm{Gravitational boundary term}+ \nn\\
			&&+ \frac1{16\pi}\int_{\partial\mathcal M} \delta\left(\sqrt{h_E^p}e^{\ga\phi} F^{\mu\nu}n_\mu\right)\le(A_\nu\ri)- \nn\\
			&&-	 \frac1{16\pi}\int_{\partial\mathcal M} \sqrt{h_E^p} \da\le(\phi\ri) n^\mu\partial_\mu\phi \,.\,.
\eear
Here, the equations of motion will be obtained when we keep the quantity $\left(\sqrt{-h} F^{\mu\nu}n_\mu\right)$ fixed on the boundary at infinity, which is precisely the usual electric conserved charge,
\be
	Q = \frac1{16\pi}\int_{S^t_\infty} \sqrt{\sigma}N e^{\ga\phi}F^{0\nu}n_\nu\,.
	\label{ElectricCharge}
\ee
This is the canonical ensemble, as thermodynamical variables are the temperature (intensive) and the electric charge (extensive parameter). The free energy in the canonical ensemble is the Helmholtz potential,
\be
	W[T,Q] = E - TS\,,
	\label{FreeEnergyCanonical}
\ee
related to the Euclidean action by $-\beta F = \tilde I -\tilde I_0$. The first law is expressed as:
\be
	\ud E =  T\ud S + \Phi\ud Q\,,
	\label{FirstLawCanonical}
\ee
or equivalently as
\be
	\ud W = \ud E -T\ud S -S\ud T = -S\ud T + \Phi\ud Q
\ee
We can then deduce the other thermodynamic quantities of the solution by taking the appropriate derivatives of the Helmholtz potential with respect to the thermodynamic variables,
\bear
	E_W&=&W-T\frac{\partial W}{\partial T}\Big|_Q \label{EnergyCanonical}\,,\\
	S_W&=&-\frac{\partial W}{\partial T}\Big|_Q \label{EntropyCanonical}\,,\\
	\Phi_W&=&\frac{\partial W}{\partial Q}\Big|_T \label{ElectricPotentialCanonical}\,.
\eear

We also define various quantities which will be useful to evaluate the thermodynamical stability of the solutions \cite{RN2}, namely the heat capacity at constant electric charge and dilaton charge,
\be
	C_Q = T\le(\frac{\partial S}{\partial T}\ri)_{Q}\,,
	\label{HeatCapacityCanonical}
\ee
and the electric permittivity,
\be
	\epsilon_T = \le(\frac{\partial Q}{\partial\Phi}\ri)_{T}\,.
	\label{ElectricPermittivityCanonical}
\ee
Stability of the solutions requires that both quantities are positive. This corresponds respectively to the fact that larger black holes should heat up and radiate more, while smaller black holes should go colder and radiate less; and that the chemical potential should increase when more charge is added to black hole \cite{RN1}, making it harder to move away from equilibrium, as expected from classical physics or more generally, from Le Chatellier's principle.

Equivalently, one can require that
\be
	\le(\frac{\partial^2W}{\partial T^2}\ri)_{Q}\leq0\,, \quad  \le(\frac{\partial^2 W}{\partial Q^2}\ri)_{T}\leq0\,,
\ee
as can be seen by varying the free energy in the canonical ensemble \eqref{FreeEnergyCanonical}.

Finally, let us note that all these boundary terms in the case of black hole space-times should be considered both on the boundary at spatial infinity, the outer boundary of space-time, but also on the horizon, the inner boundary. In the case of the Maxwell boundary term, this is not a problem: indeed, there is a gauge freedom in choosing the value of the chemical potential on the outer boundary, $\Phi$. It cannot be set to zero for a black hole space-time as the coordinate transformation required to do that would be singular on the horizon \cite{Gibbons:1976ue}, so we fix it so that it is zero on the horizon: this way, the Maxwell boundary term never contributes on the inner boundary. On the other hand, for the Gibbons-Hawking boundary term, including the inner boundary contribution would precisely mean that we do not fix the temperature of our thermodynamical system, but rather its entropy: this means going over to the microcanonical ensemble. However, in order to carry calculations in this ensemble, we would need to be able to count the available microstates needed to fix the entropy to a particular value. Unfortunately, this requires a quantum theory of gravity: some progress in the case of extremal (or so-called BPS) black holes have been made in some specific realizations of String Theory \cite{Strominger:1996sh}, but this is as far as our knowledge extends for the present. Thus, we will leave aside the microcanonical ensemble in the following.

\section{DBI dynamics and the strong interaction limit}

In the case where the U(1) symmetry arises from fundamentals, the gauge field originates from a stack of flavor D-branes. In this case the proper gauge field action is the DBI action that includes ``self-interactions" of charge.
\be S_{DBI}=-\int d^{p+1}x ~\tilde Z(\phi)\le[\sqrt{-\det(g^{\s}+F)}-\sqrt{-\det g^{\s}}\ri],
 \label{11}\ee
where we have adjusted the action so that the potential is still given by V in (\ref{al1}).
Therefore in this case the full action is $S_g+S_{DBI}$ instead of $S_g+S_F$.
The metric that enters the DBI action is the induced metric in the string frame. As we do not consider extra embedding coordinates, we can rewrite it using the Einstein metric using (\ref{ein1}).
\be S_{DBI}=-\int d^{p+1}x ~  e^{2k\phi} Z(\phi)[\sqrt{-\det(g+e^{-k\phi}F)}-\sqrt{-\det g}]\,,\qquad Z=M^{p-1}e^{{(p-3)\over 2}k\phi}\tilde Z\,.
 \label{11a}\ee
Here the coupling $Z$ has been defined such as to appear in the kanonically normalised Maxwell term, $-Z(\phi) F^2/4$.

The equation of motion for the gauge field  is
\be
\partial_{\m}\left[e^{+k\phi}Z \sqrt{-\det(g+e^{-k\phi}F)}\left((g+e^{-k\phi}F)^{\m\n}-(g+e^{-k\phi}F)^{\n\m}\right)\right]=0\,,
 \label{12}\ee
while the variation with respect to the metric gives
\be
{1\over \sqrt{-g}}{\delta S_{DBI}\over \delta g^{\m\n}}={{Z e^{2k\phi}}\over 4}\sqrt{\det(g+e^{-k\phi}F)\over \det g}\left[g_{\m a}\left(g+e^{-k\phi}F\right)^{ab}g_{b\n}+\left(\mu\leftrightarrow \n\right)\right]-{{Z e^{2k\phi}}\over 2}g_{\m\n}
\label{104}\ee
where $\left(g+e^{-k\phi}F\right)^{ab}$ is the inverse matric of $\left(g+e^{-k\phi}F\right)_{ab}$.

For a non-trivial $F_{rt}$  profile in the coordinate system (\ref{ToroidalMetrics}) we obtain
\be
-\det(g+e^{-k\phi}F)=C^{p-1}(DB-e^{-2k\phi}F_{rt}^2)\,,
\label{105}\ee
and the equation of motion (\ref{12}) becomes
\be
\pa_r\left({ZC^{p-1\over 2}F_{rt}\over \sqrt{DB-e^{-2k\phi}F_{rt}^2}}\right)=0~~\to ~~  {ZC^{p-1\over 2}F_{rt}\over \sqrt{DB-e^{-2k\phi}F_{rt}^2}}=q\,.
 \label{13}\ee
Solving we obtain
\be
F_{rt}=A_t'={q \sqrt{DB}e^{{k}\phi}\over \sqrt{q^2+e^{2k\phi}Z^2C^{(p-1)}}}\,.
 \label{14}\ee
The on-shell action is
\bea
S^{on~shell}_{DBI}&=&-\int \ud^{p+1}x~ e^{2k\phi} Z\le[\sqrt{|\det(g+e^{-k\phi}F)|}-\sqrt{\det g}\ri] \label{15}\\
									&=&-V_{p}\int \ud r e^{2k\phi}Z C^{p-1\over 2}\sqrt{DB}\left(
{e^{2k\phi}ZC^{p-1\over 2}\over \sqrt{q^2e^{2k\phi}+e^{4k\phi} Z^2C^{p-1}}}-1\right)\nn\\
									&=&V_{p}\int \ud r\left[{q^2\over 2}{\sqrt{DB}\over ZC^{(p-1)\over 2}}+{\cal O}(q^4)\right].
\eea

The last line is the result in the linearised (Maxwell) approximation.

The Einstein equations  (\ref{Einstein2})-(\ref{Einstein4})  now become

\be
\phi'^2+(p-1){C''\over C}-\left({D'\over D}+{B'\over B}+(p-2){C'\over C}\right){C'\over C}
=0\,,
\label{16b}\ee
\be
{D''\over D}-{C''\over C}+{4-p\over 2}\left({C'\over C}-{D'\over D}\right)\left({B'\over B}+{D'\over D}\right)-{q^2 e^{2k\phi}BC^{-{p-1\over 2}}\over
\sqrt{q^2e^{2k\phi}+e^{4k\phi}Z^2C^{p-1}}}=0\,,
\label{16}\ee
\be
-{1\over 2}\phi'^2-B(V+e^{2k\phi}Z)+{p-1\over 2}{C'\over C}\left({p-2\over 2}{C'\over C}+{D'\over D}\right)+{B\over C^{p-1\over 2}}\sqrt{q^2e^{2k\phi}+e^{4k\phi}Z^2C^{p-1}}=0\,.
\label{17}\ee
In particular, in  the domain wall frame, ($D=e^{2A+g}$, $B=e^{-g}$, $C=e^{2A}$), the equations above translate to
\be
(p-1)A^{\prime\prime}+{1 \over 2}\phi^{\prime 2}=0\,,\qquad
pA^{\prime}+g^{\prime}+{g^{\prime\prime} \over g^{\prime}} -
{q^2e^{-2(p-1)A} e^{-g+2k\phi} \over  g^{\prime}\sqrt{e^{4k\phi}Z^2+q^2e^{-2(p-1)A+2k\phi}}} =0\,,
\label{16a}\ee

\be
(p-1)A'(g'+pA')-{1\over 2}\phi^{\prime 2}-e^{-g}(V+e^{2k\phi}Z)
+e^{-g}\sqrt{e^{4k\phi}Z^2+q^2e^{-2(p-1)A+2k\phi}}=0\,.
\label{17a}\ee
There is still a conserved Noether charge
\be
{\cal Q}= e^{pA}g'e^g-qA_t\,,\qquad {\ud{\cal Q}\over \ud r}=0\,.
\label{18}\ee
and it vanishes at extremality.

\subsection{The weak self-interaction limit}

In a region where
\be
Z(\phi)~C^{(p-1)\over 2}e^{{k}\phi}\gg |q|\,,
\label{19}
\ee
we can linearize the DBI action and the equations (\ref{16}) and (\ref{17}) reduce to the Maxwell equations in (\ref{Einstein2})-(\ref{Einstein4}) as can be directly verified.

\subsection{The strong self-interaction limit\label{strong}}

If the strong self-interaction limit it is the opposite inequality that is satisfied
\be
Z(\phi)~C^{(p-1)\over 2}e^{{k}\phi}\ll |q|\,.
\label{20}\ee
In the domain wall coordinates, the DBI square root is approximated as
\be
\sqrt{e^{4k\phi}Z^2+q^2e^{2k\phi-2(p-1)A}}\simeq |q|e^{{k}\phi-(p-1)A}+{Z^2 e^{(p-1)A+{3k}\phi}\over 2|q|}+\cdots
\label{21}\ee
and the rest of the equations become to leading order
\be
(p-1)A^{\prime\prime}+{1 \over 2}\phi^{\prime 2}=0\,,\qquad
pA^{\prime}+g^{\prime}+{g^{\prime\prime} \over g^{\prime}} -
{|q| e^{-g+{k}\phi} \over  g^{\prime}e^{(p-1)A}} =0\,,
\label{22}\ee

\be
(p-1)A'(g'+pA')-{1\over 2}\phi^{\prime 2}-e^{-g}V
+|q|e^{-g-(p-1)A+{k}\phi}=0\,.
\label{23}\ee

Note that in this limit the gauge coupling constant function $Z(\phi)$ drops out of the equations to leading order after taking into account the inequality \eqref{20}.
Moreover the gauge field solution becomes
\be
F_{rt}=A_t'={\rm sign}(q)e^{A+{k}\phi}\left[1-{Z^2e^{2(p-1)A+2k\phi}\over 2q^2}+\cdots\right]
 \label{24}\ee
Therefore to leading order the gauge field solution is $q$-independent.

\subsection{The probe limit\label{probe}}

In this limit the presence of the gauge field solution is subleading in the gravity plus scalar equations, with equivalent condition
\be
V\gg e^{2k\phi}Z\left(
{e^{2k\phi}ZC^{p-1\over 2}\over \sqrt{ q^2e^{2k\phi}+e^{4k\phi}Z^2C^{p-1}}}-1\right),
\label{24a}\ee
In the linearized limit this becomes

\be
V\gg {q^2\over 2ZC^{p-1}}\,,
\label{25}\ee
while in the strong self-interaction limit it is
\be
V\gg e^{2k\phi} Z\,.
\label{26}\ee

\section{The Sturm-Liouville problem for spin-1 and 2 in general charged backgrounds at extremality.\label{slc}}

We will study here two special fluctuations of the system, namely spin-2 excitations and the spin-2 excitations arising from the gauge field at extremality. This will give us information on the reliability of the extremal naked singularity as to the spectral properties of the system.
If the related Sturm-Liouville problem is well-defined with a normalizable and a non-normalizable solution, then the associated physics is largely independent of the resolution of the singularity.
In the opposite case it depends in a crucial way.

\subsection{The spectrum of the Laplacian}

The spectra of the spin-two excitations are readily computable as they satisfy the standard Laplace equation.
 In the background above and for a plane wave basis for the perturbation, $h(r)e^{i\omega t+i\vec k \cdot \vec x}$ it becomes
 (in Euclidean space)
\be
h''+\left({f'\over f}+pA'\right)h'-\left({\omega^2\over f^2e^{2A}}+{\vec k^2\over fe^{2A}}\right)h=0
\label{d4}\ee
At extremality ($m=0$ in (\ref{88e})), $f\sim r^{-v}$ and
\be
{f'\over f}+pA'={a\over r}\sp a=2+{\g^2-\d^2\over 2}+{(\g-\d)^2\over 2(p-1)}=1+{w_pu\over 2(p-1)}
\label{d5}\ee
while the terms depending on frequencies or momenta scale as some inverse powers of $r$,
\be
{\omega^2\over f^2e^{2A}} \sim {\omega^2\over r^{2(2+\d(\g-\d))+ {(\g-\d)^2\over (p-1)}}} \,,\qquad
{\vec k^2\over fe^{2A}} \sim {\vec k^2\over r^{2+\d(\g-\d)+ {(\g-\d)^2\over (p-1)}}} \,.
\ee
The Gubser constraint (\ref{d3}) implies that $a>1$.

Therefore, near the singularity $r=0$,  equation (\ref{d4}) becomes
\be
h''+{a\over r}h'-{w^2\over r^{2b}}h=0\sp b=-v +{(\g-\d)^2\over 2(p-1)}>0.
\label{d6}
\ee
with $w\sim \omega$.
Equation can be solved by the modified  Bessel functions, and the two independent solutions are

\be
h_{\pm}=r^{1-a\over 2}\left\{ \begin{array}{l}
\displaystyle  I_{|\nu|}\left({w\over |1-b|}r^{1-b}\right)\\
\\
\displaystyle I_{-|\nu|}\left({w
\over |1-b|}r^{1-b}\right).
\end{array}\right.  \sp \nu={1-a\over 2(b-1)}
\label{d7}\ee

When $b<1$, the argument of the Bessel function is small near $r=0$ and the solutions behave as
\be
h_{1}\sim r^{1-a}\sp h_{2}\sim {\rm constant}
\label{d8}\ee
Using the standard norm
\be
|h|^2=\int_0 dr~ r^a~h(r)^2
\label{d9}\ee
we observe that $h_1$ is normalizable at $r=0$ when $a<3$ while $h_2$ is always normalizable as $a>1$. The Sturm-Liouville problem is therefore unacceptable when $1<a<3$.
Therefore, when
\be
b<1\sp  w_pu< (p-1)(\g-\d)^2
\label{d10}\ee
the dynamics of the low temperature system is not reliable and a resolution of the singularity is needed to access it. The relevant region is shown in figure \ref{Fig41} in the main text.

When $b>1$, the argument of the Bessel function is large near $r=0$ and the solutions behave as
\be
h_{\pm}\sim r^{b-a  \over 2}e^{\pm{w \over |b-1|r^{b-1}}} \,,
\label{d11}\ee
Therefore one is always normalizable and the other is always not normalizable.
we conclude that  when $b>1$, the Sturm-Liouville problem for spin two is always well defined.

There are a few marginal cases that need to be checked separately. In the  $a=1$ case the two linearly independent solutions
are $I_0$ and $K_0$.
For $b<1$, both solutions are normalizable and therefore this case is not reliable. For $b>1$ only one is normalizable.
The other case is $b=1$. In this case the two solutions behave as
\be
h_{\pm} \sim r^{1-a\pm \sqrt{(1-a)^2+4w^2}\over 2}
\label{d12}\ee
Since $h_{+}$ is always normalizable the constraint that $h_-$ is non-normalizable reads, $(1-a)^2+4w^2>4$.
Since $w\sim\omega$ in order for this to hold for all $\omega$, me finally must have
\be
b=1\sp a>3
\label{d13}\ee
This is still in agreement with (\ref{d10}) with $b=1$.

\subsection{The spectrum of current excitations}

The spectrum of current excitations at zero momentum $\vec{k}=0$ can be computed in the parity odd sector, in which the $A_i$ and $g_{ti}$ fluctuations decouple from the rest of the fluctuation spectrum, from eq.~\eqref{30a}. In the domain wall coordinate system used in \eqref{Sol3DW} eq.~\eqref{30a} reads in Euclidean signature
\bea\label{823}
0 &=& A_i'' + \left({Z'\over Z}+{f'\over f}+(p-2)A'\right)A_i' - \left[ \frac{\omega^2}{f^2 e^{2A}} + \frac{q^2}{f Z e^{2(p-1)A}}\right]A_i\\
&=& A_i '' + \frac{\tilde a}{r} A_i' - \left[ \frac{m_1^2}{r^{2b}} + \frac{m_2^2}{r^{2 b'}} \right]A_i\\
\tilde a &=& 2 - \frac{p(\ga-\d)^2}{2(p-1)} \leq 2 \,,\quad b = \frac{(\ga-\d)^2}{2(p-1)} - v \geq 0\,,\,\,
b' = \frac{\tilde a}{2}+\frac{p(\ga-\d)^2}{4(p-1)} = 1\,,\label{tan3} \\
m_1^2 &=& \left( \frac{ w_p u^2 e^{\d \phi_0}}{8(p-1) (-\Lambda \ell^2)} \right)^2 \omega^2 \ell^{2 b} \,,\quad m_2^2 = \frac{q^2 w_pu^2 e^{(\d-\ga)\phi_0} }{8(p-1)(-\Lambda)}
\label{d14}
\eea

For $b > 1$ the frequency term dominates and, as in the spin-2 case, the Sturm-Liouville problem is always acceptable. For $b < 1$, the charge term dominates as $r\rightarrow 0$ and the solutions behave as
\be
A_i^{\pm} \sim r^{{1-\tilde a\pm \sqrt{(1-\tilde a)^2+4m_2^2}\over 2}}\,.
\label{d15}\ee
Again the $A^{+}_i$ solution is always normalizable. For $A^{-}_i$ to be non-normalizable, we must have
\be
\sqrt{(1-\tilde a)^2+4m_2^2}>2
\label{d16}\ee
As $m_2\sim q$ this is always satisfied for sufficiently large charge density
\be
q^2>q_0^2\equiv {2(p-1)e^{(\g-\d)\phi_0}\over w_pu^2}\left(-{\Lambda\over \ell^2}\right)\left(3-{p(\g-\d)^2\over 2(p-1)}\right)\left(1+{p(\g-\d)^2\over 2(p-1)}\right)
\label{tan4}\ee
When $\tilde a<-1$, this is satisfied for any charge density. In the opposite case $2>\tilde a>-1$, the constraint (\ref{tan4}) is necessary.

For $b=b'=1$, the solutions behave as
\be
A_i^{\pm} \sim r^{{1-\tilde a\pm \sqrt{(1-\tilde a)^2+4(m_1^2+m_2^2)}\over 2}}\,.
\label{d17}\ee
Asking that this leads to an acceptable problem  for all frequencies we end up with the constraint (\ref{d16}).

To summarize, there are no constraints on $\gamma,\delta$ coming from the spin-1 Sturm-Liouville problem. However for $b<1$ there  are constraints on the charge density summarized in (\ref{d16}).

The various constraints are summarized in figure \ref{Fig42} for $p=3$ and figure \ref{Fig43} for $p=4$.

\section{The gravitational system in phase variables: fixed points and stability.\label{phasev}}

In this appendix we will formulate and study a different form of the system of equations presented in appendix
\ref{equo}. They involve three phase variables satisfying a non-linear first-order system of equations from which all the background functions
can be reconstructed by direct integration. The advantage of this formulation is that it is largely coordinate-independent and it is
useful to study local stability of fixed points
and the global profile of the solution space. It is also the most efficient formulation for finding numerical solutions.
This is a generalization of the approached developed in  \cite{ihqcd2},\cite{gkmn2}.
A related approach was developed \cite{Poletti:1994ff}, but it involves five first-order equations and is coordinate-dependent. This work investigated
the global profile of the space of solutions and we find results qualitatively similar.

We will define the phase variables first in the domain wall system (\ref{ToroidalMetrics2}, \ref{3},\ref{4}) and we will then show that they can be defined in a coordinate independent fashion.

In the domain-wall coordinate system, the equations of motion read
\begin{eqnarray}
 \label{4n}
0 &=& (p-1)A'' + {\frac{1}{2}} \phi'^2\,,\\
\label{5n}
0 &=& \frac{g''}{g'} + g' + pA' - \frac{q^2}{e^{2(p-1)A+g} Z(\phi) g'}\,,\\
\label{6n}
0 &=& (p-1)A'(g'+pA') - e^{-g} V(\phi) - {\frac{1}{2}} \phi'^2 + \frac{q^2}{2e^{2(p-1)A+g} Z(\phi)}\,.
\end{eqnarray}
along with
\be
F_{rt}=A_t'={q \over Ze^{(p-2)A}}\,.
 \label{3n}\ee
To proceed further we will need the  second derivatives
\be
A''=-{\phi'^2\over 2(p-1)}\sp g''=-g'^2-pg'A' + \frac{q^2}{e^{2(p-1)A+g} Z(\phi) }\,,
\label{7n}\ee
\be
\phi''=-(g'+pA')\phi'-e^{-g}V'-{q^2Z'\over 2e^{2(p-1)A+g} Z^2}
\label{8n}\ee
where $Z'\equiv {dZ\over d\phi}, V'\equiv {dV\over d\phi}$.

We define the three phase variables as
\be X(\ff) = {1\over \sqrt{2p(p-1)}}\frac{\ff'}{A'},\qquad Y(\ff) = \frac{g'}{pA'}\sp W(\phi)={q^2e^{-2(p-1)A}\over 2Z~V}\,.
\label{9n}\ee
In terms of these we can express $A,g$ as
\be
A=A_0+{1\over \sqrt{2p(p-1)}}\int_{\phi_0}^{\phi}{d\phi\over X}\sp  g=\sqrt{p\over 2(p-1)}\int_{\phi_0}^{\phi}{d\phi~Y\over X}\,.
\label{10n}\ee
These determine $A,g$ as functions of $\phi$.

We then use the following relations
\be
e^{-g}={p(p-1)A'^2}{X^2-Y-1\over V(W-1)}~~\to~~A'=\pm\sqrt{V(W-1)\over p(p-1)(X^2-Y-1)}~e^{-{g\over 2}}\sp W,V>0
\label{11n}\ee
to obtain
\be
{d\phi\over dr}= \pm X\sqrt{V(W-1)\over p(p-1)(X^2-Y-1)}~e^{-{g\over 2}}
\label{16n}\ee
which upon integration gives $\phi(r)$.
Therefore knowledge of $X(\phi),Y(\phi),W(\phi)$ permits the reconstruction
of $A(r),g(r),\phi(r)$ from (\ref{10n}) and (\ref{16n}) by standard integration.

Although we defined the phase variables in the domain wall coordinate system, they can be defined in the generic coordinate system (\ref{ToroidalMetrics}) as follows
\be X(\ff) = {2\over \sqrt{2p(p-1)}}\frac{d\phi}{d\log(C)}\sp Y(\ff) = {2\over p}\frac{d\log(D/C)}{d\log(C)}\sp W(\phi)={Z\over 4V}F_{\m\n}F^{\m\n}={Z\over 2BDV}(A_t')^2\,.
\label{9nn}\ee

The final ingredient in this approach is the first-order system satisfied by the phase variables that is summarized below.

\be
{dW\over d\phi}=-\left({Z'\over Z}+{V'\over V}+\sqrt{2(p-1)\over p}{1\over X}\right)W\,,
\label{70n}\ee
\be
{dX\over d\phi}=\sqrt{p\over 2(p-1)}(X^2-Y-1)\left[1-\sqrt{p-1\over 2p}{{V'\over V}+{Z'\over Z}W
\over X(W-1)}\right],
\label{71n}\ee
\be
{dY\over d\phi}={X^2-Y-1\over \sqrt{2p(p-1)}X}\left[pY+{2(p-1)W\over  W-1}\right].
\label{72n}\ee

This system has as input functions the potential $V$ and the gauge coupling constant $Z$. For the system studied in this paper, $V=V_0 e^{-\d\phi}$, $Z=e^{\g\phi}$ and the equations above become

\be
{dW\over d\phi}=-\left(\g-\d+\sqrt{2(p-1)\over p}{1\over X}\right)W\,,
\label{73n}\ee
\be
{dX\over d\phi}=\sqrt{p\over 2(p-1)}(X^2-Y-1)\left[1-\sqrt{p-1\over 2p}{\g W-\d
\over X(W-1)}\right],
\label{74n}\ee
\be
{dY\over d\phi}={X^2-Y-1\over \sqrt{2p(p-1)}X}\left[pY+{2(p-1)W\over  W-1}\right].
\label{75n}\ee

These equations have three integration constants.
The determination of $A,g$ in terms of $\phi$ in equation (\ref{10n}) has another two. The extra integration constant to determine $A(r)$ is a diffeomorphism artifact in this case although it can be traded for an initial value for the scalar $\phi_0$.
The physical set of integration constants are three: $\phi_0,T,Q$.
The two last ones are decided in the system above. The third constant of the first-order system corresponds to the scalar vev and should be tuned so as to avoid a (bad) naked singularity.

Eqs. (\ref{70n})-(\ref{72n}) can be inverted and give
\be
{d\log(W-1)\over d\phi}=-{V'\over V}+\sqrt{p\over p-1}\left(X-{Y\over 2X}\right)+{d\log(X^2-Y-1)\over d\phi}\,.
\label{17n}\ee
Integrating, we obtain
\be
W(\phi)=1-{(X^2-Y-1)\over V}~e^{-\sqrt{p\over p-1}\int_{-\phi_0}^{\phi} du \le( X-{Y \over 2 X}\ri)}.
\label{15n}\ee

The scalar invariants of the background can be expressed in terms of the phase variables. The scalar curvature is given by
\be
R=-2U+{p+1\over p-1}{U-V\over X^2-Y-1}\left[{2p\over p+1}X^2-Y-1\right].
\label{68n}\ee
while the scalar and gauge-field  kinetic terms are
\be
(\partial\phi)^2=2V(W-1){X^2\over X^2-Y-1}\sp ZF^2=4WV\,.
\label{69n}\ee
Singularities of the scalar invariants are signaled by the vanishing of $X^2-Y-1$ or the divergence of $W,V\to \infty$. The horizon is signaled by a divergence of $Y$. In particular $X^2=Y+1$ is a bad generic singularity.

The first-order equations above are equivalent to the original equations if $\phi$ is not constant.
It can be shown that only in the case $\gamma=\d$ a solution $\phi=\phi_0$ constant exists, and therefore this case must be studied separately. In what follows we assume $\g\not=\d$.

There are special solutions to the system  (\ref{73n})-(\ref{75n}) that relate to the vanishing of the right-hand sides of equations. These are fixed point solutions to the equations and as we will see they coincide with the charged and uncharged extremal solutions as well as with completely singular (and unacceptable) solutions. The rest of the solutions interpolate between various fixed points solutions.

\subsection{Uncharged solutions, $W=0$}

$W=0$ is a fixed point of (\ref{70n}) and describes uncharged black holes.

The rest of the equations become
\be
{dX\over d\phi}=\sqrt{p\over 2(p-1)}(X^2-Y-1)\left[1-\sqrt{p-1\over 2p}{\delta
\over X}\right],
\label{36n}\ee
\be
{dY\over d\phi}={X^2-Y-1\over \sqrt{2p(p-1)}X}pY\,.
\label{37n}\ee

Note that the second equation can be solved as
\be
Y(\phi)={e^{-\sqrt{p\over 2(p-1)}\int^{\phi}{du\over X(u)}}\over C+\sqrt{p\over 2(p-1)}\int^{\phi}du{1-X^2(u)\over X(u)}e^{-\sqrt{p\over 2(p-1)}\int^{u}{du'\over X(u')}}}.
\label{38n}\ee
For an extremal solution we must have $Y=0$.

The fixed point solution here has
\be
Y=0\sp X=a\sp a=\sqrt{p-1\over 2p}\delta\,.
\label{90n}\ee
This corresponds to the extremal solution in (\ref{9}) for $r_0=0$.
We can also deform $Y$ by keeping $X=a$ to obtain
\be
Y={1-a^2\over e^{p(1-a^2)(\phi-\phi_0)\over \sqrt{2p(p-1)}a}-1}\,.
\label{39n}\ee
This has a  singularity hidden by a horizon. This coincides with the solution in (\ref{9}).

In this case we can write the general solution in  the implicit form as
\be
{(X-a)^{2a}\over (X-1)^{a+1}(X+1)^{a-1} }=e^{k(\phi-\phi_0)}\sp |X|>1\,,
\label{40n}\ee
and
\be
{(X-a)^{2a}\over (1-X^2)^{a}}{X+1\over X-1}=e^{k(\phi-\phi_0)}\sp |X|<1\sp k=2\sqrt{p\over 2(p-1)}\left({p-1\over 2p}\delta^2-1\right).
\label{41n}\ee

This solution captures all the flows among the three fixed points $\pm 1,a$ and $\pm \infty$. In particular, the Gubser bound is satisfied when $|a|<1$.
The fixed points at $X=\pm 1$ are bad singularities whose nature is independent of the potentials.

The solutions for $X>1$ are unphysical due to (\ref{11n}).
Therefore all these solutions are unacceptable as they interpolate between a bad and a good singularity. They become acceptable if we introduce a UV cutoff (or change the UV potential), and we keep the flow to the good IR singularity.

\subsection{Charged solutions, $W\not= 0$}

There is a single acceptable fixed point solution given by
\be
X_c=-\sqrt{2(p-1)\over p}{1\over (\g-\d)}\sp Y_c={2(p-1)\over p}{2+\g\d-\d^2\over (\g-\d)^2}\sp
 W_c={2+\d(\g-\d)\over 2+\g(\g-\d)}\,.
    \label{60n}\ee

 This is the extremal charged solution described in (\ref{Sol3DW})-(\ref{wpu}) in section \ref{nearex} with zero temperature.
 One can also solve for a non-trivial  Y keeping $X=X_c$ and $W=W_c$ describing the BH solution:
 \be
 Y={{2(p-1)\over p}{W_c\over W_c-1}e^{{1+\d(\g-\d)\over \g-\d}(\phi-\phi_0)}   +X_c^2-1\over
 e^{{1+\d(\g-\d)\over \g-\d}(\phi-\phi_0)}  -1}\,.
    \label{59n}    \ee

 Finally there is the completely singular generic fixed point $X^2=Y+1$.
 This exhausts all fixed points of the equations.

\subsection{Summary of fixed point solutions and their properties.}

The only acceptable fixed-point (extremal) solutions to equations (\ref{73n})-(\ref{75n}) we found were the uncharged solution
\be
X_u=\sqrt{p-1\over 2p}\delta\sp W_u=0\sp Y_u=0\,,
\label{91n}\ee
and the charged solution
\be
X_c=-\sqrt{2(p-1)\over p}{1\over (\g-\d)}\sp Y_c={2(p-1)\over p}{2+\g\d-\d^2\over (\g-\d)^2}=-{2(p-1)\over p}{v\over (\g-\d)^2}\,,
\label{92n}\ee
\be W_c={2+\d(\g-\d)\over 2+\g(\g-\d)}=-{v\over u}>0\,.
    \label{599n}\ee

 We can calculate that
\be
{X_u\over X_c}-1=-{2+\d(\g-\d)\over 2}={v\over 2}<0
\label{93n}\ee
which implies that always
\be
{X_u\over X_c}<1\,.
\label{94n}\ee

In table \ref{t1} the position of the fixed points is summarized as a function of the signs of two parameters.
\TABLE{
\begin{tabular}{|c|c|c|c|c|c|c|c|c|c|}
\hline
$\d$ & \phantom{a}$\g-\d$\phantom{a} & & \phantom{a} $X_u$\phantom{a} & \phantom{a} $Y_u$\phantom{a}
&\phantom{a} $W_u$\phantom{a} &&  \phantom{a} $X_c$\phantom{a} & \phantom{a}
$Y_c$\phantom{a} &\phantom{a} $W_c$\phantom{a}\\
\hline
+ & + &   & + & 0 & 0 & & -- &  + & +\\
-- & -- &   & -- & 0 & 0 & & + &  + & +\\
-- & + &   & -- & 0 & 0 & & -- &  + & +\\
+ & -- &   & + & 0 & 0 & & + &  + & +\\
\hline
\end{tabular}
\caption{A table showing the positivity properties of the fixed point solutions as a function of the values of the parameters. We have used the fact that the charged solutions exist when $w_p>0$, $u>0$, $v<0$ in (\protect\ref{Sol3DW}). \label{t1}}
}

Another important ingredient for the study of the solutions is the behavior of the dilaton as well as the potential in such solutions.
This is tabulated in table \ref{t2}.

\TABLE{
\begin{tabular}{|c|c|c|c|c|c|c|c|}
\hline
$\d$ & \phantom{a}$\g-\d$\phantom{a} & & \phantom{a} $\phi_u$\phantom{a} &  \phantom{a} $V_u$\phantom{a}
&&\phantom{a} $\phi_c$\phantom{a} & \phantom{a} $V_c$\phantom{a} \\
\hline
+ & + &   & $\searrow$ &  $\nearrow$ & & $\nearrow$ & $\searrow$\\

- & + &   &  $\nearrow$&$\nearrow$ &   &$\nearrow$ & $\nearrow$ \\

- & - &   &$\nearrow$ & $\nearrow$  &  & $\searrow$& $\searrow$\\

+& - &   & $\searrow$ & $\nearrow$ & & $\searrow$ & $\nearrow$ \\
\hline
\end{tabular}
\caption{We indicate the behavior of the scalar $\phi$ and the potential $V$ for the uncharged and charged extremal solutions. An arrow $\nearrow$ indicates an increase towards the IR while $\searrow$ denotes a decrease towards the IR.
Note that  when $\d(\g-\d)>0$,  $V_c$ decreases towards the IR. Since in  the uncharged solution
$V_u$ always increases towards the IR, this suggests that in these cases
$\phi'$ and $X$ should vanish somewhere in between in order for an interpolation between the uncharged and the charged solution to exist, as indicated by  the table above.\label{t2}}
}

\subsection{Stability analysis of fixed-point solutions}

We will now proceed to the stability analysis around the fixed-point solutions with or without charge.

\subsubsection{Uncharged fixed-point solution}

We start around the uncharged solution, expand  $W\to W_u+w$,$X\to X_u+x$, $Y\to Y_u+y$ and linearize the equations to obtain
\be
\left(\begin{matrix}
\dot x\\ \dot y\\\dot w
\end{matrix}\right)=M_u~\left(\begin{matrix}
x\\ y\\ w
\end{matrix}\right)\sp M_u=  \left(\begin{matrix}
{(p-1)\d^2-2p\over 2(p-1)\d}  &0& {((p-1)\d^2-2p)(\g-\d)\over 2\sqrt{2p(p-1)\d}}            \\
0& {(p-1)\d^2-2p\over 2(p-1)\d} & -{(p-1)\d^2-2p\over p\d}\\
0&0& -{2+\d(\g-\d)\over \d}
\end{matrix}\right).
\label{79n}\ee

 There are two equal eigenvalues $\l_1=\l_2$ with respective left eigenvectors $\vec v_{\l}$ given by
\be
\l_1={(p-1)\d^2-2p\over 2(p-1)\d}\sp \vec v_{1}=\left(\begin{matrix}
1\\ 0\\ 0\end{matrix}\right)\sp \l_2={(p-1)\d^2-2p\over 2(p-1)\d}\sp \vec v_2=\left(\begin{matrix}
0\\ 1\\ 0\end{matrix}\right),
\label{80n}\ee
and a third eigenvalue
\be
\l_3= -{2+\d(\g-\d)\over \d}={v\over \d}\sp \vec v_{3}=\left(\begin{matrix}
-\sqrt{p-1\over 2p}{(\g-\d)((p-1)\d^2-2p)\over (p-1)\d(2\g-\d)+2(p-2)}\\
{2(p-1)\over p}{((p-1)\d^2-2p)\over (p-1)\d(2\g-\d)+2(p-2)}\\ 1\end{matrix}\right).
\label{81n}\ee

By combining the data of tables \ref{t1} and \ref{t2} we obtain that in all cases the uncharged solution is unstable to flow away from it, if $\d^2<{2p\over p-1}$. To reach this conclusion we must take into account whether $\phi$ increases or decreases towards the IR as this crucial for the sign of the appropriate eigenvalues.
 Generically, the endpoint of such a flow is the singularity $X^2=Y+1$. With appropriately-tuned initial conditions the charged fixed point solution is obtained.

Finally we would like to point out that we can perturbatively solve the charged equations of motion by perturbing in $q^2$ around the uncharged solution.
If we denote collectively by $\Phi_0$ the metric components and scalar of the extremal uncharged solution and by $\Phi_1$ the first ${\cal O}(q^2)$ correction, the domain-wall perturbations around the uncharged extremal background
\be
	A_0=\frac{2}{(p-1)\da^2)}\ln r\,,\qquad g_0=0\,,\qquad \phi_0=\frac2\delta\ln r\,.
\ee
imply that
\be
\Phi=\Phi_0+\Phi_1+\cdots\sp \Phi_1\sim q^2 ~r^{2v\over \d^2}\,.
\ee
As $v<0$ for the existence of acceptable charged extremal solutions, this indicates that near the boundary ($r\to\infty$) the charge induces a subleading perturbation to the uncharged solution.

\subsubsection{Charged fixed-point solution}
Next we expand and linearize around the charged solution to obtain
\be
\left(\begin{matrix}
\dot x\\ \dot y\\\dot w
\end{matrix}\right)=M_c~\left(\begin{matrix}
x\\ y\\ w
\end{matrix}\right)\sp M_c=  \left(\begin{matrix}
 \frac{w_pu}{2(p-1)(\g-\d)} &0&  \frac{w_pu^3}{2\sqrt{2p(p-1)}(\g-\d)^4}         \\
0&{w_pu\over 2(p-1)(\g-\d)} & -{w_pu^3\over p(\g-\d)^5}  \\
-\sqrt{p\over 2(p-1)}~{v\over u}(\g-\d)^2&0&0
\end{matrix}\right).
\label{82n}\ee

An obvious left eigenvector and eigenvalue is
\be
\l_1={w_pu\over 2(p-1)(\g-\d)}\sp \vec v_{1}=\left(\begin{matrix}
0\\ 1\\ 0\end{matrix}\right).
\label{95n}\ee
The other two eigenvalues satisfy
\be
\l^2-{w_p u\over 2(p-1)(\g-\d)}\l+{w_p u^2 v\over 4(p-1) (\g-\d)^2}=0\,,
\label{96n}\ee
with solution
\be
\l_{\pm}={w_pu\pm\sqrt{w_p^2u^2-4(p-1)w_p u^2 v}\over 4(p-1)(\g-\d)}\,.
\label{97n}\ee
The charged extremal solutions exist for $v<0$, $w_p>0,u>0$.
As $v<0$, $\l_+$ has the sign of $\g-\d$ and $\l_-$ the opposite one.

Similarly, an analysis of the stability of the charged solution taking into account the results of table \ref{t2} shows that
the solutions have two stable directions and one that is unstable.
This is in accordance with what we expect from AdS/CFT: we must ``tune" one initial condition in order to arrive at an acceptable solution in the IR.
The tuning allows precisely the avoidance of the negative eigenvalue in the IR that drives the solution to be bad there.

Therefore a ``well-tuned" solution interpolates between the uncharged extremal solution in the UV and the charged extremal solution in the IR.
In the next subsection we will express the $\g\d=1$ solutions of section as concrete such interpolations.
Numerical interpolations were found in other concrete cases in \cite{perl}, where this issue has been studied in detail, and also in \cite{kachru} where numerical interpolation between the charged extremal fixed point in the IR and AdS asymptotics in the UV is provided for the case of a flat potential, $\da=0$.

\subsection{The extremal $\g\d=1$ solution in terms of phase variables\label{gd11}}

The solution was given in (\ref{Sol1}) in section \ref{gd-1}. Its extremal version reads
\be
	\ud s^2 = - {r^2\over \ell^2}V^{2-{\frac{4(1-\da^2)}{(3-\da^2)(1+\da^2)}}}\ud t^2+ \left({r\over \ell}\right)^{2\d^2-2}
V^{{4\d^2(\d^2-1)\over (3-\d^2)(1+\d^2)}-2}\ud r^2
		+ r^2V^{\frac{2(\da^2-1)^2}{(3-\da^2)(1+\da^2)}}\le(\ud x^2+\ud y^2\ri),
\label{98n}\ee
\be
V(r) = 1-\left({r_-\over r}\right)^{3-\d^2}\sp
	r_{-}^{3-\da^2} = {\sqrt{1+\d^2}q\ell^{2-\d^2}\over 2\d(3-\d^2)}\,,
\label{99n}\ee
\be
	e^{\phi}= \le(\frac r\ell\ri)^{2\da}V^{\frac{4\da(\da^2-1)}{(3-\da^2)(1+\da^2)}}\sp
	\mathcal A= {2\d\over \sqrt{1+\d^2}}~V\ud t\,.
\label{100n}\ee

The solution can be brought in the domain wall frame $r\to u$ by
\be
e^{2A}=r^2V^{\frac{2(\da^2-1)^2}{(3-\da^2)(1+\da^2)}}\sp f={1\over \ell^2}V^{4\d^2\over (1+\d^2)}\,,
\label{101n}\ee
\be
\ud u=V^{(\d^2-1)(3+\d^2)\over (1+\d^2)(3-\d^2)}{\ud r\over \ell}.
\label{102n}\ee

A direct evaluation of the phase variables gives
\be
X={1\over 2\sqrt{3}}{d\phi\over dA}={1\over \sqrt{3}}{\d+{2\d(\d^2-1)\over
 (1+\d^2)(3-\d^2)}{rV'\over V}\over 1+{(\d^2-1)^2\over (1+\d^2)(3-\d^2)}{rV'\over V}}
 ={\d\over \sqrt{3}}{V+{2(\d^2-1)\over (\d^2+1)}(1-V)\over V+{(\d^2-1)^2\over (1+\d^2)}(1-V)}\,,
\label{103n}\ee
\be
Y={1\over 3}{{4\d^2\over (1+\d^2)}{rV'\over V}\over 1+{(\d^2-1)^2\over (3-\d^2)(1+\d^2)}{rV'\over V}}
={1\over 3}{{4\d^2(3-\d^2)\over (1+\d^2)}(1-V)\over V+{(\d^2-1)^2\over (1+\d^2)}(1-V)}\,,
\label{104n}\ee
\be
W={\d^2(3-\d^2)\over 1+\d^2}(1-V)^2\,,
\label{105n}\ee
where we used
\be
{rV'\over V}=(3-\d^2){1-V\over V}\,.
\label{106n}\ee
The solution  interpolates between $(X_u,Y_u,W_u)$ at $V=1$ and $(X_c,Y_c,W_c)$ at $V=0$.
The relation between $V$ and $\phi$ can be calculated as
\be
e^{\phi}=\left({r_-\over \ell}\right)^{2\d}(1-V)^{-{2\d\over (3-\d^2)}}~V^{\frac{4\da(\da^2-1)}{(3-\da^2)(1+\da^2)}}\,.
\label{107n}\ee

\section{UV asymptotics\label{ap-uv}}

In this appendix we will investigate the UV asymptotics of the solutions at finite density. This information is useful among other things in order to understand the spectra of charged fluctuations in such backgrounds. We will remain in $p=3$ as this is the only subtle case where such an analysis is needed.

The action we consider is (\ref{ActionLiouville}) with
\be
V={6\over \ell^2}-{m^2\over 2}\phi^2+\cdots\sp Z=1-{k\over 2}\phi^2+\cdots\,.
\label{e1}\ee
The field $\phi$ corresponds to an operator with UV dimension $\D$ so that
\be
m^2\ell^2=\D(\D-3)\,.
\label{e2}\ee
We work in the general diagonal ansatz (\ref{ToroidalMetrics})

The UV fixed point is given by $\phi=0$, and the general finite density solution is the RN black-hole given by
 \be
D_o={\ell^2\over r^2}f(r)\sp B_o={\ell^2\over r^2 f(r)}\sp C_o={\ell^2\over r^2}\sp f(r)=1-Mr^3+{q^2r^4\over 4\ell^2}\,.
\label{e3}\ee

We now consider a relevant perturbation associated with the scalar $\phi$. We turn on an infinitesimal source for the dual operator and we will work to second order in this parameter.

To leading order in the perturbation the solution is
\be
D=D_o(1+x^2d(r)+{\cal O}(x^3))\sp B=B_o(1+x^2b(r)+{\cal O}(x^3))\,,
\label{e444}\ee
$$C=C_o(1+x^2c(r)+{\cal O}(x^3))\sp \phi=xr^{\D}g(r)\,,
$$
where $x$ is a small parameter, the UV coupling of the operator dual to the scalar.
The functions have a regular expansion around the boundary $r=0$:
\be
b(r)=b_0+b_3r^3+b_4r^4+{\cal O}(r^6)\sp c(r)=c_0+c_3r^3+c_4r^4+{\cal O}(r^6)\,,
\label{e5}\ee
\be
 d(r)=d_0+d_3r^3+d_4r^4+{\cal O}(r^6)\sp g(r)=1+g_3r^3+g_4r^4+{\cal O}(r^6)\,.
\label{e6}\ee
One of the functions can be trivialized via a coordinate transformation.
We will keep it explicit and write the general solution as a function of the $b(r)$ coefficients.

Solving the equations (\ref{16bb})-(\ref{17b}), we obtain:
\be
g_3={\D\over 6}M\sp g_4=-{(\D(\D+1)-2k)q^2\over 16(2\D+1)\ell^2}
\sp g_6={\D(\D+3)^2 M^2\over 36(2\D+3)}\,,
 \label{e7}\ee
\be
c_0=-{1\over 8}-{b_0\over 2\D}\sp c_3={3Mb_0\over 4\D(2\D+3)}-{b_3\over 2\D+3}-{(4\D^2+6\D-9)M\over 48(2\D+3)}\,,
\label{e8}\ee
\be
c_4=-{b_0q^2\over 8\D(\D+2)}-{b_4\over 2(\D+2)}+{((\D^2+4\D+2)(\D-1)-2(\D+2)k)q^2\over 64(\D+2)(2\D+1)\ell^2}\,,
\label{e9}\ee

\be
d_0=-{1\over 8}-{b_0\over 2\D}\sp d_3=-{3(\D+1)Mb_0\over 2\D(2\D+3)}-{b_3\over 2\D+3}-{(2\D^2+3\D+9)M\over 24(2\D+3)}\,,
\label{e10}\ee
\be
d_4={(2\D+3)q^2b_0\over 8\D(\D+2)\ell^2}-{b_4\over 2(\D+2)}+{((\D^3+3\D^2+10\D+6)-2(\D-6)k)q^2\over 64(\D+2)(2\D+1)\ell^2}\,.
\label{e11}\ee

These formul\ae\phantom{} are sufficient to compute the UV asymptotics of effective Schr\"odiger potentials.

\addcontentsline{toc}{section}{References}

\bibliography{EHT}
\bibliographystyle{JHEP}

\end{document}